\documentclass[9pt,DIV=24,a4paper]{scrartcl}

\usepackage{amssymb}
\usepackage{amsmath}
\usepackage{array}
\usepackage{algorithm}
\usepackage{algpseudocode}
\usepackage{graphicx}
\usepackage{cite}

\usepackage{color}

\newcommand{\libxsmm}{\texttt{libxsmm} }
\newcommand{\mkl}{\texttt{Intel MKL} }
\newcommand{\jitters}{\texttt{jitters} }

\newcommand{\newtext}[1]{} 
\graphicspath{{./figs/}}

\title{The effective use of BLAS interface for implementation of finite-element ADER-DG and finite-volume ADER-WENO methods}
\author{Popov I.S.\thanks{Department of Theoretical Physics, Dostoevsky Omsk State University, Omsk, Russia {\em diphosgen@mail.ru, popovis@omsu.ru}}}

\begin{document}
\vskip 1cm
\maketitle

\begin{abstract}
\noindent
Numerical methods of the ADER family, in particular finite-element ADER-DG and finite-volume ADER-WENO methods, are among the most accurate numerical methods for solving quasilinear \newtext{hyperbolic} PDE systems. The internal structure of ADER-DG and ADER-WENO numerical methods contains a large number of basic linear algebra operations related to matrix multiplications. The main interface of software libraries for matrix multiplications for high-performance computing is BLAS. An effective method for integration the standard functions of the BLAS interface into the implementation of these numerical methods is presented. The calculated matrices are small matrices; \newtext{and this allows to use effectively} JIT technologies. The proposed approach immediately operates on AoS, which allows to efficiently calculate flux, source and non-conservative terms without transposition. The obtained computational costs demonstrated that the effective implementation, based on the use of the JIT functions of the BLAS, outperformed both the implementation based on the general BLAS functions and the vanilla implementations by several orders of magnitude. The complexity of developing an implementation based on the proposed approach does not exceed the complexity of developing a vanilla implementation. \newtext{Performance analysis using roofline partly explains the observed features of the decreasing of computational costs.}
\end{abstract}




\section*{Introduction}
\label{sec:intro}

Numerical methods of the ADER family, including finite-element discontinuous Galerkin methods (ADER-DG) and finite-volume methods based on WENO reconstruction (ADER-WENO), are modern accurate methods of arbitrarily high order intended for solving systems of quasilinear partial differential equations (PDE systems). Titarev and Toro~\cite{ader_init_1, ader_init_2} developed the ADER paradigm for finite-volume methods, which was based on the solution of the generalized Riemann problem (GRP), and the evolution of the solution at finite-volume interfaces was determined by the Cauchy-Kovalevski procedure based on a high-order expansion of the solution in a Taylor series and the use of the original PDE system to calculate the derivatives (see also~\cite{ader_init_3}). Further development of the ADER paradigm for finite-volume methods was carried out in the works of Titarev and Toro~\cite{ader_init_4, ader_init_5, ader_init_6}. Dumbser \textit{et al}~\cite{ader_stiff_1, ader_stiff_2} modified the ADER paradigm for finite-volume methods based on the use of a discrete space-time solution obtained in a separate stage of the local space-time DG-predictor (LST-DG predictor), and showed higher solution accuracy than the original version, especially for stiff problems. 

Finite-element ADER-DG methods are fundamentally linear methods for solving PDE systems, so they are subject to the well-known Godunov theorem and such methods 
\newtext{require}
limiters to preserve the monotonicity of the numerical solution. Currently, finite-element ADER-DG and finite-volume ADER-WENO methods are based on the development of the Multi-dimensional Optimal Order Detection (MOOD) paradigm~\cite{mood_par_1, mood_par_2, mood_par_3, mood_par_4}, which resulted in the development~\cite{ader_dg_dev_1, ader_dg_dev_2} of finite-element ADER-DG methods with \textit{a posteriori} correction of the solution in subcells by a finite-volume limiter, which can be the finite-volume ADER-WENO method~\cite{ader_weno_lstdg_ideal, ader_weno_lstdg_diss}. This approach made it possible to preserve the monotonicity of the numerical solution and the subgrid resolution characteristic of DG methods~\cite{ader_dg_dev_1}. However, 
the further development of purely finite-volume methods of the ADER family, 
\newtext{carried out in the works~\cite{ader_weno_new_1, ader_weno_new_2, ader_weno_new_3, ader_weno_new_4, ader_weno_new_5, ader_weno_new_6, ader_weno_new_7}, is not directly related to the ADER-DG method with a posteriori correction of the solution}.

Currently, the finite element ADER-DG methods and the finite volume ADER-WENO methods are used to solve a wide range of specific problems in the physics and mechanics of continuous media. The main application, in this case, is associated with the use of the finite-element ADER-DG methods with a posteriori correction of the solution in subcells by a finite-volume limiter, which can be the finite-volume ADER-WENO method. Among these problems, it is necessary to highlight the problems of simulating ideal and dissipative flows in both classical hydrodynamics and magnetohydrodynamics~\cite{ader_dg_ideal_flows, ader_dg_diss_flows, ader_dg_semiexpl}, as well as special and general relativistic hydrodynamics and magnetohydrodynamics~\cite{ader_dg_grmhd, ader_dg_gr_prd, ader_dg_gr_z4_2024}, the latter of which are characterized by high stiffness of dissipative terms. It is also necessary to note the works in which these numerical methods were used to solve problems of the motion of elastic deformable media~\cite{ader_dg_eff_impl, fron_phys, ader_dg_hyperelastic}, problems of seismic wave propagation~\cite{ader_dg_eff_impl, exahype, ader_dg_seiemic, ader_dg_seiemic_underwater}, solving shallow water equations~\cite{ADER_GRP_LST_DG_1, ader_dg_wb_shwater_2018, ader_dg_wb_shwater_2022}, simulating blood flow~\cite{ader_eno_fv_blood_2022}, simulating compressible
barotropic two‐fluid flows~\cite{ader_dg_barotropic} and nonlinear dispersive systems~\cite{ader_dg_dispersive}. A special place is occupied by problems associated with the study of the formation and propagation of detonation waves in reacting flows, which are characterized by abnormally high stiffness, for which the finite element ADER-DG methods and finite volume ADER-WENO methods were developed and studied in the works~\cite{popov_j_sci_comp_2023, popov_comp_fluids_2024}. 
\newtext{Application of finite element ADER-DG methods and finite volume ADER-WENO methods for solving a wide range of other problems are also presented in works~\cite{ader_dg_eff_impl, exahype}.}

Finite-element ADER-DG methods and finite-volume ADER-WENO methods are developed for structured Cartesian meshes~\cite{ader_dg_dev_1, ader_dg_dev_2, ader_dg_grmhd, ader_dg_gr_prd, ader_dg_gr_z4_2024}, using the adaptive mesh refinement paradigm~\cite{ader_weno_lstdg_ideal, ader_weno_lstdg_diss, ader_dg_ideal_flows, ader_dg_diss_flows, ader_dg_PNPM}, unstructured triangular and tetrahedral meshes~\cite{ader_dg_simple_mod, PNPM_DG_2009, PNPM_DG_2010, ader_dg_eff_impl, ader_dg_wb_shwater_2018} and moving meshes~\cite{ader_dg_ale} (see also~\cite{ader_rev_2024}). 
\newtext{A detailed discussion of the features of the mathematical and algorithmic formulations of numerical methods and their software implementation is presented in works~\cite{ader_dg_eff_impl, exahype, fron_phys, ader_dg_hpc_impl_1, ader_dg_hpc_impl_2, ader_dg_hpc_impl_3, ader_dg_hpc_impl_4}.}
\newtext{The implementation of the ADER-DG numerical method on a GPU is presented in the paper~\cite{ader_dg_impl_gpu}.}
Dumbser \textit{et al} developed \textit{well-balanced} ADER-DG method for the Euler equations in general relativity problems~\cite{ader_dg_gr_z4_2024} and for the multilayer shallow water equations~\cite{ader_dg_wb_shwater_2022}. ADER-DG methods were used to solve initial value problems for ordinary differential systems of equations~\cite{ader_stiff_1, PNPM_DG_2010, ader_dg_ode_ivp, ader_eff} and differential-algebraic systems of equations~\cite{ader_dg_dae}. Zhao \textit{et al} developed \textit{path-conservative} discontinuous Galerkin (DG) method to solve shallow water equations~\cite{ader_dg_wb_pc_shwater_2024}. 
In~\cite{dg_entropy} Gaburro \textit{et al} developed entropy preserving ADER-DG schemes (see also~\cite{dg_entropy_add}). 
\newtext{Gaburro \textit{et al} developed entropy preserving ADER-DG schemes~\cite{dg_entropy} (see also~\cite{dg_entropy_add}).}
\newtext{Based on the ADER paradigm, a SPH numerical method was created~\cite{ader_weno_sph}, which demonstrated a very high quality of modeling hydrodynamic and magnetohydrodynamic flows with discontinuities, compared to classical SPH methods.}
The modern state of development of the space-time finite-element ADER-DG is presented in the works~\cite{ader_dg_mod_1, ader_dg_mod_2, ader_dg_mod_3}.

Han Veiga \textit{et al}~\cite{dec_vs_ader_2021} showed that the numerical methods of the ADER family are significantly interconnected with numerical methods based on the deferred correction (DeC) paradigm. DeC methods have a long history and their application to initial value problems for ODE systems goes back to~\cite{dec_src_1968}, and are effectively used to solve partial differential equations~\cite{dec_abgrall_2017, dec_abgrall_2019} and ordinary differential systems of equations~\cite{dec_dutt_2000, dec_dutt_2000, dec_minion_2003, dec_shu_2008}. DeC methods, like the methods of the ADER family, allow one to obtain an arbitrarily high order and are characterized by a high accuracy of the numerical solution. The current state of research on DeC methods~\cite{dec_vs_ader_2021} shows that they compete with methods of the ADER family. 
\newtext{The features of the relationship between the ADER paradigm and the DeC paradigm, which also allows one to obtain numerical methods of an arbitrary high order, are discussed in papers~\cite{dec_vs_ader_2021, dec_vs_ader_2023};} 
\newtext{it is also noted that DeC numerical methods are easier to implement than ADER methods.}

Numerical methods of the ADER family, as accurate high-order methods, require large computational costs. It is clear that a necessary condition for the successful application of any high-order numerical method is the possibility of its effective implementation for modern computing architectures~\cite{exahype, Robey_Zamora_hpc_2021}, when the effective use of modern computing subsets of instructions of the instruction set architecture is allowed. Modern finite-element ADER-DG methods and finite-volume ADER-WENO methods also contain the LST-DG predictor for obtaining a discrete space-time solution that contains information on the local evolution of the solution for each mesh cell at a time step and is characterized by a high dimensionality~\cite{ader_dg_eff_impl, exahype}, which leads to memory requirements and locality of the data structure placement in memory~\cite{Robey_Zamora_hpc_2021}. 

Internal structure of the ADER-DG and ADER-WENO methods, especially the LST-DG predictor, contains a large number of basic operations of linear algebra --- scalar-vector, matrix-vector and matrix-matrix multiplications. At present, the main interface of basic operations of linear algebra in the high-performance computing is the BLAS interface (Basic Linear Algebra Subprograms)~\cite{blas_ref_1, blas_ref_2}, for which implementations have been developed for all computer architectures used in this field of research. An important feature of the basic operations of linear algebra in the ADER-DG and ADER-WENO methods is the use of small matrices. For operations with small matrices, special implementations of the BLAS interface have been developed at the moment, using JIT technologies, such as \libxsmm library~\cite{libxsmm_1, libxsmm_2, libxsmm_3} and JIT subsystem of the \mkl library~\cite{intel_mkl_ref_man}. 

Implementation of the LST-DG predictor for the ADER-DG and ADER-WENO methods based on the use of the BLAS interface was proposed in the work~\cite{ader_dg_eff_impl}. However, the use of the BLAS interface led to the problem of storing a discrete space-time solution --- matrix-vector and matrix-matrix multiplications of the BLAS required the SoA (structure of arrays) storage paradigm, and the calculation of flux, source and non-conservative terms required the AoS (array of structures) storage paradigm. Moreover, this problem is solved only for the LST-DG predictor, but \newtext{it} is not solved \newtext{for the correctors at all} --- the finite-element ADER-DG method and the finite-volume ADER-WENO method, which use a discrete solution both to solve the Riemann problem when calculating flux integrals at cell interfaces and to calculate integrals over cell volumes. This is a classic ``AoS~vs.~SoA problem'', well-known in computer science and high-performance computing~\cite{Robey_Zamora_hpc_2021}. To solve this problem, \newtext{the procedure of transposing the discrete space-time solution, allowing to convert AoS to SoA and SoA to AoS, were used in work~\cite{ader_dg_eff_impl} and subsequent work~\cite{exahype}.}
Despite an effective implementation of this procedure, optimized for the use of registers of AVX, AVX2 and AVX-512 instruction subsets, this led to a complication of the software implementation, and also made it initially dependent on the details of the device of the modern version of the x86 instruction set architecture.

This work proposes and presents an effective approach to using the BLAS interface for developing a software implementation of both the ADER-DG and ADER-WENO methods separately, and the finite-element ADER-DG methods with a posteriori correction of the solution in subcells by a finite-volume ADER-WENO limiter, which combines the ADER-DG and ADER-WENO methods and some additional procedures for analyzing and rebuilding the solution. The developed approach allows operating with AoS data storage directly, without using conversion between AoS and SoA. It differs significantly from the approach proposed in works~\cite{ader_dg_eff_impl, exahype}, which requires matrix transposition. The approach developed in this paper is essentially based on the detailed structure of the ADER-DG and ADER-WENO formula expressions. \newtext{Therefore,} further in the text, the formula expressions for these methods will be presented in detail, in particular --- the representation of matrices in the form of Kronecker products, the structure of indices and summations, and the algorithms implementing these formulas. \newtext{The expression of matrices in the form of Kronecker products allows to use effectively the BLAS interface operations with small matrices.} A detailed description of all the main algorithms and parameters of the used BLAS interface operations is presented. It is shown that \jitters used in BLAS implementations for small matrices, based on the use of JIT compilation, can be created before the start of calculations once and exist for the entire time of calculations.

The proposed approach immediately operates on AoS, which makes it possible to efficiently calculate flux, source and non-conservative terms. This approach simply scales in case of large values of $M$, which is typical for problems of simulating multicomponent reacting flows, in which the number of components can be large, as well as problems for simulating dynamic problems of general relativity, in particular within the framework of the conformal and covariant Z4 formalism~\cite{ader_dg_gr_prd, ader_dg_grmhd, ader_dg_gr_z4_2024}, where $M = 110$ of the fully well-balanced scheme (the state vector is composed of $59$ dynamical variables).

The text of the article is presented in three main parts: Section~\ref{sec:framework} ``Mathematical framework'', Section~\ref{sec:apps} ``Applications of the numerical method'', Section~\ref{sec:comp_costs} ``Computational costs''\newtext{, Section~\ref{sec:perf_anal} ``Performance analysis''}. The main results related to the development of the algorithmic implementation of the numerical methods and the efficient use of the BLAS interface are presented in Section~\ref{sec:framework}. The applications of the numerical method\newtext{,} presented in Section~\ref{sec:apps}, demonstrate the operability of the developed algorithmic and software implementations, with \newtext{some sample} data used to test the performance of the ADER family methods. 
The results of the computational cost estimates\newtext{,} presented in Section~\ref{sec:comp_costs},
quantitatively demonstrate the increase in computational performance. \newtext{The results of the performance analysis using roofline are presented in Section~\ref{sec:perf_anal}, which can partly explain the observed features of the decreasing of computational costs and show that there may still be some potential for optimization for the presented numerical methods.}

The efficient use of the BLAS interface for implementing the finite-element ADER-DG and the finite-volume ADER-WENO methods presented in this paper is quite native with respect to the selected numerical methods. The presented approaches for the efficient use of the BLAS interface can be used to implement other finite-volume and finite-element numerical methods if the main matrix operations in these methods are similar to the corresponding operations in the ADER family methods.

\section{Mathematical framework}
\label{sec:framework}

In this paper, the non-homogeneous non-conservative PDE system is chosen in the following form:
\begin{equation}\label{eq:pde_src}
\frac{\partial \mathbf{U}}{\partial t} + \nabla\cdot\mathbb{F}(\mathbf{U}) + 
	\mathbf{B}(\mathbf{U})\cdot\nabla\mathbf{U} = \mathbf{S}(\mathbf{U}, \mathbf{r}, t),
\end{equation}
where $\mathbf{U} \in \Omega_{U} \subseteq \mathcal{R}^{M}$ is the $M$-dimensional state vector of conserved variables, $\mathbb{F}(\mathbf{U}) = (\mathbf{F}_{x}(\mathbf{U}), \mathbf{F}_{y}(\mathbf{U}), \mathbf{F}_{z}(\mathbf{U}))$ is the flux tensor, $\mathbf{B}(\mathbf{U}) = (\mathbb{B}_{x}(\mathbf{U}), \mathbb{B}_{y}(\mathbf{U}), \mathbb{B}_{z}(\mathbf{U}))$ is a vector of matrices $M \times M$ defining non-conservative terms $\mathbf{B}(\mathbf{U})\cdot\nabla\mathbf{U}$, and $\mathbf{S} = \mathbf{S}(\mathbf{U}, \mathbf{r}, t)$ is an algebraic source term. \newtext{An arbitrary dependence of the source terms $\mathbf{S}$ on coordinates $\mathbf{r}$ and $t$ is assumed in the expressions presented here and further in the text.} However, the flux tensor  $\mathbb{F}(\mathbf{U})$ and matrices of non-conservative terms $\mathbf{B}$ can also depend on $(\mathbf{r}, t)$, this is in no way restricted to the obtained expressions --- such dependence can occur in case of curvilinear systems for coordinates and time. The system of equations can be represented in the non-conservative form $\partial\mathbf{U}/\partial t + \mathbf{A}(\mathbf{U})\cdot\nabla\mathbf{U} = \mathbf{S}(\mathbf{U}, \mathbf{r}, t)$, where $\mathbf{A}(\mathbf{U}) = \partial\mathbb{F}(\mathbf{U})/\partial\mathbf{U} + \mathbf{B}$. The PDE system (\ref{eq:pde_src}) is hyperbolic if for all vectors $\mathbf{a} \neq \mathbf{0}$ and for all vectors $\mathbf{U} \in \Omega_{U}$ the matrix $\mathbf{A}(\mathbf{U})\cdot\mathbf{a}$ has $M$ real eigenvalues and a full set of $M$ linearly independent right eigenvectors. Hyperbolicity is of fundamental importance for obtaining a solution to the Riemann problem of calculating fluxes at cell interfaces in these numerical methods. The efficient implementation of numerical methods of the ADER family developed in this paper is applicable to solving hyperbolic PDE systems; however, this approach is directly extended to solving PDE systems with parabolic terms, following work~\cite{ader_dg_diss_flows} in terms \newtext{of} calculating eigenvalues for determining the time step and in terms of solving the Riemann problem. 

The implementation of the numerical method in this paper is performed for the case of a Cartesian coordinate system. The system of equations (\ref{eq:pde_src}) in Cartesian coordinates $(x, y, z)$, which is used in this work, takes the following form:
\begin{equation}\label{eq:pde_src_cart}
\frac{\partial \mathbf{U}}{\partial t} + 
	\frac{\partial \mathbf{F}_{x}(\mathbf{U})}{\partial x} + 
	\frac{\partial \mathbf{F}_{y}(\mathbf{U})}{\partial y} + 
	\frac{\partial \mathbf{F}_{z}(\mathbf{U})}{\partial z} + 
	\mathbb{B}_{x}(\mathbf{U})\cdot\frac{\partial \mathbf{U}}{\partial x} + 
	\mathbb{B}_{y}(\mathbf{U})\cdot\frac{\partial \mathbf{U}}{\partial y} +
	\mathbb{B}_{z}(\mathbf{U})\cdot\frac{\partial \mathbf{U}}{\partial z} = 
	\mathbf{S}(\mathbf{U}, \mathbf{r}, t).
\end{equation}
The finite-element ADER-DG and finite-volume ADER-WENO methods use discretization of the space-time domain of the solution. In this paper, the computational domain $\Omega \subset \mathcal{R}^{3}$ is discretized by a spatial mesh $\Omega = \cup_{i}\Omega_{i}$ with coordinate cells $\Omega_{i} = [x_{i_{1}}, x_{i_{1}+1}]\times[y_{i_{2}}, y_{i_{2}+1}]\times[z_{i_{3}}, z_{i_{3}+1}]$, where $x_{i_{1}+1} - x_{i_{1}} = \Delta x$, $y_{i_{2}+1} - y_{i_{2}} = \Delta y$, $z_{i_{3}+1} - z_{i_{3}} = \Delta z$ are coordinate steps, and $i$ is a three-component multi-index $i = (i_{1}, i_{2}, i_{3})$. As a result of time discretization $t^{n}$ with time step $\Delta t^{n} = t^{n+1} - t^{n}$, the set of space-time elements $\Omega_{n,i} = [t^{n}, t^{n+1}]\times\Omega_{i}$ is obtained in which a local space-time coordinate system $(\tau, \boldsymbol{\xi})$, with $\boldsymbol{\xi} = (\xi, \eta, \zeta)$, is introduced in the following form:
\begin{equation}\label{eq:local_space_time_coords}
t = t^{n} + \Delta t^{n}\cdot\tau;\quad
x = x_{i_{1}} + \Delta x\cdot\xi;\quad
y = y_{i_{2}} + \Delta y\cdot\eta;\quad
z = z_{i_{3}} + \Delta z\cdot\zeta;\quad
0 \leqslant \tau,\ \xi,\ \eta,\ \zeta \leqslant 1,
\end{equation}
which maps the element $\Omega_{n,i}$ to a reference space-time element $\omega_{4} = [0, 1]\times\omega_{3} = [0, 1]^{4}$. In the coordinate system (\ref{eq:local_space_time_coords}) of the reference space-time element $\Omega_{n,i}$, the system of equations (\ref{eq:pde_src_cart}) is rewritten in the following form:
\begin{equation}\label{eq:pde_ref}
\frac{\partial \mathbf{u}}{\partial t} + 
	\frac{\partial \mathbf{f}_{\xi}(\mathbf{u})}{\partial \xi} + 
	\frac{\partial \mathbf{f}_{\eta}(\mathbf{u})}{\partial \eta} + 
	\frac{\partial \mathbf{f}_{\zeta}(\mathbf{u})}{\partial \zeta} + 
	\mathfrak{B}_{\xi}(\mathbf{u})\cdot\frac{\partial \mathbf{u}}{\partial \xi} + 
	\mathfrak{B}_{\eta}(\mathbf{u})\cdot\frac{\partial \mathbf{u}}{\partial \eta} +
	\mathfrak{B}_{\zeta}(\mathbf{u})\cdot\frac{\partial \mathbf{u}}{\partial \zeta} = 
	\mathbf{s}(\mathbf{u}, \mathbf{r}(\boldsymbol{\xi}), t(\tau)),
\end{equation}
where
\begin{equation}
\begin{array}{llll}
\mathbf{u} = \mathbf{U},&
\mathbf{f}_{\xi} = \cfrac{\Delta t^{n}}{\Delta x}\cdot\mathbf{F}_{x},&
\mathbf{f}_{\eta} = \cfrac{\Delta t^{n}}{\Delta y}\cdot\mathbf{F}_{y},&
\mathbf{f}_{\zeta} = \cfrac{\Delta t^{n}}{\Delta z}\cdot\mathbf{F}_{z},\\
\mathfrak{B}_{\xi} = \cfrac{\Delta t^{n}}{\Delta x}\cdot\mathbb{B}_{x},&
\mathfrak{B}_{\eta} = \cfrac{\Delta t^{n}}{\Delta y}\cdot\mathbb{B}_{y},&
\mathfrak{B}_{\zeta} = \cfrac{\Delta t^{n}}{\Delta z}\cdot\mathbb{B}_{z},&
\mathbf{s} = \Delta t^{n}\cdot\mathbf{S}. 
\end{array}
\end{equation}
Designations \newtext{$\mathbf{u}$, $\mathbf{f}_{\xi}$, $\mathbf{f}_{\eta}$, $\mathbf{f}_{\zeta}$, $\mathfrak{B}_{\xi}$, $\mathfrak{B}_{\eta}$, $\mathfrak{B}_{\zeta}$, $\mathbf{s}$} define the rescaled flux tensor, matrices of non-conservative terms, and source terms \newtext{respectively}.
A detailed description of the mathematical framework is based on works~\cite{ader_dg_dev_1, ader_dg_dev_2, ader_weno_lstdg_ideal, ader_weno_lstdg_diss, ader_dg_ideal_flows, ader_dg_diss_flows, ader_dg_ale, ader_dg_grmhd, ader_dg_gr_prd, ader_dg_gr_z4_2024, ader_dg_PNPM, PNPM_DG_2009, PNPM_DG_2010}, and \newtext{it} is provided primarily to derive the final formulaic apparatus for which the implementation has been developed.

The finite-element ADER-DG and finite-volume ADER-WENO methods are based on the use of a discrete space-time solution $\mathbf{q}(\tau, \boldsymbol{\xi})$ obtained by the LST-DG predictor~\cite{ader_dg_dev_1, ader_weno_lstdg_ideal, ader_weno_lstdg_diss, ader_dg_ideal_flows, ader_dg_diss_flows, ader_dg_eff_impl}. The discrete solution $\mathbf{q}$ determines the solution of the problem in the small, describing the dynamic evolution of the solution in the space-time element $\Omega_{n,i}$ without taking into account the interaction with other mesh cells, and is represented in the form:
\begin{equation}\label{eq:lst_dg_repr}
\mathbf{q}(\tau, \boldsymbol{\xi}) = \sum\limits_{p} \hat{\mathbf{q}}_{p}\Theta_{p}(\tau, \boldsymbol{\xi}),
\end{equation}
where $\Theta_{p}(\tau, \boldsymbol{\xi})$ is the basis functions, $\hat{\mathbf{q}}_{p}$ is the representation coefficients, $p = (p_{0}, p_{1}, p_{2}, p_{3})$ is the four-component multi-index. The basis functions $\Theta_{p}(\tau, \boldsymbol{\xi})$ are chosen in the form of tensor products of the nodal basis functions $\Theta_{p}(\tau, \boldsymbol{\xi}) = \varphi_{p_{0}}(\tau)\varphi_{p_{1}}(\xi)\varphi_{p_{2}}(\eta)\varphi_{p_{3}}(\zeta)$, where $\varphi_{k}(\xi)$ are the nodal basis functions, which are Lagrange interpolation polynomials of degree $N$, with nodal points at the roots $\xi_{l}$ of the Legendre polynomials $P_{N+1}(\xi)$: $\varphi_{k}(\xi_{l}) = \delta_{kl}$, and $0 \leqslant p_{0}, p_{1}, p_{2}, p_{3} \leqslant N$. However, the roots of Radau polynomials, Lobatto polynomials and other polynomials can be chosen as nodal points~\cite{rev_bases_2024}. In the finite-element ADER-DG method, the solution $\mathbf{u}^{n}(\boldsymbol{\xi})$ in the cell $\Omega_{i}$ at the time step $t^{n}$ is represented in the form:
\begin{equation}\label{eq:dg_repr}
\mathbf{u}^{n}(\boldsymbol{\xi}) = \sum\limits_{k} \hat{\mathbf{u}}_{k}^{n}\Phi_{k}(\boldsymbol{\xi}),
\end{equation}
where $\Phi_{k}(\boldsymbol{\xi})$ is the basis functions, $\hat{\mathbf{u}}_{k}^{n}$ is the representation coefficients, $k = (k_{1}, k_{2}, k_{3})$ is the three-component multi-index, $0 \leqslant k_{1}, k_{2}, k_{3} \leqslant N$. The basis functions $\Phi_{k}(\boldsymbol{\xi})$ are also chosen in the form of tensor products of the nodal basis functions $\Phi_{k}(\tau, \boldsymbol{\xi}) = \varphi_{k_{1}}(\xi)\varphi_{k_{2}}(\eta)\varphi_{k_{3}}(\zeta)$. In the finite-volume ADER-WENO method, the solution in the cell $\Omega_{i}$ at the time step $t^{n}$ is represented by a finite-volume average value $\mathbf{v}_{i}^{n}$:
\begin{equation}\label{eq:fv_repr}
\mathbf{v}_{i}^{n} = \int\limits_{0}^{1}d\xi \int\limits_{0}^{1}d\eta \int\limits_{0}^{1}d\zeta \cdot \mathbf{u}(\mathbf{r}(\boldsymbol{\xi}), t^{n}).
\end{equation}
Using the values $\mathbf{v}_{j}^{n}$ in the set of cells $\Omega_{j}$ surrounding the cell $\Omega_{i}$, WENO-reconstruction of the solution is performed, allowing to obtain a highly accurate representation of the solution $\mathbf{w}^{n}(\boldsymbol{\xi})$ at the time step $t^{n}$, which is presented in the form:
\begin{equation}\label{eq:weno_reconstr}
\mathbf{w}^{n}(\boldsymbol{\xi}) = \sum\limits_{k} \hat{\mathbf{w}}_{k}^{n}\Phi_{k}(\boldsymbol{\xi}),
\end{equation}
which is technically similar to the DG representation (\ref{eq:dg_repr}) of the solution, however, 
the coefficients $\hat{\mathbf{w}}_{k}^{n}$ that are calculated in the WENO-reconstruction of the solution are used 
\newtext{instead of the coefficients $\hat{\mathbf{u}}_{k}^{n}$ in the expression.}

This work proposes and presents an effective approach to using the BLAS interface for developing a software implementation of both the ADER-DG and ADER-WENO methods separately, and the finite-element ADER-DG methods with a posteriori correction of the solution in subcells by a finite-volume ADER-WENO limiter, which combines the ADER-DG and ADER-WENO methods and some additional procedures for analyzing and recalculation the solution. The adaptive finite-element ADER-DG method is characterized by very high accuracy and resolution of the smooth components of the solution. Non-physical anomalies of the numerical solution arising due to the fundamental linearity of the ADER-DG method, which are explained by the well-known Godunov theorem, are corrected by an a posteriori limiter, for which the high-precision finite-volume ADER-WENO method is chosen. Correction of the solution obtained by the ADER-DG method is carried out in subcells, for which the representations transformations in subcell correction by finite-volume method are used. Each time step of the space-time adaptive ADER-DG finite-element method with LST-DG predictor and a posteriori sub-cell ADER-WENO finite-volume limiting involved a sequence of steps~\cite{ader_dg_dev_1, ader_dg_dev_2, ader_weno_lstdg_ideal, ader_weno_lstdg_diss, ader_dg_ideal_flows, ader_dg_diss_flows, ader_dg_grmhd, ader_dg_gr_prd, ader_dg_PNPM, PNPM_DG_2009, PNPM_DG_2010, ader_dg_ale}:
\begin{itemize}
\setlength\itemsep{-2.5pt}
	\item a LST-DG predictor, using which a discrete space-time solution in the small is obtained;
	\item a pure finite-element ADER-DG scheme, using which a candidate high accuracy solution is obtained;
	\item a determination of physical and numerical admissibility of the obtained high accuracy candidate solution and identification of ``troubled'' cells;
	\item a recalculation of the solution in ``troubled'' cells by a stable ADER-WENO finite-volume limiter.
\end{itemize}
Numerical ADER-DG and ADER-WENO methods separately involved a sequence of steps of the LST-DG predictor and a corrector, which is a completely one-step finite-element ADER-DG scheme in the first case and a completely one-step finite-volume ADER-WENO scheme in the second case. An excellent detailed description of this computational scheme is given in the basic works~\cite{ader_dg_dev_1, ader_dg_dev_2, ader_weno_lstdg_ideal, ader_weno_lstdg_diss, ader_dg_ideal_flows, ader_dg_diss_flows, ader_dg_grmhd, ader_dg_gr_prd, ader_dg_PNPM, PNPM_DG_2009, PNPM_DG_2010, ader_dg_ale} 
\newtext{by} the developers of this method. Details of the internal structure of the ADER-DG are presented in the works~\cite{ader_dg_dev_1, ader_dg_dev_2, PNPM_DG_2009, PNPM_DG_2010}. Details of the internal structure of the finite-element ADER-DG and finite-volume ADER-WENO methods, in which the LST-DG prediction method is used, are presented in the works~\cite{ader_weno_lstdg_ideal, ader_weno_lstdg_diss}. Peculiarities of mathematical formulation and efficient software implementation are discussed in~\cite{ader_dg_eff_impl, fron_phys} and~\cite{exahype, ader_dg_hpc_impl_1, ader_dg_hpc_impl_2, ader_dg_hpc_impl_3, ader_dg_hpc_impl_4}. The modern state of development of the space-time adaptive ADER finite-element DG method with a posteriori correction technique of solutions on sub-cells by the finite-volume limiter using AMR for use on unstructured meshes and using the ALE approach is presented in the works~\cite{ader_dg_mod_1, ader_dg_mod_2}.

Further in the text of this paper, a detailed mathematical description of each individual step of the methods is presented, which is necessary for a detailed introduction of the algorithmic implementation of the effective use of BLAS procedures. After each mathematical description, a detailed description of the corresponding algorithmic implementation follows, for which a pseudocode and the main quantitative parameters of the called functions are presented.

\newtext{It should be noted that in this paper the presented numerical methods, its algorithmic and software implementations are used to solve hyperbolic systems of equations. However, in the presented form these methods can also be considered suitable for elliptic problems.}

\subsection{LST-DG predictor}
\label{sec:framework:lst_dg_predictor}

\paragraph{General description}
\label{sec:framework:lst_dg_predictor:descr}

The system of nonlinear algebraic equations of the LST-DG predictor is obtained as a result of substitution $\mathbf{u} \mapsto \mathbf{q}$ into the system of equations (\ref{eq:pde_ref}) and using the condition for the $L_{2}$ projection of the residual to be zero on the test functions. 
$\Theta_{p}(\tau, \boldsymbol{\xi})$ are chosen \newtext{for the test functions}, integrating by parts over time $\tau$ in the time term, and using the initial condition $\mathbf{q}(0, \boldsymbol{\xi}) = \mathbf{u}(\boldsymbol{\xi})$ (\ref{eq:dg_repr}). \newtext{All this} leads to the following expression~\cite{ader_weno_lstdg_ideal, ader_weno_lstdg_diss, ader_dg_dev_1, ader_dg_ideal_flows, ader_dg_diss_flows, ader_dg_eff_impl}:
\begin{equation}
\begin{split}
\left[\Theta_{p}, \mathbf{q}\right](1) -
\left[\Theta_{p}, \mathbf{u}\right](0) &+
\left\langle\frac{\partial\Theta_{p}}{\partial\tau}, \mathbf{q}\right\rangle +
\left\langle
	\Theta_{p}, 
	\left(
		\frac{\partial \mathbf{f}_{\xi}(\mathbf{q})}{\partial \xi} + 
		\frac{\partial \mathbf{f}_{\eta}(\mathbf{q})}{\partial \eta} + 
		\frac{\partial \mathbf{f}_{\zeta}(\mathbf{q})}{\partial \zeta}
	\right)
\right\rangle\\
&+\left\langle
	\Theta_{p}, 
	\left(
		\mathfrak{B}_{\xi}(\mathbf{q})\cdot\frac{\partial \mathbf{q}}{\partial \xi} + 
		\mathfrak{B}_{\eta}(\mathbf{q})\cdot\frac{\partial \mathbf{q}}{\partial \eta} +
		\mathfrak{B}_{\zeta}(\mathbf{q})\cdot\frac{\partial \mathbf{q}}{\partial \zeta} -
		\mathbf{s}(\mathbf{q}, \mathbf{r}(\boldsymbol{\xi}), t(\tau))
	\right)
\right\rangle = 0,
\end{split}
\end{equation}
where \newtext{the} following integral operators for $L_{2}$ scalar products are introduced to simplify the notation:
\begin{equation}
\begin{split}
&\left[f(\tau, \boldsymbol{\xi}), g(\tau, \boldsymbol{\xi})\right](\tau) = 
	\int\limits_{0}^{1}d\xi \int\limits_{0}^{1}d\eta \int\limits_{0}^{1}d\zeta \cdot
	f(\tau, \boldsymbol{\xi}) g(\tau, \boldsymbol{\xi});\\
&\left\langle f(\tau, \boldsymbol{\xi}), g(\tau, \boldsymbol{\xi}) \right\rangle = 
	\int\limits_{0}^{1}d\tau \cdot \left[f(\tau, \boldsymbol{\xi}), g(\tau, \boldsymbol{\xi})\right](\tau) = 
	\int\limits_{0}^{1}d\tau \int\limits_{0}^{1}d\xi \int\limits_{0}^{1}d\eta \int\limits_{0}^{1}d\zeta \cdot
	f(\tau, \boldsymbol{\xi}) g(\tau, \boldsymbol{\xi}).
\end{split}
\end{equation}
This form of the system of predictor equations is written for ADER-DG method, but for finite-volume ADER-WENO method it is enough to select the initial condition in the form $\mathbf{q}(0, \boldsymbol{\xi}) = \mathbf{w}(\boldsymbol{\xi})$ of a reconstructed function (\ref{eq:weno_reconstr}). 
Gaussian-Legendre quadrature formula
\begin{equation}\label{eq:gl_rule}
\int\limits_{0}^{1}d\xi \cdot f(\xi) \approx \sum\limits_{k = 0}^{N} w_{k} f(\xi_{k}),\qquad
w_{k} = \int\limits_{0}^{1}d\xi \cdot \varphi_{k}^{2}(\xi),\qquad
\sum_{k = 0}^{N} w_{k} = 1,
\end{equation}
\newtext{enables approximate point-wise evaluation~\cite{Jackson_2017, Zanotti_lectures_2016} of the physical fluxes and sources:}
\begin{equation}
\begin{split}
&\mathbf{f}_{\xi}(\mathbf{q}) \approx \sum\limits_{p} \Theta_{p}\cdot\mathbf{f}_{\xi}(\hat{\mathbf{q}}_{p}),\quad
\mathbf{f}_{\eta}(\mathbf{q}) \approx \sum\limits_{p} \Theta_{p}\cdot\mathbf{f}_{\eta}(\hat{\mathbf{q}}_{p}),\quad
\mathbf{f}_{\zeta}(\mathbf{q}) \approx \sum\limits_{p} \Theta_{p}\cdot\mathbf{f}_{\zeta}(\hat{\mathbf{q}}_{p}),\\
&\mathfrak{B}_{\xi}(\mathbf{q})\cdot\frac{\partial \mathbf{q}}{\partial \xi} \approx
\sum\limits_{p} \Theta_{p}\cdot \left(
	\mathfrak{B}_{\xi}(\hat{\mathbf{q}}_{p}) \sum\limits_{r} \frac{\partial\Theta_{r}(\tau_{p}, \boldsymbol{\xi}_{p})}{\partial\xi}\cdot\hat{\mathbf{q}}_{r}
\right),\quad
\mathfrak{B}_{\eta}(\mathbf{q})\cdot\frac{\partial \mathbf{q}}{\partial \eta} \approx
\sum\limits_{p} \Theta_{p}\cdot \left(
	\mathfrak{B}_{\eta}(\hat{\mathbf{q}}_{p}) \sum\limits_{r} \frac{\partial\Theta_{r}(\tau_{p}, \boldsymbol{\xi}_{p})}{\partial\eta}\cdot\hat{\mathbf{q}}_{r}
\right),\\
&\mathfrak{B}_{\zeta}(\mathbf{q})\cdot\frac{\partial \mathbf{q}}{\partial \zeta} \approx
\sum\limits_{p} \Theta_{p}\cdot \left(
	\mathfrak{B}_{\zeta}(\hat{\mathbf{q}}_{p}) \sum\limits_{r} \frac{\partial\Theta_{r}(\tau_{p}, \boldsymbol{\xi}_{p})}{\partial\zeta}\cdot\hat{\mathbf{q}}_{r}
\right),\quad
\mathbf{s}(\mathbf{q}, \mathbf{r}(\boldsymbol{\xi}), t(\tau)) \approx 
\sum\limits_{p} \Theta_{p}\cdot\mathbf{s}(\hat{\mathbf{q}}_{p}, \mathbf{r}(\boldsymbol{\xi}_{p}), t(\tau_{p})).
\end{split}
\end{equation}
The approximation by point-wise evaluation can be proved by the following consideration. The representation of the original term is considered as an expansion in basis functions in the form:
\begin{equation}
\mathbf{s}(\mathbf{q}, \mathbf{r}(\boldsymbol{\xi}), t(\tau)) = 
\sum\limits_{p} \Theta_{p}\cdot\mathbf{s}_{p},\qquad
\mathbf{s}_{p} = 
	\frac{\left\langle\Theta_{p}, \mathbf{s}(\mathbf{q}, \mathbf{r}(\boldsymbol{\xi}), t(\tau))\right\rangle}
	{\left\langle\Theta_{p}, \Theta_{p}\right\rangle}.
\end{equation}
As a result of using the Gauss-Legendre quadrature formula and the definition of Lagrange interpolation polynomials $\varphi_{p}(\xi_{k}) = \delta_{pk}$, the following expression is obtained:
\begin{equation}
\left\langle\Theta_{p}, \mathbf{s}(\mathbf{q}, \mathbf{r}(\boldsymbol{\xi}), t(\tau))\right\rangle \approx 
	w_{p_{0}}w_{p_{1}}w_{p_{2}}w_{p_{3}}\cdot\mathbf{s}(\mathbf{q}_{p}, \mathbf{r}(\boldsymbol{\xi}_{p}), t(\tau_{p})),\qquad
\left\langle\Theta_{p}, \Theta_{p}\right\rangle = w_{p_{0}}w_{p_{1}}w_{p_{2}}w_{p_{3}}.
\end{equation}
This expression leads to the point-wise evolution for the source term $\mathbf{s}_{p} \approx \mathbf{s}(\mathbf{q}_{p}, \mathbf{r}(\boldsymbol{\xi}_{p}), t(\tau_{p}))$. Similar transformations can be carried out for all conservative and non-conservative flow terms, which also can be proved for the point-wise evaluation of the physical fluxes. As a result of the approximation by point-wise evaluation, the system of equations of the LST-DG predictor takes the following form:
\begin{equation}
\begin{split}
&\sum\limits_{q} \left[\Theta_{p}, \Theta_{q}\right](1)\cdot\hat{\mathbf{q}}_{q} -
\sum\limits_{q} \left\langle\frac{\partial\Theta_{p}}{\partial\tau}, \Theta_{q}\right\rangle\cdot\hat{\mathbf{q}}_{q} +
\sum\limits_{q} \left\langle
	\Theta_{p}, 
	\frac{\partial\Theta_{q}}{\partial\xi}\cdot\mathbf{f}_{\xi}(\hat{\mathbf{q}}_{q}) +
	\frac{\partial\Theta_{q}}{\partial\eta}\cdot\mathbf{f}_{\eta}(\hat{\mathbf{q}}_{q}) +
	\frac{\partial\Theta_{q}}{\partial\zeta}\cdot\mathbf{f}_{\zeta}(\hat{\mathbf{q}}_{q})
\right\rangle\\
&-\sum\limits_{q} \left\langle\Theta_{p}, \Theta_{q}\right\rangle\left(
	\mathbf{s}(\hat{\mathbf{q}}_{q}, \mathbf{r}(\boldsymbol{\xi}_{q}), t(\tau_{q})) - 
	\sum\limits_{r} \left(
		\mathfrak{B}_{\xi}(\hat{\mathbf{q}}_{q}) \frac{\partial\Theta_{r}(\tau_{q}, \boldsymbol{\xi}_{q})}{\partial\xi} +
		\mathfrak{B}_{\eta}(\hat{\mathbf{q}}_{q}) \frac{\partial\Theta_{r}(\tau_{q}, \boldsymbol{\xi}_{q})}{\partial\eta} +
		\mathfrak{B}_{\zeta}(\hat{\mathbf{q}}_{q}) \frac{\partial\Theta_{r}(\tau_{q}, \boldsymbol{\xi}_{q})}{\partial\zeta}
	\right)\cdot\hat{\mathbf{q}}_{r}
\right)\\
&=\sum\limits_{k} \left[\Theta_{p}, \Phi_{k}\right](0)\cdot\hat{\mathbf{u}}_{k},
\end{split}
\end{equation}
where the left side of the system of equations contains functional structures \newtext{with} the desired quantities $\hat{\mathbf{q}}_{p}$, and the right side contains known coefficients $\hat{\mathbf{u}}_{k}$ for representing the initial conditions specified at the time step $t^{n}$. To simplify the final form of the system of equations, the following notation is introduced:
\begin{equation}
\begin{split}
&\hat{\mathbf{f}}^{\xi}_{p} = \mathbf{f}_{\xi}(\hat{\mathbf{q}}_{p});\quad
\hat{\mathbf{f}}^{\eta}_{p} = \mathbf{f}_{\eta}(\hat{\mathbf{q}}_{p});\quad
\hat{\mathbf{f}}^{\zeta}_{p} = \mathbf{f}_{\zeta}(\hat{\mathbf{q}}_{p});\quad
\hat{\mathbf{s}}_{p} = \mathbf{s}(\hat{\mathbf{q}}_{p}, \mathbf{r}(\boldsymbol{\xi}_{p}), t(\tau_{p})),\\
&\hat{\mathbf{b}}_{p} = 
	\sum\limits_{r} \left(
		\mathfrak{B}_{\xi}(\hat{\mathbf{q}}_{p}) \frac{\partial\Theta_{r}(\tau_{p}, \boldsymbol{\xi}_{p})}{\partial\xi} +
		\mathfrak{B}_{\eta}(\hat{\mathbf{q}}_{p}) \frac{\partial\Theta_{r}(\tau_{p}, \boldsymbol{\xi}_{p})}{\partial\eta} +
		\mathfrak{B}_{\zeta}(\hat{\mathbf{q}}_{p}) \frac{\partial\Theta_{r}(\tau_{p}, \boldsymbol{\xi}_{p})}{\partial\zeta}
	\right)\cdot\hat{\mathbf{q}}_{r};
\end{split}
\end{equation}
and notation for the matrices is also introduced:
\begin{equation}\label{eq:lst_dg_predictor_src_matrices}
\begin{split}
&\mathbb{K}^{\tau}_{pq} = \left[\Theta_{p}, \Theta_{q}\right](1) - \left\langle\frac{\partial\Theta_{p}}{\partial\tau}, \Theta_{q}\right\rangle;\quad
\mathbb{M}_{pq} = \left\langle\Theta_{p}, \Theta_{q}\right\rangle;\quad
\mathbb{F}^{0}_{pk} = \left[\Theta_{p}, \Phi_{k}\right](0);\\
&\mathbb{K}^{\xi}_{pq} = \left\langle\Theta_{p}, \frac{\partial\Theta_{q}}{\partial\xi}\right\rangle;\quad
\mathbb{K}^{\eta}_{pq} = \left\langle\Theta_{p}, \frac{\partial\Theta_{q}}{\partial\eta}\right\rangle;\quad
\mathbb{K}^{\zeta}_{pq} = \left\langle\Theta_{p}, \frac{\partial\Theta_{q}}{\partial\zeta}\right\rangle;
\end{split}
\end{equation}
where it is necessary to take into account that $p = (p_{0}, p_{1}, p_{2}, p_{3})$ and $k = (k_{1}, k_{2}, k_{3})$ are multi-indices for which a \newtext{pass-through} index can be introduced (multi-indices $q$ and $r$ are similar in structure to $p$):
\begin{equation}\label{eq:through_indices}
\begin{array}{lll}
p = p_{0}\cdot(N+1)^{3} + p_{1}\cdot(N+1)^{2} + p_{2}\cdot(N+1) + p_{3};& 
0 \leqslant p_{0}, p_{1}, p_{2}, p_{3} \leqslant N;& 
0 \leqslant p \leqslant (N+1)^{4};\\
k = k_{1}\cdot(N+1)^{2} + k_{2}\cdot(N+1) + k_{3};&
0 \leqslant k_{1}, k_{2}, k_{3} \leqslant N;&
0 \leqslant k \leqslant (N+1)^{3}.
\end{array}
\end{equation}
In simplifying notation, the system of predictor equations can be rewritten in the following form:
\begin{equation}
\sum\limits_{q} \left(
	\mathbb{K}^{\tau}_{pq} \hat{\mathbf{q}}_{q} +
	\mathbb{K}^{\xi}_{pq} \hat{\mathbf{f}}^{\xi}_{q} +
	\mathbb{K}^{\eta}_{pq} \hat{\mathbf{f}}^{\eta}_{q} +
	\mathbb{K}^{\zeta}_{pq} \hat{\mathbf{f}}^{\zeta}_{q} -
	\mathbb{M}_{pq}\left(\hat{\mathbf{s}}_{q} - \hat{\mathbf{b}}_{q}\right)
\right) =
\sum\limits_{k} \mathbb{F}^{0}_{pk}\hat{\mathbf{u}}_{k},
\end{equation}
which after multiplication on the left by the inverse matrix $[\mathbb{K}^{\tau}]^{-1}$ takes the following form~\cite{Zanotti_lectures_2016}:
\begin{equation}
\hat{\mathbf{q}}_{p} - \sum\limits_{q} \mathbb{L}^{s}_{pq}\left(\hat{\mathbf{s}}_{q} - \hat{\mathbf{b}}_{q}\right) =
\sum\limits_{k} \mathbb{L}^{0}_{pk}\hat{\mathbf{u}}_{k} -
\sum\limits_{q} \left(
	\mathbb{L}^{\xi}_{pq} \hat{\mathbf{f}}^{\xi}_{q} +
	\mathbb{L}^{\eta}_{pq} \hat{\mathbf{f}}^{\eta}_{q} +
	\mathbb{L}^{\zeta}_{pq} \hat{\mathbf{f}}^{\zeta}_{q}
\right),
\end{equation}
where the following notation is introduced:
\begin{equation}\label{eq:lst_dg_predictor_res_matrices}
\begin{split}
&\mathbb{L}^{\xi}_{pq} = \sum\limits_{r} \left[\mathbb{K}^{\tau}\right]^{-1}_{pr} \mathbb{K}^{\xi}_{rq},\qquad
\mathbb{L}^{\eta}_{pq} = \sum\limits_{r} \left[\mathbb{K}^{\tau}\right]^{-1}_{pr} \mathbb{K}^{\eta}_{rq},\qquad
\mathbb{L}^{\zeta}_{pq} = \sum\limits_{r} \left[\mathbb{K}^{\tau}\right]^{-1}_{pr} \mathbb{K}^{\zeta}_{rq},\\
&\mathbb{L}^{\sigma}_{pq} = \sum\limits_{r} \left[\mathbb{K}^{\tau}\right]^{-1}_{pr} \mathbb{M}_{rq},\qquad
\mathbb{L}^{0}_{pk} = \sum\limits_{r} \left[\mathbb{K}^{\tau}\right]^{-1}_{pr} \mathbb{F}_{rk}.
\end{split}
\end{equation}
\newtext{It is strictly proved~\cite{Jackson_2017}} that the eigenvalues of the matrices $\mathbb{L}^{\xi}$, $\mathbb{L}^{\eta}$ and $\mathbb{L}^{\zeta}$ are zero, which was previously stated in~\cite{Zanotti_lectures_2016}. Therefore, the Picard iterative process is of the following form:
\begin{equation}\label{eq:lst_dg_predictor_picard}
\hat{\mathbf{q}}_{p}^{(l+1)} - \sum\limits_{q} \mathbb{L}^{\sigma}_{pq}\left(\hat{\mathbf{s}}_{q}^{(l+1)} - \hat{\mathbf{b}}_{q}^{(l+1)}\right) =
\sum\limits_{k} \mathbb{L}^{0}_{pk}\hat{\mathbf{u}}_{k} -
\sum\limits_{q} \left(
	\mathbb{L}^{\xi}_{pq} \hat{\mathbf{f}}^{\xi, (l)}_{q} +
	\mathbb{L}^{\eta}_{pq} \hat{\mathbf{f}}^{\eta, (l)}_{q} +
	\mathbb{L}^{\zeta}_{pq} \hat{\mathbf{f}}^{\zeta, (l)}_{q}
\right),
\end{equation}
and \newtext{it} strictly converges to the solution of the problem in case of zero terms $\mathbf{B} \equiv 0$ and $\mathbf{S} \equiv 0$ in (\ref{eq:pde_src}). The matrix $\mathbb{L}^{0}$ has an interesting property:
\begin{equation}\label{eq:ind_prop}
\mathbb{L}^{0}_{pk} \equiv \mathbb{L}^{0}_{p_{0}p_{1}p_{2}p_{3}, k_{1}k_{2}k_{3}} = \delta_{p_{1}k_{1}}\delta_{p_{2}k_{2}}\delta_{p_{3}k_{3}},
\end{equation}
which is due to the fact that in case of zero fluxes $\mathbb{F} \equiv 0$ and terms $\mathbf{B} \equiv 0$ and $\mathbf{S} \equiv 0$ in (\ref{eq:pde_src}) the solution $\mathbf{q}(\tau, \boldsymbol{\xi})$ is a constant in time $\tau$, therefore the condition $\hat{\mathbf{q}}_{p_{0}p_{1}p_{2}p_{3}} = \hat{\mathbf{u}}_{p_{1}p_{2}p_{3}}$ must be satisfied\newtext{,} that is equivalent to the condition (\ref{eq:ind_prop}). This condition is applicable in software implementation --- the calculation of the first term on the right side of the system of equations (\ref{eq:lst_dg_predictor_picard}) is reduced to initializing each subarray $\hat{\mathbf{q}}_{p_{0}}$ by array $\hat{\mathbf{u}}$.

\paragraph{Using the BLAS interface}
\label{sec:framework:lst_dg_predictor:blas}

The matrices (\ref{eq:lst_dg_predictor_res_matrices}) included in the system of nonlinear algebraic equations (\ref{eq:lst_dg_predictor_picard}) of the LST-DG predictor can be expressed through the matrices of mass $\Lambda$ and stiffness $\Xi$ determined by the basis functions $\{\varphi_{k}(\xi)\}$:
\begin{equation}\label{eq:m_and_kappa_matrices}
\Lambda_{kl} = \int\limits_{0}^{1} \varphi_{k}\varphi_{l} d\xi = \Lambda_{k}\delta_{kl};\quad
\Xi_{kl} = \int\limits_{0}^{1} \varphi_{k}\frac{d\varphi_{l}}{d\xi} d\xi;\quad
\Upsilon_{kl} = \varphi_{k}(1)\varphi_{l}(1) - \Xi_{lk},
\end{equation}
where the orthogonality property of the functional basis $\varphi_{k}(\xi)$ is taken into account and an expression for the matrix $\Upsilon$ is introduced. The expressions (\ref{eq:lst_dg_predictor_src_matrices}) for the matrices can be written as follows:
\begin{equation}
\begin{array}{llll}
&\mathbb{K}^{\tau} = \Upsilon \otimes \Lambda \otimes \Lambda \otimes \Lambda;
&\mathbb{M} = \Lambda \otimes \Lambda \otimes \Lambda \otimes \Lambda;&\\
&\mathbb{K}^{\xi} = \Lambda \otimes \Xi \otimes \Lambda \otimes \Lambda;
&\mathbb{K}^{\eta} = \Lambda \otimes \Lambda \otimes \Xi \otimes \Lambda;
&\mathbb{K}^{\zeta} = \Lambda \otimes \Lambda \otimes \Lambda \otimes \Xi,
\end{array}
\end{equation}
where pass-through indices $p$ and $q$ (\ref{eq:through_indices}) are used, and $\otimes$ denotes the Kronecker product. Using these expressions and introducing additional matrices $\tilde{\ae} = \Upsilon^{-1}\cdot\Lambda$ and $\tilde{k} = \Lambda^{-1}\cdot\Xi$, expressions for the LST-DG predictor matrices (\ref{eq:lst_dg_predictor_res_matrices}) are obtained:
\begin{equation}
\begin{array}{lll}
\mathbb{L}^{\sigma} &\hspace{-3mm}= \left[\Upsilon \otimes \Lambda \otimes \Lambda \otimes \Lambda\right]^{-1}\cdot
\left[\Lambda \otimes \Lambda \otimes \Lambda \otimes \Lambda\right] &\hspace{-2mm}= 
\tilde{\ae} \otimes \mathbb{I} \otimes \mathbb{I} \otimes \mathbb{I};\\
\mathbb{L}^{\xi} &\hspace{-3mm}= \left[\Upsilon \otimes \Lambda \otimes \Lambda \otimes \Lambda\right]^{-1}\cdot
\left[\Lambda \otimes \Xi \otimes \Lambda \otimes \Lambda\right] &\hspace{-2mm}= 
\tilde{\ae} \otimes \tilde{k} \otimes \mathbb{I} \otimes \mathbb{I};\\
\mathbb{L}^{\eta} &\hspace{-3mm}= \left[\Upsilon \otimes \Lambda \otimes \Lambda \otimes \Lambda\right]^{-1}\cdot
\left[\Lambda \otimes \Lambda \otimes \Xi \otimes \Lambda\right] &\hspace{-2mm}= 
\tilde{\ae} \otimes \mathbb{I} \otimes \tilde{k} \otimes \mathbb{I};\\
\mathbb{L}^{\zeta} &\hspace{-3mm}= \left[\Upsilon \otimes \Lambda \otimes \Lambda \otimes \Lambda\right]^{-1}\cdot
\left[\Lambda \otimes \Lambda \otimes \Lambda \otimes \Xi\right] &\hspace{-2mm}= 
\tilde{\ae} \otimes \mathbb{I} \otimes \mathbb{I} \otimes \tilde{k},
\end{array}
\end{equation}
where $\mathbb{I}$ is the $(N+1)\times(N+1)$ identity matrix. Calculation of non-conservative terms $\hat{\mathbf{b}}_{q}$ uses multiplication by gradient matrices for basis functions $\Theta_{p}$:
\begin{equation}
\mathbb{D}^{\xi}_{pq} = \frac{\partial\Theta_{q}(\tau_{p}, \boldsymbol{\xi}_{p})}{\partial\xi};\qquad
\mathbb{D}^{\eta}_{pq} = \frac{\partial\Theta_{q}(\tau_{p}, \boldsymbol{\xi}_{p})}{\partial\eta};\qquad
\mathbb{D}^{\zeta}_{pq} = \frac{\partial\Theta_{q}(\tau_{p}, \boldsymbol{\xi}_{p})}{\partial\zeta},
\end{equation}
for which the following representations are obtained:
\begin{equation}
\mathbb{D}^{\xi} = \mathbb{I} \otimes [\mathbb{D}^{\varphi}]^{T} \otimes \mathbb{I} \otimes \mathbb{I};\qquad
\mathbb{D}^{\eta} = \mathbb{I} \otimes \mathbb{I} \otimes [\mathbb{D}^{\varphi}]^{T} \otimes \mathbb{I};\qquad
\mathbb{D}^{\zeta} = \mathbb{I} \otimes \mathbb{I} \otimes \mathbb{I} \otimes [\mathbb{D}^{\varphi}]^{T},
\end{equation}
where $\mathbb{D}^{\varphi} = ||\varphi_{k}'(\xi_{l})||$ is the $(N+1)\times(N+1)$ matrix of derivatives of the basis functions $\varphi_{k}(\xi)$ in nodes $\xi_{l}$.

\begin{table}[h!]
\begin{center}
\caption{\label{tab:lst_dg_predictor_blas}
The main parameters of the BLAS procedures \texttt{gemm} used to calculate the discrete solution $\mathbf{q}$ by the LST-DG predictor.
}
\begin{tabular}{|l|ccc|ccc|ll|}
\hline
kernel name &  $m$ & $n$ & $k$ & \texttt{lda} & \texttt{ldb} & \texttt{ldc} & $\alpha$ & $\beta$\\
\hline
\texttt{k\_ksi\_kernel} & $N+1$ & $M$ & $N+1$ & $N+1$ & $(N+1)^{2} M$ & $(N+1)^{2} M$ & $1$ & $0$\\
\texttt{k\_eta\_kernel} & $N+1$ & $M$ & $N+1$ & $N+1$ & $(N+1) M$ & $(N+1) M$ & $1$ & $1$\\
\texttt{k\_zeta\_kernel} & $N+1$ & $M$ & $N+1$ & $N+1$ & $M$ & $M$ & $1$ & $1$\\
\texttt{kappa\_kernel} & $N+1$ & $M$ & $N+1$ & $N+1$ & $(N+1)^{3} M$ & $(N+1)^{3} M$ & $1$ & $1$\\
\texttt{d\_ksi\_kernel} & $N+1$ & $M$ & $N+1$ & $N+1$ & $(N+1)^{2} M$ & $(N+1)^{2} M$ & $1$ & $0$\\
\texttt{d\_eta\_kernel} & $N+1$ & $M$ & $N+1$ & $N+1$ & $(N+1) M$ & $(N+1) M$ & $1$ & $0$\\
\texttt{d\_zeta\_kernel} & $N+1$ & $M$ & $N+1$ & $N+1$ & $M$ & $M$ & $1$ & $0$\\
\texttt{b\_ksi\_kernel} & $M$ & $M$ & $M$ & $M$ & $M$ & $M$ & $1$ & $0$\\
\texttt{b\_eta\_kernel} & $M$ & $M$ & $M$ & $M$ & $M$ & $M$ & $1$ & $1$\\
\texttt{b\_zeta\_kernel} & $M$ & $M$ & $M$ & $M$ & $M$ & $M$ & $1$ & $1$\\
\hline
\end{tabular}
\end{center}
\end{table}

\begin{algorithm}[h!]
\caption{\label{alg:lst_dg_predictor}
The basic general structure of using BLAS interface procedure calls in the implementation of the LST-DG predictor.
}
\begin{algorithmic}[1]
\State $\mathtt{q\_init}$, $\mathtt{q\_prev}$, $\mathtt{q\_next}$, $\mathtt{kf\_sum}$, 
$\mathtt{q\_int\_prev}$, $\mathtt{q\_int\_next}$, $\mathtt{s}$, $\mathtt{d}$: \texttt{array}[$(N+1)^{4} \cdot M$]
\State $\mathtt{s\_args}$: \texttt{array}[$4\cdot(N+1)^{4}$]
\Function{get\_cell\_local\_solution}{$\mathtt{u\_prev}$}
\ForAll{$p_{0} \in [0, N]$}
\State $\mathtt{q\_init}[p_{0}] \gets \mathtt{u\_prev}$
\EndFor
\State $\mathtt{q\_prev} \gets$ \Call{init\_iter}{$\mathtt{q\_init},\, \mathtt{dt\_div\_dr}$}
\Repeat
\State $\mathtt{q\_next} \gets \mathtt{q\_init}$\Comment{added only for symmetry}
\State \Call{set\_flux\_terms\_sum}{$\mathtt{kf\_sum}$}
\State $\mathtt{q\_int\_prev} \gets \mathtt{q\_prev}$
\Repeat
\State $\mathtt{q\_int\_next} \gets \mathtt{q\_init}$
\ForAll{$(p_{0}, p_{1}, p_{2}, p_{3}) \in [0, N]^{4}$}
\State $\mathtt{idx} \gets (p_{0}\cdot(N+1)^{3} + p_{1}\cdot(N+1)^{2} + p_{2}\cdot(N+1) + p_{3}) \cdot M$
\State \Call{set\_rescaled\_source\_terms}{$\mathtt{s}[\mathtt{idx}],\, \mathtt{q\_int\_prev}[\mathtt{idx}],\, \mathtt{s\_args}[\mathtt{idx}],\, \mathtt{dt}$}
\EndFor
\State \Call{set\_non\_conservative\_terms\_sum}{$\mathtt{b},\, \mathtt{q\_int\_prev},\, \mathtt{dt\_div\_dr}$}
\State $\mathtt{s} \gets \mathtt{s} - \mathtt{b} - \mathtt{kf\_sum}$
\ForAll{$(p_{1}, p_{2}, p_{3}) \in [0, N]^{3}$}
\State $\mathtt{idx} \gets (p_{1}\cdot(N+1)^{2} + p_{2}\cdot(N+1) + p_{3}) \cdot M$
\State \Call{kappa\_kernel}{$\mathtt{q\_int\_next}[\mathtt{idx}],\, \mathtt{kappa\_tilde},\, \mathtt{s}[\mathtt{idx}]$}
\EndFor
\State $\varepsilon_{\mathtt{inner}} \gets \|\mathtt{q\_int\_next} - \mathtt{q\_int\_prev}\|$
\State $\mathtt{q\_int\_prev} \gets \mathtt{q\_int\_next}$
\Until{$\varepsilon_{\mathtt{inner}} \leqslant \varepsilon_{0}$}
\State $\mathtt{q\_next} \gets \mathtt{q\_int\_next}$
\State $\varepsilon_{\mathtt{outer}} \gets \|\mathtt{q\_next} - \mathtt{q\_prev}\|$
\State $\mathtt{q\_prev} \gets \mathtt{q\_next}$
\Until{$\varepsilon_{\mathtt{outer}} \leqslant \varepsilon_{0}$}
\State \Return $\mathtt{q\_next}$
\EndFunction
\end{algorithmic}
\end{algorithm}

\begin{algorithm}[h!]
\caption{\label{alg:lst_dg_predictor:supp_1}
Procedure to Algorithm~\ref{alg:lst_dg_predictor} for calculating the sum of flux terms multiplied by matrices $\mathbb{L}^{\xi}$, $\mathbb{L}^{\eta}$, $\mathbb{L}^{\zeta}$.
}
\begin{algorithmic}[1]
\vspace{2.5mm}
\State $\mathtt{f\_ksi}$, $\mathtt{f\_eta}$, $\mathtt{f\_zeta}$, $\mathtt{kf\_sum}$, $\mathtt{q\_prev}$: \texttt{array}[$(N+1)^{4} \cdot M$]
\Procedure{set\_flux\_terms\_sum}{$\mathtt{kf\_sum},\, \mathtt{q\_prev},\, \mathtt{dt\_div\_dr}$}
\ForAll{$(p_{0}, p_{1}, p_{2}, p_{3}) \in [0, N]^{4}$}
\State $\mathtt{idx} \gets (p_{0}\cdot(N+1)^{3} + p_{1}\cdot(N+1)^{2} + p_{2}\cdot(N+1) + p_{3}) \cdot M$
\State \Call{set\_rescaled\_flux\_terms\_3d}%
	{$\mathtt{f\_ksi}[\mathtt{idx}],\, \mathtt{f\_eta}[\mathtt{idx}],\, \mathtt{f\_zeta}[\mathtt{idx}],\, \mathtt{q\_prev}[\mathtt{idx}],\, \mathtt{dt\_div\_dr}$}
\EndFor
\ForAll{$(p_{0}, p_{2}, p_{3}) \in [0, N]^{3}$}
\State $\mathtt{idx} \gets (p_{0}\cdot(N+1)^{3} + p_{2}\cdot(N+1) + p_{3}) \cdot M$
\State \Call{k\_ksi\_kernel}{$\mathtt{kf\_sum}[\mathtt{idx}],\, \mathtt{k\_tilde},\, \mathtt{f\_ksi}[\mathtt{idx}]$}
\EndFor
\ForAll{$(p_{0}, p_{1}, p_{3}) \in [0, N]^{3}$}
\State $\mathtt{idx} \gets (p_{0}\cdot(N+1)^{3} + p_{1}\cdot(N+1)^{2} + p_{3}) \cdot M$
\State \Call{k\_eta\_kernel}{$\mathtt{kf\_sum}[\mathtt{idx}],\, \mathtt{k\_tilde},\, \mathtt{f\_eta}[\mathtt{idx}]$}
\EndFor
\ForAll{$(p_{0}, p_{1}, p_{2}) \in [0, N]^{3}$}
\State $\mathtt{idx} \gets (p_{0}\cdot(N+1)^{3} + p_{1}\cdot(N+1)^{2} + p_{2}\cdot(N+1)) \cdot M$
\State \Call{k\_zeta\_kernel}{$\mathtt{kf\_sum}[\mathtt{idx}],\, \mathtt{k\_tilde},\, \mathtt{f\_zeta}[\mathtt{idx}]$}
\EndFor
\EndProcedure
\vspace{2.5mm}
\end{algorithmic}
\end{algorithm}
\begin{algorithm}[h!]
\caption{\label{alg:lst_dg_predictor:supp_2}
Procedure to Algorithm~\ref{alg:lst_dg_predictor} for calculating the sum of non-conservative terms $\hat{\mathbf{b}}_{p}$.
}
\begin{algorithmic}[1]
\vspace{2.5mm}
\State $\mathtt{b}$, $\mathtt{q\_int\_prev}$, $\mathtt{dq\_ksi}$, $\mathtt{dq\_eta}$, $\mathtt{dq\_zeta}$: \texttt{array}[$(N+1)^{4} \cdot M$]
\State $\mathtt{B\_ksi}$, $\mathtt{B\_eta}$, $\mathtt{B\_zeta}$: \texttt{array}[$M^{2}$]
\Procedure{set\_non\_conservative\_terms\_sum}{$\mathtt{b},\, \mathtt{q\_int\_prev},\, \mathtt{dt\_div\_dr}$}
\ForAll{$(p_{0}, p_{2}, p_{3}) \in [0, N]^{3}$}
\State $\mathtt{idx} \gets (p_{0}\cdot(N+1)^{3} + p_{2}\cdot(N+1) + p_{3}) \cdot M$
\State \Call{d\_ksi\_kernel}{$\mathtt{dq\_ksi}[\mathtt{idx}],\, \mathtt{D\_trans},\, \mathtt{q\_int\_prev}[\mathtt{idx}]$}
\EndFor
\ForAll{$(p_{0}, p_{1}, p_{3}) \in [0, N]^{3}$}
\State $\mathtt{idx} \gets (p_{0}\cdot(N+1)^{3} + p_{1}\cdot(N+1)^{2} + p_{3}) \cdot M$
\State \Call{d\_eta\_kernel}{$\mathtt{dq\_eta}[\mathtt{idx}],\, \mathtt{D\_trans},\, \mathtt{q\_int\_prev}[\mathtt{idx}]$}
\EndFor
\ForAll{$(p_{0}, p_{1}, p_{2}) \in [0, N]^{3}$}
\State $\mathtt{idx} \gets (p_{0}\cdot(N+1)^{3} + p_{1}\cdot(N+1)^{2} + p_{2}\cdot(N+1)) \cdot M$
\State \Call{d\_zeta\_kernel}{$\mathtt{dq\_zeta}[\mathtt{idx}],\, \mathtt{D\_trans},\, \mathtt{q\_int\_prev}[\mathtt{idx}]$}
\EndFor
\ForAll{$(p_{0}, p_{1}, p_{2}, p_{3}) \in [0, N]^{4}$}
\State $\mathtt{idx} \gets (p_{0}\cdot(N+1)^{3} + p_{1}\cdot(N+1)^{2} + p_{2}\cdot(N+1) + p_{3}) \cdot M$
\State \Call{set\_rescaled\_non\_conservative\_matrices}
	{$\mathtt{B\_ksi},\, \mathtt{B\_eta},\, \mathtt{B\_zeta},\, \mathtt{q\_int\_prev}[\mathtt{idx}],\, \mathtt{dt\_div\_dr}$}
\State \Call{b\_ksi\_kernel}{$\mathtt{b}[\mathtt{idx}],\, \mathtt{B\_ksi},\, \mathtt{dq\_ksi}[\mathtt{idx}]$}
\State \Call{b\_eta\_kernel}{$\mathtt{b}[\mathtt{idx}],\, \mathtt{B\_eta},\, \mathtt{dq\_eta}[\mathtt{idx}]$}
\State \Call{b\_zeta\_kernel}{$\mathtt{b}[\mathtt{idx}],\, \mathtt{B\_zeta},\, \mathtt{dq\_zeta}[\mathtt{idx}]$}
\EndFor
\EndProcedure
\vspace{2.5mm}
\end{algorithmic}
\end{algorithm}

The numerical method for solving the system (\ref{eq:lst_dg_predictor_picard}) of nonlinear algebraic equations of the LST-DG predictor can be based on the Picard method and the Newton method~\cite{Jackson_2017}. It is noted~\cite{Jackson_2017} that the solution of the system of nonlinear algebraic equations of the predictor in the case of the presence of stiff terms is preferably performed using Newton's method. \newtext{The use of Newton's method can only be justified in the case of essentially stiff terms, otherwise it will only lead to an increase in computational costs, especially those associated with calculating the inverse of a sufficiently large matrix.} However, it should be noted that in case of sufficiently high stiffness, Newton's method can be used for internal iterations~\cite{Zanotti_lectures_2016}, and local time steps can also be used~\cite{popov_comp_fluids_2024}. This work proposes an implementation of the Picard iterative method for solving a system of equations (\ref{eq:lst_dg_predictor_picard}), which is organized in the form of two nested Picard iterative processes:
\begin{equation}\label{eq:lst_dg_predictor_nested_picards}
\hat{\mathbf{q}}_{p}^{(l+1, m+1)} =
\sum\limits_{q} \mathbb{L}^{\sigma}_{pq}\left(\hat{\mathbf{s}}_{q}^{(l+1, m)} - \hat{\mathbf{b}}_{q}^{(l+1, m)}\right) +
\sum\limits_{k} \mathbb{L}^{0}_{pk}\hat{\mathbf{u}}_{k} -
\sum\limits_{q} \left(
	\mathbb{L}^{\xi}_{pq} \hat{\mathbf{f}}^{\xi, (l, m)}_{q} +
	\mathbb{L}^{\eta}_{pq} \hat{\mathbf{f}}^{\eta, (l, m)}_{q} +
	\mathbb{L}^{\zeta}_{pq} \hat{\mathbf{f}}^{\zeta, (l, m)}_{q}
\right),
\end{equation}
where in the internal iterative process on index $m$ convergence is achieved for the stiff terms associated with algebraic sources and non-conservative terms, and in the external iterative process on index $l$ a general convergence is achieved, associated with flux terms. Further, \newtext{we describe how to use} the BLAS interface to implement a LST-DG predictor of this form. Taking into account the recommendations of the work~\cite{Jackson_2017}, it can be added that the internal iterative process can be implemented using Newton method, rather than Picard method, in the case of the presence of stiff source terms in the system of equations.

All terms in the expression (\ref{eq:lst_dg_predictor_nested_picards}) are calculated using the \texttt{gemm} function of the BLAS interface. The main parameters of the standard function \texttt{gemm} determine the storage method --- \texttt{layout}, operations with matrices --- \texttt{transa} and \texttt{transb}, matrices sizes $n$, $m$ and $k$, leading dimensions \texttt{lda}, \texttt{ldb} and \texttt{lds}, and the factors $\alpha$ and $\beta$. The \texttt{row\_major} layout for storing matrices is chosen, and \texttt{notrans} is selected for the operations \texttt{transa} and \texttt{transb}. The values of all parameters used are presented in Table~\ref{tab:lst_dg_predictor_blas}. The last two lines of Table~\ref{tab:lst_dg_predictor_blas} are the same, which is done for reasons of syntactically clear introduction of functions \texttt{b\_kernel} for each individual direction \texttt{ksi}, \texttt{eta} and \texttt{zeta} (in software implementation this can be an alias, if this is important). The basic general structure of the algorithm for calculating the expansion coefficients $\mathbf{q}_{p}$ of a local discrete solution is presented in Algorithm~\ref{alg:lst_dg_predictor}, where, to simplify the pseudocode, two support procedures are designed separately, presented in Algorithm~\ref{alg:lst_dg_predictor:supp_1} and Algorithm~\ref{alg:lst_dg_predictor:supp_2}. Algorithm~\ref{alg:lst_dg_predictor} receives as input an array of coefficients $\mathbf{u}$, as well as the time step $\Delta t$ and coordinate steps $(\Delta x, \Delta y, \Delta z)$ (as $(\Delta t/\Delta x, \Delta t/\Delta y, \Delta t/\Delta z)$ ratios). The procedures \texttt{set\_rescaled\_flux\_terms\_3d}, \texttt{set\_rescaled\_non\_conservative\_matrices} and \texttt{set\_rescaled\_source\_terms} calculate fluxes, source terms, and matrices of non-conservative terms, respectively, calculated immediately in rescaled variables. The matrices \texttt{k\_tilde}, \texttt{kappa\_tilde}, \texttt{D\_trans} used for calculations are corresponded to matrices $\tilde{k}$, $\tilde{\ae}$, $[\mathbb{D}^{\varphi}]^{T}$ and are pre-computed. Array \texttt{s\_args} stores the values $(t(\tau_{p}), \mathbf{r}(\boldsymbol{\xi})_{p})$ for each cell, with the values $t(\tau_{p})$ being recalculated for each time step. The function \texttt{init\_iter} can use simple initialization $\hat{\mathbf{q}}_{p_{0}p_{1}p_{2}p_{3}} = \hat{\mathbf{u}}_{p_{1}p_{2}p_{3}}$, based on property (\ref{eq:ind_prop}), or more advanced methods for obtaining initial iterations of the solution~\cite{ader_dg_eff_impl}. Internal and external iterations are implemented in loops, the exit conditions of which are determined by decreasing the error $\varepsilon$ to a value of $\varepsilon_{0}$, which 
\newtext{is} chosen equal to $10^{-12}$-$10^{-14}$ for the case of using the floating point number type \texttt{double} and $10^{-4}$-$10^{-6}$ for the case of using the floating point number type \texttt{float}. The pseudocode of Algorithm~\ref{alg:lst_dg_predictor} does not provide any methods for error handling and infinite loop protection of the iterative process. It should be noted that all parameters of the \texttt{gemm} functions are known in advance and are constants, except for the arrays used. Therefore, this implementation satisfies all the requirements for the use of \texttt{JIT} functions for small matrices computations, in particular, the \texttt{jit gemm} from the \texttt{Intel MKL} library.

The similar implementation was proposed and discussed in the work~\cite{ader_dg_eff_impl}. However, the method of use of the BLAS interface proposed in the work~\cite{ader_dg_eff_impl} involves application of additional procedures associated with the calculation of flux, source and non-conservative terms. In general, the local discrete solution $\mathbf{q}(\tau, \boldsymbol{\xi})$ is represented as an expansion (\ref{eq:lst_dg_repr}) in basis functions with expansion coefficients $\mathbf{q}_{p}$. Each expansion coefficient $\mathbf{q}_{p}$ is an array of $M$ elements. \newtext{Then we should choose a method for $5$d array storage.} If we choose conserved variables as an index $r$ of an element of a homogeneous structure, then storage in format $\mathbf{q}_{p_{0}p_{1}p_{2}p_{3},r}$ is AoS, and in format $\mathbf{q}_{r, p_{0}p_{1}p_{2}p_{3}}$ is SoA, which is the classic AoS vs SoA problem, known in computer science. \newtext{SoA is applied in the works~\cite{ader_dg_eff_impl, exahype}}, \newtext{that}
allows the use of small matrices, well suited for the \texttt{lixsmm} library of small matrices computations. However, this choice requires complicated calculations of flux, source and non-conservative terms --- direct calculation requires transposing the matrices $\mathbf{q}_{r,p} \mapsto \mathbf{q}_{p, r}$. This work proposes an approach different from that proposed in the works~\cite{ader_dg_eff_impl, exahype}. The proposed approach immediately operates on AoS $\mathbf{q}_{p, r}$, which makes it possible to efficiently calculate flux, source and non-conservative terms. This approach can be simply scaled in case of large values of $M$, which is typical for problems of simulating multicomponent reacting flows, in which the number of components can be large, as well as problems for simulating dynamic problems of general relativity, in particular within the framework of the conformal and covariant Z4 formalism~\cite{ader_dg_gr_prd, ader_dg_grmhd, ader_dg_gr_z4_2024}, where $M = 110$ of the fully well-balanced scheme (the state vector is composed of $59$ dynamical variables).

\subsection{ADER-DG method}
\label{sec:framework:ader_dg_method}

\paragraph{General description}
\label{sec:framework:ader_dg_method:descr}

The one-step discrete ADER-DG scheme is obtained using the condition of zero $L_{2}$ projection in reference space-time element $\omega_{4}$ of the residual (\ref{eq:pde_ref}) onto the test functions $\Phi_{k}(\boldsymbol{\xi})$~\cite{ader_weno_lstdg_ideal, ader_weno_lstdg_diss, ader_dg_dev_1, ader_dg_ideal_flows, ader_dg_diss_flows, ader_dg_eff_impl}:
\begin{equation}
\left\langle 
	\Phi_{k}, \frac{\partial \mathbf{u}}{\partial \tau} + 
	\frac{\partial \mathbf{f}_{\xi}(\mathbf{u})}{\partial \xi} + 
	\frac{\partial \mathbf{f}_{\eta}(\mathbf{u})}{\partial \eta} + 
	\frac{\partial \mathbf{f}_{\zeta}(\mathbf{u})}{\partial \zeta} + 
	\mathfrak{B}_{\xi}(\mathbf{u})\cdot\frac{\partial \mathbf{u}}{\partial \xi} + 
	\mathfrak{B}_{\eta}(\mathbf{u})\cdot\frac{\partial \mathbf{u}}{\partial \eta} +
	\mathfrak{B}_{\zeta}(\mathbf{u})\cdot\frac{\partial \mathbf{u}}{\partial \zeta} -
	\mathbf{s}(\mathbf{u}, \mathbf{r}(\boldsymbol{\xi}), t(\tau))
\right\rangle = 0.
\end{equation}
\newtext{Integration by parts with the selection of complete divergence in the flux terms, and substitution of the DG representation (\ref{eq:dg_repr}) $\mathbf{u}^{n}$ and $\mathbf{u}^{n+1}$ in the time term and the discrete solution $\mathbf{q}$ (\ref{eq:lst_dg_repr}) instead of $\mathbf{u}$ in the integral expressions leads to the following form:}
\begin{equation}
\begin{split}
\sum\limits_{l} \left\langle \Phi_{k}, \Phi_{l} \right\rangle \left(\mathbf{u}^{n+1}_{l} - \mathbf{u}^{n}_{l}\right) &+
\left\{ \Phi_{k}, \mathfrak{G}\left(\mathbf{q}^{(-)}, \mathbf{q}^{(+)}\right) \right\} -
\left(
	\left\langle \frac{\partial\Phi_{k}}{\partial\xi}, \mathbf{f}_{\xi}(\mathbf{q}) \right\rangle +
	\left\langle \frac{\partial\Phi_{k}}{\partial\eta}, \mathbf{f}_{\eta}(\mathbf{q}) \right\rangle +
	\left\langle \frac{\partial\Phi_{k}}{\partial\zeta}, \mathbf{f}_{\zeta}(\mathbf{q}) \right\rangle 
\right)\\
&+\left(
	\left\langle \Phi_{k}, \mathfrak{B}_{\xi}(\mathbf{q})\cdot\frac{\partial \mathbf{q}}{\partial \xi} \right\rangle +
	\left\langle \Phi_{k}, \mathfrak{B}_{\eta}(\mathbf{q})\cdot\frac{\partial \mathbf{q}}{\partial \eta} \right\rangle +
	\left\langle \Phi_{k}, \mathfrak{B}_{\zeta}(\mathbf{q})\cdot\frac{\partial \mathbf{q}}{\partial \zeta} \right\rangle 
\right)\\
&=\left\langle \Phi_{k}, \mathbf{s}(\mathbf{q}, \mathbf{r}(\boldsymbol{\xi}), t(\tau)) \right\rangle.
\end{split}
\end{equation}
\newtext{Here,} $\mathfrak{G} = (\boldsymbol{\mathfrak{G}}_{\xi}, \boldsymbol{\mathfrak{G}}_{\eta}, \boldsymbol{\mathfrak{G}}_{\zeta})$ is the Riemann solver, taking into account non-conservative terms in the system of equations and expressed in the rescaled coordinate system (\ref{eq:local_space_time_coords}) of the reference space-time element $\omega_{4}$, $\mathbf{q}^{(+)}$ and $\mathbf{q}^{(-)}$ are the local discrete solutions in cells in the outer normal direction $\mathbf{n}$ of the cell $\Omega_{i}$ and in the opposite direction.
\newtext{Also} the notation for the integral operator is introduced for simplicity:\vspace{-1mm}
\begin{equation}
\begin{split}
\left\{f(\tau, \boldsymbol{\xi}), \mathbf{g}(\tau, \boldsymbol{\xi})\right\} 
&=\int\limits_{0}^{1}d\tau \oint\limits_{\partial\omega_{3}} dS_{\xi} \cdot f(\tau, \boldsymbol{\xi}) \mathbf{g}(\tau, \boldsymbol{\xi}) \cdot \mathbf{n}\\
&=\int\limits_{0}^{1}d\tau \int\limits_{0}^{1}d\eta \int\limits_{0}^{1}d\zeta \cdot  
\left(f(\tau, 1, \eta, \zeta) g_{\xi}(\tau, 1, \eta, \zeta) - f(\tau, 0, \eta, \zeta) g_{\xi}(\tau, 0, \eta, \zeta)\right)\\
&+\int\limits_{0}^{1}d\tau \int\limits_{0}^{1}d\xi \int\limits_{0}^{1}d\zeta \cdot  
\left(f(\tau, \xi, 1, \zeta) g_{\eta}(\tau, \xi, 1, \zeta) - f(\tau, \xi, 0, \zeta) g_{\eta}(\tau, \xi, 0, \zeta)\right)\\
&+\int\limits_{0}^{1}d\tau \int\limits_{0}^{1}d\xi \int\limits_{0}^{1}d\eta \cdot  
\left(f(\tau, \xi, \eta, 1) g_{\zeta}(\tau, \xi, \eta, 1) - f(\tau, \xi, \eta, 0) g_{\zeta}(\tau, \xi, \eta, 0)\right),\vspace{-1mm}
\end{split}
\end{equation}
where $\mathbf{g} = (g_{\xi}, g_{\eta}, g_{\zeta})$ is a vector field. Substituting the representation (\ref{eq:lst_dg_repr}) of the discrete solution $\mathbf{q}$ into the resulting expression for the ADER-DG scheme and using the Gauss-Legendre quadrature formulas (\ref{eq:gl_rule}) used in case of the selected set of nodal basis functions $\varphi_{k}(\xi)$ leads to the following expression:\vspace{-1mm}
\begin{equation}\label{eq:ader_dg_method_full}
\begin{split}
&w_{k_{1}}w_{k_{2}}w_{k_{3}}\left(\mathbf{u}^{n+1}_{k_{1}k_{2}k_{3}} - \mathbf{u}^{n}_{k_{1}k_{2}k_{3}}\right) \\
&= \sum\limits_{l_{0}=0}^{N} w_{l_{0}} \Bigg[
	\sum\limits_{l_{1}=0}^{N} w_{l_{1}}w_{k_{2}}w_{k_{3}} \mathbb{D}^{\varphi}_{k_{1}l_{1}} \mathbf{f}_{\xi}(\hat{\mathbf{q}}_{l_{0}l_{1}k_{2}k_{3}}) +
	\sum\limits_{l_{2}=0}^{N} w_{k_{1}}w_{l_{2}}w_{k_{3}} \mathbb{D}^{\varphi}_{k_{2}l_{2}} \mathbf{f}_{\eta}(\hat{\mathbf{q}}_{l_{0}k_{1}l_{2}k_{3}}) +
	\sum\limits_{l_{3}=0}^{N} w_{k_{1}}w_{k_{2}}w_{l_{3}} \mathbb{D}^{\varphi}_{k_{3}l_{3}} \mathbf{f}_{\zeta}(\hat{\mathbf{q}}_{l_{0}k_{1}k_{2}l_{3}})
\Bigg]\\
&-\sum\limits_{l_{0}=0}^{N} w_{l_{0}} w_{k_{2}}w_{k_{3}} \left[
	\tilde{\psi}_{k_{1}}\boldsymbol{\mathfrak{G}}_{\xi}\left(
		\sum\limits_{p_{1}=0}^{N} \tilde{\psi}_{p_{1}} \hat{\mathbf{q}}_{l_{0}p_{1}k_{2}k_{3}},
		\sum\limits_{p_{1}=0}^{N} \psi_{p_{1}} \hat{\mathbf{q}}^{(R, \xi)}_{l_{0}p_{1}k_{2}k_{3}}
	\right) - \psi_{k_{1}}\boldsymbol{\mathfrak{G}}_{\xi}\left(
		\sum\limits_{p_{1}=0}^{N} \tilde{\psi}_{p_{1}} \hat{\mathbf{q}}^{(L, \xi)}_{l_{0}p_{1}k_{2}k_{3}},
		\sum\limits_{p_{1}=0}^{N} \psi_{p_{1}} \hat{\mathbf{q}}_{l_{0}p_{1}k_{2}k_{3}}
	\right)
\right]\\
&-\sum\limits_{l_{0}=0}^{N} w_{l_{0}} w_{k_{1}}w_{k_{3}} \left[
	\tilde{\psi}_{k_{2}}\boldsymbol{\mathfrak{G}}_{\eta}\left(
		\sum\limits_{p_{2}=0}^{N} \tilde{\psi}_{p_{2}} \hat{\mathbf{q}}_{l_{0}k_{1}p_{2}k_{3}},
		\sum\limits_{p_{2}=0}^{N} \psi_{p_{2}} \hat{\mathbf{q}}^{(R, \eta)}_{l_{0}k_{1}p_{2}k_{3}}
	\right) - \psi_{k_{2}}\boldsymbol{\mathfrak{G}}_{\eta}\left(
		\sum\limits_{p_{2}=0}^{N} \tilde{\psi}_{p_{2}} \hat{\mathbf{q}}^{(L, \eta)}_{l_{0}k_{1}p_{2}k_{3}},
		\sum\limits_{p_{2}=0}^{N} \psi_{p_{2}} \hat{\mathbf{q}}_{l_{0}k_{1}p_{2}k_{3}}
	\right)
\right]\\
&-\sum\limits_{l_{0}=0}^{N} w_{l_{0}} w_{k_{1}}w_{k_{2}} \left[
	\tilde{\psi}_{k_{3}}\boldsymbol{\mathfrak{G}}_{\zeta}\left(
		\sum\limits_{p_{3}=0}^{N} \tilde{\psi}_{p_{3}} \hat{\mathbf{q}}_{l_{0}k_{1}k_{2}p_{3}},
		\sum\limits_{p_{3}=0}^{N} \psi_{p_{3}} \hat{\mathbf{q}}^{(R, \zeta)}_{l_{0}k_{1}k_{2}p_{3}}
	\right) - \psi_{k_{3}}\boldsymbol{\mathfrak{G}}_{\zeta}\left(
		\sum\limits_{p_{3}=0}^{N} \tilde{\psi}_{p_{3}} \hat{\mathbf{q}}^{(L, \zeta)}_{l_{0}k_{1}k_{2}p_{3}},
		\sum\limits_{p_{3}=0}^{N} \psi_{p_{3}} \hat{\mathbf{q}}_{l_{0}k_{1}k_{2}p_{3}}
	\right)
\right]\\
&-\sum\limits_{l_{0}=0}^{N} w_{l_{0}}w_{k_{1}}w_{k_{2}}w_{k_{3}} \Bigg[
	\mathfrak{B}_{\xi}(\hat{\mathbf{q}}_{l_{0}k_{1}k_{2}k_{3}}) \sum\limits_{r_{1}=0}^{N} 
	\mathbb{D}^{\varphi}_{r_{1}k_{1}} \hat{\mathbf{q}}_{l_{0}r_{1}k_{2}k_{3}} +
	\mathfrak{B}_{\eta}(\hat{\mathbf{q}}_{l_{0}k_{1}k_{2}k_{3}}) \sum\limits_{r_{2}=0}^{N} 
	\mathbb{D}^{\varphi}_{r_{2}k_{2}} \hat{\mathbf{q}}_{l_{0}k_{1}r_{2}k_{3}}\\
	&\qquad\qquad\qquad\qquad\quad\: +
	\mathfrak{B}_{\zeta}(\hat{\mathbf{q}}_{l_{0}k_{1}k_{2}k_{3}}) \sum\limits_{r_{3}=0}^{N} 
	\mathbb{D}^{\varphi}_{r_{3}k_{3}} \hat{\mathbf{q}}_{l_{0}k_{1}k_{2}r_{3}} -
	\mathbf{s}(\hat{\mathbf{q}}_{l_{0}k_{1}k_{2}k_{3}}, \mathbf{r}(\boldsymbol{\xi}_{k_{1}k_{2}k_{3}}), t(\tau_{l_{0}}))
\Bigg],
\end{split}
\end{equation}
where $\psi_{k} = \varphi_{k}(0)$, $\tilde{\psi}_{k} = \varphi_{k}(1)$, and upper indices $(L, \xi)$, $(L, \eta)$, $(L, \zeta)$ and $(R, \xi)$, $(R, \eta)$, $(R, \zeta)$ denote the discrete space-time solution $\mathbf{q}$ in the left and right cells with respect to the selected cell in directions $\xi$, $\eta$ and $\zeta$, respectively.

\begin{table}[h!]
\begin{center}
\caption{\label{tab:ader_dg_blas}
The main parameters of the BLAS procedures \texttt{gemm} used in the one-step discrete ADER-DG scheme.
}
\begin{tabular}{|l|ccc|ccc|ll|}
\hline
kernel name & $m$ & $n$ & $k$ & \texttt{lda} & \texttt{ldb} & \texttt{ldc} & $\alpha$ & $\beta$\\
\hline
\texttt{interface::psi\_ksi\_conv\_kernel} & $1$ & $M$ & $N+1$ & $N+1$ & $(N+1)^{2} M$ & $M$ & $1$ & $0$\\
\texttt{interface::psi\_eta\_conv\_kernel} & $1$ & $M$ & $N+1$ & $N+1$ & $(N+1) M$ & $M$ & $1$ & $0$\\
\texttt{interface::psi\_zeta\_conv\_kernel} & $1$ & $M$ & $N+1$ & $N+1$ & $M$ & $M$ & $1$ & $0$\\
\texttt{interface::w\_conv\_kernel} & $1$ & $M$ & $N+1$ & $N+1$ & $(N+1)^{2} M$ & $M$ & $1$ & $0$\\
\texttt{volume::w\_conv\_kernel} & $1$ & $M$ & $N+1$ & $N+1$ & $(N+1)^{3} M$ & $(N+1)^{2} M$ & $1$ & $0$\\
\texttt{volume::d\_tilde\_f\_ksi\_kernel} & $N+1$ & $M$ & $N+1$ & $N+1$ & $(N+1)^{2} M$ & $(N+1)^{2} M$ & $1$ & $0$\\
\texttt{volume::d\_tilde\_f\_eta\_kernel} & $N+1$ & $M$ & $N+1$ & $N+1$ & $(N+1) M$ & $(N+1) M$ & $1$ & $0$\\
\texttt{volume::d\_tilde\_f\_zeta\_kernel} & $N+1$ & $M$ & $N+1$ & $N+1$ & $M$ & $M$ & $1$ & $0$\\
\texttt{volume::d\_q\_ksi\_kernel} & $N+1$ & $M$ & $N+1$ & $N+1$ & $(N+1)^{2} M$ & $(N+1)^{2} M$ & $1$ & $0$\\
\texttt{volume::d\_q\_eta\_kernel} & $N+1$ & $M$ & $N+1$ & $N+1$ & $(N+1) M$ & $(N+1) M$ & $1$ & $0$\\
\texttt{volume::d\_q\_zeta\_kernel} & $N+1$ & $M$ & $N+1$ & $N+1$ & $M$ & $M$ & $1$ & $0$\\
\texttt{volume::b\_ksi\_kernel} & $M$ & $M$ & $M$ & $M$ & $M$ & $M$ & $1$ & $0$\\
\texttt{volume::b\_eta\_kernel} & $M$ & $M$ & $M$ & $M$ & $M$ & $M$ & $1$ & $1$\\
\texttt{volume::b\_zeta\_kernel} & $M$ & $M$ & $M$ & $M$ & $M$ & $M$ & $1$ & $1$\\
\hline
\end{tabular}
\end{center}
\end{table}

\begin{algorithm}[h!]
\caption{\label{alg:ader_dg_method}
The basic general structure of using BLAS interface procedure calls in the implementation of the one-step discrete ADER-DG scheme.
}
\begin{algorithmic}[1]
\Function{get\_dg\_solution}{$\mathtt{u\_prev}$}
\State $\mathtt{u\_next} \gets \mathtt{u\_prev}$
\State \Call{update\_volume\_terms}{$\mathtt{u\_next}$}
\State \Call{update\_interface\_terms}{$\mathtt{u\_next}$}
\State \Return $\mathtt{u\_next}$
\EndFunction
\end{algorithmic}
\end{algorithm}

\begin{algorithm}[h!]
\caption{\label{alg:ader_dg_method:supp_volume_terms}
Procedure to Algorithm~\ref{alg:ader_dg_method} responsible for calculating the contributions from the volume terms 
in the expression (\ref{eq:ader_dg_method_full}) and updating them to the preliminary solution $\mathtt{u\_next\_upd}$ at a next time step $t^{n+1}$.
}
\begin{algorithmic}[1]
\vspace{2.5mm}
\State $\mathtt{f\_ksi}$, $\mathtt{f\_eta}$, $\mathtt{f\_zeta}$, $\mathtt{s}$, $\mathtt{b}$: \texttt{array}[$(N+1)^{4} \cdot M$]
\State $\mathtt{w\_conv\_f\_ksi}$, $\mathtt{w\_conv\_f\_eta}$, $\mathtt{w\_conv\_f\_zeta}$: \texttt{array}[$(N+1)^{3} \cdot M$]
\State $\mathtt{D\_mult\_ww\_conv\_f\_ksi}$, $\mathtt{D\_mult\_ww\_conv\_f\_eta}$, $\mathtt{D\_mult\_ww\_conv\_f\_zeta}$: \texttt{array}[$(N+1)^{3} \cdot M$]
\Procedure{update\_volume\_terms}{$\mathtt{u\_next\_upd}$}
\ForAll{$(p_{0}, p_{1}, p_{2}, p_{3}) \in [0, N]^{4}$}
\State $\mathtt{idx} \gets (p_{0}\cdot(N+1)^{3} + p_{1}\cdot(N+1)^{2} + p_{2}\cdot(N+1) + p_{3}) \cdot M$
\State \Call{set\_rescaled\_flux\_terms\_3d}%
	{$\mathtt{f\_ksi}[\mathtt{idx}],\, \mathtt{f\_eta}[\mathtt{idx}],\, \mathtt{f\_zeta}[\mathtt{idx}],\, \mathtt{q}[\mathtt{idx}],\, \mathtt{dt\_div\_dr}$}
\State \Call{set\_rescaled\_source\_terms}{$\mathtt{s}[\mathtt{idx}],\, \mathtt{q}[\mathtt{idx}],\, \mathtt{s\_args}[\mathtt{idx}],\, \mathtt{dt}$}
\EndFor
\State \Call{volume::set\_non\_conservative\_terms\_sum}{$\mathtt{b},\, \mathtt{q},\, \mathtt{dt\_div\_dr}$}
\State $\mathtt{s} \gets \mathtt{s} - \mathtt{b}$
\ForAll{$(k_{1}, k_{2}, k_{3}) \in [0, N]^{3}$}
\State $\mathtt{idx} \gets (k_{1}\cdot(N+1)^{2} + k_{2}\cdot(N+1) + k_{3}) \cdot M$
\State \Call{volume::w\_conv\_kernel}{$\mathtt{w\_conv\_f\_ksi}[\mathtt{idx}],\, \mathtt{w},\, \mathtt{f\_ksi}[\mathtt{idx}]$}
\State \Call{volume::w\_conv\_kernel}{$\mathtt{w\_conv\_f\_eta}[\mathtt{idx}],\, \mathtt{w},\, \mathtt{f\_eta}[\mathtt{idx}]$}
\State \Call{volume::w\_conv\_kernel}{$\mathtt{w\_conv\_f\_zeta}[\mathtt{idx}],\, \mathtt{w},\, \mathtt{f\_zeta}[\mathtt{idx}]$}
\State \Call{volume::w\_conv\_kernel}{$\mathtt{w\_conv\_s}[\mathtt{idx}],\, \mathtt{w},\, \mathtt{s}[\mathtt{idx}]$}
\EndFor
\ForAll{$(k_{2}, k_{3}) \in [0, N]^{2}$}
\State $\mathtt{idx} \gets (k_{2}\cdot(N+1) + k_{3}) \cdot M$
\State \Call{volume::d\_tilde\_f\_ksi\_kernel}
	{$\mathtt{D\_mult\_ww\_conv\_f\_ksi}[\mathtt{idx}],\, \mathtt{D\_tilde},\, \mathtt{w\_conv\_f\_ksi}[\mathtt{idx}]$}
\EndFor
\ForAll{$(k_{1}, k_{3}) \in [0, N]^{2}$}
\State $\mathtt{idx} \gets (k_{1}\cdot(N+1)^{2} + k_{3}) \cdot M$
\State \Call{volume::d\_tilde\_f\_eta\_kernel}
	{$\mathtt{D\_mult\_ww\_conv\_f\_eta}[\mathtt{idx}],\, \mathtt{D\_tilde},\, \mathtt{w\_conv\_f\_eta}[\mathtt{idx}]$}
\EndFor
\ForAll{$(k_{1}, k_{2}) \in [0, N]^{2}$}
\State $\mathtt{idx} \gets (k_{1}\cdot(N+1)^{2} + k_{2}\cdot(N+1)) \cdot M$
\State \Call{volume::d\_tilde\_f\_zeta\_kernel}
	{$\mathtt{D\_mult\_ww\_conv\_f\_zeta}[\mathtt{idx}],\, \mathtt{D\_tilde},\, \mathtt{w\_conv\_f\_zeta}[\mathtt{idx}]$}
\EndFor
\ForAll{$(k_{1}, k_{2}, k_{3}) \in [0, N]^{3}$}
\State $\mathtt{idx} \gets (k_{1}\cdot(N+1)^{2} + k_{2}\cdot(N+1) + k_{3}) \cdot M$
\State $\mathtt{D\_mult\_ww\_conv\_f\_ksi}[\mathtt{idx}] \gets \mathtt{D\_mult\_ww\_conv\_f\_ksi}[\mathtt{idx}] \cdot \mathtt{w}[k_{2}]\cdot\mathtt{w}[k_{3}]$
\State $\mathtt{D\_mult\_ww\_conv\_f\_eta}[\mathtt{idx}] \gets \mathtt{D\_mult\_ww\_conv\_f\_eta}[\mathtt{idx}] \cdot \mathtt{w}[k_{1}]\cdot\mathtt{w}[k_{3}]$
\State $\mathtt{D\_mult\_ww\_conv\_f\_zeta}[\mathtt{idx}] \gets \mathtt{D\_mult\_ww\_conv\_f\_zeta}[\mathtt{idx}] \cdot \mathtt{w}[k_{1}]\cdot\mathtt{w}[k_{2}]$
\State $\mathtt{w\_conv\_s}[\mathtt{idx}] \gets \mathtt{w\_conv\_s}[\mathtt{idx}] \cdot \mathtt{w}[k_{1}]\cdot\mathtt{w}[k_{2}]\cdot\mathtt{w}[k_{3}]$
\EndFor
\ForAll{$(k_{1}, k_{2}, k_{3}) \in [0, N]^{3}$}
\State $\mathtt{idx} \gets (k_{1}\cdot(N+1)^{2} + k_{2}\cdot(N+1) + k_{3}) \cdot M$
\State $\mathtt{upd\_sum} \gets \mathtt{D\_mult\_ww\_conv\_f\_ksi}[\mathtt{idx}] + \mathtt{D\_mult\_ww\_conv\_f\_eta}[\mathtt{idx}]$
\State $\mathtt{upd\_sum} \gets \mathtt{upd\_sum} + \mathtt{D\_mult\_ww\_conv\_f\_zeta}[\mathtt{idx}] + \mathtt{w\_conv\_s}[\mathtt{idx}]$
\State $\mathtt{upd\_sum} \gets \mathtt{upd\_sum} / (\mathtt{w}[k_{1}]\cdot\mathtt{w}[k_{2}]\cdot\mathtt{w}[k_{3}])$
\State $\mathtt{u\_next\_upd}[\mathtt{idx}] \gets \mathtt{u\_next\_upd}[\mathtt{idx}] + \mathtt{upd\_sum}$
\EndFor
\EndProcedure
\vspace{2.5mm}
\end{algorithmic}
\end{algorithm}
\begin{algorithm}[h!]
\caption{\label{alg:ader_dg_method:supp_interface_terms}
Procedure to Algorithm~\ref{alg:ader_dg_method} responsible for calculating the contributions from the interface flux terms 
in the expression (\ref{eq:ader_dg_method_full}) and updating them to the preliminary solution $\mathtt{u\_next\_upd}$ at a next time step $t^{n+1}$.
\textit{Note}: before calling this procedure, it is necessary to call Algorithm~\ref{alg:ader_dg_method:set_flux_terms} 
responsible for calculating interface flux terms for all mesh interfaces.
}
\begin{algorithmic}[1]
\vspace{2.5mm}
\Procedure{update\_interface\_terms}{$\mathtt{u\_next\_upd}$}
\ForAll{$(k_{1}, k_{2}, k_{3}) \in [0, N]^{3}$}
\State $\mathtt{idx\_u} \gets (k_{1}\cdot(N+1)^{2} + k_{2}\cdot(N+1) + k_{3}) \cdot M$
\State $\mathtt{idx\_f\_ksi} \gets (k_{2}\cdot(N+1) + k_{3}) \cdot M$
\State $\mathtt{idx\_f\_eta} \gets (k_{1}\cdot(N+1) + k_{3}) \cdot M$
\State $\mathtt{idx\_f\_zeta} \gets (k_{1}\cdot(N+1) + k_{2}) \cdot M$
\State $\mathtt{flux\_term\_ksi} \gets \mathtt{psi\_1}[k_{1}] \cdot \mathtt{f\_rp\_ksi\_R}[\mathtt{idx\_f\_ksi}] - \mathtt{psi\_0}[k_{1}] \cdot \mathtt{f\_rp\_ksi\_L}[\mathtt{idx\_f\_ksi}]$
\State $\mathtt{flux\_term\_eta} \gets \mathtt{psi\_1}[k_{2}] \cdot \mathtt{f\_rp\_eta\_R}[\mathtt{idx\_f\_eta}] - \mathtt{psi\_0}[k_{2}] \cdot \mathtt{f\_rp\_eta\_L}[\mathtt{idx\_f\_eta}]$
\State $\mathtt{flux\_term\_zeta} \gets \mathtt{psi\_1}[k_{3}] \cdot \mathtt{f\_rp\_zeta\_R}[\mathtt{idx\_f\_zeta}] - \mathtt{psi\_0}[k_{3}] \cdot \mathtt{f\_rp\_zeta\_L}[\mathtt{idx\_f\_zeta}]$
\State $\mathtt{upd\_sum} \gets -(\mathtt{flux\_term\_ksi} + \mathtt{flux\_term\_eta} + \mathtt{flux\_term\_zsi}) / (\mathtt{w}[k_{1}]\cdot\mathtt{w}[k_{2}]\cdot\mathtt{w}[k_{3}])$
\State $\mathtt{u\_next\_upd}[\mathtt{idx\_u}] \gets \mathtt{u\_next\_upd}[\mathtt{idx\_u}] + \mathtt{upd\_sum}$
\EndFor
\EndProcedure
\vspace{2.5mm}
\end{algorithmic}
\end{algorithm}

\begin{algorithm}[h!]
\caption{\label{alg:ader_dg_method:set_flux_terms}
The procedure associated with Algorithm~\ref{alg:ader_dg_method} responsible for computing 
the interface flux terms in the expression (\ref{eq:ader_dg_method_full}).
}
\begin{algorithmic}[1]
\vspace{2.5mm}
\State $\mathtt{psi\_conv\_q\_L}$, $\mathtt{psi\_conv\_q\_R}$, $\mathtt{f\_rp\_ksi}$, $\mathtt{f\_rp\_eta}$, $\mathtt{f\_rp\_zeta}$: \texttt{array}[$(N+1)^{3} \cdot M$]
\State $\mathtt{f\_rp\_ksi\_interface}$, $\mathtt{f\_rp\_eta\_interface}$, $\mathtt{f\_rp\_zeta\_interface}$: \texttt{array}[$(N+1)^{2} \cdot M$]
\Procedure{set\_interfaces\_terms\_ksi}{}
\ForAll{$(k_{0}, k_{1}, k_{3}) \in [0, N]^{3}$}
\State $\mathtt{idx\_f} \gets (k_{0}\cdot(N+1)^{2} + k_{2}\cdot(N+1) + k_{3}) \cdot M$
\State $\mathtt{idx\_q} \gets (k_{0}\cdot(N+1)^{3} + k_{2}\cdot(N+1) + k_{3}) \cdot M$
\State \Call{interface::psi\_ksi\_conv\_kernel}{$\mathtt{psi\_conv\_q\_L}[0],\, \mathtt{psi\_1},\, \mathtt{q\_L\_ksi}[\mathtt{idx\_q}]$}
\State \Call{interface::psi\_ksi\_conv\_kernel}{$\mathtt{psi\_conv\_q\_R}[0],\, \mathtt{psi\_0},\, \mathtt{q\_R\_ksi}[\mathtt{idx\_q}]$}
\State \Call{rs\_rescaled\_flux\_ksi}
	{$\mathtt{f\_rp\_ksi}[\mathtt{idx\_f}],\, \mathtt{psi\_conv\_q\_L},\, \mathtt{psi\_conv\_q\_R}[0],\, \mathtt{dt\_div\_dx}$}
\EndFor
\ForAll{$(k_{2}, k_{3}) \in [0, N]^{2}$}
\State $\mathtt{idx} \gets (k_{1}\cdot(N+1) + k_{3}) \cdot M$
\State \Call{interface::w\_conv\_kernel}{$\mathtt{f\_rp\_ksi\_interface}[\mathtt{idx}],\, \mathtt{w},\, \mathtt{f\_rp\_ksi}[\mathtt{idx}]$}
\State $\mathtt{f\_rp\_ksi\_interface}[\mathtt{idx}] \gets \mathtt{f\_rp\_ksi\_interface}[\mathtt{idx}] \cdot \mathtt{w}[k_{2}]\cdot\mathtt{w}[k_{3}]$
\EndFor
\EndProcedure
\Procedure{set\_interfaces\_terms\_eta}{}
\ForAll{$(k_{0}, k_{1}, k_{3}) \in [0, N]^{3}$}
\State $\mathtt{idx\_f} \gets (k_{0}\cdot(N+1)^{2} + k_{1}\cdot(N+1) + k_{3}) \cdot M$
\State $\mathtt{idx\_q} \gets (k_{0}\cdot(N+1)^{3} + k_{1}\cdot(N+1)^{2} + k_{3}) \cdot M$
\State \Call{interface::psi\_eta\_conv\_kernel}{$\mathtt{psi\_conv\_q\_L}[0],\, \mathtt{psi\_1},\, \mathtt{q\_L\_eta}[\mathtt{idx\_q}]$}
\State \Call{interface::psi\_eta\_conv\_kernel}{$\mathtt{psi\_conv\_q\_R}[0],\, \mathtt{psi\_0},\, \mathtt{q\_R\_eta}[\mathtt{idx\_q}]$}
\State \Call{rs\_rescaled\_flux\_eta}
	{$\mathtt{f\_rp\_eta}[\mathtt{idx\_f}],\, \mathtt{psi\_conv\_q\_L},\, \mathtt{psi\_conv\_q\_R}[0],\, \mathtt{dt\_div\_dy}$}
\EndFor
\ForAll{$(k_{1}, k_{3}) \in [0, N]^{2}$}
\State $\mathtt{idx} \gets (k_{1}\cdot(N+1) + k_{3}) \cdot M$
\State \Call{interface::w\_conv\_kernel}{$\mathtt{f\_rp\_eta\_interface}[\mathtt{idx}],\, \mathtt{w},\, \mathtt{f\_rp\_eta}[\mathtt{idx}]$}
\State $\mathtt{f\_rp\_eta\_interface}[\mathtt{idx}] \gets \mathtt{f\_rp\_eta\_interface}[\mathtt{idx}] \cdot \mathtt{w}[k_{1}]\cdot\mathtt{w}[k_{3}]$
\EndFor
\EndProcedure
\Procedure{set\_interfaces\_terms\_eta}{}
\ForAll{$(k_{0}, k_{1}, k_{2}) \in [0, N]^{3}$}
\State $\mathtt{idx\_f} \gets (k_{0}\cdot(N+1)^{2} + k_{1}\cdot(N+1) + k_{2}) \cdot M$
\State $\mathtt{idx\_q} \gets (k_{0}\cdot(N+1)^{3} + k_{1}\cdot(N+1)^{2} + k_{2}\cdot(N+1)) \cdot M$
\State \Call{interface::psi\_zeta\_conv\_kernel}{$\mathtt{psi\_conv\_q\_L}[0],\, \mathtt{psi\_1},\, \mathtt{q\_L\_zeta}[\mathtt{idx\_q}]$}
\State \Call{interface::psi\_zeta\_conv\_kernel}{$\mathtt{psi\_conv\_q\_R}[0],\, \mathtt{psi\_0},\, \mathtt{q\_R\_zeta}[\mathtt{idx\_q}]$}
\State \Call{rs\_rescaled\_flux\_zeta}
	{$\mathtt{f\_rp\_zeta}[\mathtt{idx\_f}],\, \mathtt{psi\_conv\_q\_L},\, \mathtt{psi\_conv\_q\_R}[0],\, \mathtt{dt\_div\_dz}$}
\EndFor
\ForAll{$(k_{1}, k_{2}) \in [0, N]^{2}$}
\State $\mathtt{idx} \gets (k_{2}\cdot(N+1) + k_{3}) \cdot M$
\State \Call{interface::w\_conv\_kernel}{$\mathtt{f\_rp\_zeta\_interface}[\mathtt{idx}],\, \mathtt{w},\, \mathtt{f\_rp\_zeta}[\mathtt{idx}]$}
\State $\mathtt{f\_rp\_zeta\_interface}[\mathtt{idx}] \gets \mathtt{f\_rp\_zeta\_interface}[\mathtt{idx}] \cdot \mathtt{w}[k_{1}]\cdot\mathtt{w}[k_{2}]$
\EndFor
\EndProcedure
\vspace{2.5mm}
\end{algorithmic}
\end{algorithm}

\paragraph{Using the BLAS interface}
\label{sec:framework:ader_dg_method:blas}

The resulting expression (\ref{eq:ader_dg_method_full}) for the one-step discrete ADER-DG scheme contains basic linear algebra operations, which are implemented using the BLAS interface. The general paradigm for data storage and using the matrix-matrix multiplication operation are chosen in a form similar to the implementation of the LST-DG predictor, which is presented in Subsection~\ref{sec:framework:lst_dg_predictor:blas}. Only the \texttt{gemm} procedure of the 3rd level of the BLAS interface is used. In this case, only the row major data storage format is used. Table~\ref{tab:ader_dg_blas} presents the main parameters of all \texttt{gemm} procedures used. Algorithm~\ref{alg:ader_dg_method} presents the general form of the algorithm for calculating the solution $\mathtt{u\_next}$ at the next time step $t^{n+1}$, while the calculation of the solution is separated into two main procedures presented in Algorithms~\ref{alg:ader_dg_method:supp_volume_terms} and~\ref{alg:ader_dg_method:supp_interface_terms}, as well as in an additional procedures in Algorithm~\ref{alg:ader_dg_method:set_flux_terms} for computing the interface flux terms \texttt{f\_rp\_ksi\_interface}, \texttt{f\_rp\_eta\_interface}, \texttt{f\_rp\_zeta\_interface}. The calculation of interface flux terms is carried out in three separate procedures \texttt{set\_interfaces\_terms\_ksi}, \texttt{set\_interfaces\_terms\_eta} and \texttt{set\_interfaces\_terms\_zeta}, which is associated with the use of Cartesian meshes --- each interface contains terms only for the projection orthogonal to its plane; therefore, for each mesh interface one does not need to call all three procedures, just one is enough. It should be noted that the parameters of the procedures \texttt{d\_q\_ksi\_kernel}, \texttt{d\_q\_eta\_kernel} and \texttt{d\_q\_zeta\_kernel} in Table~\ref{tab:ader_dg_blas} from the name space \texttt{volume} are identical to the parameters procedures \texttt{d\_tilde\_f\_ksi\_kernel}, \texttt{d\_tilde\_f\_eta\_kernel} and \texttt{d\_tilde\_f\_zeta\_kernel}, which is associated with the same sizes of matrices and leading dimensions; however, the calculation of non-conservative terms requires the use of a transposed matrix $[\mathbb{D}^{\varphi}]^{T}$, which can be implemented by the parameter \texttt{transa}=\texttt{trans}, or by storing two pre-computed matrices --- the original $\mathbb{D}^{\varphi}$ and the transposed $[\mathbb{D}^{\varphi}]^{T}$; in the implementation presented in Algorithm~\ref{alg:ader_dg_method:supp_volume_terms}, the method with the transposed matrix $[\mathbb{D}^{\varphi}]^{T}$ is used. The \texttt{D\_tilde} matrix used in Algorithm~\ref{alg:ader_dg_method:supp_volume_terms} contains elements $w_{l} \mathbb{D}^{\varphi}_{kl}$, which is convenient according to the expression (\ref{eq:ader_dg_method_full}). \newtext{The} matrices \texttt{psi\_1}, \texttt{psi\_2} and \texttt{w} contain values $\psi_{k}$, $\tilde{\psi}_{k}$ and $w_{k}$. The procedures \texttt{rs\_rescaled\_flux\_ksi}, \texttt{rs\_rescaled\_flux\_eta} and \texttt{rs\_rescaled\_flux\_zeta} calculate the solution to the Riemann problem, where fluxes expressed immediately in the rescaled variables, for which the values \texttt{dt\_div\_dx}, \texttt{dt\_div\_dy} and \texttt{dt\_div\_dz} are passed as formal parameters to procedures. Procedure \texttt{volume::set\_non\_conservative\_terms\_sum}, responsible for calculating non-conservative terms, is structured similarly to the procedure presented in Algorithm~\ref{alg:lst_dg_predictor:supp_2} for the LST-DG predictor --- matrix calculations are performed using the procedures \texttt{d\_q\_ksi\_kernel}, \texttt{d\_q\_eta\_kernel} and \texttt{d\_q\_zeta\_kernel} from the name space \texttt{volume}. In general, the presented implementation allows the solution $\mathbf{u}^{n+1}$ to be computed at the next time step $t^{n+1}$ using BLAS interface procedures, without the use of transpose $\mathbf{q}_{r,p} \mapsto \mathbf{q}_{p, r}$~\cite{ader_dg_eff_impl}. The parameters of the procedure are known in advance, so the proposed approach is well suited to the use of JIT functions available in modern implementations of the BLAS interface for calculations with small matrices.

\subsection{ADER-WENO-FV method}
\label{sec:framework:ader_weno_fv_method}

\paragraph{General description}
\label{sec:framework:ader_weno_fv_method:descr}

The one-step discrete finite-volume ADER-WENO scheme is obtained by integrating the system of equations (\ref{eq:pde_ref}) over the reference space-time element $\omega_{4}$, which is technically equivalent to the $L_{2}$ projection onto ``test function'' $1$, resulting in the following expression~\cite{ader_weno_lstdg_ideal, ader_weno_lstdg_diss, ader_dg_eff_impl}:
\begin{equation}
\left\langle 
	\frac{\partial \mathbf{u}}{\partial \tau} + 
	\frac{\partial \mathbf{f}_{\xi}(\mathbf{u})}{\partial \xi} + 
	\frac{\partial \mathbf{f}_{\eta}(\mathbf{u})}{\partial \eta} + 
	\frac{\partial \mathbf{f}_{\zeta}(\mathbf{u})}{\partial \zeta} + 
	\mathfrak{B}_{\xi}(\mathbf{u})\cdot\frac{\partial \mathbf{u}}{\partial \xi} + 
	\mathfrak{B}_{\eta}(\mathbf{u})\cdot\frac{\partial \mathbf{u}}{\partial \eta} +
	\mathfrak{B}_{\zeta}(\mathbf{u})\cdot\frac{\partial \mathbf{u}}{\partial \zeta} -
	\mathbf{s}(\mathbf{u}, \mathbf{r}(\boldsymbol{\xi}), t(\tau))
\right\rangle = 0,
\end{equation}
where $\langle.\rangle \equiv \langle1, .\rangle$. 

\begin{table}[h!]
\begin{center}
\caption{\label{tab:ader_weno_fv_blas}
The main parameters of the BLAS procedures \texttt{gemm} used in the one-step discrete finite-volume ADER-WENO scheme.
}
\begin{tabular}{|l|ccc|ccc|ll|}
\hline
kernel name & $m$ & $n$ & $k$ & \texttt{lda} & \texttt{ldb} & \texttt{ldc} & $\alpha$ & $\beta$\\
\hline
\texttt{psi\_ksi\_conv\_kernel} & $1$ & $M$ & $N+1$ & $N+1$ & $(N+1)^{2} M$ & $M$ & $1$ & $0$\\
\texttt{psi\_eta\_conv\_kernel} & $1$ & $M$ & $N+1$ & $N+1$ & $(N+1) M$ & $M$ & $1$ & $0$\\
\texttt{psi\_zeta\_conv\_kernel} & $1$ & $M$ & $N+1$ & $N+1$ & $M$ & $M$ & $1$ & $0$\\
\texttt{w\_conv\_kernel} & $1$ & $M$ & $N+1$ & $N+1$ & $M$ & $M$ & $1$ & $0$\\
\texttt{d\_q\_ksi\_kernel} & $N+1$ & $M$ & $N+1$ & $N+1$ & $(N+1)^{2} M$ & $(N+1)^{2} M$ & $1$ & $0$\\
\texttt{d\_q\_eta\_kernel} & $N+1$ & $M$ & $N+1$ & $N+1$ & $(N+1) M$ & $(N+1) M$ & $1$ & $0$\\
\texttt{d\_q\_zeta\_kernel} & $N+1$ & $M$ & $N+1$ & $N+1$ & $M$ & $M$ & $1$ & $0$\\
\texttt{b\_ksi\_kernel} & $M$ & $M$ & $M$ & $M$ & $M$ & $M$ & $1$ & $0$\\
\texttt{b\_eta\_kernel} & $M$ & $M$ & $M$ & $M$ & $M$ & $M$ & $1$ & $1$\\
\texttt{b\_zeta\_kernel} & $M$ & $M$ & $M$ & $M$ & $M$ & $M$ & $1$ & $1$\\
\hline
\end{tabular}
\end{center}
\end{table}

The substitution of the finite-volume average values (\ref{eq:fv_repr}) $\mathbf{v}^{n}$ and $\mathbf{v}^{n+1}$ in the time term and the discrete solution $\mathbf{q}$ (\ref{eq:lst_dg_repr}) instead of $\mathbf{u}$ in the integrals leads to an expression of the form:
\begin{equation}
\begin{split}
\mathbf{v}^{n+1} = \mathbf{v}^{n} &-
\left\{ \mathfrak{G}\left(\mathbf{q}^{(-)}, \mathbf{q}^{(+)}\right) \right\}
-\left(
	\left\langle \mathfrak{B}_{\xi}(\mathbf{q})\cdot\frac{\partial \mathbf{q}}{\partial \xi} \right\rangle +
	\left\langle \mathfrak{B}_{\eta}(\mathbf{q})\cdot\frac{\partial \mathbf{q}}{\partial \eta} \right\rangle +
	\left\langle \mathfrak{B}_{\zeta}(\mathbf{q})\cdot\frac{\partial \mathbf{q}}{\partial \zeta} \right\rangle 
\right)
+\left\langle \mathbf{s}(\mathbf{q}, \mathbf{r}(\boldsymbol{\xi}), t(\tau)) \right\rangle,
\end{split}
\end{equation}
where $\mathfrak{G} = (\boldsymbol{\mathfrak{G}}_{\xi}, \boldsymbol{\mathfrak{G}}_{\eta}, \boldsymbol{\mathfrak{G}}_{\zeta})$ is the Riemann solver, taking into account non-conservative terms in the system of equations and expressed in the rescaled coordinate system (\ref{eq:local_space_time_coords}) of the reference space-time element $\omega_{4}$, $\mathbf{q}^{(+)}$ and $\mathbf{q}^{(-)}$ are the local discrete solutions in cells in the outer normal direction $\mathbf{n}$ of the cell $\Omega_{i}$ and in the opposite direction, and $\left\{.\right\} = \left\{1, .\right\}$. 

\begin{algorithm}[h!]
\caption{\label{alg:ader_weno_fv_method}
The basic general structure of using BLAS interface procedure calls in the implementation of the one-step discrete finite-volume ADER-WENO scheme.
}
\begin{algorithmic}[1]
\Function{get\_fv\_solution}{$\mathtt{v\_prev}$}
\State $\mathtt{v\_next} \gets \mathtt{v\_prev}$
\State \Call{update\_volume\_terms}{$\mathtt{v\_next}$}
\State \Call{update\_interface\_terms}{$\mathtt{v\_next}$}
\State \Return $\mathtt{v\_next}$
\EndFunction
\end{algorithmic}
\end{algorithm}

\begin{algorithm}[h!]
\caption{\label{alg:ader_weno_fv_method:supp_volume_terms}
Procedure to Algorithm~\ref{alg:ader_weno_fv_method} responsible for calculating the contributions from the volume terms 
in the expression (\ref{eq:ader_weno_fv_method_full}) and updating them to the preliminary solution $\mathtt{v\_next\_upd}$ at a next time step $t^{n+1}$.
}
\begin{algorithmic}[1]
\State $\mathtt{f\_ksi}$, $\mathtt{f\_eta}$, $\mathtt{f\_zeta}$, $\mathtt{s}$, $\mathtt{b}$: \texttt{array}[$(N+1)^{4} \cdot M$]
\State $\mathtt{w\_conv\_f\_ksi}$, $\mathtt{w\_conv\_f\_eta}$, $\mathtt{w\_conv\_f\_zeta}$: \texttt{array}[$(N+1)^{3} \cdot M$]
\State $\mathtt{D\_mult\_ww\_conv\_f\_ksi}$, $\mathtt{D\_mult\_ww\_conv\_f\_eta}$, $\mathtt{D\_mult\_ww\_conv\_f\_zeta}$: \texttt{array}[$(N+1)^{3} \cdot M$]
\Procedure{update\_volume\_terms}{$\mathtt{v\_next\_upd}$}
\ForAll{$(p_{0}, p_{1}, p_{2}, p_{3}) \in [0, N]^{4}$}
\State $\mathtt{idx} \gets (p_{0}\cdot(N+1)^{3} + p_{1}\cdot(N+1)^{2} + p_{2}\cdot(N+1) + p_{3}) \cdot M$
\State \Call{set\_rescaled\_source\_terms}{$\mathtt{s}[\mathtt{idx}],\, \mathtt{q}[\mathtt{idx}],\, \mathtt{s\_args}[\mathtt{idx}],\, \mathtt{dt}$}
\EndFor
\State \Call{volume::set\_non\_conservative\_terms\_sum}{$\mathtt{b},\, \mathtt{q},\, \mathtt{dt\_div\_dr}$}
\State $\mathtt{s} \gets \mathtt{s} - \mathtt{b}$
\ForAll{$(p_{0}, p_{1}, p_{2}) \in [0, N]^{3}$}
\State $\mathtt{idx\_s} \gets (p_{0}\cdot(N+1)^{3} + p_{1}\cdot(N+1)^{2} + p_{2}\cdot(N+1)) \cdot M$
\State $\mathtt{idx\_d} \gets (p_{0}\cdot(N+1)^{2} + p_{1}\cdot(N+1) + p_{2}) \cdot M$
\State \Call{w\_conv\_kernel}{$\mathtt{w\_p3\_conv\_s}[\mathtt{idx\_d}],\, \mathtt{w},\, \mathtt{s}[\mathtt{idx\_s}]$}
\EndFor
\ForAll{$(p_{0}, p_{1}) \in [0, N]^{2}$}
\State $\mathtt{idx\_s} \gets (p_{0}\cdot(N+1)^{2} + p_{1}\cdot(N+1)) \cdot M$
\State $\mathtt{idx\_d} \gets (p_{0}\cdot(N+1) + p_{1}) \cdot M$
\State \Call{w\_conv\_kernel}{$\mathtt{w\_p2p3\_conv\_s}[\mathtt{idx\_d}],\, \mathtt{w},\, \mathtt{w\_p3\_conv\_s}[\mathtt{idx\_s}]$}
\EndFor
\ForAll{$p_{0} \in [0, N]$}
\State $\mathtt{idx\_s} \gets (p_{0}\cdot(N+1) + p_{1}) \cdot M$
\State $\mathtt{idx\_d} \gets p_{0} \cdot M$
\State \Call{w\_conv\_kernel}{$\mathtt{w\_p1p2p3\_conv\_s}[\mathtt{idx\_d}],\, \mathtt{w},\, \mathtt{w\_p2p3\_conv\_s}[\mathtt{idx\_s}]$}
\EndFor
\State \Call{w\_conv\_kernel}{$\mathtt{w\_p0p1p2p3\_conv\_s}[0],\, \mathtt{w},\, \mathtt{w\_p1p2p3\_conv\_s}[0]$}
\State $\mathtt{v\_next\_upd} \gets \mathtt{v\_next\_upd} + \mathtt{w\_p0p1p2p3\_conv\_s}$
\EndProcedure
\end{algorithmic}
\end{algorithm}

\begin{algorithm}[h!]
\caption{\label{alg:ader_weno_fv_method:set_flux_terms}
The procedure associated with Algorithm~\ref{alg:ader_weno_fv_method} responsible for computing 
the interface flux terms in the expression (\ref{eq:ader_weno_fv_method_full}).
}
\begin{algorithmic}[1]
\State $\mathtt{psi\_conv\_q\_L}$, $\mathtt{psi\_conv\_q\_R}$, $\mathtt{f\_rp\_ksi}$, $\mathtt{f\_rp\_eta}$, $\mathtt{f\_rp\_zeta}$: \texttt{array}[$(N+1)^{3} \cdot M$]
\State $\mathtt{w\_k3\_conv\_f\_rp\_ksi}$, $\mathtt{w\_k3\_conv\_f\_rp\_eta}$, $\mathtt{w\_k3\_conv\_f\_rp\_zeta}$: \texttt{array}[$(N+1)^{2} \cdot M$]
\State $\mathtt{w\_k2k3\_conv\_f\_rp\_ksi}$, $\mathtt{w\_k2k3\_conv\_f\_rp\_eta}$, $\mathtt{w\_k2k3\_conv\_f\_rp\_zeta}$: \texttt{array}[$(N+1) \cdot M$]
\State $\mathtt{f\_rp\_ksi\_interface}$, $\mathtt{f\_rp\_eta\_interface}$, $\mathtt{f\_rp\_zeta\_interface}$: \texttt{array}[$M$]
\Procedure{set\_interfaces\_terms\_ksi}{}
\ForAll{$(k_{0}, k_{2}, k_{3}) \in [0, N]^{3}$}
\State $\mathtt{idx\_f} \gets (k_{0}\cdot(N+1)^{2} + k_{2}\cdot(N+1) + k_{3}) \cdot M$
\State $\mathtt{idx\_q} \gets (k_{0}\cdot(N+1)^{3} + k_{2}\cdot(N+1) + k_{3}) \cdot M$
\State \Call{interface::psi\_ksi\_conv\_kernel}{$\mathtt{psi\_conv\_q\_L}[0],\, \mathtt{psi\_1},\, \mathtt{q\_L\_ksi}[\mathtt{idx\_q}]$}
\State \Call{interface::psi\_ksi\_conv\_kernel}{$\mathtt{psi\_conv\_q\_R}[0],\, \mathtt{psi\_0},\, \mathtt{q\_R\_ksi}[\mathtt{idx\_q}]$}
\State \Call{rs\_rescaled\_flux\_ksi}
	{$\mathtt{f\_rp\_ksi}[\mathtt{idx\_f}],\, \mathtt{psi\_conv\_q\_L},\, \mathtt{psi\_conv\_q\_R}[0],\, \mathtt{dt\_div\_dx}$}
\EndFor
\ForAll{$(k_{0}, k_{2}) \in [0, N]^{2}$}
\State $\mathtt{idx\_s} \gets (k_{0}\cdot(N+1)^{2} + k_{2}) \cdot M$
\State $\mathtt{idx\_d} \gets (k_{0}\cdot(N+1) + k_{2}) \cdot M$
\State \Call{w\_conv\_kernel}{$\mathtt{w\_k3\_conv\_f\_rp\_ksi}[\mathtt{idx\_d}],\, \mathtt{w},\, \mathtt{f\_rp\_ksi}[\mathtt{idx\_s}]$}
\EndFor
\ForAll{$k_{0} \in [0, N]^{2}$}
\State $\mathtt{idx\_s} \gets k_{0}\cdot(N+1) M$
\State $\mathtt{idx\_d} \gets k_{0}\cdot M$
\State \Call{w\_conv\_kernel}{$\mathtt{w\_k2k3\_conv\_f\_rp\_ksi}[\mathtt{idx\_d}],\, \mathtt{w},\, \mathtt{w\_k3\_conv\_f\_rp\_ksi}[\mathtt{idx\_s}]$}
\EndFor
\State \Call{w\_conv\_kernel}{$\mathtt{f\_rp\_ksi\_interface}[0],\, \mathtt{w},\, \mathtt{w\_k2k3\_conv\_f\_rp\_ksi}[0]$}
\EndProcedure
\Procedure{set\_interfaces\_terms\_eta}{}
\ForAll{$(k_{0}, k_{1}, k_{3}) \in [0, N]^{3}$}
\State $\mathtt{idx\_f} \gets (k_{0}\cdot(N+1)^{2} + k_{1}\cdot(N+1) + k_{3}) \cdot M$
\State $\mathtt{idx\_q} \gets (k_{0}\cdot(N+1)^{3} + k_{1}\cdot(N+1)^{2} + k_{3}) \cdot M$
\State \Call{interface::psi\_eta\_conv\_kernel}{$\mathtt{psi\_conv\_q\_L}[0],\, \mathtt{psi\_1},\, \mathtt{q\_L\_eta}[\mathtt{idx\_q}]$}
\State \Call{interface::psi\_eta\_conv\_kernel}{$\mathtt{psi\_conv\_q\_R}[0],\, \mathtt{psi\_0},\, \mathtt{q\_R\_eta}[\mathtt{idx\_q}]$}
\State \Call{rs\_rescaled\_flux\_eta}
	{$\mathtt{f\_rp\_eta}[\mathtt{idx\_f}],\, \mathtt{psi\_conv\_q\_L},\, \mathtt{psi\_conv\_q\_R}[0],\, \mathtt{dt\_div\_dy}$}
\EndFor
\ForAll{$(k_{0}, k_{1}) \in [0, N]^{2}$}
\State $\mathtt{idx\_s} \gets (k_{0}\cdot(N+1)^{2} + k_{1}\cdot(N+1)) \cdot M$
\State $\mathtt{idx\_d} \gets (k_{0}\cdot(N+1) + k_{1}) \cdot M$
\State \Call{w\_conv\_kernel}{$\mathtt{w\_k3\_conv\_f\_rp\_eta}[\mathtt{idx\_d}],\, \mathtt{w},\, \mathtt{f\_rp\_eta}[\mathtt{idx\_s}]$}
\EndFor
\ForAll{$k_{0} \in [0, N]^{2}$}
\State $\mathtt{idx\_s} \gets k_{0}\cdot(N+1) M$
\State $\mathtt{idx\_d} \gets k_{0}\cdot M$
\State \Call{w\_conv\_kernel}{$\mathtt{w\_k2k3\_conv\_f\_rp\_eta}[\mathtt{idx\_d}],\, \mathtt{w},\, \mathtt{w\_k3\_conv\_f\_rp\_eta}[\mathtt{idx\_s}]$}
\EndFor
\State \Call{w\_conv\_kernel}{$\mathtt{f\_rp\_eta\_interface}[0],\, \mathtt{w},\, \mathtt{w\_k2k3\_conv\_f\_rp\_eta}[0]$}
\EndProcedure
\algstore{alg:ader_weno_fv_method:set_flux_terms:store}
\end{algorithmic}
\end{algorithm}
\begin{algorithm}[h!]
\caption{Continuation of Algorithm~\ref{alg:ader_weno_fv_method:set_flux_terms}.}
\begin{algorithmic}[1]
\algrestore{alg:ader_weno_fv_method:set_flux_terms:store}
\Procedure{set\_interfaces\_terms\_zeta}{}
\ForAll{$(k_{0}, k_{1}, k_{2}) \in [0, N]^{3}$}
\State $\mathtt{idx\_f} \gets (k_{0}\cdot(N+1)^{2} + k_{1}\cdot(N+1) + k_{2}) \cdot M$
\State $\mathtt{idx\_q} \gets (k_{0}\cdot(N+1)^{3} + k_{1}\cdot(N+1)^{2} + k_{2}\cdot(N+1)) \cdot M$
\State \Call{interface::psi\_zeta\_conv\_kernel}{$\mathtt{psi\_conv\_q\_L}[0],\, \mathtt{psi\_1},\, \mathtt{q\_L\_zeta}[\mathtt{idx\_q}]$}
\State \Call{interface::psi\_zeta\_conv\_kernel}{$\mathtt{psi\_conv\_q\_R}[0],\, \mathtt{psi\_0},\, \mathtt{q\_R\_zeta}[\mathtt{idx\_q}]$}
\State \Call{rs\_rescaled\_flux\_zeta}
	{$\mathtt{f\_rp\_zeta}[\mathtt{idx\_f}],\, \mathtt{psi\_conv\_q\_L},\, \mathtt{psi\_conv\_q\_R}[0],\, \mathtt{dt\_div\_dz}$}
\EndFor
\ForAll{$(k_{0}, k_{1}) \in [0, N]^{2}$}
\State $\mathtt{idx\_s} \gets (k_{0}\cdot(N+1)^{2} + k_{1}\cdot(N+1)) \cdot M$
\State $\mathtt{idx\_d} \gets (k_{0}\cdot(N+1) + k_{1}) \cdot M$
\State \Call{w\_conv\_kernel}{$\mathtt{w\_k3\_conv\_f\_rp\_zeta}[\mathtt{idx\_d}],\, \mathtt{w},\, \mathtt{f\_rp\_zeta}[\mathtt{idx\_s}]$}
\EndFor
\ForAll{$k_{0} \in [0, N]^{2}$}
\State $\mathtt{idx\_s} \gets k_{0}\cdot(N+1) M$
\State $\mathtt{idx\_d} \gets k_{0}\cdot M$
\State \Call{w\_conv\_kernel}{$\mathtt{w\_k2k3\_conv\_f\_rp\_zeta}[\mathtt{idx\_d}],\, \mathtt{w},\, \mathtt{w\_k3\_conv\_f\_rp\_zeta}[\mathtt{idx\_s}]$}
\EndFor
\State \Call{w\_conv\_kernel}{$\mathtt{f\_rp\_zeta\_interface}[0],\, \mathtt{w},\, \mathtt{w\_k2k3\_conv\_f\_rp\_zeta}[0]$}
\EndProcedure
\end{algorithmic}
\end{algorithm}

Substituting the representation (\ref{eq:lst_dg_repr}) of the discrete solution $\mathbf{q}$ into the resulting expression for the ADER-DG scheme and using the Gauss-Legendre quadrature formulas (\ref{eq:gl_rule}) used in case of the selected set of nodal basis functions $\varphi_{k}(\xi)$ leads to the following expression:
\begin{equation}\label{eq:ader_weno_fv_method_full}
\begin{split}
\mathbf{v}^{n+1} = \mathbf{v}^{n} 
&-\sum\limits_{l_{0}=0}^{N}\sum\limits_{l_{2}=0}^{N}\sum\limits_{l_{3}=0}^{N} w_{l_{0}} w_{l_{2}}w_{l_{3}} \left[
	\boldsymbol{\mathfrak{G}}_{\xi}\left(
		\sum\limits_{p_{1}=0}^{N} \tilde{\psi}_{p_{1}} \hat{\mathbf{q}}_{l_{0}p_{1}l_{2}l_{3}},
		\sum\limits_{p_{1}=0}^{N} \psi_{p_{1}} \hat{\mathbf{q}}^{(R, \xi)}_{l_{0}p_{1}l_{2}l_{3}}
	\right) - \boldsymbol{\mathfrak{G}}_{\xi}\left(
		\sum\limits_{p_{1}=0}^{N} \tilde{\psi}_{p_{1}} \hat{\mathbf{q}}^{(L, \xi)}_{l_{0}p_{1}l_{2}l_{3}},
		\sum\limits_{p_{1}=0}^{N} \psi_{p_{1}} \hat{\mathbf{q}}_{l_{0}p_{1}l_{2}l_{3}}
	\right)
\right]\\
&-\sum\limits_{l_{0}=0}^{N}\sum\limits_{l_{1}=0}^{N}\sum\limits_{l_{3}=0}^{N} w_{l_{0}} w_{l_{1}}w_{l_{3}} \left[
	\boldsymbol{\mathfrak{G}}_{\eta}\left(
		\sum\limits_{p_{2}=0}^{N} \tilde{\psi}_{p_{2}} \hat{\mathbf{q}}_{l_{0}l_{1}p_{2}l_{3}},
		\sum\limits_{p_{2}=0}^{N} \psi_{p_{2}} \hat{\mathbf{q}}^{(R, \eta)}_{l_{0}l_{1}p_{2}l_{3}}
	\right) - \boldsymbol{\mathfrak{G}}_{\eta}\left(
		\sum\limits_{p_{2}=0}^{N} \tilde{\psi}_{p_{2}} \hat{\mathbf{q}}^{(L, \eta)}_{l_{0}l_{1}p_{2}l_{3}},
		\sum\limits_{p_{2}=0}^{N} \psi_{p_{2}} \hat{\mathbf{q}}_{l_{0}l_{1}p_{2}l_{3}}
	\right)
\right]\\
&-\sum\limits_{l_{0}=0}^{N}\sum\limits_{l_{1}=0}^{N}\sum\limits_{l_{2}=0}^{N} w_{l_{0}} w_{l_{1}}w_{l_{2}} \left[
	\boldsymbol{\mathfrak{G}}_{\zeta}\left(
		\sum\limits_{p_{3}=0}^{N} \tilde{\psi}_{p_{3}} \hat{\mathbf{q}}_{l_{0}l_{1}l_{2}p_{3}},
		\sum\limits_{p_{3}=0}^{N} \psi_{p_{3}} \hat{\mathbf{q}}^{(R, \zeta)}_{l_{0}l_{1}l_{2}p_{3}}
	\right) - \boldsymbol{\mathfrak{G}}_{\zeta}\left(
		\sum\limits_{p_{3}=0}^{N} \tilde{\psi}_{p_{3}} \hat{\mathbf{q}}^{(L, \zeta)}_{l_{0}l_{1}l_{2}p_{3}},
		\sum\limits_{p_{3}=0}^{N} \psi_{p_{3}} \hat{\mathbf{q}}_{l_{0}l_{1}l_{2}p_{3}}
	\right)
\right]\\
&-\sum\limits_{l_{0}=0}^{N}\sum\limits_{l_{1}=0}^{N}\sum\limits_{l_{2}=0}^{N}\sum\limits_{l_{3}=0}^{N} w_{l_{0}} w_{l_{1}}w_{l_{2}}w_{l_{3}} \Bigg[
	\mathfrak{B}_{\xi}(\hat{\mathbf{q}}_{l_{0}l_{1}l_{2}l_{3}}) 
	\sum\limits_{r_{1}=0}^{N}\mathbb{D}^{\varphi}_{r_{1}l_{1}} \hat{\mathbf{q}}_{l_{0}r_{1}l_{2}l_{3}} +
	\mathfrak{B}_{\eta}(\hat{\mathbf{q}}_{l_{0}l_{1}l_{2}l_{3}}) 
	\sum\limits_{r_{2}=0}^{N}\mathbb{D}^{\varphi}_{r_{2}l_{2}} \hat{\mathbf{q}}_{l_{0}l_{1}r_{2}l_{3}}\\
	&\qquad\qquad\qquad\quad\qquad\qquad\qquad\quad\,\,+\mathfrak{B}_{\zeta}(\hat{\mathbf{q}}_{l_{0}l_{1}l_{2}l_{3}}) 
	\sum\limits_{r_{3}=0}^{N}\mathbb{D}^{\varphi}_{r_{2}l_{2}} \hat{\mathbf{q}}_{l_{0}l_{1}l_{2}r_{3}} -
	\mathbf{s}(\hat{\mathbf{q}}_{l_{0}l_{1}l_{2}l_{3}}, \mathbf{r}(\boldsymbol{\xi}_{l_{1}l_{2}l_{3}}), t(\tau_{l_{0}}))
\Bigg],
\end{split}
\end{equation}
where $\psi_{k} = \varphi_{k}(0)$, $\tilde{\psi}_{k} = \varphi_{k}(1)$, and upper indices $(L, \xi)$, $(L, \eta)$, $(L, \zeta)$ and $(R, \xi)$, $(R, \eta)$, $(R, \zeta)$ denote the discrete space-time solution $\mathbf{q}$ in the left and right cells with respect to the selected cell in directions $\xi$, $\eta$ and $\zeta$, respectively.

\paragraph{Using the BLAS interface}
\label{sec:framework:ader_weno_fv_method:blas}

The resulting expression (\ref{eq:ader_weno_fv_method_full}) for the one-step discrete finite-volume ADER-WENO scheme contains basic linear algebra operations, which are implemented using the BLAS interface. The general paradigm for data storage and using the matrix-matrix multiplication operation are chosen in a form similar to the implementation of the LST-DG predictor and the finite-element ADER-DG scheme, which is presented in Subsections~\ref{sec:framework:lst_dg_predictor:blas} and~\ref{sec:framework:ader_dg_method:blas}. Only the \texttt{gemm} procedure of the 3rd level of the BLAS interface is used. In this case, only the row major data storage format is used. Table~\ref{tab:ader_weno_fv_blas} presents the main parameters of all \texttt{gemm} procedures used. Algorithm~\ref{alg:ader_weno_fv_method} presents the general form of the algorithm for calculating the solution $\mathtt{v\_next}$ at the next time step $t^{n+1}$, while the calculation of the solution is separated into two main procedures, one of which presented in Algorithm~\ref{alg:ader_weno_fv_method:supp_volume_terms}, the second procedure \texttt{update\_interface\_terms} updates the solution with differences in interface flux terms, which is a trivial operation and is essentially similar to Algorithm~\ref{alg:ader_dg_method:supp_interface_terms} for the ADER-DG method. Also the additional procedures in Algorithm~\ref{alg:ader_dg_method:set_flux_terms} for computing the interface flux terms \texttt{f\_rp\_ksi\_interface}, \texttt{f\_rp\_eta\_interface}, \texttt{f\_rp\_zeta\_interface} are separated. The calculation of interface flux terms is carried out in three separate procedures \texttt{set\_interfaces\_terms\_ksi}, \texttt{set\_interfaces\_terms\_eta} and \texttt{set\_interfaces\_terms\_zeta}, which is associated with the use of Cartesian meshes. The matrices \texttt{psi\_1}, \texttt{psi\_2} and \texttt{w} contain values $\psi_{k}$, $\tilde{\psi}_{k}$ and $w_{k}$. The procedures calculating the solution to the Riemann problem \texttt{rs\_rescaled\_flux\_ksi}, \texttt{rs\_rescaled\_flux\_eta} and \texttt{rs\_rescaled\_flux\_zeta}, calculated fluxes immediately in the rescaled variables, for which the values \texttt{dt\_div\_dx}, \texttt{dt\_div\_dy} and \texttt{dt\_div\_dz} are passed as formal parameters to procedures. Procedure \texttt{volume::set\_non\_conservative\_terms\_sum}, responsible for calculating non-conservative terms, is structured similarly to the procedure presented in Algorithms~\ref{alg:lst_dg_predictor:supp_2} for the LST-DG predictor and~\ref{alg:ader_dg_method:supp_volume_terms} for finite-element ADER-DG scheme. In general, the presented implementation allows the solution $\mathbf{u}^{n+1}$ to be computed at the next time step $t^{n+1}$ using BLAS interface procedures, without the use of transpose $\mathbf{q}_{r,p} \mapsto \mathbf{q}_{p, r}$~\cite{ader_dg_eff_impl}. The parameters of the procedure are known in advance, so the proposed approach is well suited to the use of JIT functions available in modern implementations of the BLAS interface for calculations with small matrices.

\subsection{Representations transformations in subcell correction by finite-volume method}
\label{sec:framework:repr_transf}

\paragraph{General description}
\label{sec:framework:repr_transf:descr}

The finite-element ADER-DG methods are the high order linear methods, therefore, the fundamental property of monotonicity of the numerical solution obtained using them may be violated, which is explained by the well-known Godunov theorem. Therefore, ADER-DG methods are used in conjunction with high stable finite-volume limiters, which avoid violation of monotonicity and the occurrence of non-physical artifacts of the numerical solution~\cite{ader_dg_dev_1, ader_dg_ideal_flows, ader_dg_diss_flows, ader_dg_eff_impl}. The finite-volume limiters can use both the classical Godunov and TVD methods~\cite{ader_dg_diss_flows}, as well as the high-precision finite-volume ADER-WENO method~\cite{ader_dg_dev_1, ader_dg_ideal_flows}. To maintain high accuracy and subgrid resolution of the numerical solution obtained by ADER-DG methods, it is customary to use the method of subcell correction of the solution~\cite{ader_dg_dev_1, ader_dg_ideal_flows}, when a finite-volume limiter, which does not have subgrid resolution, is called in the subgrid created in the cell. A sub-grid is created within the cell $\Omega_{i}$ containing $N_{s}$ subcells with the spatial step $h/N_{s}$ in each coordinate direction --- a total of $N_{s}^{3}$ sub-cells in the 3D case. In subcells $\Omega_{i, j}$, an alternative data representation $\mathbf{v}_{j}$ is determined by finite-volume average values (\ref{eq:fv_repr}) for each subcell $\Omega_{i, j}$ from the original high-precision DG representation $\mathbf{u}(\boldsymbol{\xi})$ of the solution. It should be noted that the subgrid finite-volume alternative data representation is used not only for the purpose of subcell correction of the solution: calculating the time step $\Delta t^{n}$ requires calculating the maximum values of the signal speed $\lambda$ in the solution in the entire spatial mesh, which is usually difficult to do mathematically strictly due to the high computational cost of calculating the extrema of multidimensional polynomials degrees in each cell --- instead of strict extrema, maximum values can be calculated using a finite-volume subcell representation $\mathbf{v}_{j}$; a similar approach is used to test the numerical admissibility detector~\cite{ader_dg_ideal_flows}.

Using a subgrid solution correction requires the use of two subgrid operators --- the suitable piecewise-constant projection operator $\hat{\mathbb{P}}$ and the suitable high order accurate reconstruction operator $\hat{\mathbb{R}}$. The suitable piecewise-constant projection operator $\hat{\mathbb{P}}$ calculates finite-volume average values $\mathbf{v}$ for each subcells $\Omega$ using the known high-precision DG representation $\mathbf{u}(\boldsymbol{\xi})$ of the solution:
\begin{equation}\label{eq:p_operator}
\mathbf{v} = \hat{\mathbb{P}}\cdot\mathbf{u},
\end{equation}
and the suitable high order accurate reconstruction operator $\hat{\mathbb{R}}$ calculates the inverse transformation, calculating a high-precision DG representation of the solution using the known subgrid finite-volume representation:
\begin{equation}\label{eq:r_operator}
\mathbf{u} = \hat{\mathbb{R}}\cdot\mathbf{v}.
\end{equation}
The operators $\hat{\mathbb{P}}$ and $\hat{\mathbb{R}}$ have one-side reversibility: $\hat{\mathbb{R}}\circ\hat{\mathbb{P}} = 1$. In case $N_{s} = N$, the amount of information about the solution in the finite-volume subcell representation $\mathbf{v}$ and in the high-precision DG representation $\mathbf{u}$ is the same, and the operators have the property of complete reversibility --- $\hat{\mathbb{P}}\circ\hat{\mathbb{R}} = 1$. However, usually the value $N_{s} = 2N+1$, which is due to the requirement that the time step $\Delta t^{n}$ coincide for finite-volume numerical methods and DG methods~\cite{ader_dg_dev_1}, therefore reversibility in a different order of action of the operators does not occur --- $\hat{\mathbb{P}}\circ\hat{\mathbb{R}} \neq 1$, in the chosen case $N_{s} = 2N+1 > N+1$ for $N \geqslant 1$.

\begin{table}[h!]
\begin{center}
\caption{\label{tab:repr_trasforms_blas}
The main parameters of the BLAS procedures \texttt{gemm} used in the suitable piecewise-constant projection operator $\hat{\mathbb{P}}$ and the suitable high order accurate reconstruction operator $\hat{\mathbb{R}}$.
}
\begin{tabular}{|l|ccc|ccc|ll|}
\hline
kernel name & $m$ & $n$ & $k$ & \texttt{lda} & \texttt{ldb} & \texttt{ldc} & $\alpha$ & $\beta$\\
\hline
\texttt{p\_inner\_stage\_kernel} & $N_{s}$ & $M$ & $N+1$ & $N+1$ & $M$ & $(N+1)^{2} M$ & $1$ & $0$\\
\texttt{p\_mider\_stage\_kernel} & $N_{s}$ & $M$ & $N+1$ & $N+1$ & $M$ & $(N+1) N_{s} M$ & $1$ & $0$\\
\texttt{p\_outer\_stage\_kernel} & $N_{s}$ & $M$ & $N+1$ & $N+1$ & $M$ & $N_{s}^{2} M$ & $1$ & $0$\\
\texttt{r\_inner\_stage\_kernel} & $N+1$ & $M$ & $N_{s}$ & $N_{s}$ & $M$ & $N_{s}^{2} M$ & $1$ & $0$\\
\texttt{r\_mider\_stage\_kernel} & $N+1$ & $M$ & $N_{s}$ & $N_{s}$ & $M$ & $(N+1) N_{s} M$ & $1$ & $0$\\
\texttt{r\_outer\_stage\_kernel} & $N+1$ & $M$ & $N_{s}$ & $N_{s}$ & $M$ & $(N+1)^{2} M$ & $1$ & $0$\\
\hline
\end{tabular}
\end{center}
\end{table}

\begin{algorithm}[h!]
\caption{\label{alg:repr_trasforms:p_operator}
The basic general structure of using BLAS interface procedure calls in the suitable piecewise-constant projection operator $\hat{\mathbb{P}}$.
}
\begin{algorithmic}[1]
\State $\mathtt{u}$: \texttt{array}[$(N+1)^{3} \cdot M$]
\State $\mathtt{temp\_matrix\_inner}$: \texttt{array}[$(N+1)^{2} \cdot N_{s} \cdot M$]
\State $\mathtt{temp\_matrix\_mider}$: \texttt{array}[$(N+1) \cdot N_{s}^{2} \cdot M$]
\State $\mathtt{v}$: \texttt{array}[$N_{s}^{3} \cdot M$]
\Function{p\_operator}{$\mathtt{u}$}
\ForAll{$(p_{1}, p_{2}) \in [0, N]\times[0, N]$}
\State $\mathtt{idx\_d} \gets (p_{1}\cdot(N+1) + p_{2}) \cdot M$
\State $\mathtt{idx\_s} \gets (p_{1}\cdot(N+1)^{2} + p_{2}\cdot(N+1)) \cdot M$
\State \Call{p\_inner\_stage\_kernel}{$\mathtt{temp\_matrix\_inner}[\mathtt{idx\_d}],\, \mathtt{p\_matrix},\, \mathtt{u}[\mathtt{idx\_s}]$}
\EndFor
\ForAll{$(k_{3}, p_{1}) \in [0, N_{s}-1]\times[0, N]$}
\State $\mathtt{idx\_d} \gets (k_{3}\cdot(N+1) + p_{1}) \cdot M$
\State $\mathtt{idx\_s} \gets (k_{3}\cdot(N+1)^{2} + p_{1}\cdot(N+1)) \cdot M$
\State \Call{p\_mider\_stage\_kernel}{$\mathtt{temp\_matrix\_mider}[\mathtt{idx\_d}],\, \mathtt{p\_matrix},\, \mathtt{temp\_matrix\_inner}[\mathtt{idx\_s}]$}
\EndFor
\ForAll{$(k_{2}, p_{3}) \in [0, N_{s}-1]\times[0, N_{s}-1]$}
\State $\mathtt{idx\_d} \gets (k_{2} \cdot N_{s} + k_{3}) \cdot M$
\State $\mathtt{idx\_s} \gets (k_{2} \cdot (N+1) N_{s} + k_{3}\cdot(N+1)) \cdot M$
\State \Call{p\_outer\_stage\_kernel}{$\mathtt{v}[\mathtt{idx\_d}],\, \mathtt{p\_matrix},\, \mathtt{temp\_matrix\_mider}[\mathtt{idx\_s}]$}
\EndFor
\State \Return $\mathtt{v}$
\EndFunction
\end{algorithmic}
\end{algorithm}

\begin{algorithm}[h!]
\caption{\label{alg:repr_trasforms:r_operator}
The basic general structure of using BLAS interface procedure calls in the suitable high order accurate reconstruction operator $\hat{\mathbb{R}}$.
}
\begin{algorithmic}[1]
\State $\mathtt{v}$: \texttt{array}[$N_{s}^{3} \cdot M$]
\State $\mathtt{temp\_matrix\_inner}$: \texttt{array}[$N_{s}^{2} \cdot (N+1) \cdot M$]
\State $\mathtt{temp\_matrix\_mider}$: \texttt{array}[$N_{s} \cdot (N+1)^{2} \cdot M$]
\State $\mathtt{u}$: \texttt{array}[$(N+1)^{3} \cdot M$]
\Function{r\_operator}{$\mathtt{v}$}
\ForAll{$(k_{1}, k_{2}) \in [0, N_{s}-1]\times[0, N_{s}-1]$}
\State $\mathtt{idx\_d} \gets (k_{1} \cdot N_{s} + k_{2}) \cdot M$
\State $\mathtt{idx\_s} \gets (k_{1} \cdot N_{s}^{2} + k_{2} \cdot N_{s}) \cdot M$
\State \Call{p\_inner\_stage\_kernel}{$\mathtt{temp\_matrix\_inner}[\mathtt{idx\_d}],\, \mathtt{p\_matrix},\, \mathtt{v}[\mathtt{idx\_s}]$}
\EndFor
\ForAll{$(p_{3}, k_{1}) \in [0, N]\times[0, N_{s}-1]$}
\State $\mathtt{idx\_d} \gets (p_{3} \cdot N_{s} + k_{1}) \cdot M$
\State $\mathtt{idx\_s} \gets (p_{3} \cdot N_{s}^{2} + k_{1} \cdot N_{s}) \cdot M$
\State \Call{p\_mider\_stage\_kernel}{$\mathtt{temp\_matrix\_mider}[\mathtt{idx\_d}],\, \mathtt{p\_matrix},\, \mathtt{temp\_matrix\_inner}[\mathtt{idx\_s}]$}
\EndFor
\ForAll{$(p_{2}, p_{3}) \in [0, N]\times[0, N]$}
\State $\mathtt{idx\_d} \gets (p_{2} \cdot (N+1) + p_{3}) \cdot M$
\State $\mathtt{idx\_s} \gets (p_{2} \cdot N_{s}^{2} + p_{3} \cdot N_{s}) \cdot M$
\State \Call{p\_outer\_stage\_kernel}{$\mathtt{u}[\mathtt{idx\_d}],\, \mathtt{p\_matrix},\, \mathtt{temp\_matrix\_mider}[\mathtt{idx\_s}]$}
\EndFor
\State \Return $\mathtt{u}$
\EndFunction
\end{algorithmic}
\end{algorithm}

The derivation of the formula expression for operator $\mathbb{P}$ is based on the following considerations. A high-precision DG representation of the solution $\mathbf{u} = \mathbf{u}(\boldsymbol{\xi})$ (\ref{eq:dg_repr}) is averaged over each subcell $\Omega_{i, j}$ of the subgrid in cell $\Omega_{i}$, resulting in the following expression (\ref{eq:fv_repr}):
\begin{equation}
\mathbf{v}_{j} = N_{s}^{3} \int\limits_{\Omega_{i, j}} d\boldsymbol{\xi} \cdot \mathbf{u}(\boldsymbol{\xi}) 
= N_{s}^{3} \int\limits_{\frac{j_{1}}{N_{s}}}^{\frac{j_{1}+1}{N_{s}}}d\xi \int\limits_{\frac{j_{2}}{N_{s}}}^{\frac{j_{2}+1}{N_{s}}}d\eta
\int\limits_{\frac{j_{3}}{N_{s}}}^{\frac{j_{3}+1}{N_{s}}}d\zeta \cdot \sum\limits_{k} \mathbf{u}_{k} \Phi_{k}(\xi, \eta, \zeta) =
\sum\limits_{k} \mathbb{P}_{j,k} \mathbf{u}_{k},
\end{equation}
where $k = (k_{1}, k_{2}, k_{3})$, $j = (j_{1}, j_{2}, j_{3})$ is the three-component multi-indices: $0 \leqslant k_{1}, k_{2}, k_{3} \leqslant N$, $0 \leqslant j_{1}, j_{2}, j_{3} < N_{s}$; it should be noted that these multi-indices define different index ranges and can be unambiguously converted into \newtext{pass-through} indices (\ref{eq:through_indices}):
\begin{equation}
\begin{split}
&k = k_{1}\cdot(N+1)^{2} + k_{2}\cdot(N+1) + k_{3};\\
&j = j_{1} \cdot N_{s}^{2} + j_{2} \cdot N_{s} + j_{3};
\end{split}
\end{equation}
which take ranges of $0 \leqslant k < (N+1)^{3}$ and $0 \leqslant j \leqslant N_{s}^{3}$. In an application-interesting~\cite{ader_dg_dev_1, ader_dg_ideal_flows, ader_dg_diss_flows, ader_dg_eff_impl} case $N_{s} = 2N+1 > N+1$, the inverse transformation is defined by suitable high order accurate reconstruction operator $\hat{\mathbb{R}}$ using a pseudo-inverse matrix:
\begin{equation}
\mathbf{u}_{k} = \sum\limits_{j}\mathbb{R}_{k,j}\mathbf{v}_{j} = 
\sum\limits_{j}\left\{\left[\mathbb{P}^{T}\mathbb{P}\right]^{-1}\cdot\mathbb{P}\right\}_{k,j}\mathbf{v}_{j},
\end{equation}
where expression $[\mathbb{P}^{T}\mathbb{P}]^{-1}\mathbb{P}$ defines the pseudo-inverse matrix. The presented expressions for the suitable piecewise-constant projection operator $\hat{\mathbb{P}}$ and the suitable high order accurate reconstruction operator $\hat{\mathbb{R}}$ allowed to conclude that the matrices of these operators can be represented in the form of Kronecker products:
\begin{equation}
\begin{split}
&\mathbb{P} = \mathrm{P}\otimes\mathrm{P}\otimes\mathrm{P};\\
&\mathbb{R} = \mathrm{R}\otimes\mathrm{R}\otimes\mathrm{R};
\end{split}
\end{equation}
which can be pre-computed in code, and where $\mathrm{P}$ is a $N_{s}\times(N+1)$ matrix containing elements of the following form:
\begin{equation}
\mathrm{P}_{j,k} = N_{s} \int\limits_{\frac{j}{N_{s}}}^{\frac{j+1}{N_{s}}}d\xi \cdot \varphi_{k}(\xi),
\end{equation}
where indices take ranges of $0 \leqslant j < N_{s}$ and $0 \leqslant k \leqslant N$, and $\mathrm{R}$ is the pseudo-inverse of matrix $\mathrm{P}$.

\paragraph{Using the BLAS interface}
\label{sec:framework:repr_transf:blas}

The resulting expressions for the suitable piecewise-constant projection operator $\hat{\mathbb{P}}$ (\ref{eq:p_operator}) and the suitable high order accurate reconstruction operator $\hat{\mathbb{R}}$ (\ref{eq:r_operator}) are implemented using the \texttt{gemm} function of the BLAS interface. In this case, only the row major data storage format is used, and \texttt{notrans} is selected for the operations \texttt{transa} and \texttt{transb}. The parameters of the \texttt{gemm} function that are used in the implementation are presented in Table~\ref{tab:repr_trasforms_blas}. The algorithmic description of the implementation of the suitable piecewise-constant projection operator $\hat{\mathbb{P}}$ is presented in Algorithm~\ref{alg:repr_trasforms:p_operator}; the algorithmic description of the implementation of the suitable high order accurate reconstruction operator $\hat{\mathbb{R}}$ is presented in Algorithm~\ref{alg:repr_trasforms:r_operator}. The parameters of the procedure are known in advance, so the proposed approach is well suited to the use of JIT functions available in modern implementations of the BLAS interface for calculations with small matrices.

\subsection{Time step constraint}
\label{sec:framework:cfl_condition}

The one-step discrete ADER-DG scheme is explicit, despite the presence of a locally implicit LST-DG predictor. The Courant-Friedrichs-Lewy stability criterion is imposed on the time step $\Delta t^{n}$~\cite{ADER_DG_time_step_1, ADER_DG_time_step_2}:
\begin{equation}
\Delta t^{n} = \mathtt{CFL} \cdot \frac{1}{d}\cdot\frac{1}{2N+1} \cdot \min\limits_{k = 1,\ldots,d}\left[\frac{h_{k}}{\left|\lambda_{k}^{max}\right|}\right],
\end{equation}
where $\mathtt{CFL} \leqslant 1$ is the Courant number, $d$ is the spatial dimension of the problem, $N$ is the degrees of polynomials used in the DG representation (\ref{eq:dg_repr}), $h_{k}$ is the spatial mesh step in the $k$-direction, $|\lambda_{k}^{max}|$ is the maximum signal speed in the $k$-direction for which the expression $|\lambda_{k}^{max}| = |u_{k}|+c$ is used, where $u_{k}$ is the flow velocity in the $k$-direction, $c$ is the sound speed. This form differs from the classical form of the Courant-Friedrichs-Lewy stability criterion, used in finite-volume numerical methods, in particular --- for the finite-volume ADER-WENO method, by the presence of the expression $2N+1$ in the denominator. This is used to choose effectively the spatial step of the subgrid in which the a posteriori limiting of the solution is performed~\cite{ader_dg_dev_1, ader_dg_dev_2, ader_dg_ideal_flows}.

\subsection{Admissibility criteria}
\label{sec:framework:adm_criteria}

The one-step discrete ADER-DG scheme is fundamentally a linear scheme of arbitrarily high order, so the numerical solution $\mathbf{u}^{n+1}$ may contain a violation of monotonicity, which is determined by the well-known Godunov theorem, which shows that there are no linear monotonic numerical schemes above first order for hyperbolic equations. Violation of the monotonicity of the numerical solution $\mathbf{u}^{n+1}$ leads to the inadmissibility of the numerical solution, in particular, non-physical oscillations of the numerical solution arise, leading to negative values of density $\rho$, internal energy density $e$ and pressure $p$ for classical hydrodynamical problems. Therefore, the numerical solution $\mathbf{u}^{n+1}$ obtained as a result of using the one-step discrete ADER-DG scheme is only preliminary --- the so-called candidate solution $\mathbf{u}^{*}$. 

Within the framework of the space-time adaptive ADER-DG finite element method with LST-DG predictor and a posteriori sub-cell ADER-WENO finite-volume limiting, the resulting numerical candidate solution $\mathbf{u}_{h}^{*}$ is checked for admissibility, for which admissibility criteria are used: the physical admissibility detector (PAD) and the numerical admissibility detector (NAD), which are widely used~\cite{ader_dg_dev_1, ader_dg_dev_2, ader_weno_lstdg_ideal, ader_weno_lstdg_diss, ader_dg_ideal_flows, ader_dg_diss_flows, ader_dg_grmhd, ader_dg_gr_prd, ader_dg_PNPM, PNPM_DG_2009, PNPM_DG_2010, ader_dg_ale} in numerical methods for solving quasi-linear equations with a posteriori correction of the numerical solution. 

The physical admissibility detector checks the candidate numerical solution for the admissibility of the main physical assumptions of the problem. In this work, for non-stationary Euler equations these were the condition of positivity of the density $\rho$ and internal energy density $e$. The positiveness of the pressure $p$ automatically follow from the fulfillment of these admissibility conditions. In this work the numerical admissibility detector is chosen in a cell representation. It is based on the use of the relaxed discrete maximum principle (DMP) in the polynomial sense, and is expressed by the following inequality~\cite{ader_dg_dev_2, ader_dg_ideal_flows, ader_dg_diss_flows}:
\begin{equation}\label{eq:nad_ineq}
\begin{split}
\min\limits_{\mathbf{r}' \in V_{i}} \left(\mathbf{u}(\mathbf{r}', t^{n})\right) - \boldsymbol{\delta}
\leqslant \mathbf{u}^{*}(\mathbf{r}, t^{n+1}) \leqslant
\max\limits_{\mathbf{r}' \in V_{i}} \left(\mathbf{u}(\mathbf{r}', t^{n})\right) + \boldsymbol{\delta},\quad
\forall \mathbf{r} \in \Omega_{i},
\end{split}
\end{equation}
where the maximum and minimum are taken over the set $V_{i}$ that contains this cell $\Omega_{i}$ and its Voronov neighboring cells. An additional small vector quantity $\boldsymbol{\delta}$, which is given by the expression
\begin{equation}
\boldsymbol{\delta} = \max\left[\mathbf{\delta}_{0}, \epsilon_{0} \cdot \left(
	\max\limits_{\mathbf{r}' \in V_{i}} \left(\mathbf{u}(\mathbf{r}', t^{n})\right) -
	\min\limits_{\mathbf{r}' \in V_{i}} \left(\mathbf{u}(\mathbf{r}', t^{n})\right)
\right)\right],
\end{equation}
determines the tolerance of the criterion --- the real calculation in software implementation is carried out not on the polynomial representation $\mathbf{u}$ (\ref{eq:dg_repr}) of the solution, but on the basis of a finite-volume sub-cell representation $\mathbf{v}_{j}$ (\ref{eq:p_operator}), which is formed in a subgrid of the cell; the use of exact extrema of the solution representation would require significant computational costs, especially in the case of high degrees of $N$ and two- and three-dimensional problems, so an approach is chosen with the analysis of the finite-volume representation $\mathbf{v}_{j}$ and a small expansion of the criterion admissibility window by $\boldsymbol{\delta}$. The values $\delta_{0} = 10^{-4}$ and $\epsilon = 10^{-3}$ are chosen in accordance with the recommendations of the works~\cite{ader_dg_dev_2, ader_dg_ideal_flows}. The inequality (\ref{eq:nad_ineq}) is stated in vector form --- feasibility is checked for each individual component of the vector of conservative variables, and the final conclusion of feasibility is determined by the $\land$ logical operation for all components of the vector. 

Subcell forms of the numerical admissibility detector (the so-called SubNAD), used in particular in methods with flux reconstruction~\cite{DG_FR_2019, DG_FR_2023}, are not used in this work. In general, the choice of criteria for the admissibility of a candidate numerical solution corresponds to the approaches proposed in the works~\cite{ader_dg_dev_2, ader_dg_ideal_flows, ader_dg_diss_flows}.

\section{Applications of the numerical method}
\label{sec:apps}

Applications of the implementations of the numerical methods of the ADER family presented in this work \newtext{are demonstrated} using the example of classical gas dynamics problems described by a system of non-stationary Euler equations:
\begin{equation}\label{eq:system_of_equations}
\frac{\partial}{\partial t}\left[
\begin{array}{c}
\rho\\
\rho\mathbf{v}\\
\varepsilon
\end{array}
\right] + \nabla\left[
\begin{array}{c}
\rho\mathbf{v}\\
\rho\mathbf{v}\otimes\mathbf{v} + p\mathbf{I}\\
(\varepsilon + p) \mathbf{v}
\end{array}
\right] = \mathbf{S},
\end{equation}
where $\rho$ is the mass density; $\mathbf{v} = (u, v, w)$ is the velocity; $p$ is the pressure; $\varepsilon$ is the total energy density including the thermal $e$ and the kinetic contributions $\varepsilon = e + \frac{1}{2} \rho v^{2}$; $\mathbf{v}\otimes\mathbf{v} \equiv \mathbf{v}\mathbf{v}^{T}$ is the tensor products. The source term $\mathbf{S} = \mathbf{S}(\mathbf{U};\, \mathbf{r},\, t)$ allows arbitrary dependence on conserved variables $\mathbf{U}$, spatial coordinates $\mathbf{r} = (x, y, z)$ and time $t$.

The presented applications of the numerical method demonstrate the performance of the developed algorithmic and software implementation. Four groups of tests are presented: tests for accuracy and convergence, classical one-dimensional tests of Riemann problems (the calculations were carried out by a two-dimensional software implementation), a Kelvin-Helmholtz instability test, and tests with cylindrical and spherical explosion problems. The two-dimensional and three-dimensional sine wave advection problems characterized by a continuous solution are then used to test the performance of the ADER-DG finite-element method and the ADER-WENO finite volume method separately. The cylindrical and spherical explosion problems are then used to test the performance of the finite-element ADER-DG methods with a posteriori correction of the solution in subcells by a finite-volume ADER-WENO limiter.

\subsection{Accuracy and convergence}
\label{sec:apps:acc_conv}

\paragraph{PDE system problem}
Accuracy and convergence of the space-time adaptive ADER-DG numerical method with a posteriori sub-cell ADER-WENO finite-volume limiting are tested based on a numerical solution of the two-dimensional problem of the advection of a sine wave in a periodic spatial domain~\cite{DG_FR_2019}. The accuracy and convergence for two-dimensional and three-dimensional problems are studied. The use of a sine wave advection problem is standard in the field of research of numerical methods.

The initial conditions are chosen in the form of a stationary ambient gas flow with the parameters: density $\rho_{\infty} = 1$, velocity $(u_{\infty}, v_{\infty}, w_{\infty})$, pressure $p_{\infty} = 1$. In the two-dimensional case, the velocity is chosen as $(u_{\infty}, v_{\infty}, w_{\infty}) = (1, 1, 0)$, in the three-dimensional case, the velocity is chosen as $(u_{\infty}, v_{\infty}, w_{\infty}) = (1, 1, 1)$. The local density perturbation $\delta\rho$ is imposed on the flow, expressed in the form of a sine wave, which in the two-dimensional case is given in the form $\delta\rho = 0.5\cdot\sin(2\pi(x+y))$, in the three-dimensional case --- in form $\delta\rho = 0.5\cdot\sin(2\pi(x+y+z))$. The adiabatic index $\gamma = 1.4$. 

\begin{figure}[h!]
\centering
\includegraphics[width=0.49\textwidth]{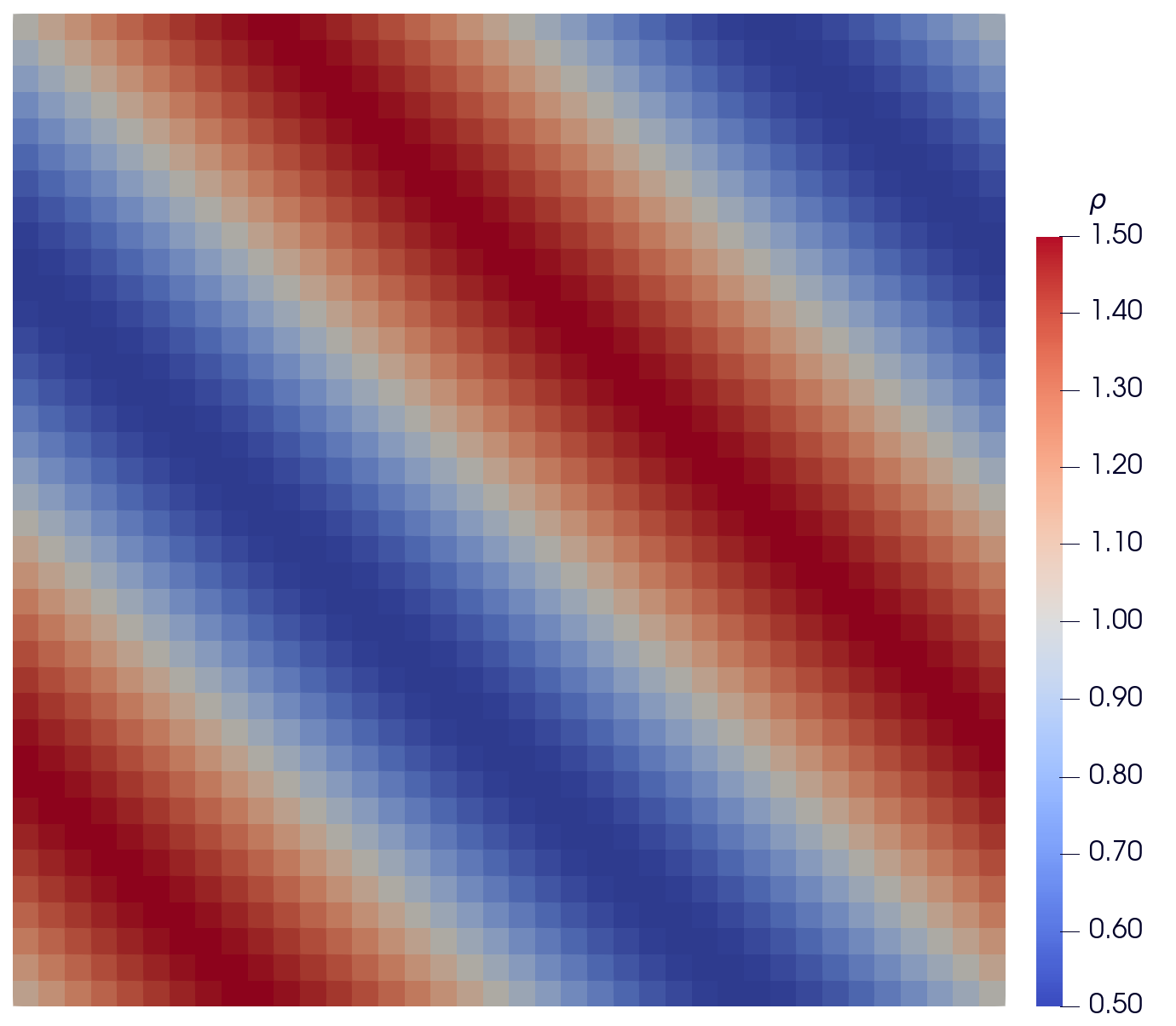}
\includegraphics[width=0.49\textwidth]{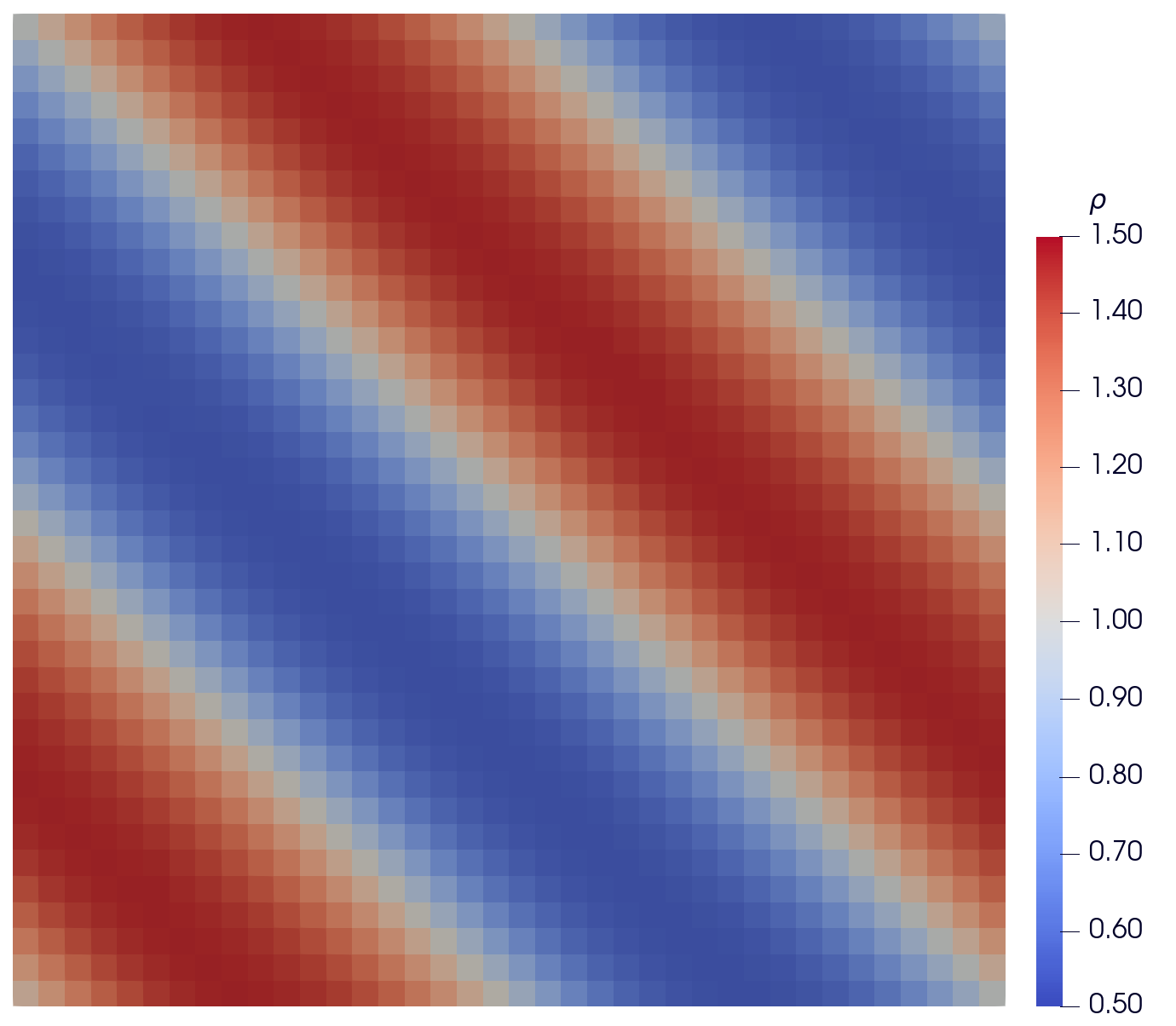}
\caption{\label{fig:sine_wave_adv_2d_dg_9_2x2}%
Numerical solution of the two-dimensional problem of a sine wave advection 
(a detailed statement of the problem is presented in the text)
obtained using the ADER-DG-$\mathbb{P}_{9}$ method on a $2 \times 2$ mesh
at the final time $t_\mathrm{final} = 1.0$:
subcells finite-volume representation of exact (left) and numerical (right) solutions for density $\rho$.
}
\end{figure}

\begin{figure}[h!]
\centering
\includegraphics[width=0.49\textwidth]{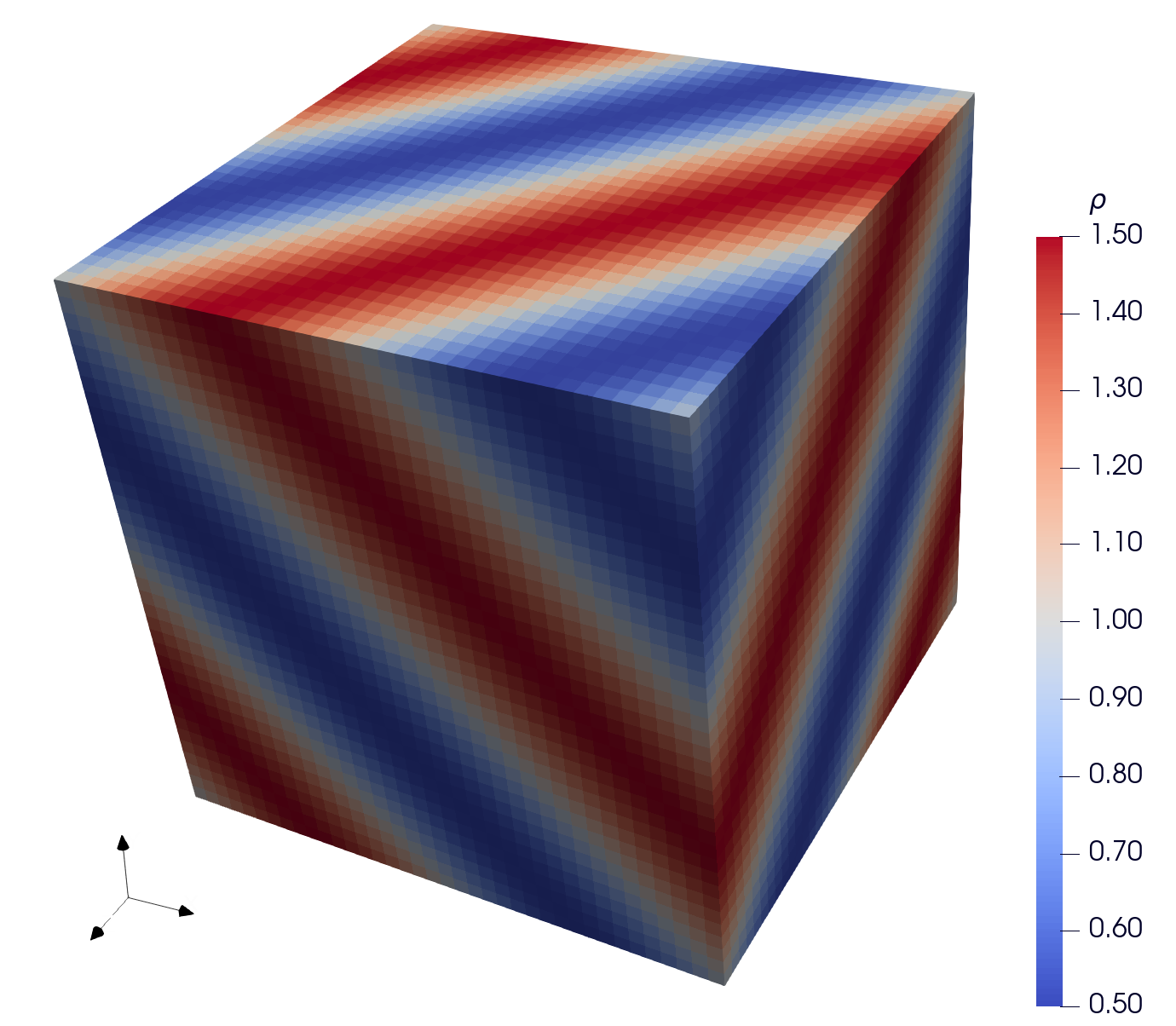}
\includegraphics[width=0.49\textwidth]{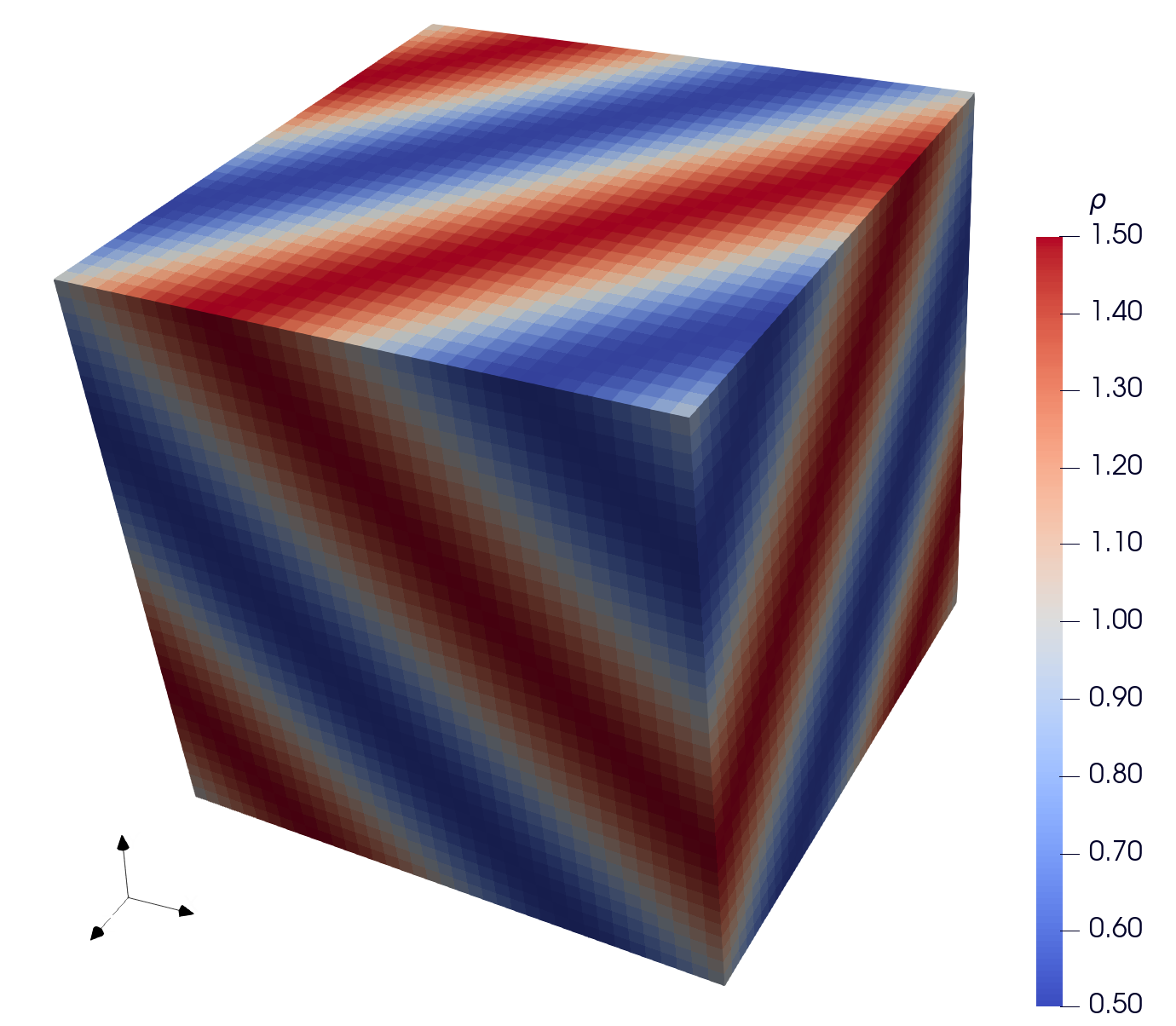}
\caption{\label{fig:sine_wave_adv_3d_dg_9_2x2}%
Numerical solution of the three-dimensional problem of a sine wave advection 
(a detailed statement of the problem is presented in the text)
obtained using the ADER-DG-$\mathbb{P}_{9}$ method on a $2 \times 2 \times 2$ mesh
at the final time $t_\mathrm{final} = 1.0$:
subcells finite-volume representation of exact (left) and numerical (right) solutions for density $\rho$.
}
\end{figure}

The coordinate domain is chosen in the form of a unit square $\Omega = [0, 1]\times[0, 1]$ in the two-dimensional case and \newtext{as} a unit cube $\Omega = [0, 1]\times[0, 1]\times[0, 1]$ in the three-dimensional case. Periodic boundary conditions are chosen. The exact analytical solution of the problem represents the process of simple advection of a sine wave, which can be expressed by a function $\rho = \rho_{0}(\mathbf{r} - \mathbf{v}_{\infty}t)$ for the density, constant pressure $p = p_{\infty}$ and constant flow velocity $(u, v, w) = (u_{\infty}, v_{\infty}, w_{\infty})$. With the selected initial conditions, coordinate domain and periodic boundary conditions, the sine wave returns to its original coordinate position, and the exact solution returns to the initial conditions, after a period of time $\Delta t = 1$. Therefore the final time $t_\mathrm{final} = 1.0$ is chosen. The exact solutions are shown in Figure~\ref{fig:sine_wave_adv_2d_dg_9_2x2} (left) in the two-dimensional case and Figure~\ref{fig:sine_wave_adv_3d_dg_9_2x2} (left) in the three-dimensional case.


\begin{table*}[h!]
\begin{center}
\caption{\label{tab:conv_orders_pde_2d}
$L_{1}$, $L_{2}$ and $L_{\infty}$ norms of errors $\epsilon$, by density $\rho$,
and convergence orders $p$ for ADER-DG-$\mathbb{P}_{N}$ method with a posteriori ADER-WENO2 finite-volume limiter, 
for two-dimensional sine wave advection problem;
$p_\mathrm{theor.} = N+1$ is the theoretical accuracy order of the ADER-DG-$\mathbb{P}_{N}$ method.
}
\begin{tabular}{|r|r|ccc|ccc|l|}
\hline
& cells 
& $\quad \epsilon_{L_{1}} \quad$ & $\quad \epsilon_{L_{2}} \quad$ & $\quad \epsilon_{L_{\infty}} \quad$
& $\quad p_{L_{1}} \quad$ & $\quad p_{L_{2}} \quad$ & $\quad p_{L_{\infty}} \quad$ & theor. \\
\hline
DG-$\mathbb{P}_{1}$		&	$10^{2}$	&	5.48E$-$03	&	7.33E$-$03	&	3.29E$-$02	&	--		&	--		&	--		&	2\\
						&	$15^{2}$	&	2.44E$-$03	&	3.26E$-$03	&	1.46E$-$02	&	1.99	&	1.99	&	2.00	&	\\
						&	$20^{2}$	&	1.37E$-$03	&	1.84E$-$03	&	8.16E$-$03	&	2.01	&	2.00	&	2.02	&	\\
						&	$25^{2}$	&	8.78E$-$04	&	1.18E$-$03	&	5.25E$-$03	&	2.00	&	2.00	&	1.98	&	\\\hline
DG-$\mathbb{P}_{2}$		&	$10^{2}$	&	2.87E$-$04	&	3.89E$-$04	&	1.98E$-$03	&	--		&	--		&	--		&	3\\
						&	$15^{2}$	&	8.44E$-$05	&	1.16E$-$04	&	6.04E$-$04	&	3.02	&	2.99	&	2.93	&	\\
						&	$20^{2}$	&	3.55E$-$05	&	4.88E$-$05	&	2.54E$-$04	&	3.01	&	3.00	&	3.02	&	\\
						&	$25^{2}$	&	1.81E$-$05	&	2.50E$-$05	&	1.31E$-$04	&	3.00	&	3.00	&	2.97	&	\\\hline
DG-$\mathbb{P}_{3}$		&	$10^{2}$	&	1.13E$-$05	&	1.54E$-$05	&	9.07E$-$05	&	--		&	--		&	--		&	4\\
						&	$15^{2}$	&	2.22E$-$06	&	3.05E$-$06	&	1.79E$-$05	&	4.00	&	4.00	&	3.99	&	\\
						&	$20^{2}$	&	7.02E$-$07	&	9.66E$-$07	&	5.65E$-$06	&	4.01	&	4.00	&	4.02	&	\\
						&	$25^{2}$	&	2.87E$-$07	&	3.96E$-$07	&	2.33E$-$06	&	4.00	&	4.00	&	3.97	&	\\\hline
DG-$\mathbb{P}_{4}$		&	$10^{2}$	&	3.58E$-$07	&	4.87E$-$07	&	3.04E$-$06	&	--		&	--		&	--		&	5\\
						&	$15^{2}$	&	4.67E$-$08	&	6.42E$-$08	&	4.15E$-$07	&	5.02	&	5.00	&	4.91	&	\\
						&	$20^{2}$	&	1.10E$-$08	&	1.52E$-$08	&	9.77E$-$08	&	5.01	&	5.00	&	5.03	&	\\
						&	$25^{2}$	&	3.61E$-$09	&	5.00E$-$09	&	3.23E$-$08	&	5.00	&	5.00	&	4.96	&	\\\hline
DG-$\mathbb{P}_{5}$		&	$ 5^{2}$	&	6.09E$-$07	&	8.12E$-$07	&	5.50E$-$06	&	--		&	--		&	--		&	6\\
						&	$10^{2}$	&	9.33E$-$09	&	1.28E$-$08	&	8.89E$-$08	&	6.03	&	5.99	&	5.95	&	\\
						&	$15^{2}$	&	8.18E$-$10	&	1.12E$-$09	&	7.80E$-$09	&	6.00	&	6.00	&	6.00	&	\\
						&	$20^{2}$	&	1.45E$-$10	&	2.00E$-$10	&	1.38E$-$09	&	6.01	&	6.00	&	6.02	&	\\\hline
DG-$\mathbb{P}_{6}$		&	$ 5^{2}$	&	2.75E$-$08	&	3.66E$-$08	&	2.69E$-$07	&	--		&	--		&	--		&	7\\
						&	$10^{2}$	&	2.12E$-$10	&	2.88E$-$10	&	2.05E$-$09	&	7.02	&	6.99	&	7.04	&	\\
						&	$15^{2}$	&	1.23E$-$11	&	1.69E$-$11	&	1.25E$-$10	&	7.03	&	7.00	&	6.90	&	\\
						&	$20^{2}$	&	1.63E$-$12	&	2.25E$-$12	&	1.65E$-$11	&	7.01	&	7.00	&	7.03	&	\\\hline
DG-$\mathbb{P}_{7}$		&	$ 5^{2}$	&	1.08E$-$09	&	1.44E$-$09	&	1.08E$-$08	&	--		&	--		&	--		&	8\\
						&	$10^{2}$	&	4.13E$-$12	&	5.66E$-$12	&	4.39E$-$11	&	8.03	&	7.99	&	7.95	&	\\
						&	$15^{2}$	&	1.61E$-$13	&	2.21E$-$13	&	1.71E$-$12	&	8.01	&	8.00	&	8.00	&	\\
						&	$20^{2}$	&	1.60E$-$14	&	2.22E$-$14	&	1.71E$-$13	&	8.01	&	8.00	&	8.00	&	\\\hline
DG-$\mathbb{P}_{8}$		&	$ 2^{2}$	&	1.49E$-$07	&	1.84E$-$07	&	7.36E$-$07	&	--		&	--		&	--		&	9\\
						&	$ 4^{2}$	&	2.77E$-$10	&	3.74E$-$10	&	2.29E$-$09	&	9.08	&	8.94	&	8.33	&	\\
						&	$ 6^{2}$	&	7.37E$-$12	&	9.79E$-$12	&	7.03E$-$11	&	8.94	&	8.98	&	8.59	&	\\
						&	$ 8^{2}$	&	5.39E$-$13	&	7.37E$-$13	&	5.58E$-$12	&	9.09	&	8.99	&	8.81	&	\\\hline
DG-$\mathbb{P}_{9}$		&	$ 2^{2}$	&	1.09E$-$08	&	1.45E$-$08	&	1.19E$-$07	&	--		&	--		&	--		&	10\\
						&	$ 4^{2}$	&	1.09E$-$11	&	1.47E$-$11	&	9.27E$-$11	&	9.97	&	9.95	&	10.33	&	\\
						&	$ 6^{2}$	&	1.88E$-$13	&	2.57E$-$13	&	2.14E$-$12	&	10.00	&	9.98	&	9.30	&	\\
						&	$ 8^{2}$	&	1.06E$-$14	&	1.45E$-$14	&	1.15E$-$13	&	10.01	&	9.99	&	10.17	&	\\
\hline
\end{tabular}
\end{center}
\end{table*}

\begin{table*}[h!]
\begin{center}
\caption{\label{tab:conv_orders_pde_3d}
$L_{1}$, $L_{2}$ and $L_{\infty}$ norms of errors $\epsilon$, by density $\rho$,
and convergence orders $p$ for ADER-DG-$\mathbb{P}_{N}$ method with a posteriori ADER-WENO2 finite-volume limiter, 
for three-dimensional sine wave advection problem;
$p_\mathrm{theor.} = N+1$ is the theoretical accuracy order of the ADER-DG-$\mathbb{P}_{N}$ method.
}
\begin{tabular}{|r|r|ccc|ccc|l|}
\hline
& cells 
& $\quad \epsilon_{L_{1}} \quad$ & $\quad \epsilon_{L_{2}} \quad$ & $\quad \epsilon_{L_{\infty}} \quad$
& $\quad p_{L_{1}} \quad$ & $\quad p_{L_{2}} \quad$ & $\quad p_{L_{\infty}} \quad$ & theor. \\
\hline
DG-$\mathbb{P}_{1}$		&	$10^{3}$	&	5.24E$-$02	&	5.83E$-$02	&	1.75E$-$01	&	--		&	--		&	--		&	2\\
						&	$15^{3}$	&	1.08E$-$02	&	1.30E$-$02	&	5.23E$-$02	&	3.89	&	3.71	&	2.98	&	\\
						&	$20^{3}$	&	4.62E$-$03	&	5.70E$-$03	&	2.99E$-$02	&	2.95	&	2.86	&	1.94	&	\\
						&	$25^{3}$	&	2.82E$-$03	&	3.53E$-$03	&	2.18E$-$02	&	2.21	&	2.15	&	1.42	&	\\\hline
DG-$\mathbb{P}_{2}$		&	$10^{3}$	&	7.69E$-$04	&	9.29E$-$04	&	6.07E$-$03	&	--		&	--		&	--		&	3\\
						&	$15^{3}$	&	2.22E$-$04	&	2.72E$-$04	&	1.92E$-$03	&	3.06	&	3.03	&	2.85	&	\\
						&	$20^{3}$	&	9.29E$-$05	&	1.15E$-$04	&	8.20E$-$04	&	3.03	&	3.01	&	2.95	&	\\
						&	$25^{3}$	&	4.74E$-$05	&	5.86E$-$05	&	4.20E$-$04	&	3.02	&	3.00	&	3.00	&	\\\hline
DG-$\mathbb{P}_{3}$		&	$10^{3}$	&	2.17E$-$05	&	2.78E$-$05	&	2.46E$-$04	&	--		&	--		&	--		&	4\\
						&	$15^{3}$	&	4.28E$-$06	&	5.48E$-$06	&	4.89E$-$05	&	4.00	&	4.01	&	3.99	&	\\
						&	$20^{3}$	&	1.35E$-$06	&	1.73E$-$06	&	1.56E$-$05	&	4.01	&	4.01	&	3.97	&	\\
						&	$25^{3}$	&	5.54E$-$07	&	7.08E$-$07	&	6.42E$-$06	&	4.00	&	4.00	&	3.97	&	\\\hline
DG-$\mathbb{P}_{4}$		&	$10^{3}$	&	7.69E$-$07	&	9.96E$-$07	&	8.96E$-$06	&	--		&	--		&	--		&	5\\
						&	$15^{3}$	&	1.03E$-$07	&	1.34E$-$07	&	1.26E$-$06	&	4.95	&	4.95	&	4.84	&	\\
						&	$20^{3}$	&	2.48E$-$08	&	3.21E$-$08	&	2.99E$-$07	&	4.97	&	4.97	&	4.99	&	\\
						&	$25^{3}$	&	8.16E$-$09	&	1.06E$-$08	&	9.68E$-$08	&	4.97	&	4.98	&	5.05	&	\\\hline
DG-$\mathbb{P}_{5}$		&	$ 5^{3}$	&	1.11E$-$06	&	1.45E$-$06	&	1.36E$-$05	&	--		&	--		&	--		&	6\\
						&	$10^{3}$	&	1.67E$-$08	&	2.19E$-$08	&	2.26E$-$07	&	6.05	&	6.05	&	5.90	&	\\
						&	$15^{3}$	&	1.46E$-$09	&	1.92E$-$09	&	1.97E$-$08	&	6.01	&	6.01	&	6.02	&	\\
						&	$20^{3}$	&	2.60E$-$10	&	3.41E$-$10	&	3.52E$-$09	&	6.00	&	6.00	&	5.99	&	\\\hline
DG-$\mathbb{P}_{6}$		&	$ 5^{3}$	&	5.24E$-$08	&	6.83E$-$08	&	6.55E$-$07	&	--		&	--		&	--		&	7\\
						&	$10^{3}$	&	4.36E$-$10	&	5.74E$-$10	&	5.37E$-$09	&	6.91	&	6.90	&	6.93	&	\\
						&	$15^{3}$	&	2.59E$-$11	&	3.42E$-$11	&	3.39E$-$10	&	6.96	&	6.96	&	6.82	&	\\
						&	$20^{3}$	&	3.49E$-$12	&	4.59E$-$12	&	4.55E$-$11	&	6.98	&	6.98	&	6.98	&	\\\hline
DG-$\mathbb{P}_{7}$		&	$ 5^{3}$	&	1.90E$-$09	&	2.52E$-$09	&	2.43E$-$08	&	--		&	--		&	--		&	8\\
						&	$10^{3}$	&	7.14E$-$12	&	9.50E$-$12	&	1.01E$-$10	&	8.05	&	8.05	&	7.90	&	\\
						&	$15^{3}$	&	2.78E$-$13	&	3.70E$-$13	&	3.97E$-$12	&	8.00	&	8.01	&	7.99	&	\\
						&	$20^{3}$	&	2.74E$-$14	&	3.72E$-$14	&	4.45E$-$13	&	8.06	&	7.99	&	7.61	&	\\\hline
DG-$\mathbb{P}_{8}$		&	$ 2^{3}$	&	2.27E$-$07	&	2.94E$-$07	&	2.17E$-$06	&	--		&	--		&	--		&	9\\
						&	$ 4^{3}$	&	5.08E$-$10	&	6.67E$-$10	&	5.01E$-$09	&	8.81	&	8.78	&	8.76	&	\\
						&	$ 6^{3}$	&	1.39E$-$11	&	1.84E$-$11	&	1.58E$-$10	&	8.88	&	8.86	&	8.52	&	\\
						&	$ 8^{3}$	&	1.07E$-$12	&	1.42E$-$12	&	1.28E$-$11	&	8.91	&	8.90	&	8.73	&	\\\hline
DG-$\mathbb{P}_{9}$		&	$ 2^{3}$	&	2.20E$-$08	&	2.87E$-$08	&	2.56E$-$07	&	--		&	--		&	--		&	10\\
						&	$ 4^{3}$	&	1.98E$-$11	&	2.63E$-$11	&	1.89E$-$10	&	10.12	&	10.09	&	10.40	&	\\
						&	$ 6^{3}$	&	3.30E$-$13	&	4.40E$-$13	&	4.36E$-$12	&	10.10	&	10.08	&	9.30	&	\\
						&	$ 8^{3}$	&	1.91E$-$14	&	2.53E$-$14	&	2.73E$-$13	&	9.91	&	9.93	&	9.62	&	\\
\hline
\end{tabular}
\end{center}
\end{table*}

Figure~\ref{fig:sine_wave_adv_2d_dg_9_2x2} and Figure~\ref{fig:sine_wave_adv_3d_dg_9_2x2} shows numerical solution to this problem in the two-dimensional and the three-dimensional cases obtained using the ADER-DG-$\mathbb{P}_{9}$ method on $2 \times 2$ and $2 \times 2 \times 2$ meshes, respectively. The presented results demonstrate that the numerical solution differs slightly from the exact solution visually, while in the three-dimensional case these differences are practically not visible. However, when the $2 \times 2$ size of the spatial mesh is taken into account, this result demonstrates the unusually high accuracy of the ADER-DG-$\mathbb{P}_{N}$ method.

The error $\epsilon$ of the numerical solution is calculated in three functional norms $L_{1}$, $L_{2}$, $L_{\infty}$ for the density $\rho$:
\begin{equation}\label{eq:epsilons}
\begin{split}
\epsilon_{L_{1}} = \int\limits_{\Omega} \left|\rho(\mathbf{r}, t_\mathrm{final}) - \rho_\mathrm{exact}(\mathbf{r})\right| dV;\quad
\epsilon_{L_{2}}^{2} = \int\limits_{\Omega} \left[\rho(\mathbf{r}, t_\mathrm{final}) - \rho_\mathrm{exact}(\mathbf{r})\right]^{2} dV;\quad
\epsilon_{L_{\infty}} = \operatorname{ess}\sup\limits_{\hspace{-5mm}\mathbf{r}\in\Omega} 
						 \left|\rho(\mathbf{r}, t_\mathrm{final}) - \rho_\mathrm{exact}(\mathbf{r})\right|;
\end{split}
\end{equation}
where the integrals are calculated as the sum of the integrals for each finite-element cell $\Omega_{k}$ over the function of the DG representation of the solution $\mathbf{u}$, and the calculation is carried out using the Gauss-Legendre quadrature formula based on polynomials of degree $25$ for each coordinate direction; the calculation of the supremum to determine the norm of error $\epsilon_{L_{\infty}}$ is performed using a finite-volume subcell representation of the solution. The errors $\epsilon$ are obtained for a set of mesh coordinate steps $h$, from which the empirical values of the convergence orders $p$ are calculated: $\epsilon \sim h^{p}$, therefore $p = \ln(\epsilon(h_{1})/\epsilon(h_{2}))/\ln(h_{1}/h_{2})$.

The calculated empirical convergence orders $p_{1}$, $p_{2}$, $p_{\infty}$ for errors $\epsilon_{L_{1}}$, $\epsilon_{L_{2}}$, $\epsilon_{L_{\infty}}$ for ADER-DG-$\mathbb{P}_{N}$ method with a posteriori limitation of the solution by a ADER-WENO2 finite-volume limiter are presented in Table~\ref{tab:conv_orders_pde_2d} for two-dimensional problem and in Table~\ref{tab:conv_orders_pde_3d} for three-dimensional problem. The results are obtained for degrees $N = 1, \ldots, 9$. It should be noted that the solution to this test problem is smooth, and the limiter is not called in all cases of the solution obtained from the use of a limiter --- the admissibility criteria are not activated. The expected theoretical values of the convergence orders $p_\mathrm{theor.} = N+1$~\cite{ader_dg_dev_1} are presented in Tables~\ref{tab:conv_orders_pde_2d} and~\ref{tab:conv_orders_pde_3d} for comparison. The presented results show that in all cases of orders of polynomials $N \leqslant 9$ in the DG-representation, there is a good correspondence between the empirical $p$ and theoretical $p_\mathrm{theor.}$ convergence orders. It should be noted, in order to obtain correct values of empirical orders of convergence in the case of large degrees $N = 8, 9$, a special choice of the mesh coordinate step $h$ is necessary --- in the region of small $h$ there is a significant increase in round-off errors, which leads to an increase in error $\epsilon$. 

The ADER-DG-$\mathbb{P}_{N}$ numerical method demonstrates very high accuracy of the numerical solution even on very coarse meshes, which is especially well observed for polynomial degrees $N = 7$-$9$. In the case of ADER-DG-$\mathbb{P}_{7}$ method, the round-off error of double-precision floating-point numbers is practically achieved on $15^{2}$-$20^{2}$ and $15^{3}$-$20^{3}$ meshes for two-dimensional and three-dimensional problems, respectively, and in the case of ADER-DG-$\mathbb{P}_{9}$ method, such a low error of the numerical solution is achieved already on $6^{2}$-$8^{2}$ and $6^{3}$-$8^{3}$ meshes.

The presented results on the convergence orders for ADER-DG-$\mathbb{P}_{N}$ method with a posteriori limitation of the solution by finite-volume limiter are in good agreement with the results of the basic works~\cite{ader_dg_dev_1, ader_dg_ideal_flows}. This allows us to conclude that the software implementation of the ADER-DG-$\mathbb{P}_{N}$ method is correct.

\paragraph{ODE system problem}
It should be noted separately that the numerical method applied to initial value problems (IVP) for ordinary differential equation systems (ODE systems) demonstrates superconvergence with convergence orders $2N+1$, which is obtained when considering the solution at time nodes $\mathbf{u}^{n}$ (\ref{eq:dg_repr}). This type of superconvergence for DG methods for solving ODE systems is well known~\cite{Delfour_1981, Delfour_1986} and has been studied in many works~\cite{Baccouch_dg_ivp_ode_1, Baccouch_dg_ivp_ode_2, Baccouch_dg_ivp_ode_3, Baccouch_dg_ivp_ode_4, Baccouch_dg_ivp_ode_5, Baccouch_dg_ivp_ode_6}. In work~\cite{popov_j_sci_comp_2023}, the superconvergence with convergence orders $2N+1$ was shown specifically for the ADER-DG method as applied to ODE systems. However, the convergence of the discrete space-time solution $\mathbf{q}$ (\ref{eq:lst_dg_repr}) (in the case of ODE systems, this is already a local discrete \textit{time} solution, because the sought function depends only on one argument with respect to the ODE system problem) demonstrates only the convergence order $N+1$, which is typical for the ADER-DG method when solving PDE systems. 

In this work, the convergence of the ADER-DG-$\mathbb{P}_{N}$ method is also investigated in solving ``ODE systems'', which are PDE systems from the point of view of the formula statement of the problem, and ODE systems taking into account the coordinate dependence of the initial conditions and the selected boundary conditions. In the selected problem, the initial conditions are completely free of dependence on space variables, making the coordinate derivatives of the flux terms zero, and the boundary conditions are chosen to be consistent with the selected initial conditions. As a result of this choice of initial and boundary conditions for the PDE system, the dynamic evolution of the solution is determined only by the source term $\mathbf{S}$, which was also chosen to be independent of the coordinates.

\begin{table*}[h!]
\begin{center}
\caption{\label{tab:conv_orders_global_ode}
Global $L_{1}$, $L_{2}$ and $L_{\infty}$ norms of errors $\epsilon$
and global convergence orders $p$ for ADER-DG-$\mathbb{P}_{N}$ method for IVP ODE system problem;
$p_\mathrm{theor.} = 2N+1$ is the theoretical global accuracy order of the ADER-DG-$\mathbb{P}_{N}$ method for IVP ODE system problem.
}\vspace{2mm}
\begin{tabular}{|r|c|ccc|ccc|l|}
\hline
& time steps 
& $\quad \epsilon_{L_{1}} \quad$ & $\quad \epsilon_{L_{2}} \quad$ & $\quad \epsilon_{L_{\infty}} \quad$
& $\quad p_{L_{1}} \quad$ & $\quad p_{L_{2}} \quad$ & $\quad p_{L_{\infty}} \quad$ & theor. \\
\hline
DG-$\mathbb{P}_{1}$	& $10$	&	1.56E$+$00	&	5.15E$-$01	&	2.53E$-$01		&	--	&	--	&	--			&	$3$	\\
	& $20$	&	2.35E$-$01	&	7.81E$-$02	&	4.06E$-$02		&	$2.73$	&	$2.72$	&	$2.64$			&		\\
	& $30$	&	7.16E$-$02	&	2.37E$-$02	&	1.25E$-$02		&	$2.93$	&	$2.94$	&	$2.90$			&		\\
	& $40$	&	3.05E$-$02	&	1.01E$-$02	&	5.34E$-$03		&	$2.97$	&	$2.98$	&	$2.96$			&		\\
	& $50$	&	1.56E$-$02	&	5.15E$-$03	&	2.75E$-$03		&	$2.99$	&	$2.99$	&	$2.98$			&		\\\hline
DG-$\mathbb{P}_{2}$	& $10$	&	2.86E$-$02	&	9.60E$-$03	&	4.95E$-$03		&	--	&	--	&	--			&	$5$	\\
	& $20$	&	9.56E$-$04	&	3.18E$-$04	&	1.67E$-$04		&	$4.90$	&	$4.92$	&	$4.89$			&		\\
	& $30$	&	1.27E$-$04	&	4.19E$-$05	&	2.23E$-$05		&	$4.98$	&	$5.00$	&	$4.97$			&		\\
	& $40$	&	3.03E$-$05	&	9.98E$-$06	&	5.31E$-$06		&	$4.98$	&	$4.99$	&	$4.98$			&		\\
	& $50$	&	9.91E$-$06	&	3.26E$-$06	&	1.74E$-$06		&	$5.01$	&	$5.01$	&	$4.99$			&		\\\hline
DG-$\mathbb{P}_{3}$	& $10$	&	2.35E$-$04	&	7.92E$-$05	&	4.13E$-$05		&	--	&	--	&	--			&	$7$	\\
	& $20$	&	1.94E$-$06	&	6.44E$-$07	&	3.39E$-$07		&	$6.92$	&	$6.94$	&	$6.93$			&		\\
	& $30$	&	1.14E$-$07	&	3.75E$-$08	&	2.00E$-$08		&	$7.00$	&	$7.01$	&	$6.98$			&		\\
	& $40$	&	1.53E$-$08	&	5.03E$-$09	&	2.68E$-$09		&	$6.97$	&	$6.99$	&	$6.99$			&		\\
	& $50$	&	3.19E$-$09	&	1.05E$-$09	&	5.63E$-$10		&	$7.01$	&	$7.01$	&	$6.99$			&		\\\hline
DG-$\mathbb{P}_{4}$	& $10$	&	1.16E$-$06	&	3.93E$-$07	&	2.05E$-$07		&	--	&	--	&	--			&	$9$	\\
	& $20$	&	2.37E$-$09	&	7.87E$-$10	&	4.14E$-$10		&	$8.94$	&	$8.96$	&	$8.95$			&		\\
	& $30$	&	6.16E$-$11	&	2.04E$-$11	&	1.09E$-$11		&	$9.00$	&	$9.01$	&	$8.98$			&		\\
	& $40$	&	4.65E$-$12	&	1.53E$-$12	&	8.17E$-$13		&	$8.98$	&	$8.99$	&	$8.99$			&		\\
	& $50$	&	6.23E$-$13	&	2.05E$-$13	&	1.10E$-$13		&	$9.01$	&	$9.02$	&	$9.00$			&		\\\hline
DG-$\mathbb{P}_{5}$	& $3$	&	1.51E$-$03	&	5.35E$-$04	&	2.53E$-$04		&	--	&	--	&	--			&	$11$	\\
	& $4$	&	8.19E$-$05	&	2.83E$-$05	&	1.30E$-$05		&	$10.13$	&	$10.22$	&	$10.31$			&		\\
	& $5$	&	7.09E$-$06	&	2.45E$-$06	&	1.23E$-$06		&	$10.97$	&	$10.97$	&	$10.59$			&		\\
	& $6$	&	1.01E$-$06	&	3.49E$-$07	&	1.73E$-$07		&	$10.67$	&	$10.68$	&	$10.73$			&		\\
	& $7$	&	1.93E$-$07	&	6.68E$-$08	&	3.27E$-$08		&	$10.78$	&	$10.74$	&	$10.81$			&		\\\hline
DG-$\mathbb{P}_{6}$	& $3$	&	4.14E$-$05	&	1.48E$-$05	&	7.09E$-$06		&	--	&	--	&	--			&	$13$	\\
	& $4$	&	1.25E$-$06	&	4.31E$-$07	&	1.98E$-$07		&	$12.17$	&	$12.29$	&	$12.43$			&		\\
	& $5$	&	6.73E$-$08	&	2.33E$-$08	&	1.18E$-$08		&	$13.08$	&	$13.07$	&	$12.66$			&		\\
	& $6$	&	6.69E$-$09	&	2.30E$-$09	&	1.14E$-$09		&	$12.66$	&	$12.69$	&	$12.78$			&		\\
	& $7$	&	9.24E$-$10	&	3.21E$-$10	&	1.58E$-$10		&	$12.84$	&	$12.79$	&	$12.84$			&		\\\hline
DG-$\mathbb{P}_{7}$	& $3$	&	8.37E$-$07	&	3.01E$-$07	&	1.46E$-$07		&	--	&	--	&	--			&	$15$	\\
	& $4$	&	1.41E$-$08	&	4.87E$-$09	&	2.24E$-$09		&	$14.20$	&	$14.34$	&	$14.52$			&		\\
	& $5$	&	4.78E$-$10	&	1.66E$-$10	&	8.41E$-$11		&	$15.16$	&	$15.15$	&	$14.71$			&		\\
	& $6$	&	3.30E$-$11	&	1.14E$-$11	&	5.65E$-$12		&	$14.66$	&	$14.70$	&	$14.81$			&		\\
	& $7$	&	3.33E$-$12	&	1.16E$-$12	&	5.71E$-$13		&	$14.89$	&	$14.83$	&	$14.87$			&		\\\hline
DG-$\mathbb{P}_{8}$	& $1$	&	1.53E$-$01	&	6.11E$-$02	&	2.44E$-$02		&	--	&	--	&	--			&	$17$	\\
	& $2$	&	1.00E$-$05	&	3.65E$-$06	&	1.60E$-$06		&	$13.90$	&	$14.03$	&	$13.90$			&		\\
	& $3$	&	1.32E$-$08	&	4.76E$-$09	&	2.31E$-$09		&	$16.35$	&	$16.38$	&	$16.13$			&		\\
	& $4$	&	1.23E$-$10	&	4.25E$-$11	&	1.96E$-$11		&	$16.26$	&	$16.40$	&	$16.58$			&		\\
	& $5$	&	2.65E$-$12	&	9.19E$-$13	&	4.66E$-$13		&	$17.19$	&	$17.18$	&	$16.75$			&		\\\hline
DG-$\mathbb{P}_{9}$	& $1$	&	2.49E$-$02	&	9.91E$-$03	&	3.96E$-$03		&	--	&	--	&	--			&	$19$	\\
	& $2$	&	2.95E$-$07	&	1.07E$-$07	&	4.69E$-$08		&	$16.36$	&	$16.50$	&	$16.36$			&		\\
	& $3$	&	1.66E$-$10	&	5.98E$-$11	&	2.89E$-$11		&	$18.45$	&	$18.48$	&	$18.23$			&		\\
	& $4$	&	8.55E$-$13	&	2.95E$-$13	&	1.36E$-$13		&	$18.32$	&	$18.46$	&	$18.63$			&		\\
\hline
\end{tabular}
\end{center}
\end{table*}

\begin{table*}[h!]
\begin{center}
\caption{\label{tab:conv_orders_local_ode}
Local $L_{1}$, $L_{2}$ and $L_{\infty}$ norms of errors $\varepsilon$
and local convergence orders $p$ for ADER-DG-$\mathbb{P}_{N}$ method for IVP ODE system problem;
$p_\mathrm{theor.} = N+1$ is the theoretical local accuracy order of the ADER-DG-$\mathbb{P}_{N}$ method for IVP ODE system problem.
}\vspace{2mm}
\begin{tabular}{|r|c|ccc|ccc|l|}
\hline
& time steps 
& $\quad \varepsilon_{L_{1}} \quad$ & $\quad \varepsilon_{L_{2}} \quad$ & $\quad \varepsilon_{L_{\infty}} \quad$
& $\quad p_{L_{1}} \quad$ & $\quad p_{L_{2}} \quad$ & $\quad p_{L_{\infty}} \quad$ & theor. \\
\hline
DG-$\mathbb{P}_{1}$	& $10$	&	1.87E$+$00	&	2.49E$+$00	&	3.09E$-$01		&	--	&	--	&	--			&	$2$	\\
	& $20$	&	3.55E$-$01	&	4.84E$-$01	&	6.04E$-$02		&	$2.40$	&	$2.37$	&	$2.36$			&		\\
	& $30$	&	1.38E$-$01	&	1.90E$-$01	&	2.78E$-$02		&	$2.34$	&	$2.31$	&	$1.91$			&		\\
	& $40$	&	7.21E$-$02	&	1.01E$-$01	&	1.60E$-$02		&	$2.24$	&	$2.20$	&	$1.92$			&		\\
	& $50$	&	4.43E$-$02	&	6.28E$-$02	&	1.03E$-$02		&	$2.18$	&	$2.12$	&	$1.95$			&		\\\hline
DG-$\mathbb{P}_{2}$	& $10$	&	1.09E$-$01	&	1.54E$-$01	&	3.08E$-$02		&	--	&	--	&	--			&	$3$	\\
	& $20$	&	1.24E$-$02	&	1.95E$-$02	&	4.07E$-$03		&	$3.13$	&	$2.98$	&	$2.92$			&		\\
	& $30$	&	3.64E$-$03	&	5.84E$-$03	&	1.22E$-$03		&	$3.03$	&	$2.98$	&	$2.98$			&		\\
	& $40$	&	1.53E$-$03	&	2.47E$-$03	&	5.11E$-$04		&	$3.02$	&	$2.99$	&	$3.02$			&		\\
	& $50$	&	7.82E$-$04	&	1.27E$-$03	&	2.64E$-$04		&	$3.00$	&	$2.98$	&	$2.96$			&		\\\hline
DG-$\mathbb{P}_{3}$	& $10$	&	7.75E$-$03	&	1.23E$-$02	&	2.89E$-$03		&	--	&	--	&	--			&	$4$	\\
	& $20$	&	4.84E$-$04	&	7.99E$-$04	&	1.84E$-$04		&	$4.00$	&	$3.95$	&	$3.97$			&		\\
	& $30$	&	9.53E$-$05	&	1.59E$-$04	&	3.65E$-$05		&	$4.01$	&	$3.98$	&	$3.99$			&		\\
	& $40$	&	3.00E$-$05	&	5.04E$-$05	&	1.15E$-$05		&	$4.02$	&	$4.00$	&	$4.03$			&		\\
	& $50$	&	1.23E$-$05	&	2.08E$-$05	&	4.74E$-$06		&	$3.98$	&	$3.97$	&	$3.96$			&		\\\hline
DG-$\mathbb{P}_{4}$	& $10$	&	4.87E$-$04	&	8.16E$-$04	&	2.02E$-$04		&	--	&	--	&	--			&	$5$	\\
	& $20$	&	1.53E$-$05	&	2.65E$-$05	&	6.43E$-$06		&	$4.99$	&	$4.95$	&	$4.97$			&		\\
	& $30$	&	2.01E$-$06	&	3.51E$-$06	&	8.50E$-$07		&	$5.01$	&	$4.98$	&	$4.99$			&		\\
	& $40$	&	4.75E$-$07	&	8.34E$-$07	&	2.00E$-$07		&	$5.02$	&	$5.00$	&	$5.03$			&		\\
	& $50$	&	1.56E$-$07	&	2.75E$-$07	&	6.62E$-$08		&	$4.98$	&	$4.97$	&	$4.95$			&		\\\hline
DG-$\mathbb{P}_{5}$	& $3$	&	3.32E$-$02	&	4.85E$-$02	&	1.27E$-$02		&	--	&	--	&	--			&	$6$	\\
	& $4$	&	6.60E$-$03	&	1.04E$-$02	&	2.53E$-$03		&	$5.61$	&	$5.34$	&	$5.63$			&		\\
	& $5$	&	1.63E$-$03	&	2.65E$-$03	&	6.95E$-$04		&	$6.28$	&	$6.14$	&	$5.78$			&		\\
	& $6$	&	5.51E$-$04	&	9.22E$-$04	&	2.39E$-$04		&	$5.94$	&	$5.79$	&	$5.85$			&		\\
	& $7$	&	2.18E$-$04	&	3.72E$-$04	&	9.63E$-$05		&	$6.00$	&	$5.89$	&	$5.90$			&		\\\hline
DG-$\mathbb{P}_{6}$	& $3$	&	5.15E$-$03	&	7.81E$-$03	&	2.14E$-$03		&	--	&	--	&	--			&	$7$	\\
	& $4$	&	7.76E$-$04	&	1.26E$-$03	&	3.13E$-$04		&	$6.58$	&	$6.35$	&	$6.69$			&		\\
	& $5$	&	1.53E$-$04	&	2.54E$-$04	&	6.82E$-$05		&	$7.29$	&	$7.17$	&	$6.82$			&		\\
	& $6$	&	4.32E$-$05	&	7.35E$-$05	&	1.95E$-$05		&	$6.92$	&	$6.80$	&	$6.88$			&		\\
	& $7$	&	1.47E$-$05	&	2.53E$-$05	&	6.70E$-$06		&	$7.01$	&	$6.91$	&	$6.92$			&		\\\hline
DG-$\mathbb{P}_{7}$	& $3$	&	6.95E$-$04	&	1.10E$-$03	&	3.09E$-$04		&	--	&	--	&	--			&	$8$	\\
	& $4$	&	7.88E$-$05	&	1.32E$-$04	&	3.33E$-$05		&	$7.57$	&	$7.37$	&	$7.74$			&		\\
	& $5$	&	1.23E$-$05	&	2.12E$-$05	&	5.78E$-$06		&	$8.31$	&	$8.20$	&	$7.85$			&		\\
	& $6$	&	2.91E$-$06	&	5.10E$-$06	&	1.37E$-$06		&	$7.92$	&	$7.81$	&	$7.90$			&		\\
	& $7$	&	8.47E$-$07	&	1.50E$-$06	&	4.03E$-$07		&	$8.01$	&	$7.92$	&	$7.93$			&		\\\hline
DG-$\mathbb{P}_{8}$	& $1$	&	8.40E$-$01	&	8.43E$-$01	&	1.99E$-$01		&	--	&	--	&	--			&	$9$	\\
	& $2$	&	3.00E$-$03	&	4.52E$-$03	&	1.24E$-$03		&	$8.13$	&	$7.54$	&	$7.32$			&		\\
	& $3$	&	8.21E$-$05	&	1.35E$-$04	&	3.89E$-$05		&	$8.87$	&	$8.66$	&	$8.54$			&		\\
	& $4$	&	6.95E$-$06	&	1.21E$-$05	&	3.11E$-$06		&	$8.58$	&	$8.39$	&	$8.78$			&		\\
	& $5$	&	8.71E$-$07	&	1.55E$-$06	&	4.31E$-$07		&	$9.31$	&	$9.21$	&	$8.87$			&		\\\hline
DG-$\mathbb{P}_{9}$	& $1$	&	2.70E$-$01	&	2.98E$-$01	&	8.27E$-$02		&	--	&	--	&	--			&	$10$	\\
	& $2$	&	4.99E$-$04	&	7.56E$-$04	&	2.14E$-$04		&	$9.08$	&	$8.62$	&	$8.60$			&		\\
	& $3$	&	9.00E$-$06	&	1.48E$-$05	&	4.36E$-$06		&	$9.90$	&	$9.70$	&	$9.60$			&		\\
	& $4$	&	5.74E$-$07	&	9.90E$-$07	&	2.60E$-$07		&	$9.57$	&	$9.40$	&	$9.80$			&		\\
\hline
\end{tabular}
\end{center}
\end{table*}

The original system of Euler equations (\ref{eq:system_of_equations}), extended by the system of two convection-reaction equations, taking into account the chosen reaction mechanism in a two-component medium, takes the following form:
\begin{equation}\label{eq:pdes_2_comps}
\frac{\partial}{\partial t}\left[
\begin{array}{c}
\rho\\
\rho\mathbf{v}\\
\varepsilon\\
\rho c_{1}\\
\rho c_{2}
\end{array}
\right] + \nabla\left[
\begin{array}{c}
\rho\mathbf{v}\\
\rho\mathbf{v}\otimes\mathbf{v} + p\mathbf{I}\\
(\varepsilon + p) \mathbf{v}\\
\rho c_{1}\mathbf{v}\\
\rho c_{2}\mathbf{v}
\end{array}
\right] = \left[
\begin{array}{c}
0\\
\mathbf{0}\\
0\\
+\rho c_{2}\\
-\omega^{2}\rho c_{1}
\end{array}
\right],
\end{equation}
where $c_{1}$ is the mass concentration of the reaction reagent $A$, $c_{2}$ is the mass concentration of the reaction reagent $B$, $[\rho c_{2}, -\omega^{2}\rho c_{1}]^{T}$ is the ``reaction rates'' (in real reacting flows, it is clear that such a reaction mechanism does not exist --- the equations allow negative concentrations $c_{1}$ and $c_{2}$ of components), $\omega$ is the constant.

The initial conditions are presented in the form of spatially uniform stationary gas flow with parameters $\rho_{\infty} = 1$, $(u_{\infty}, v_{\infty}, w_{\infty}) = (1, 1, 1)$, $p_{\infty} = 1$. The initial concentration values are chosen as follows: $c_{1} = 1$, $c_{2} = 0$. The periodic boundary conditions are chosen. The coordinate domain $\Omega = [0, 1]\times[0, 1]\times[0, 1]$. Taking into account such initial and boundary conditions, the problem (\ref{eq:pdes_2_comps}) is reduced to the following IVP for ODE system:
\begin{equation}\label{eq:harm_osc}
\begin{cases}
\cfrac{du_{1}}{dt} = u_{2};\qquad &u_{1}(0) = 1;\\
\cfrac{du_{2}}{dt} = -\omega^{2} u_{1};\qquad &u_{2}(0) = 0;
\end{cases}
\end{equation}
which is a system of equations of a one-dimensional linear harmonic oscillator with frequency $\omega$, where $u_{1} \equiv c_{1}$ and $u_{2} \equiv c_{2}$. The exact analytical solution of this problem is: $u_{1}^\mathrm{ex}(t) = \cos(\omega t)$, $u_{2}^\mathrm{ex}(t) = -\omega\sin(\omega t)$. In the calculations it is assumed that $\omega = 1$. The final time is chosen as $t_\mathrm{final} = 2\pi$. The spatial mesh is selected as a single cell. The values of the time step $\Delta t^{n}$ are not determined from the Courant condition, but they are explicitly specified by the amount of time steps, chosen uniformly in the range $[0, t_\mathrm{final}]$.

The convergence orders are calculated separately for the solution at the nodes $\mathbf{u}^{n}$, for which the phenomenon of superconvergence $p = 2N+1$ is expected, and for the local discrete solution $\mathbf{q}$. The errors for solution at time nodes $\epsilon$ and for the local discrete solution $\varepsilon$ are calculated based on the following expressions:
\begin{equation}\label{eq:epsilons_q}
\begin{split}
\epsilon_{L_{1}} = \sum_{n} \Delta t^{n} \int\limits_{\Omega} \left|\mathbf{u}^{n}(\boldsymbol{\xi}(\mathbf{r})) - \mathbf{u}_\mathrm{exact}(\mathbf{r}, t^{n})\right| dV;\qquad\,\;
&\varepsilon_{L_{1}} = \int\limits_{0}^{t_\mathrm{final}}\int\limits_{\Omega} \left|\mathbf{q}(\boldsymbol{\xi}(\mathbf{r}), \tau(t)) - \mathbf{u}_\mathrm{exact}(\mathbf{r}, t)\right| dt dV;\\
\epsilon_{L_{2}}^{2} = \sum_{n} \Delta t^{n} \int\limits_{\Omega} \left|\mathbf{u}^{n}(\boldsymbol{\xi}(\mathbf{r})) - \mathbf{u}_\mathrm{exact}(\mathbf{r}, t^{n})\right|^{2} dV;\qquad
&\varepsilon_{L_{2}}^{2} = \int\limits_{0}^{t_\mathrm{final}}\int\limits_{\Omega} 
							\left|\mathbf{q}(\boldsymbol{\xi}(\mathbf{r}), \tau(t)) - \mathbf{u}_\mathrm{exact}(\mathbf{r}, t)\right|^{2} dt dV;\\[3mm]
\epsilon_{L_{\infty}} = \max\limits_{n}\operatorname{ess}\sup\limits_{\hspace{-5mm}\mathbf{r}\in\Omega} 
						\left|\mathbf{u}^{n}(\boldsymbol{\xi}(\mathbf{r})) - \mathbf{u}_\mathrm{exact}(\mathbf{r}, t^{n})\right|;\qquad\,\,\,\,\,
&\varepsilon_{L_{\infty}} = \operatorname{ess}\hspace{-3mm}\sup\limits_{\hspace{-5mm}\mathbf{r},\, t\, \in\, \Omega\times[0, t_\mathrm{final}]} 
						 	 \left|\mathbf{q}(\boldsymbol{\xi}(\mathbf{r}), \tau(t)) - \mathbf{u}_\mathrm{exact}(\mathbf{r}, t)\right|;
\end{split}
\end{equation}
where the integrals are calculated as an integral by single finite-element cell $\Omega_{k}$ and as a sum for each time range $[t^{n}, t^{n+1}] \subseteq [0, t_\mathrm{final}]$ over the function of the DG representation of the solution $\mathbf{u}$, and the calculation is carried out using the Gauss-Legendre quadrature formula based on polynomials of degree $25$ for each coordinate direction and by time; the calculation of the supremum to determine the norm of error $\epsilon_{L_{\infty}}$ is performed using a finite-volume subcell representation of the solution for coordinate directions and obtaining the maximum value of $1000$ equally spaced nodes in each time range $[t^{n}, t^{n+1}]$. The operation $|\ldots|$ denotes taking the maximum absolute value of the components of a vector. The errors $\epsilon$ are obtained for a set of time steps amount with time step $\Delta t$, from which the empirical values of the convergence orders $p$ are calculated: $\epsilon \sim {\Delta t}^{p}$, therefore $p = \ln(\epsilon({\Delta t}_{1})/\epsilon({\Delta t}_{2}))/\ln({\Delta t}_{1}/{\Delta t}_{2})$.

The calculated empirical convergence orders $p_{1}$, $p_{2}$, $p_{\infty}$ by errors $\epsilon_{L_{1}}$, $\epsilon_{L_{2}}$, $\epsilon_{L_{\infty}}$ for solution at time nodes and by $\varepsilon_{L_{1}}$, $\varepsilon_{L_{2}}$, $\varepsilon_{L_{\infty}}$ for local solution are presented in Tables~\ref{tab:conv_orders_global_ode} and~\ref{tab:conv_orders_local_ode} for ADER-DG-$\mathbb{P}_{N}$ method. The results have been obtained for degrees $N = 1, \ldots, 9$. The expected theoretical values of the convergence orders $p_\mathrm{theor.} = N+1$ for solution at time nodes $\mathbf{u}^{n}$ and $p_\mathrm{theor.} = 2N+1$ for the local solution $\mathbf{q}$ are presented in Tables~\ref{tab:conv_orders_global_ode} and~\ref{tab:conv_orders_local_ode} for comparison. The presented results show that in all cases of orders of polynomials $N \leqslant 9$ in the DG-representation, there is a good correspondence between the empirical $p$ and theoretical $p_\mathrm{theor.}$ convergence orders. The numerical method demonstrates well the expected superconvergence with convergence order $2N+1$ for the solution at time nodes $\mathbf{u}^{n}$. It should be noted that in order to obtain correct values of empirical orders of convergence in the case of large degrees $N = 8, 9$, a special choice of the time step is necessary --- in the region of small $\Delta t$ there is a significant increase in round-off errors, which leads to an increase in errors $\epsilon$ and $\varepsilon$. For the local solution $\mathbf{q}$, the errors do not reach the scale of round-off errors for double-precision floating-point numbers, however, this is achieved for the solution at time nodes $\mathbf{u}^{n}$, which of course affected the local solution, for which the node solution acts as the initial condition at each time step.

The presented results on the convergence orders for ADER-DG-$\mathbb{P}_{N}$ method are in good agreement with the results~\cite{popov_j_sci_comp_2023}. The numerical method demonstrates well the expected superconvergence with convergence order $2N+1$ for the solution at time nodes $\mathbf{u}^{n}$. This allows us to conclude that the software implementation of the ADER-DG-$\mathbb{P}_{N}$ method is correct.

\subsection{Classical Riemann problems}
\label{sec:apps_cgd_problems:problems_1d}

Obtaining a correct solution to the problem of advection of a sine wave made it possible to determine the correctness of the software implementation of the ADER-DG-$\mathbb{P}_{N}$ method. However, numerical simulation of detonation waves requires clarification of the correctness of the full software implementation of the ADER-DG-$\mathbb{P}_{N}$ method with a posteriori sub-cell ADER-WENO finite-volume limiting for simulation of flows with discontinuities in the solution.

\begin{table}[h!]
\caption{%
Data for one-dimensional Riemann problem tests.
The parameter values $(\rho_{L}, u_{L}, p_{L})$ correspond to the state of the flow to the left of the discontinuity; 
the parameter values $(\rho_{R}, u_{R}, p_{R})$ correspond to the state of the flow to the right of the discontinuity.
}
\label{tab:classical_tests_1d}
\centering
\begin{tabular}{|c||c|c|c||c|c|c|}
\hline
Test & $\qquad \rho_{L} \qquad$ & $\qquad u_{L} \qquad$ & $\qquad p_{L} \qquad$ & 
	   $\qquad \rho_{R} \qquad$ & $\qquad u_{R} \qquad$ & $\qquad p_{R} \qquad$ \\
\hline
1 & $1.000$ & $0.000$ & $1.000$ & $0.125$ & $0.000$ & $0.100$ \\
\hline
2 & $0.445$ & $0.698$ & $3.528$ & $0.500$ & $0.000$ & $0.571$ \\
\hline
3 & $1.000$ & $+1.000$ & $1.000$ & $1.000$ & $-1.000$ & $1.000$ \\
\hline
4 & $1.000$ & $-1.000$ & $1.000$ & $1.000$ & $+1.000$ & $1.000$ \\
\hline
\end{tabular}
\end{table}

This subsection presents the results of calculating classical test cases based on exactly solvable classical one-dimensional Riemann problems~\cite{Toro_solvers_2009}. Verification and testing of the developed software implementation in this work is carried out on four gas-dynamic tests~\cite{Toro_solvers_2009}: the classical Sod and Lax problems 
\newtext{(Tests $1$ and $2$)}, and the problem with two shock waves \newtext{(Test $3$)}, the problem with two strong rarefaction waves \newtext{(Test $4$)}. The spatial domain of the flow is chosen as $\Omega = [0, 1]^{2}$. The initial discontinuity is located at the vertical line with coordinate $x_\mathrm{c} = 0.5$. The initial conditions in the Riemann problems is chosen in the following form:
\begin{equation}\label{eq:rp_1d_init}
(\rho, u, p)(x, y, t = 0) = \left\{
\begin{array}{ll}
(\rho_{L}, u_{L}, p_{L}), & \mathrm{if}\ x \leqslant 0.5; \\
(\rho_{R}, u_{R}, p_{R}), & \mathrm{if}\ x >\, 0.5;
\end{array}
\right.
\end{equation}
where the parameter values $(\rho_{L}, u_{L}, p_{L})$ and $(\rho_{R}, u_{R}, p_{R})$ correspond to the state of the flow to the left and \newtext{to the} right of the discontinuity. Data for the parameter values $(\rho, u, p)$ of these four Riemann problem tests are presented in Table~\ref{tab:classical_tests_1d}. The boundary conditions are chosen as free outflow conditions. The final time of the simulation is chosen as $t_\mathrm{final} = 0.15$ for all four tests. The adiabatic index $\gamma = 1.4$. The Courant number $\mathtt{CFL} = 0.4$.

\begin{figure*}[h!]
\centering
\includegraphics[width=0.245\textwidth]{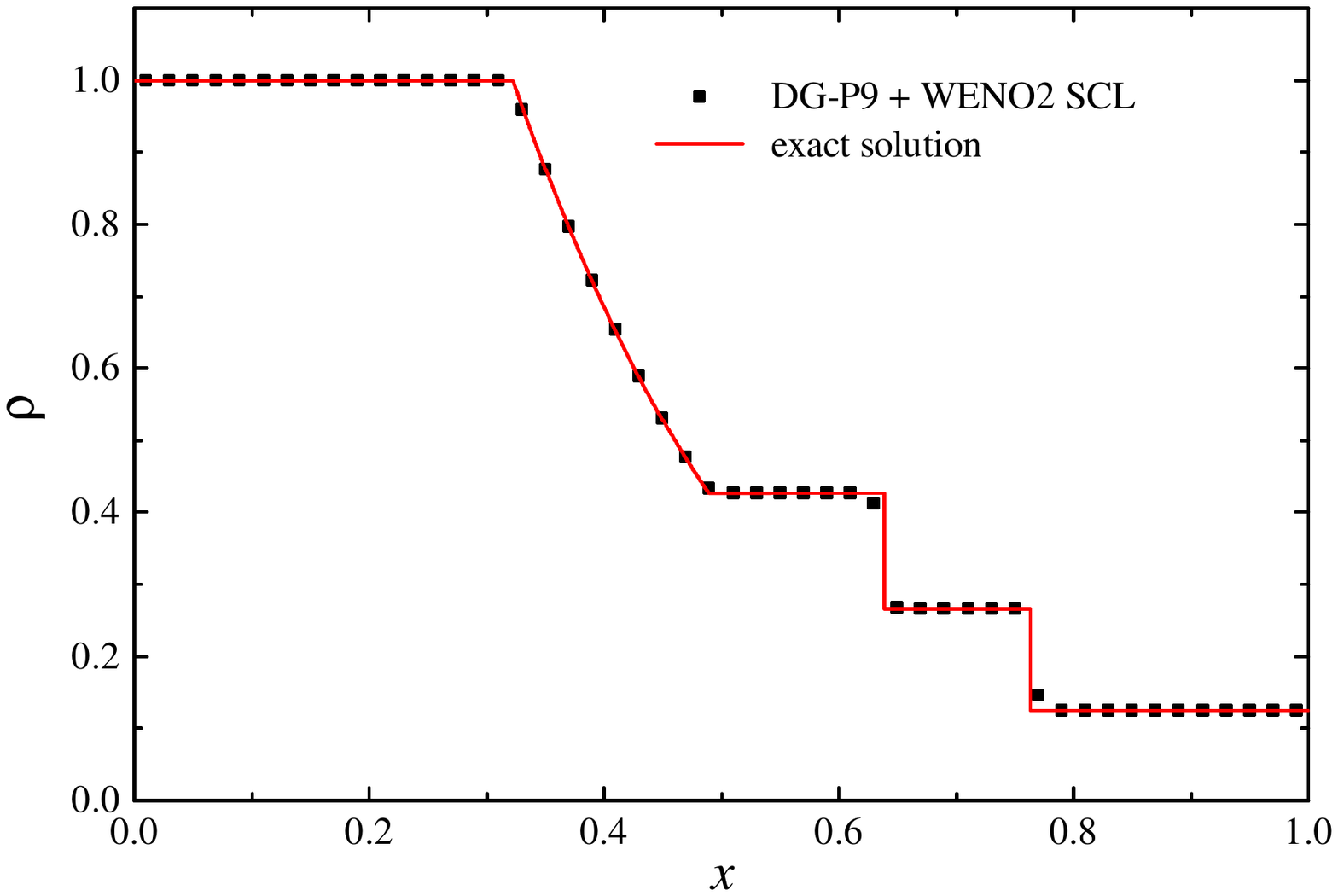}
\includegraphics[width=0.245\textwidth]{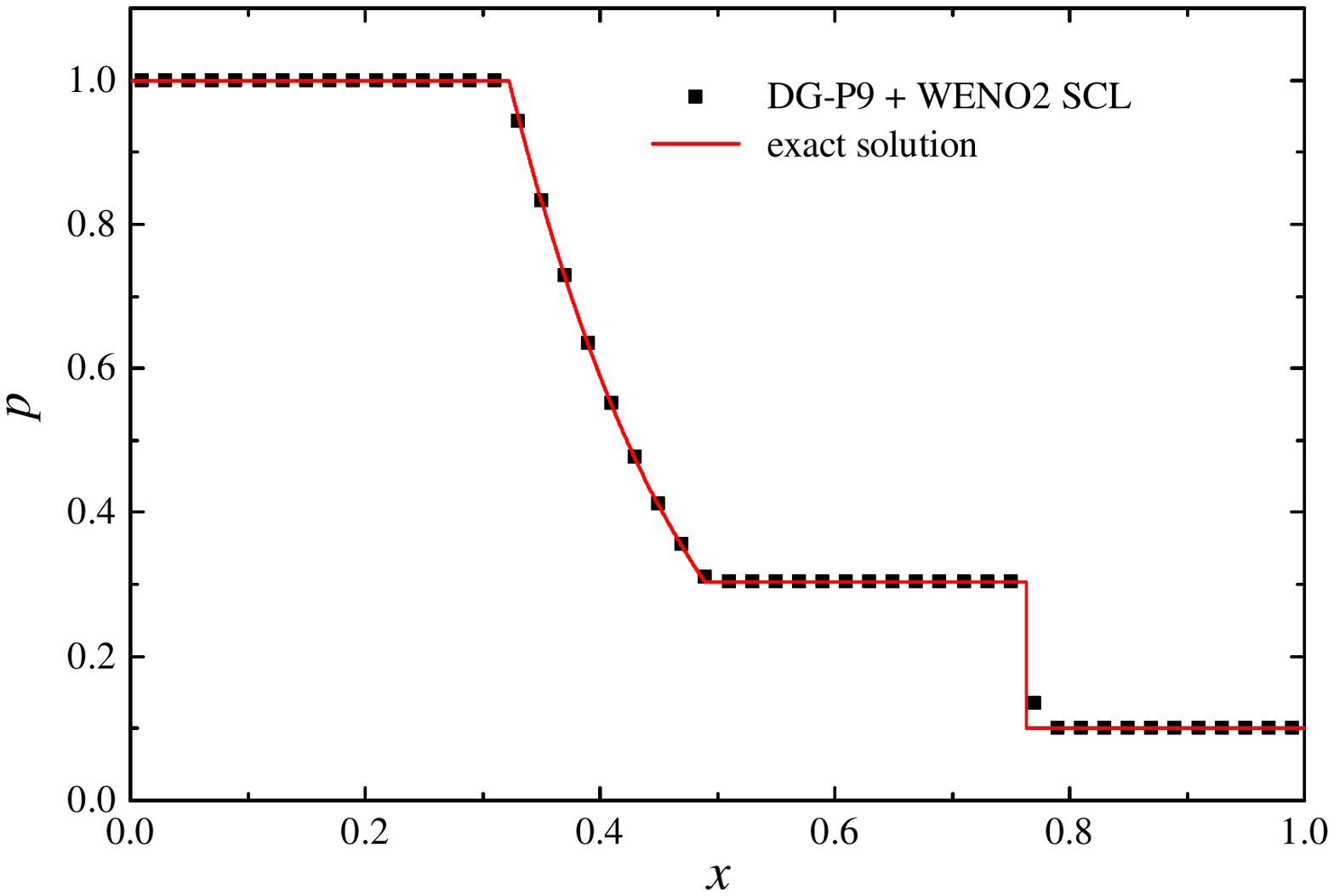}
\includegraphics[width=0.245\textwidth]{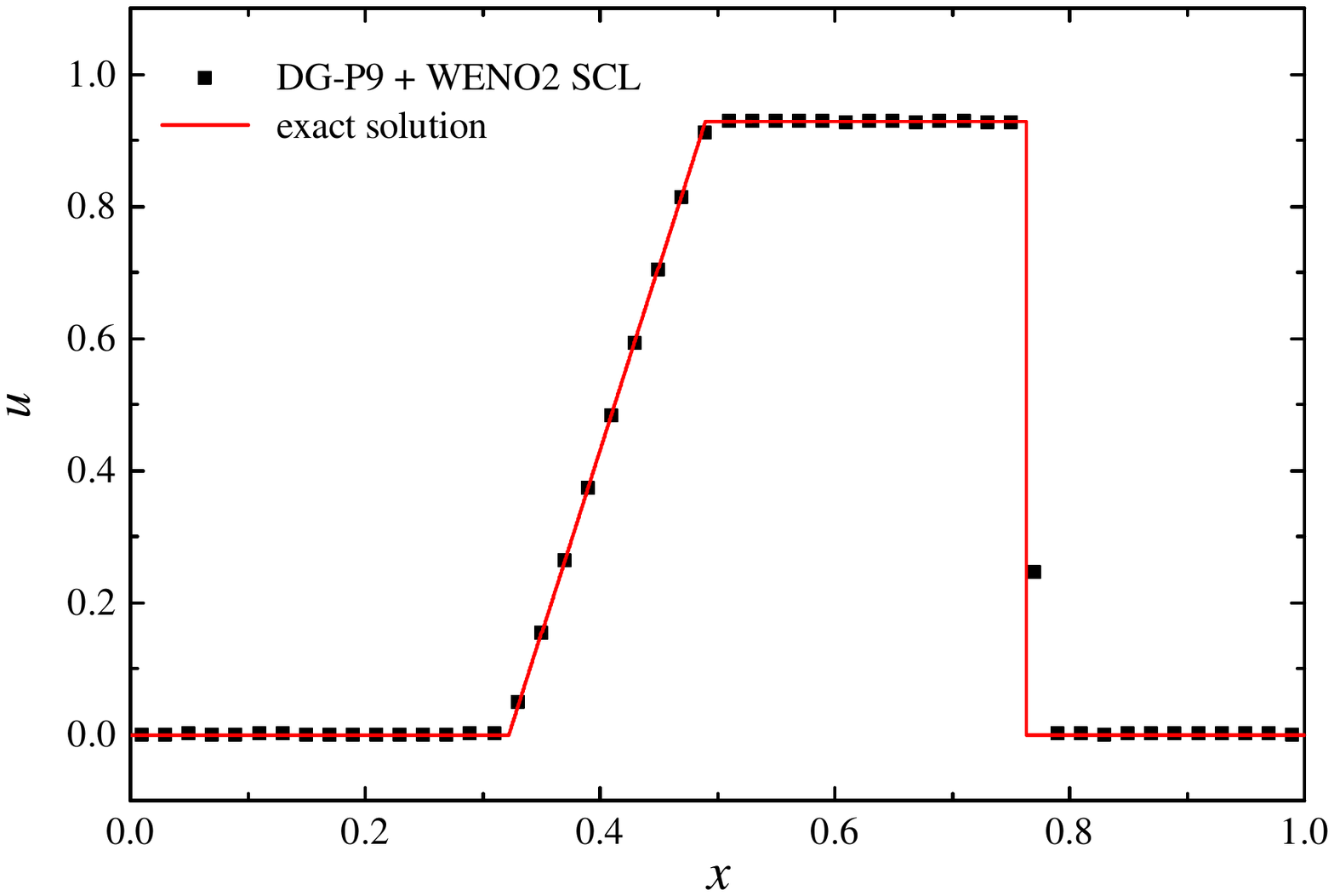}
\includegraphics[width=0.245\textwidth]{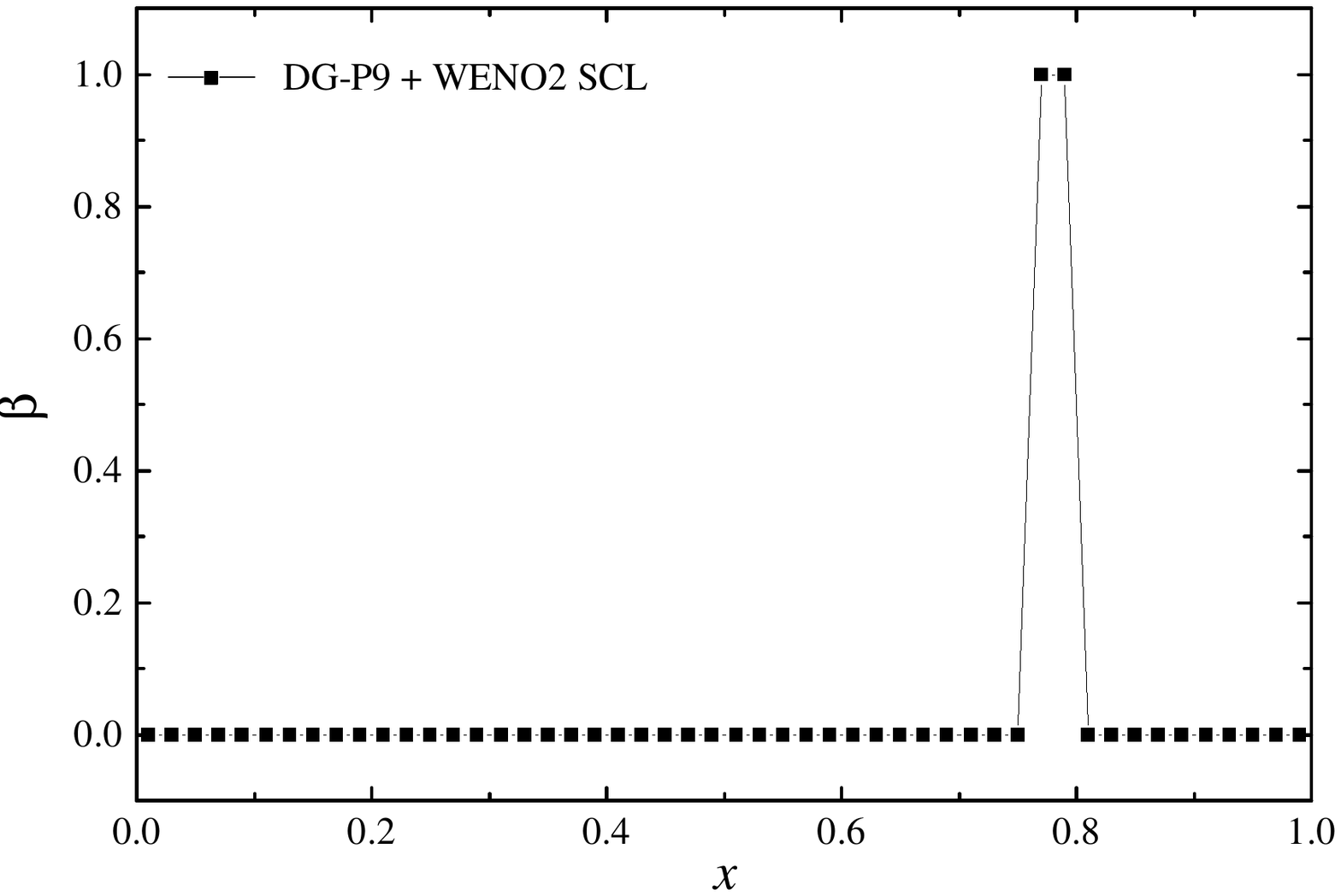}\\
\includegraphics[width=0.245\textwidth]{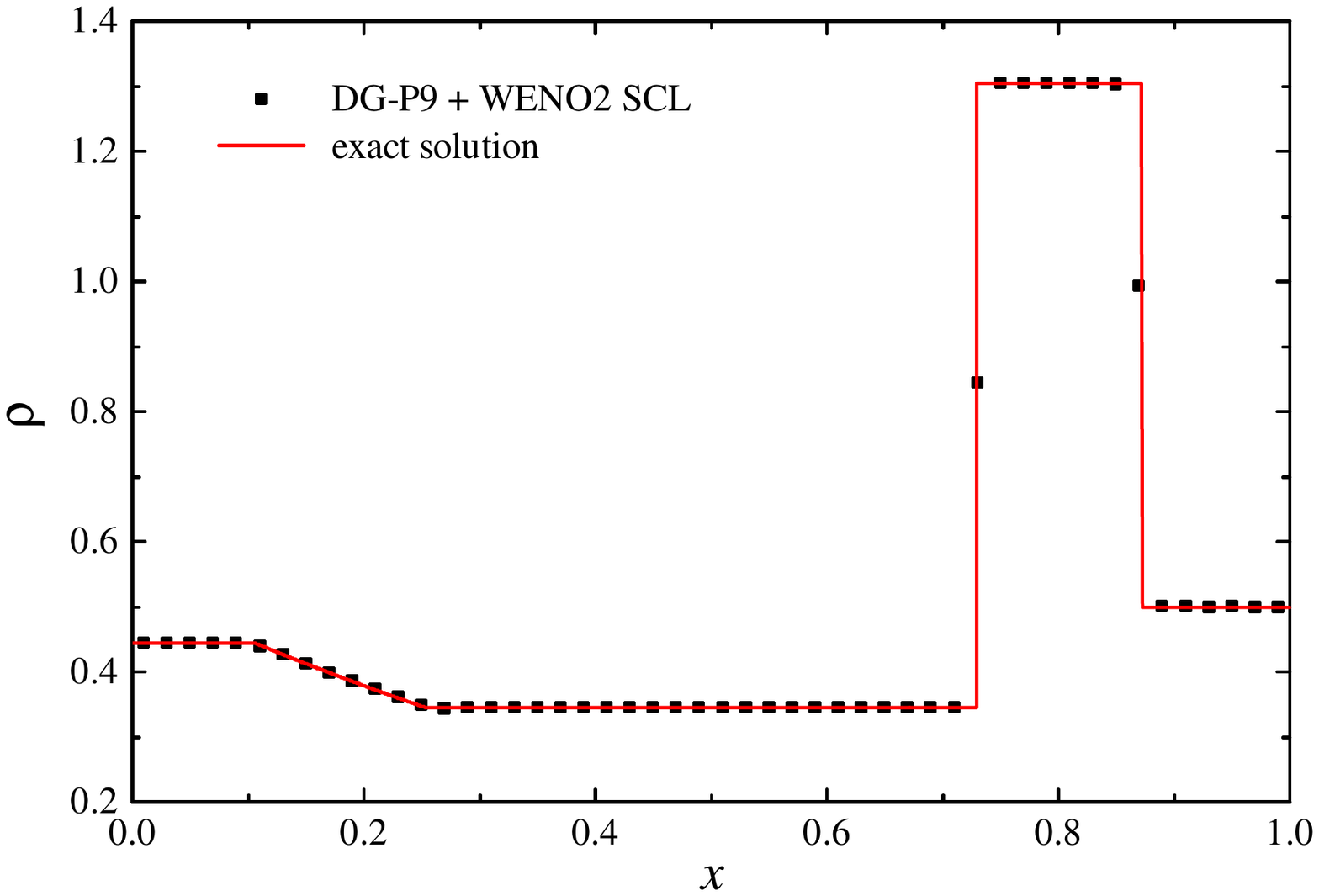}
\includegraphics[width=0.245\textwidth]{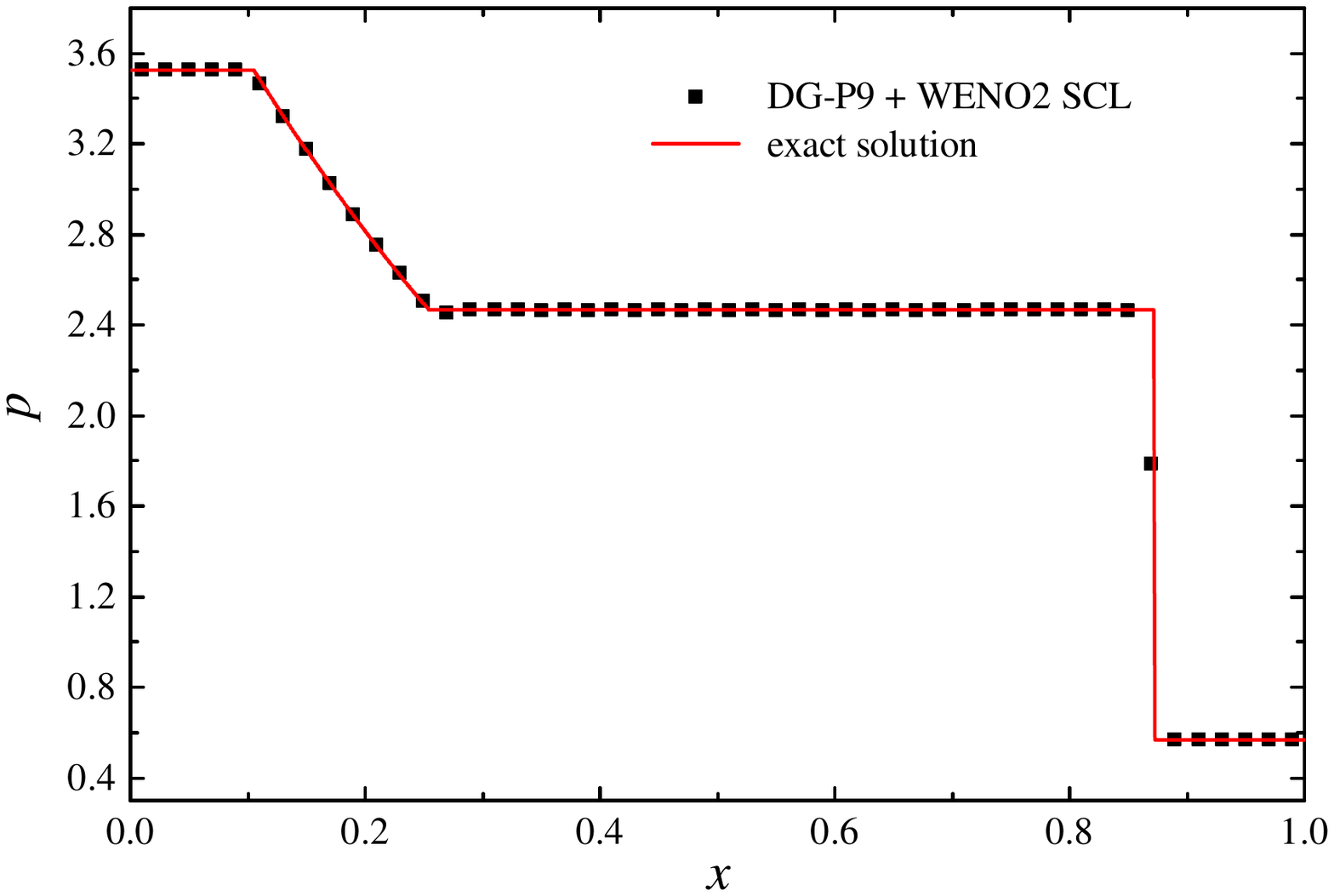}
\includegraphics[width=0.245\textwidth]{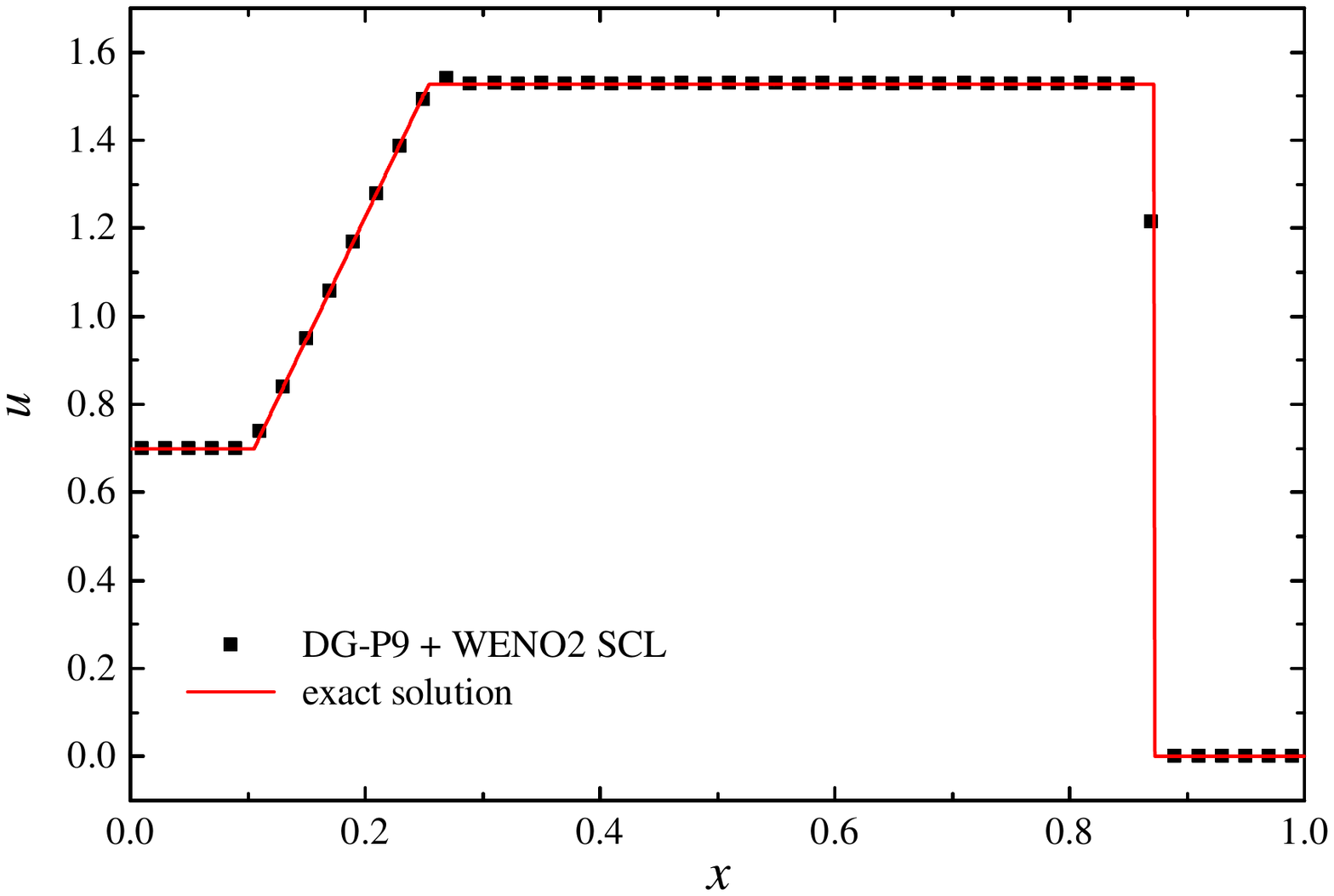}
\includegraphics[width=0.245\textwidth]{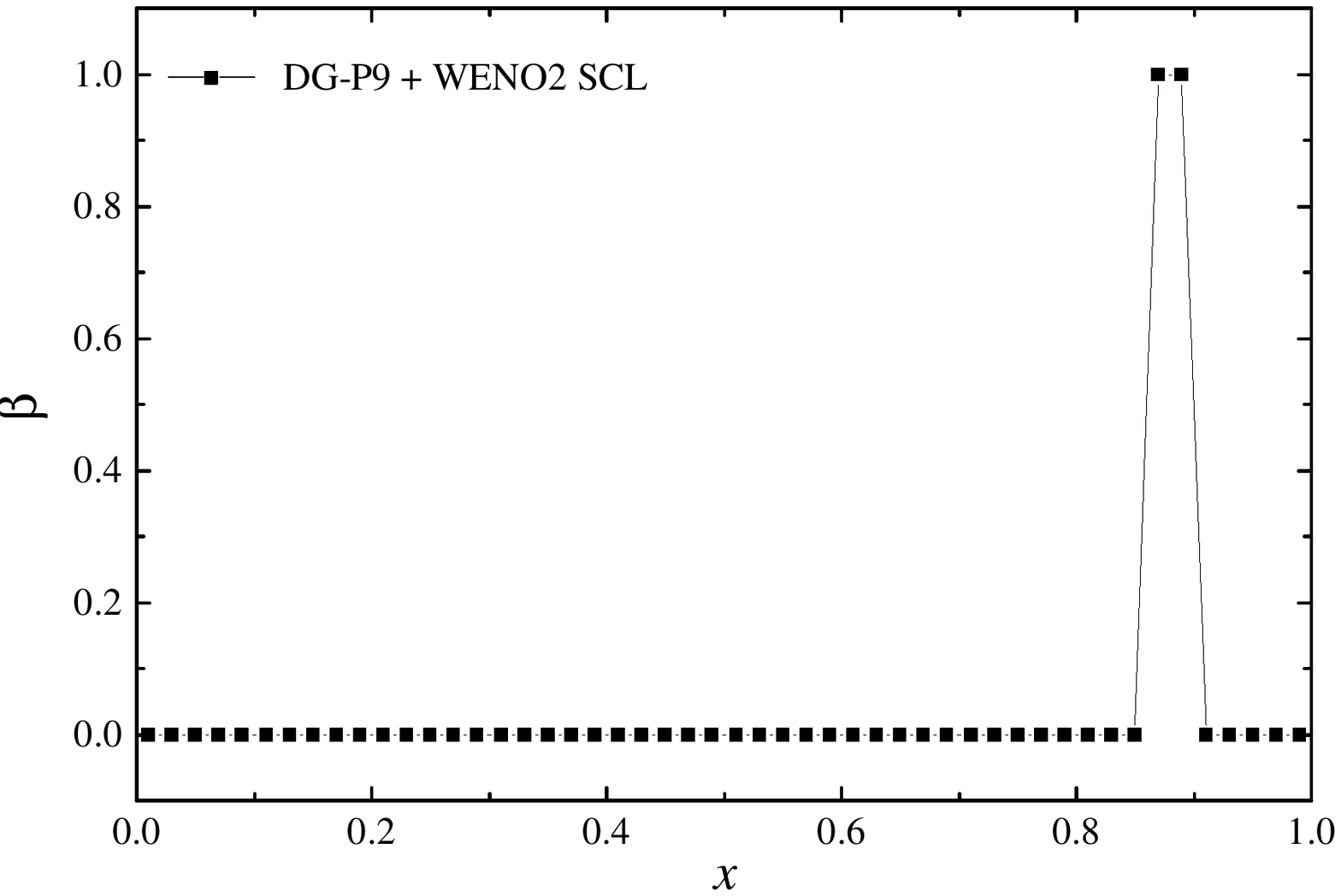}\\
\includegraphics[width=0.245\textwidth]{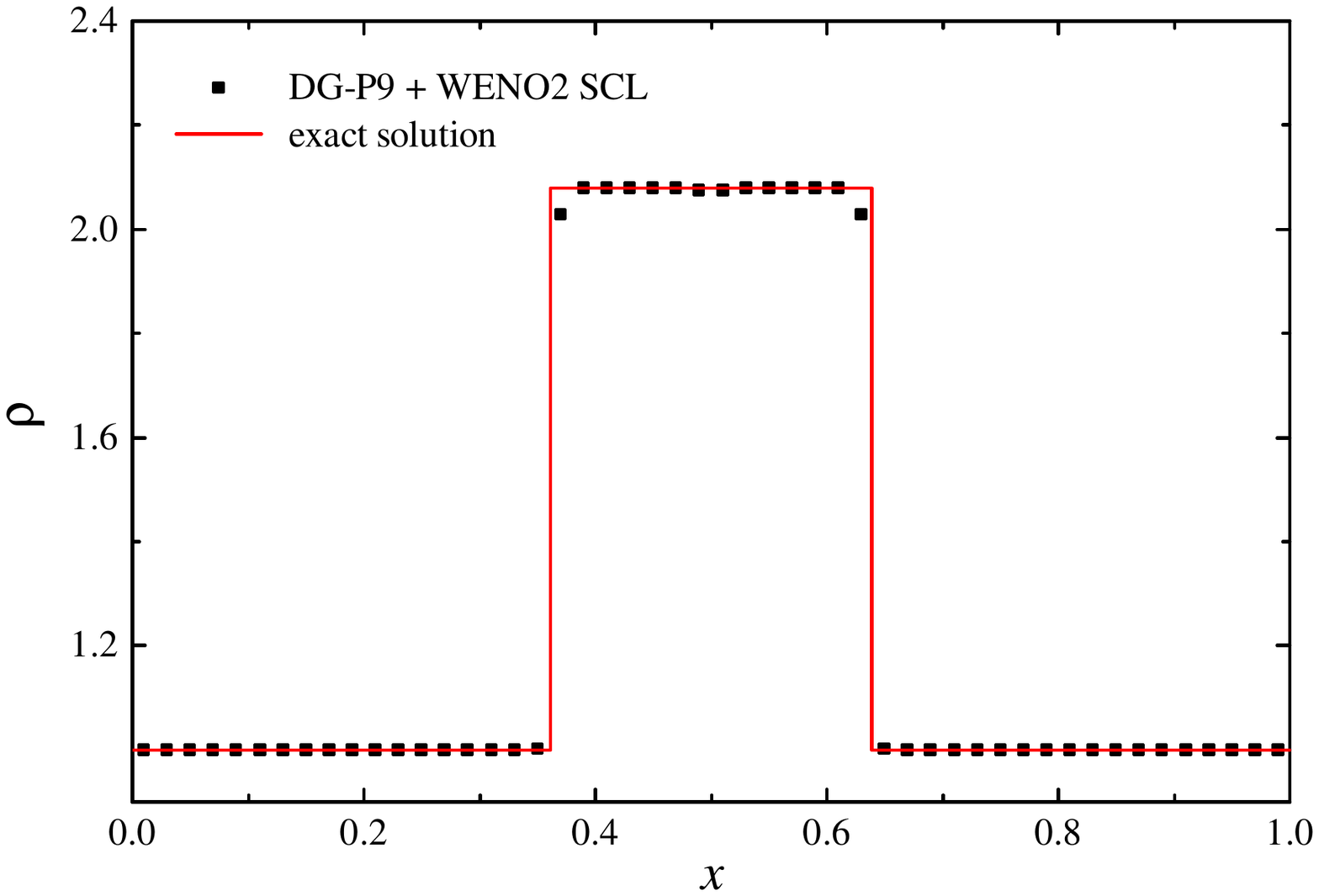}
\includegraphics[width=0.245\textwidth]{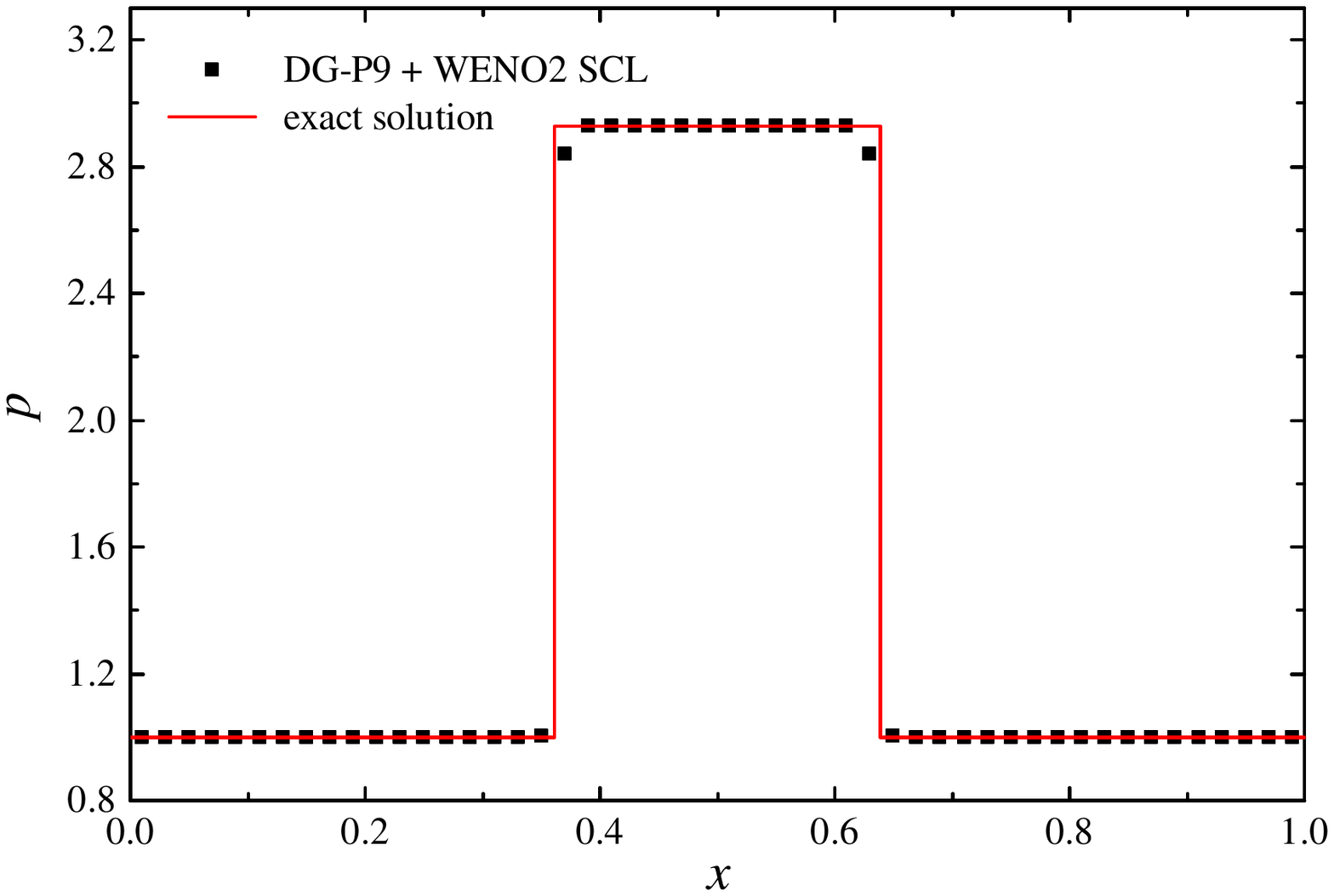}
\includegraphics[width=0.245\textwidth]{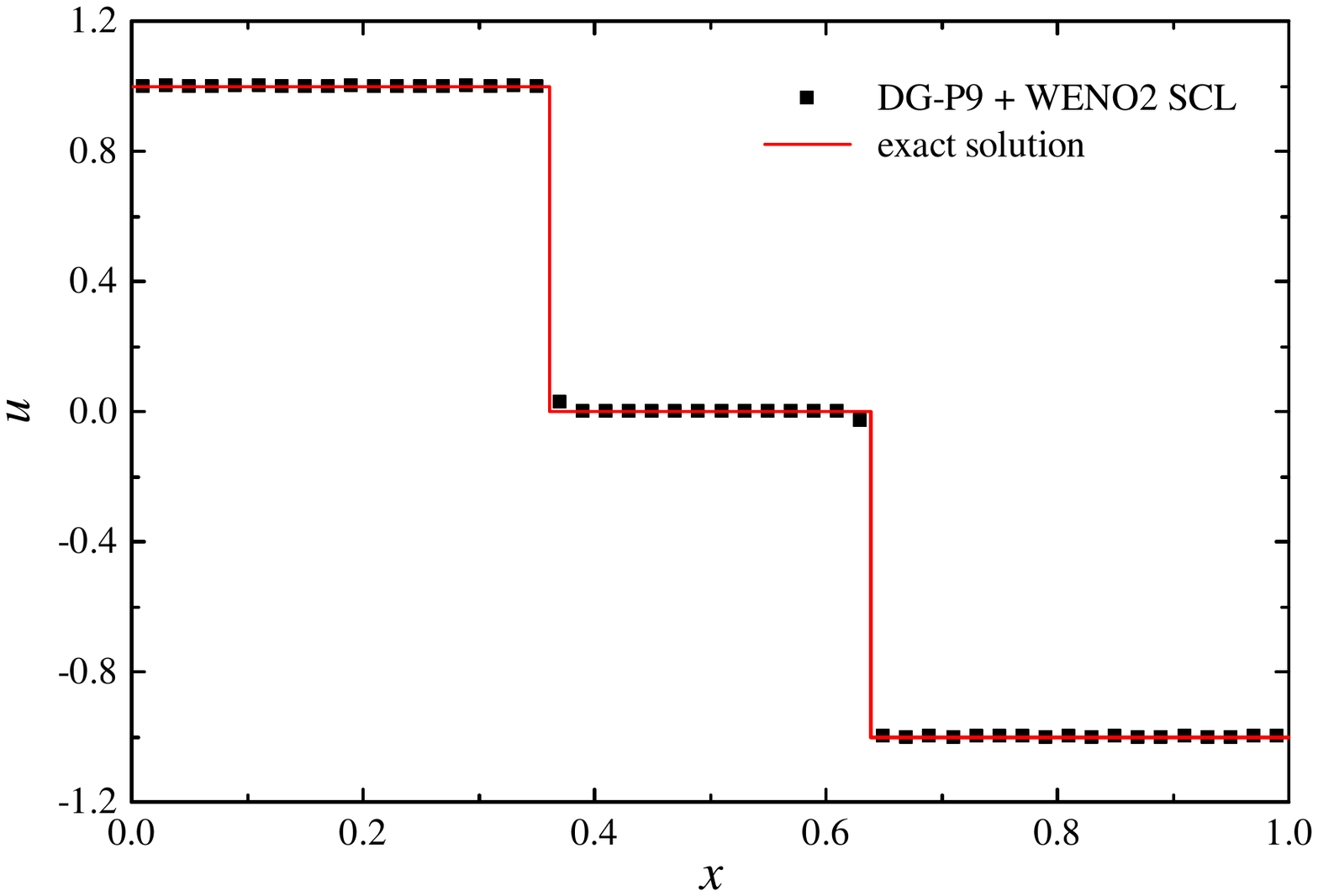}
\includegraphics[width=0.245\textwidth]{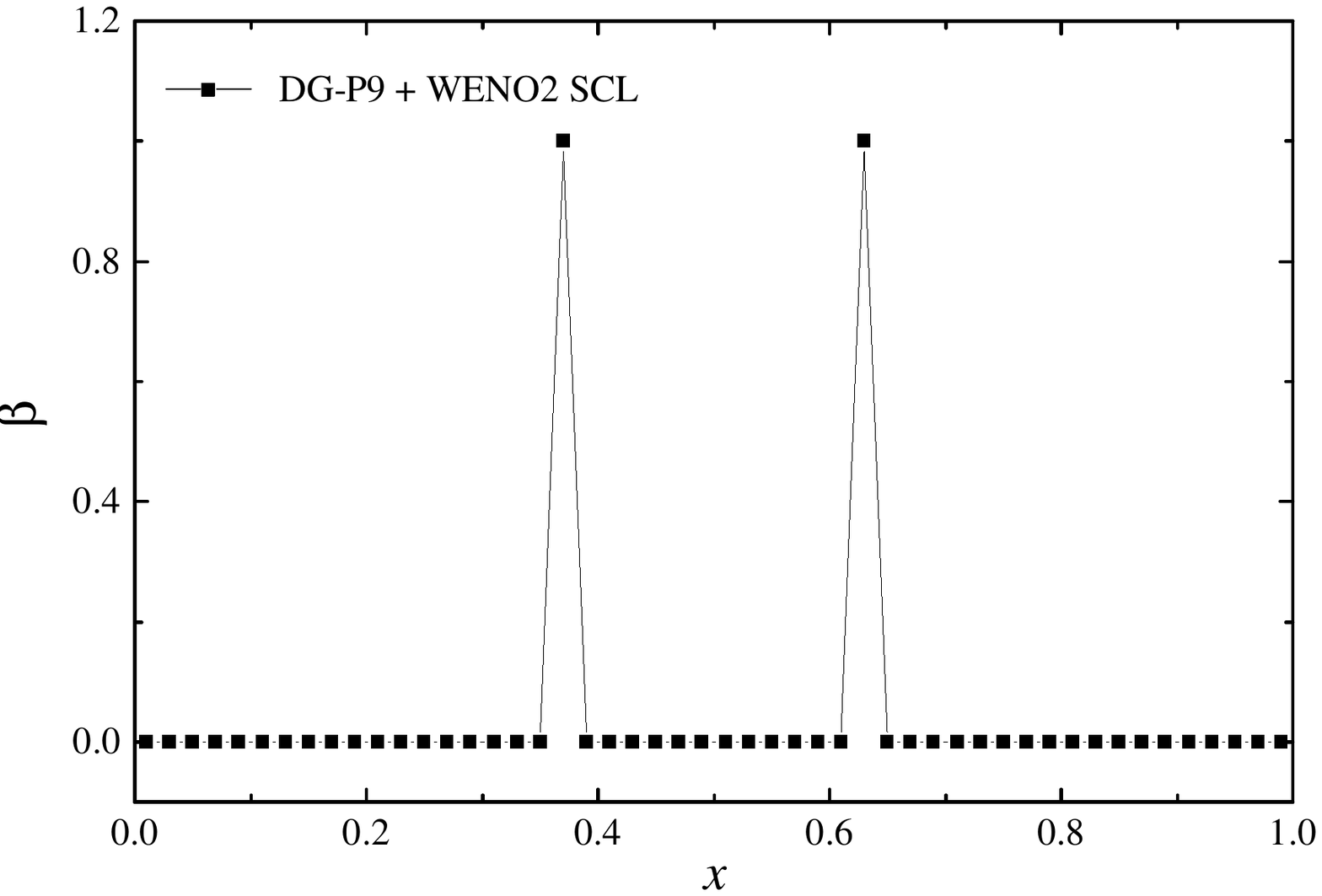}\\
\includegraphics[width=0.245\textwidth]{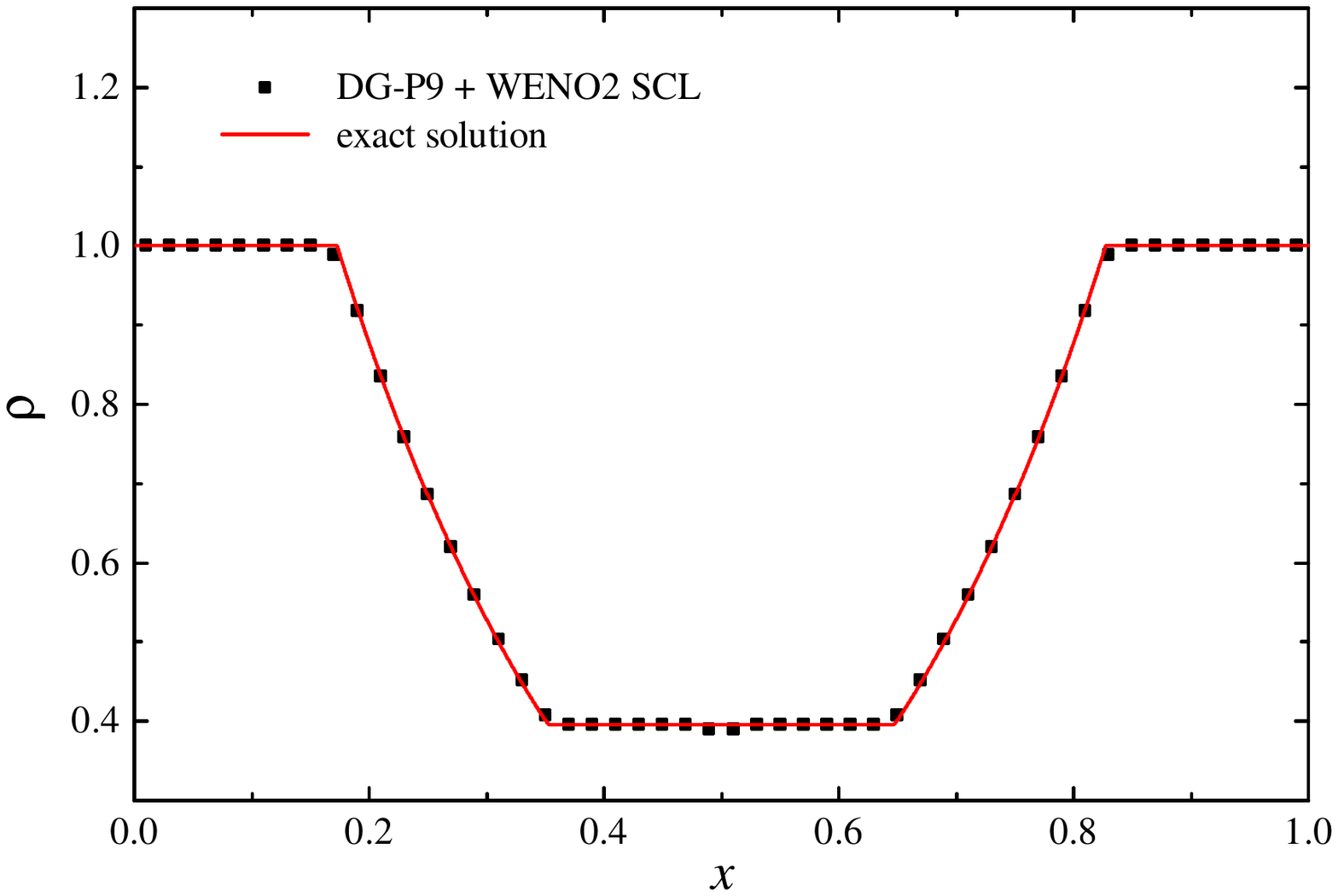}
\includegraphics[width=0.245\textwidth]{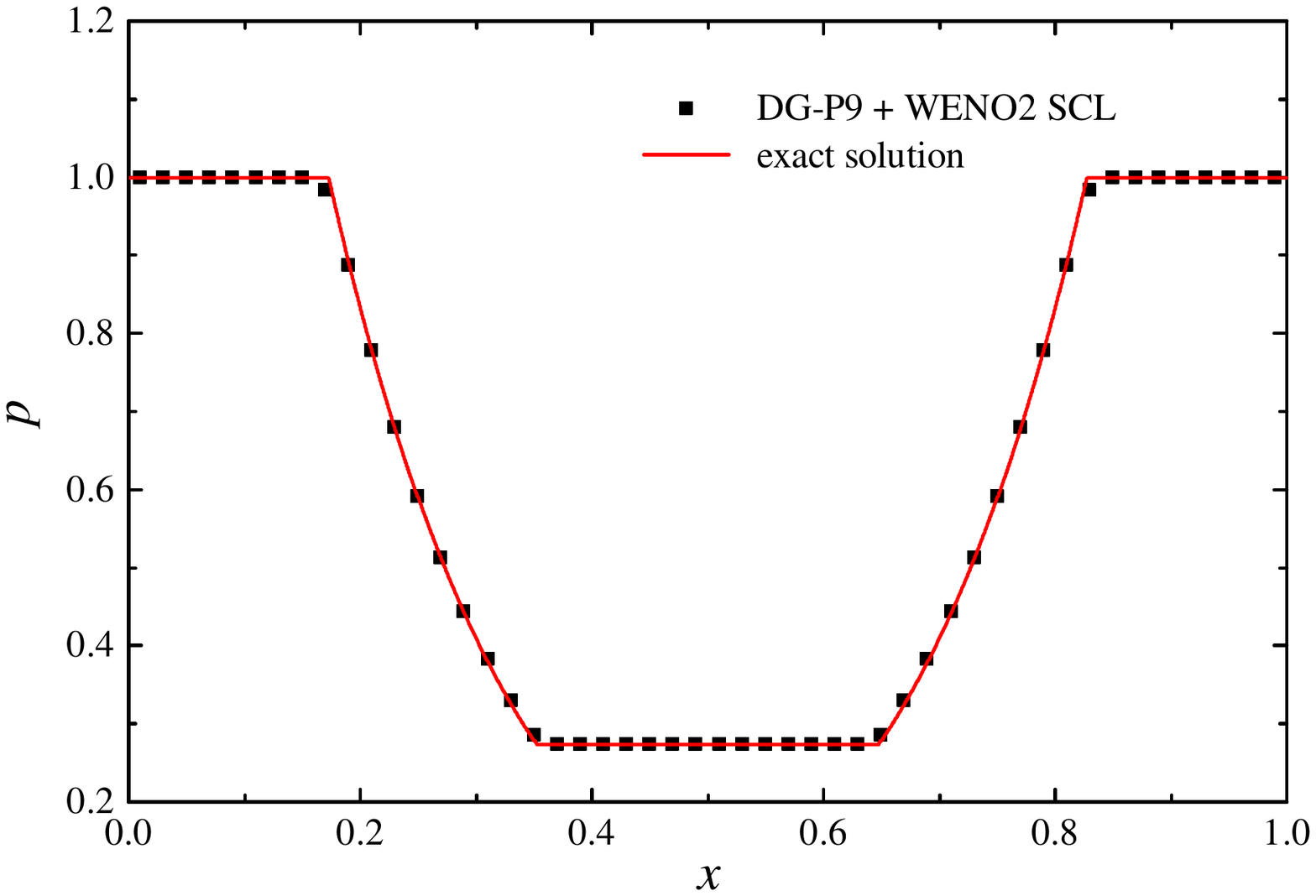}
\includegraphics[width=0.245\textwidth]{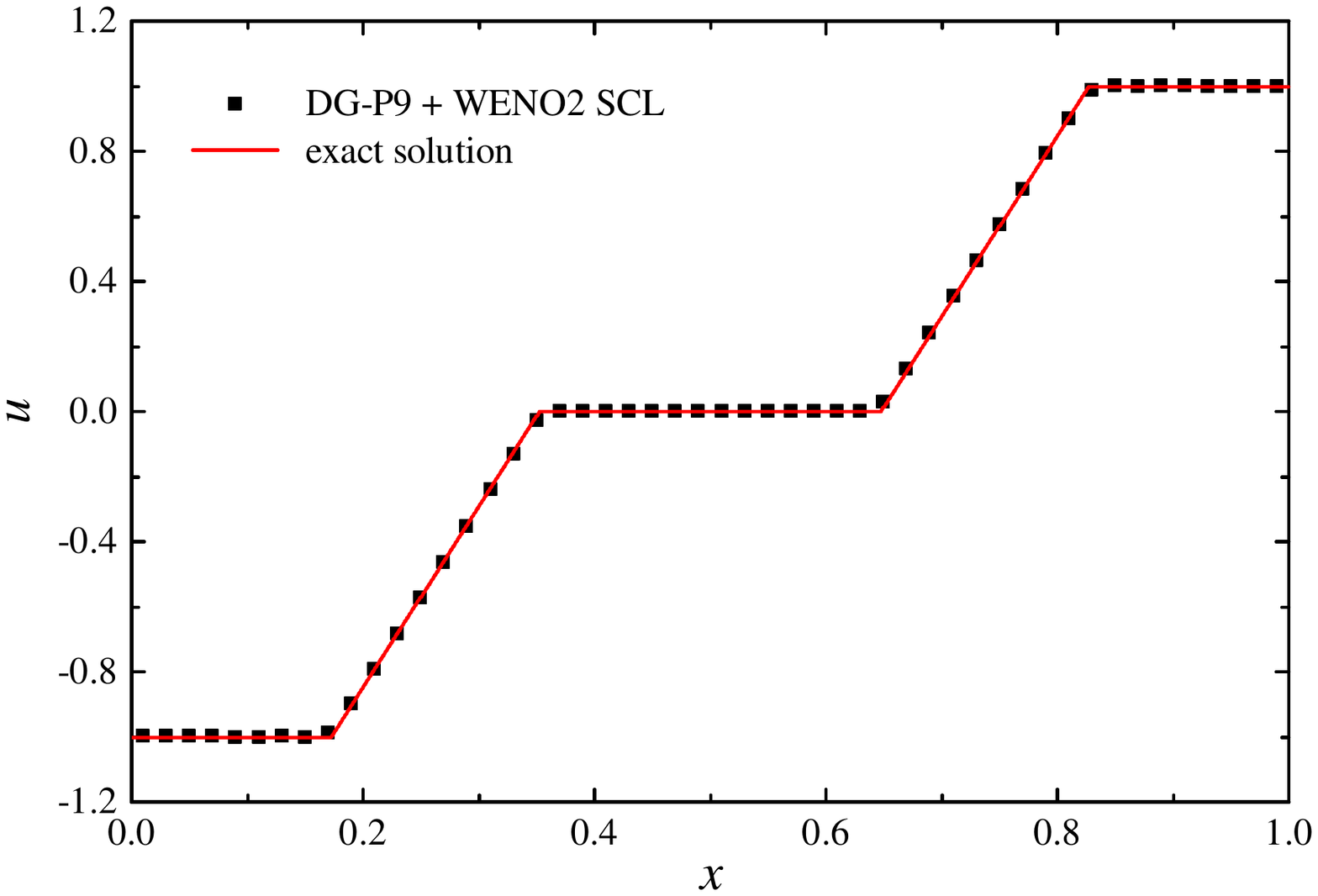}
\includegraphics[width=0.245\textwidth]{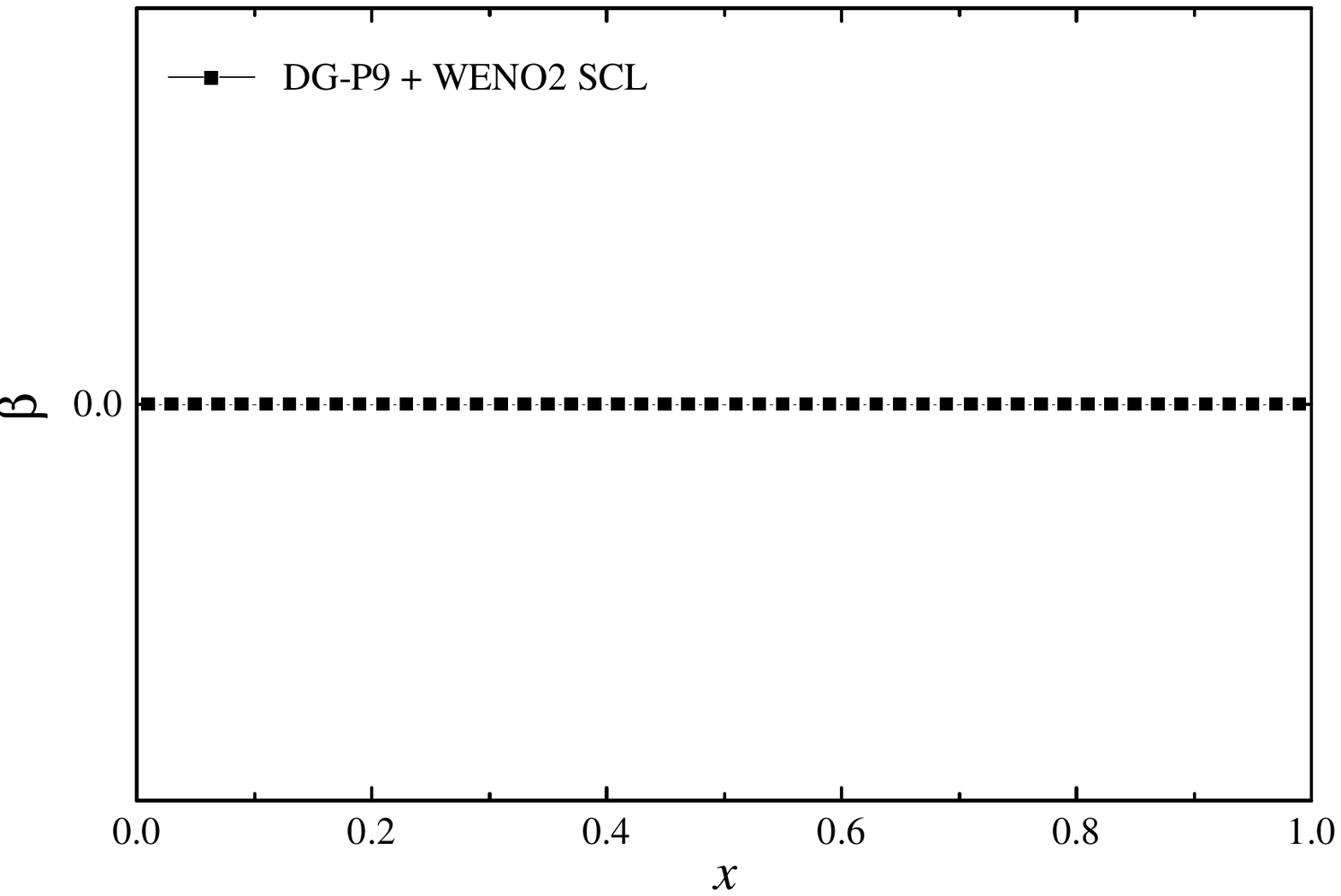}
\caption{\label{fig:classical_tests_1d}%
Numerical solution of the classical Sod, Lax, two shock waves problems and two rarefaction waves (from top to bottom)
obtained using the ADER-DG-$\mathbb{P}_{9}$ method with a posteriori limitation of the solution by a ADER-WENO2 finite-volume limiter
on mesh with $10\times50$ cells (a detailed statement of the problem is presented in the text).
The graphs show the coordinate dependencies of density $\rho$, pressure $p$, flow velocity $u$ 
and troubled cells indicator $\beta$ (from left to right) at the final time $t_\mathrm{final} = 0.15$.
The black square symbols represent the numerical solution; 
the red solid lines represents the exact analytical solution of the problem.
}
\end{figure*}

Numerical solution of the classical Sod, Lax, two shock waves and two rarefaction waves problems obtained using the ADER-DG-$\mathbb{P}_{9}$ method with a posteriori limitation of the solution by a ADER-WENO2 finite-volume limiter on two-dimensional mesh with $10\times50$ cells is presented in Figure~\ref{fig:classical_tests_1d}.

The one-dimensional flow that occurs in the Sod problem contains a shock wave, a contact discontinuity, and a rarefaction wave. The obtained coordinate dependencies show that the shock front and contact discontinuity are resolved in the solution with an accuracy of one mesh cell. In this case, the width of the contact discontinuity does not increase over time, which is often characteristic of first- and second-order methods. The contact discontinuity does not appear in any way in the coordinate dependence of pressure $p$ in the form of non-physical artifacts. The one-dimensional flow that arises in the Lax problem is also characterized by a shock wave, a contact discontinuity and a rarefaction wave, however, compared to the Sod problem, it is much more difficult to solve this problem by high-order methods. The obtained coordinate dependencies demonstrate properties of the numerical solution similar to the solution to the Sod problem --- the shock wave and contact discontinuity are resolved with an accuracy of one mesh cell. An important feature of the numerical solution of the Sod and Lax problems is that in the solution only two troubled cells arise at each time step grid, as can be seen from the presented results for the troubled cells indicator $\beta$. Troubled cells are formed only in the coordinate vicinity of the shock wave front, while in the vicinity of the contact discontinuity, as well as the boundary characteristics of the rarefaction wave, troubled cells are not formed.

The one-dimensional flow that occurs in Test 3 contains two symmetrical shock waves. Shock waves are resolved in a numerical solution with an accuracy of one mesh cell, while two troubled cells are formed in the vicinity of the shock wave fronts. The features of shock wave resolution in Test 3 are, in general, not different from the shock wave resolution in the Sod and Lax problems. The one-dimensional flow that occurs in Test 4 contains two symmetrical rarefaction waves. The presented results show that the main features of a flow with two rarefaction waves are resolved quite accurately in the numerical solution. It should also be noted that in this test case the limiter is not called --- there are no grid cells, the troubled cells indicator $\beta = 0$ is everywhere. It can also be noted that on the coordinate dependence of the density at the point of the initial position of the discontinuity, one can observe a small area of convexity downwards, the size of which is two cells. The features of the resolution of rarefaction waves in Test 4, in general, are not different from the resolution of rarefaction waves in the Sod and Lax problems.

As a result of the analysis, we can conclude that the developed software implementation of the ADER-DG-$\mathbb{P}_{N}$ method with a posteriori limitation of the solution by an ADER-WENO finite-volume limiter makes it possible to obtain numerical solutions to one-dimensional problems that are characterized by the appearance of discontinuous components in the solution. The properties of the numerical solution correspond to the results presented in the basic works~\cite{ader_dg_dev_1, ader_dg_ideal_flows}.

\subsection{Kelvin-Helmholtz instability problem}
\label{sec:apps_cgd_problems:khi_2d}

The evolution of the Kelvin-Helmholtz instability has been considered as a two-dimensional problem. The formation of a vortex street in a shear layer significantly depends on the dissipative properties of the numerical method used. In the one-dimensional Riemann problems discussed above, it is shown that contact discontinuities are resolved very well by the method, while the contact discontinuity region does not form troubles for mesh cells. Two-dimensional tangential discontinuities in gas dynamics exhibit instability to small perturbations of the discontinuity, accompanied by the generation of vortices that form an irregular structure of multiple vortices with fractal properties, which in dynamic evolution form a regular structure of large vortices in the form of a shifted sequence~\cite{Springel_mnras_2010}.

\begin{figure*}[h!]
\centering
\includegraphics[width=0.49\textwidth]{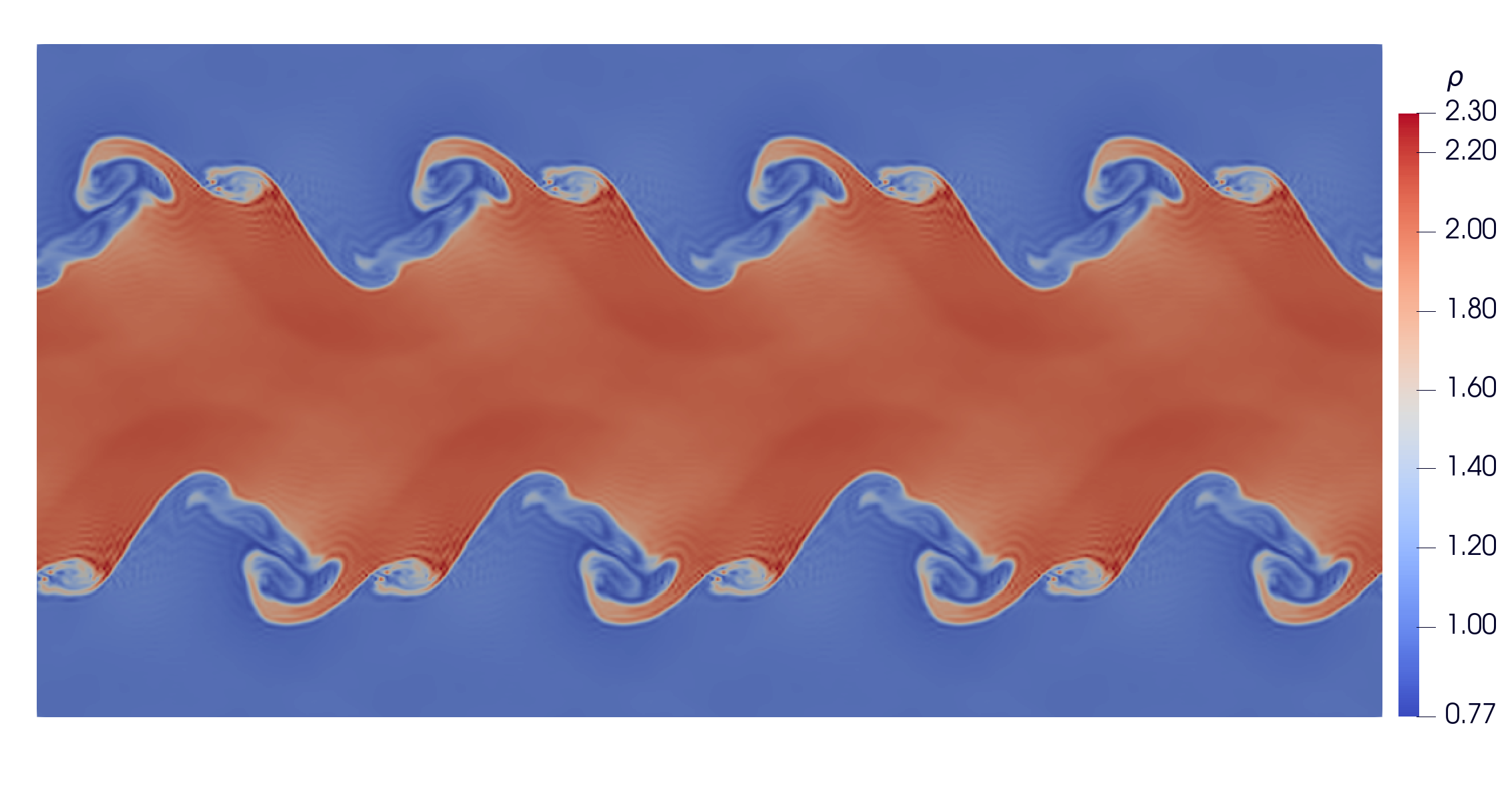}
\includegraphics[width=0.49\textwidth]{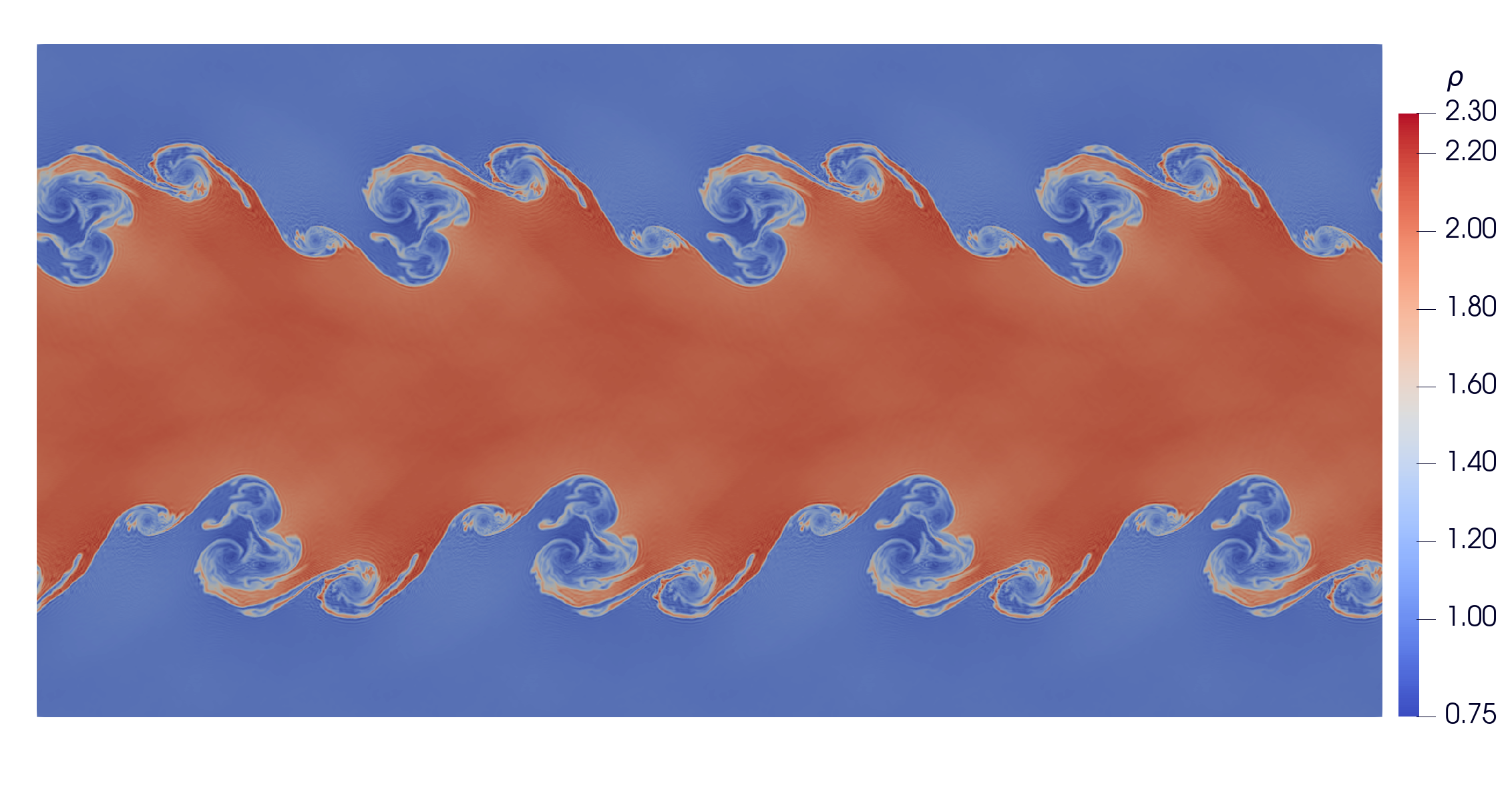}\\
\includegraphics[width=0.49\textwidth]{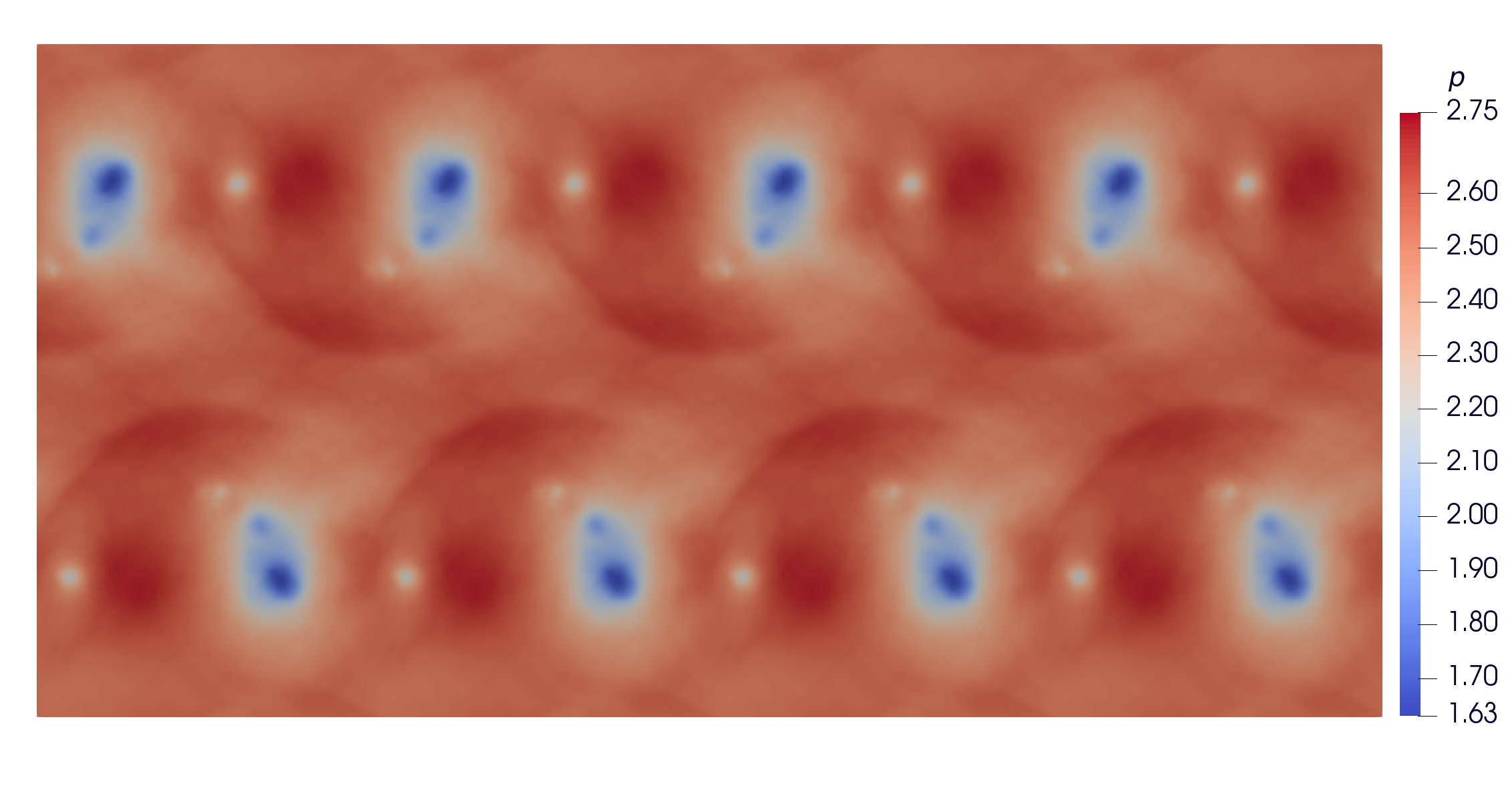}
\includegraphics[width=0.49\textwidth]{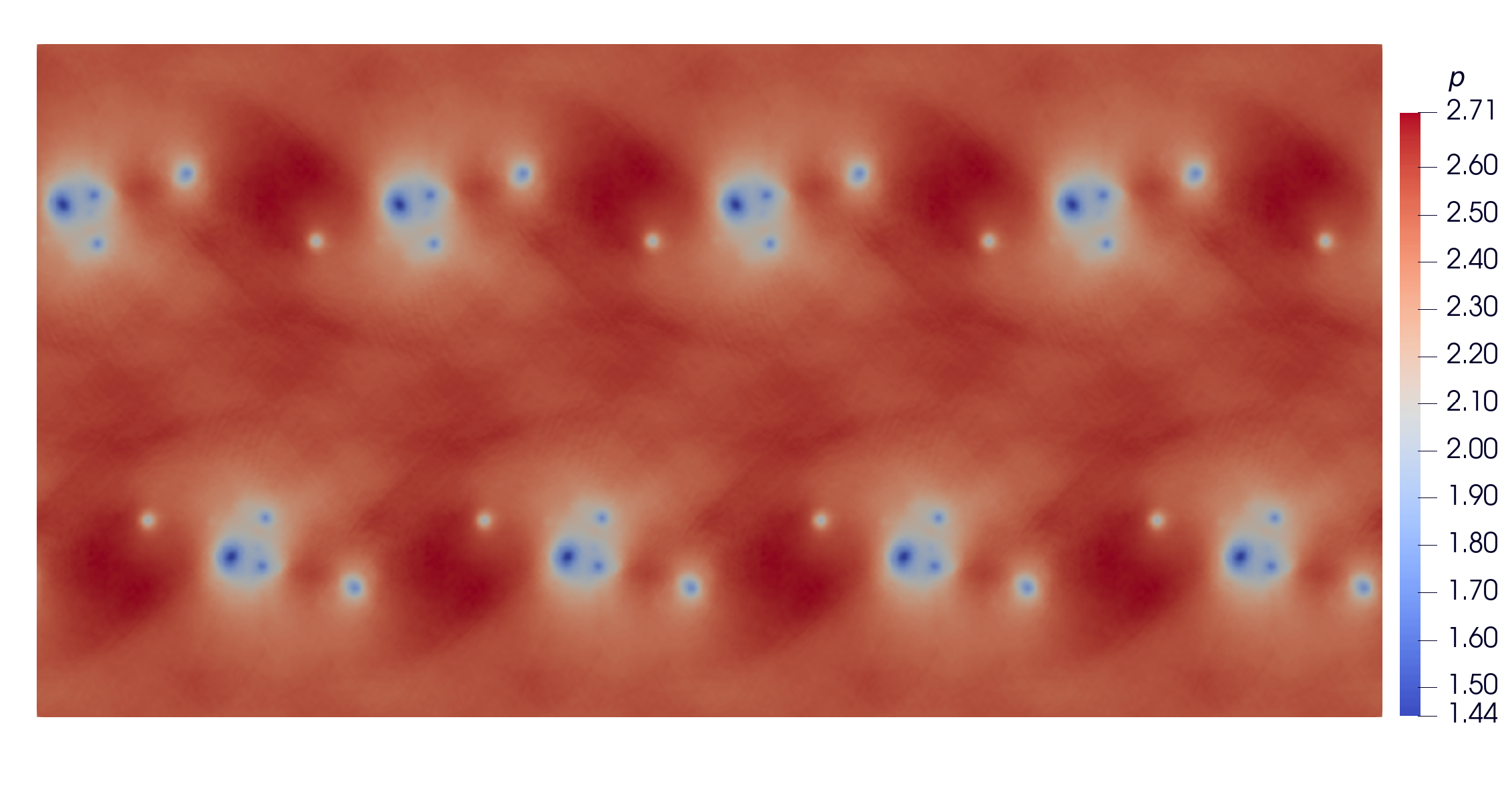}\\
\includegraphics[width=0.49\textwidth]{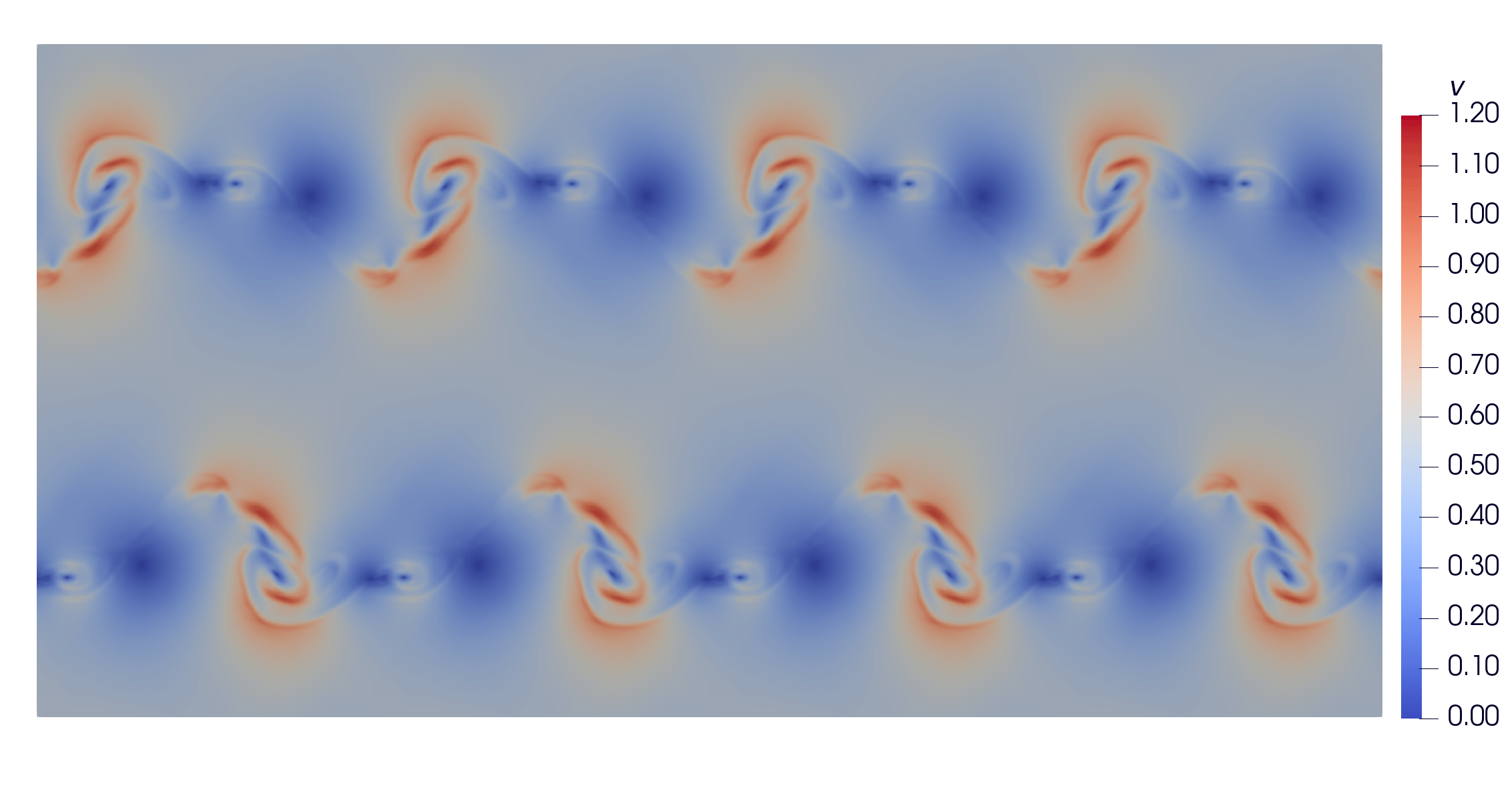}
\includegraphics[width=0.49\textwidth]{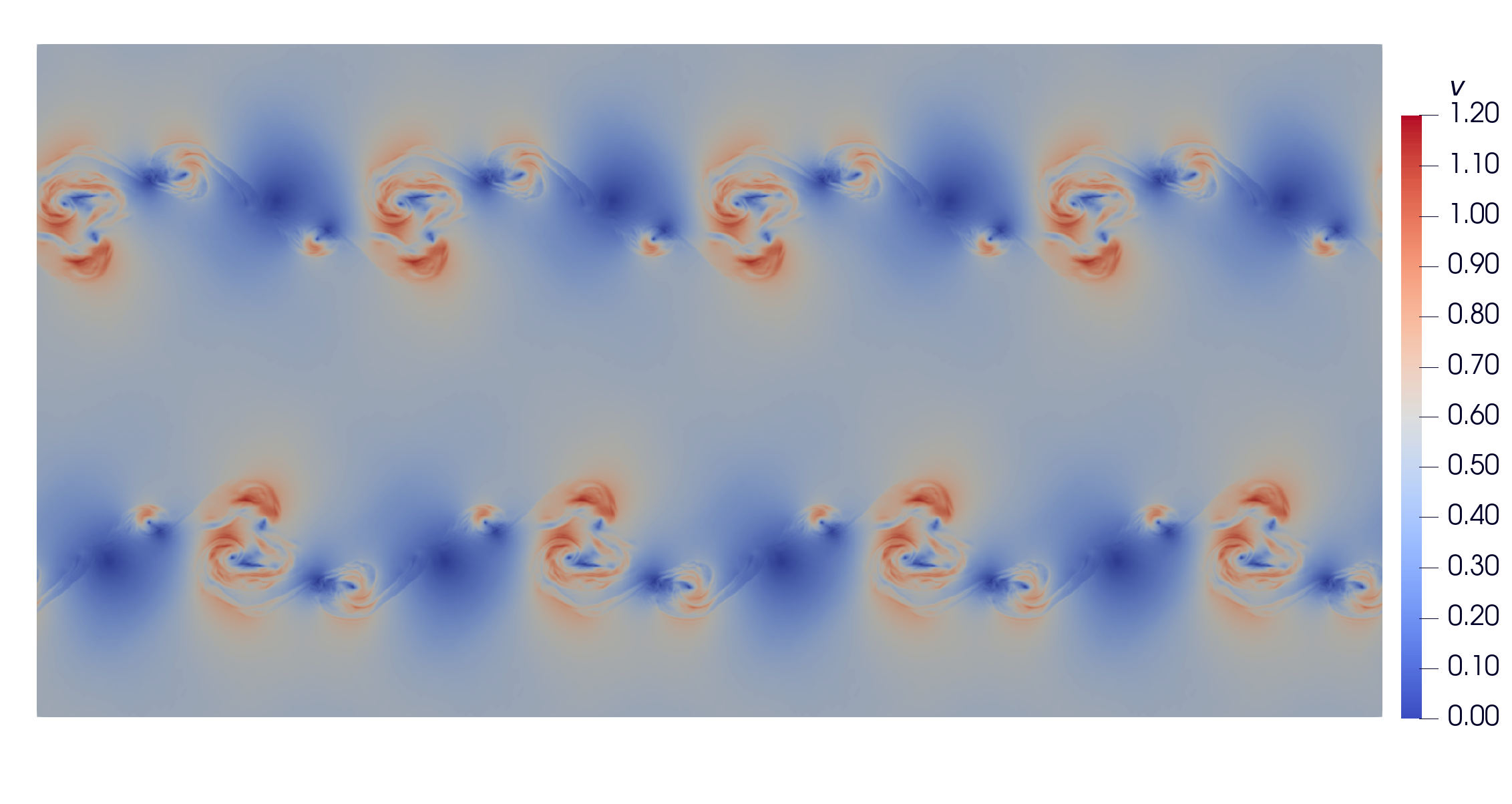}\\
\includegraphics[width=0.49\textwidth]{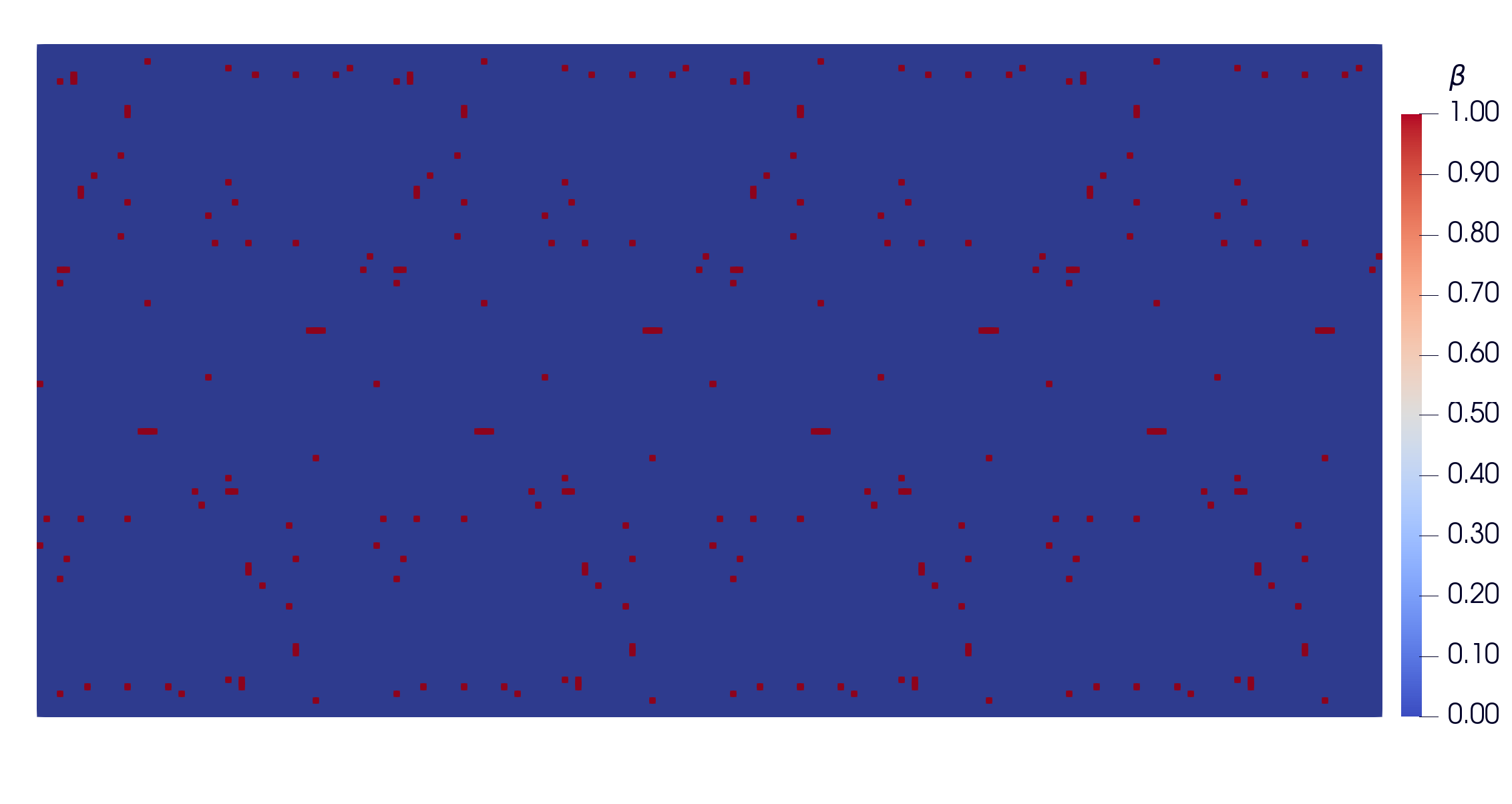}
\includegraphics[width=0.49\textwidth]{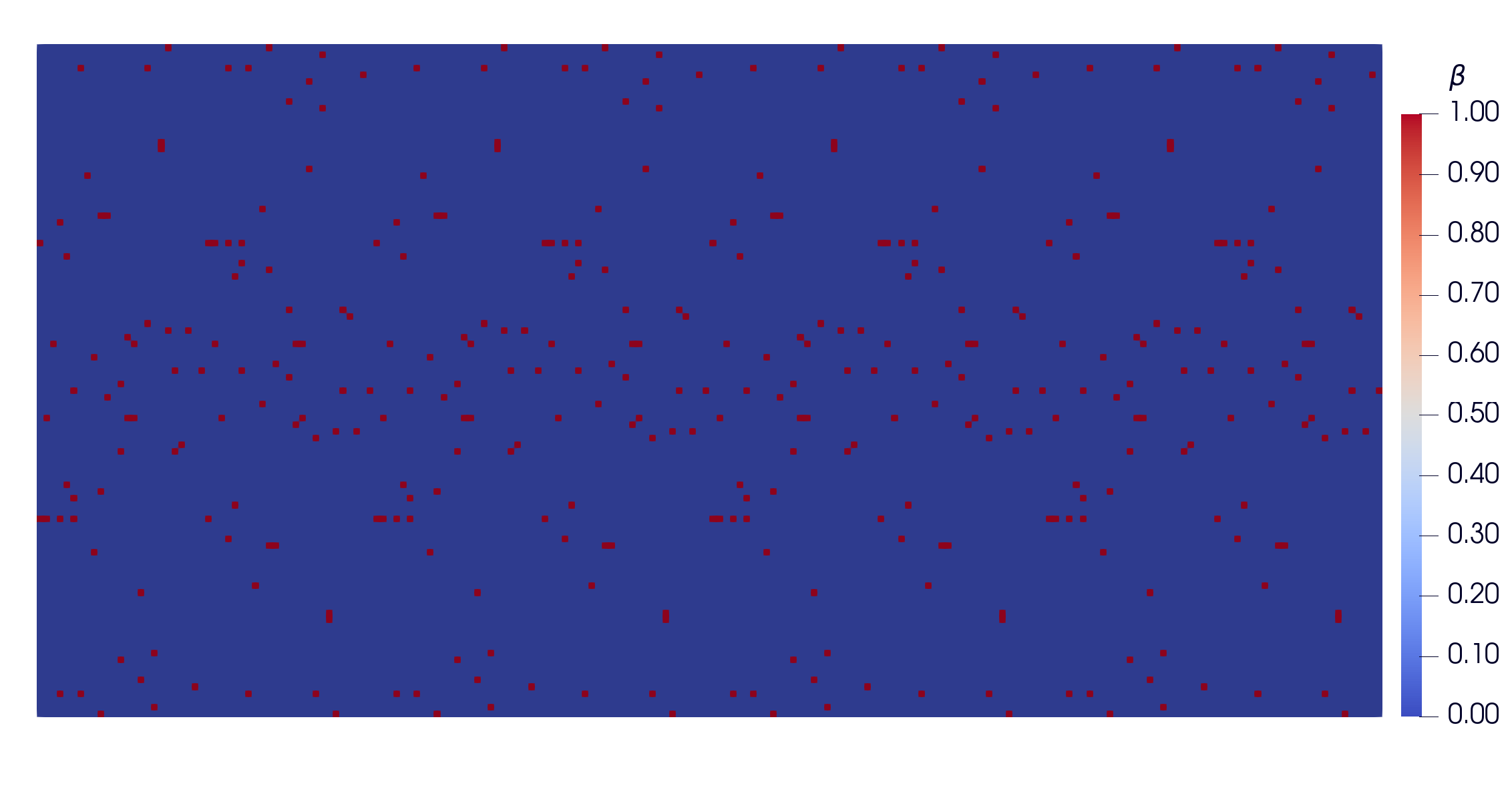}
\caption{\label{fig:khi_2d_08}%
Numerical solution of the two-dimensional Kelvin-Helmholtz instability problem (a detailed statement of the problem is presented in the text)
obtained using the ADER-DG-$\mathbb{P}_{2}$ (left) and ADER-DG-$\mathbb{P}_{5}$ (right) methods on mesh with $200 \times 100$ cells 
at the time $t = 0.8$. 
The graphs show the coordinate dependencies of the subcells finite-volume representation of density $\rho$, pressure $p$,
flow velocity magnitude $v$ and troubled cells indicator $\beta$ (bottom).
}
\end{figure*}
\begin{figure*}[h!]
\centering
\includegraphics[width=0.49\textwidth]{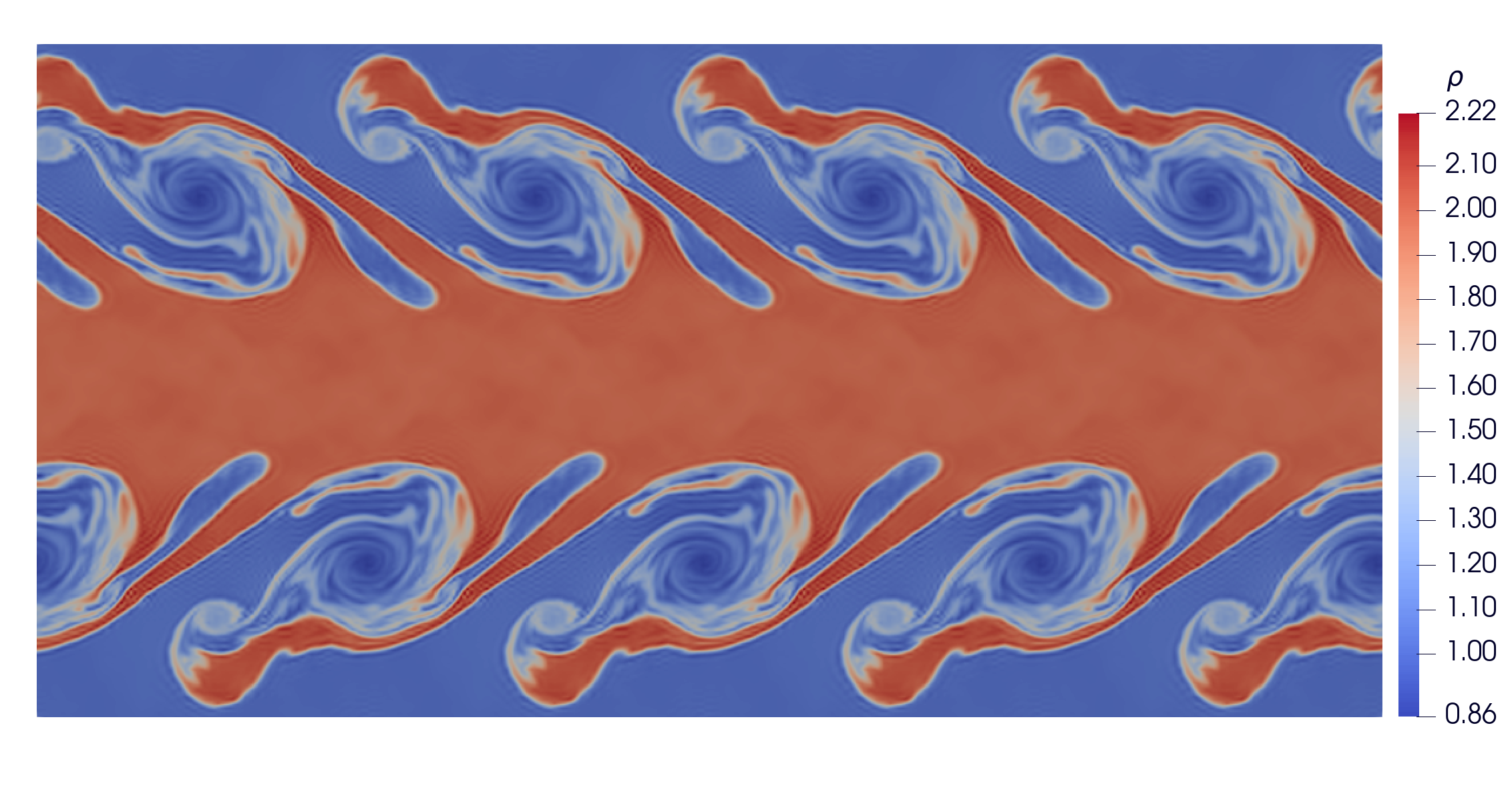}
\includegraphics[width=0.49\textwidth]{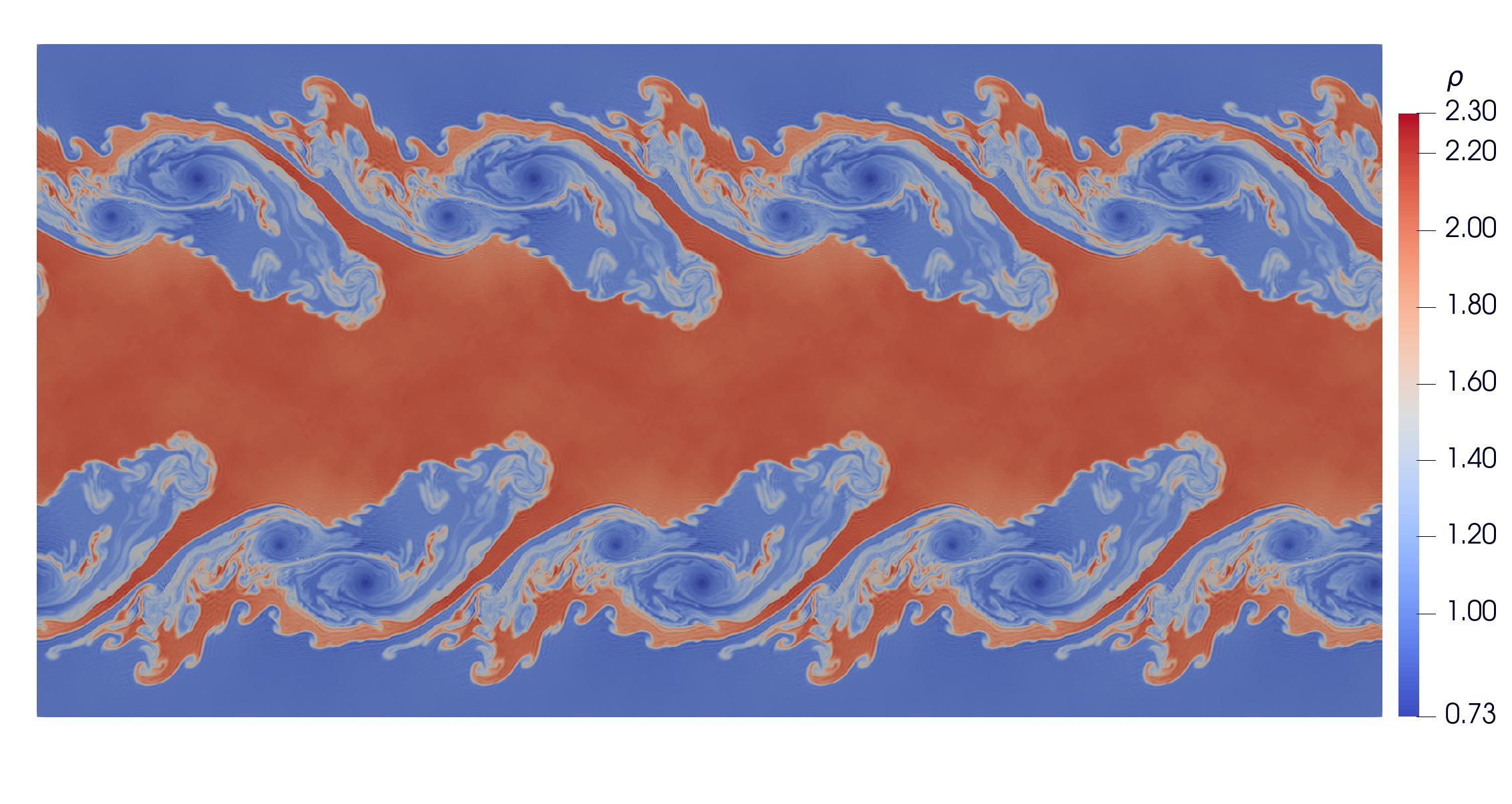}\\
\includegraphics[width=0.49\textwidth]{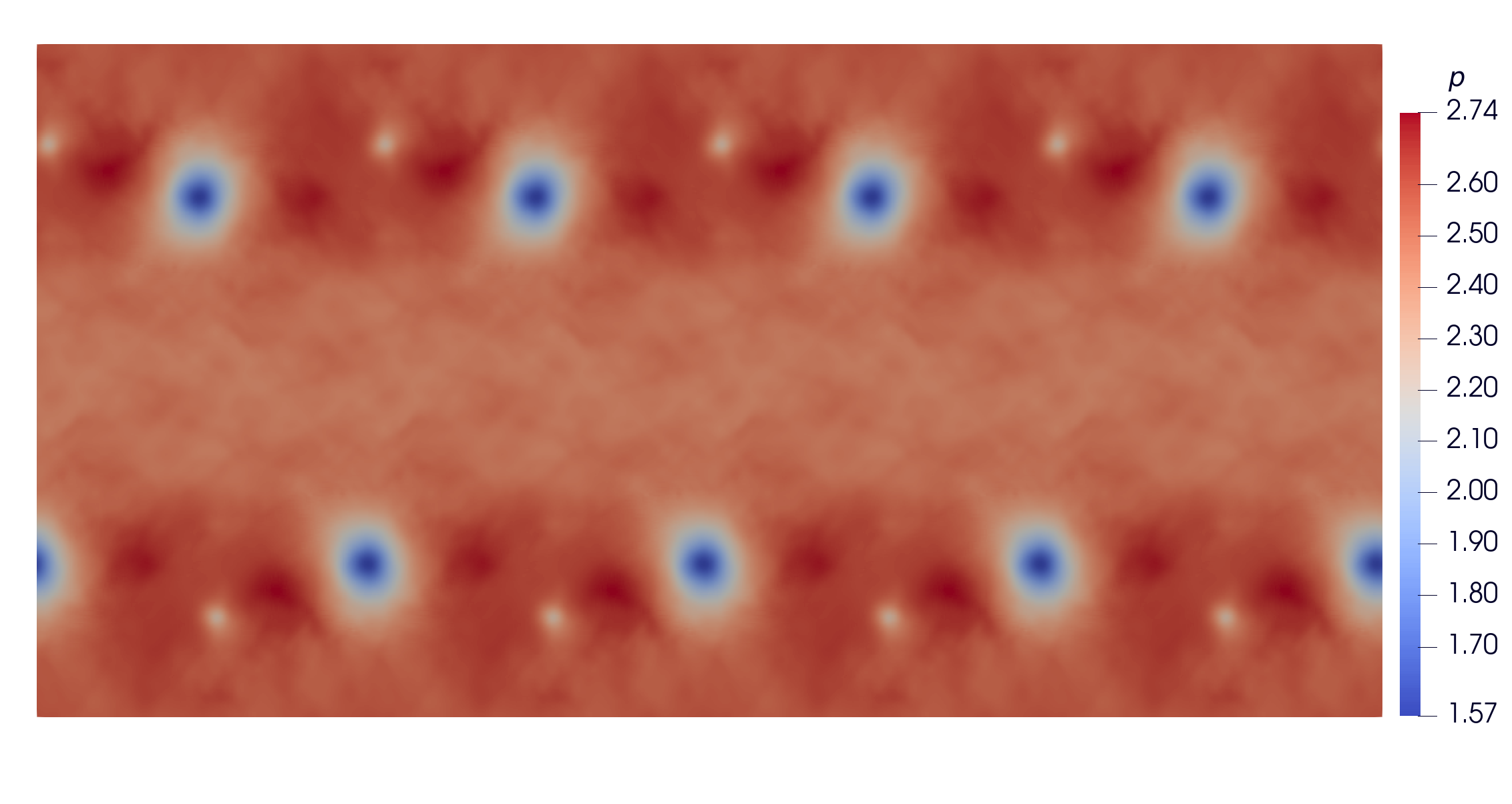}
\includegraphics[width=0.49\textwidth]{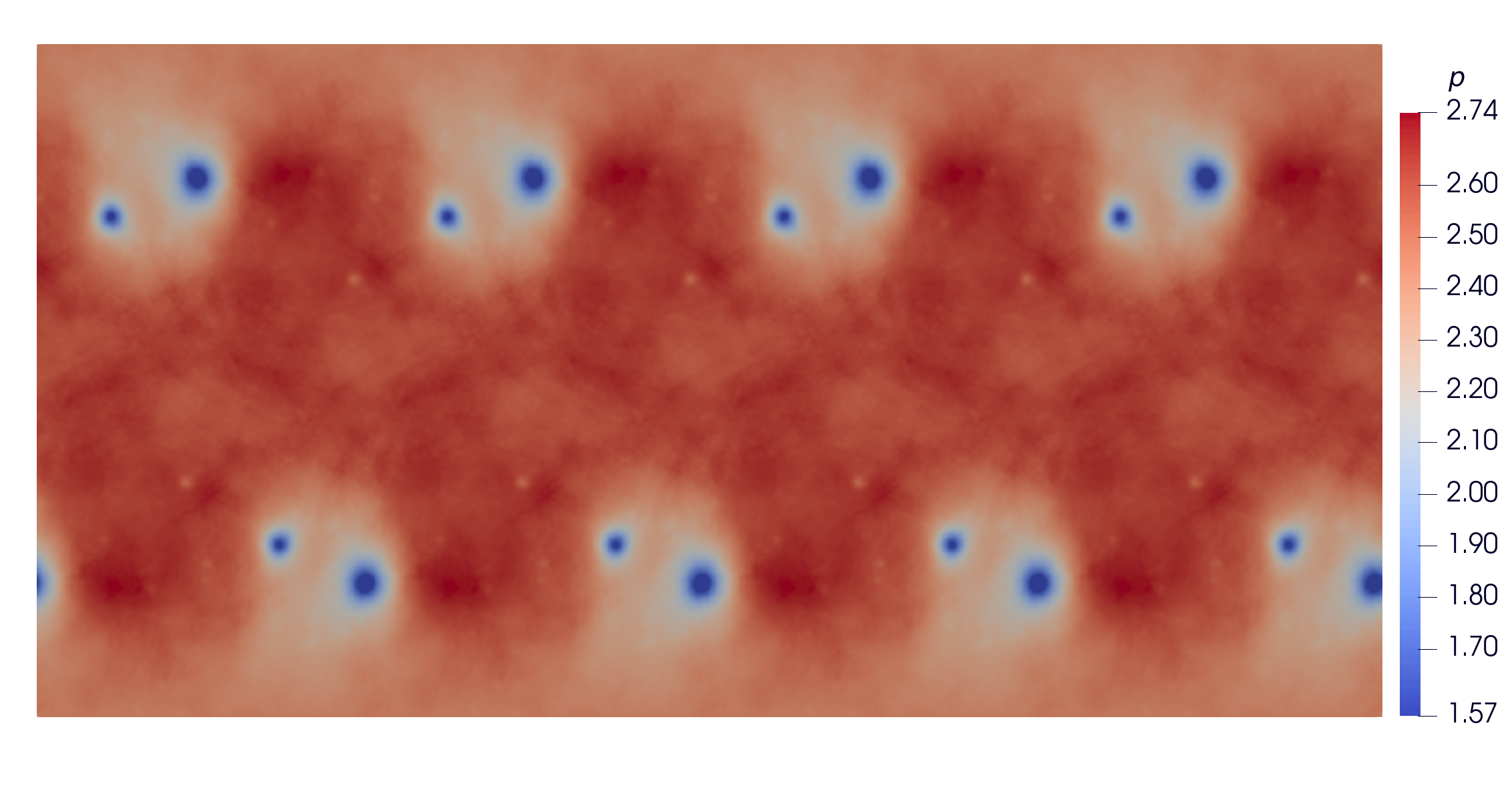}\\
\includegraphics[width=0.49\textwidth]{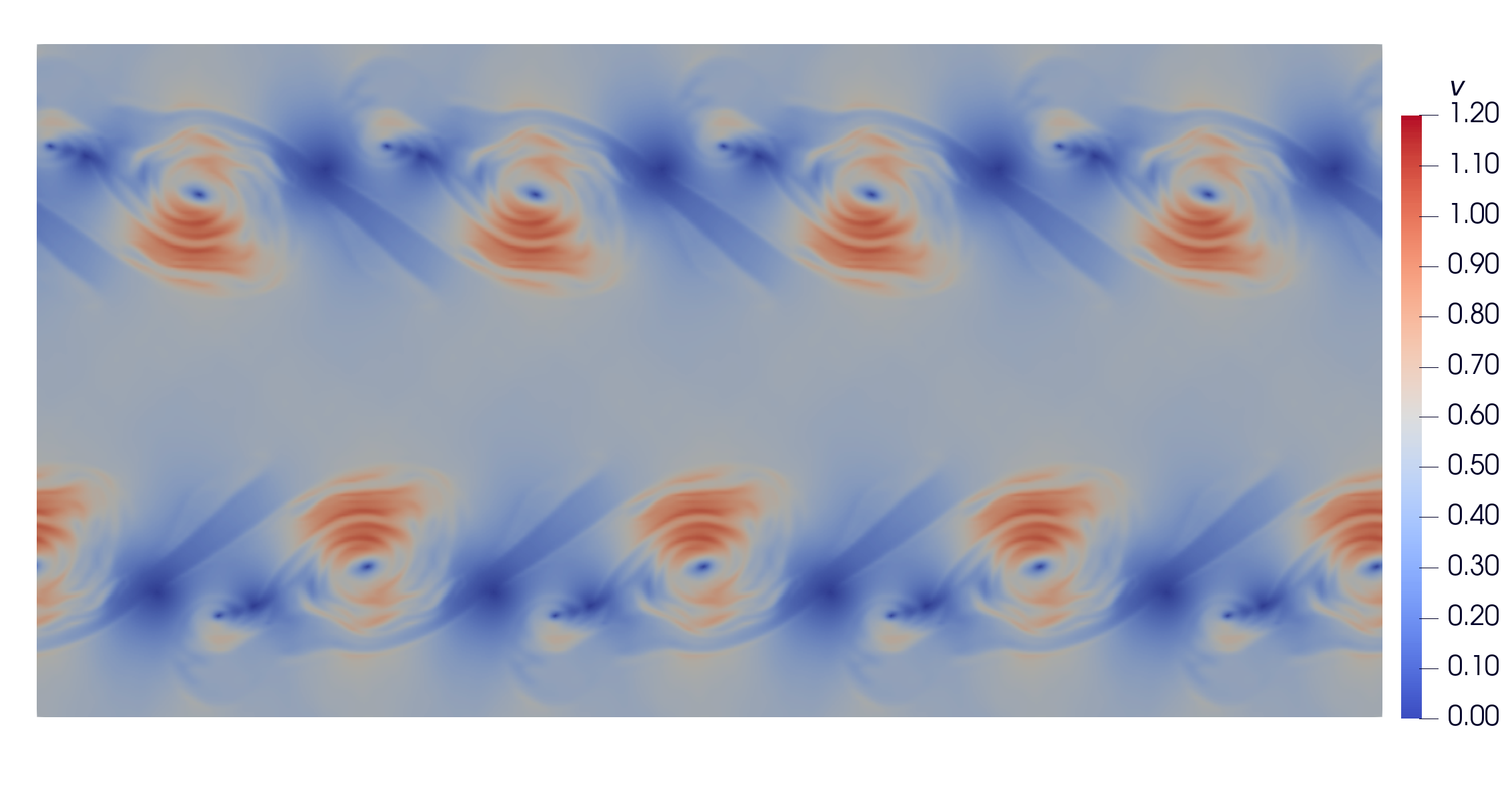}
\includegraphics[width=0.49\textwidth]{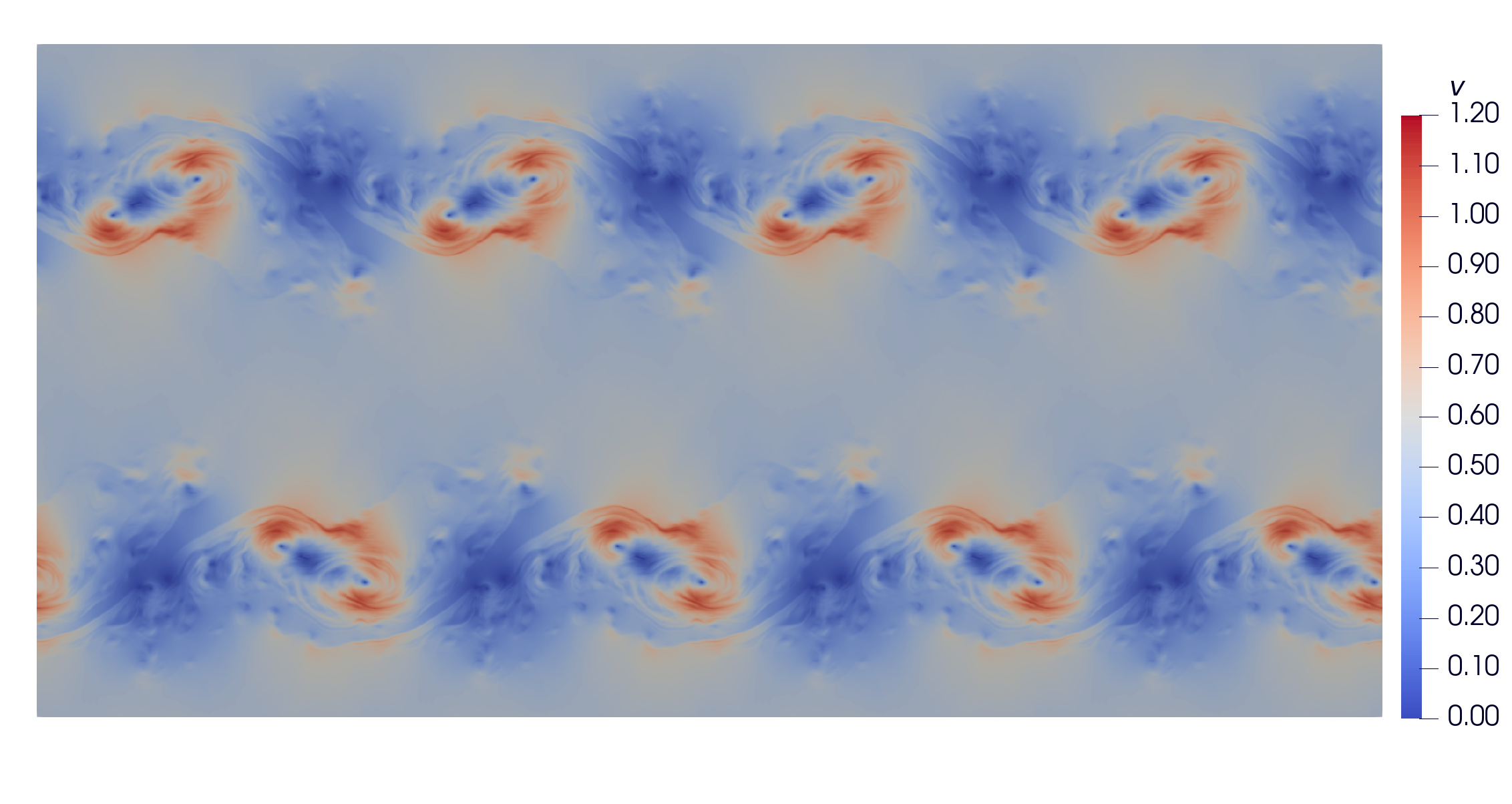}\\
\includegraphics[width=0.49\textwidth]{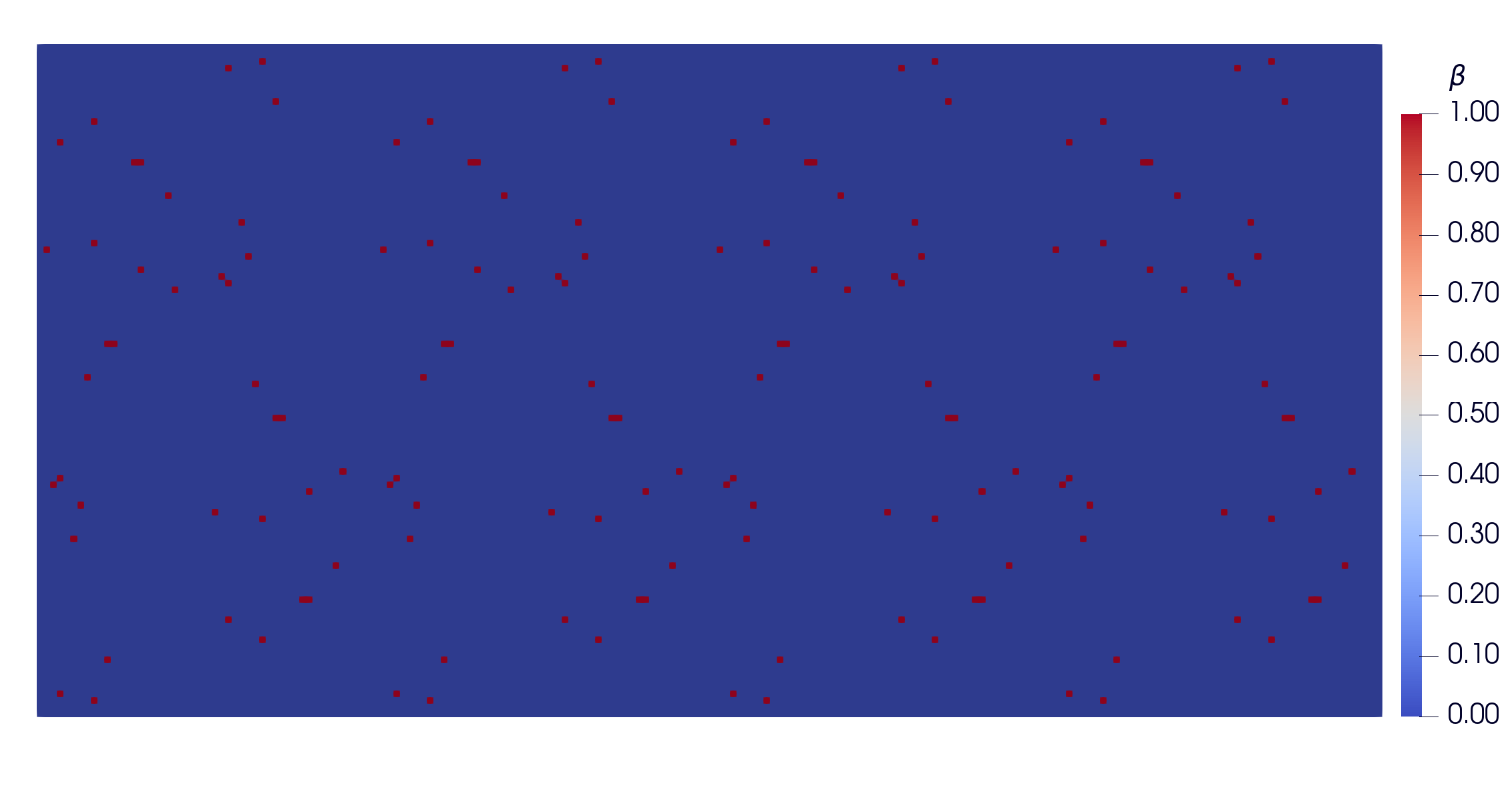}
\includegraphics[width=0.49\textwidth]{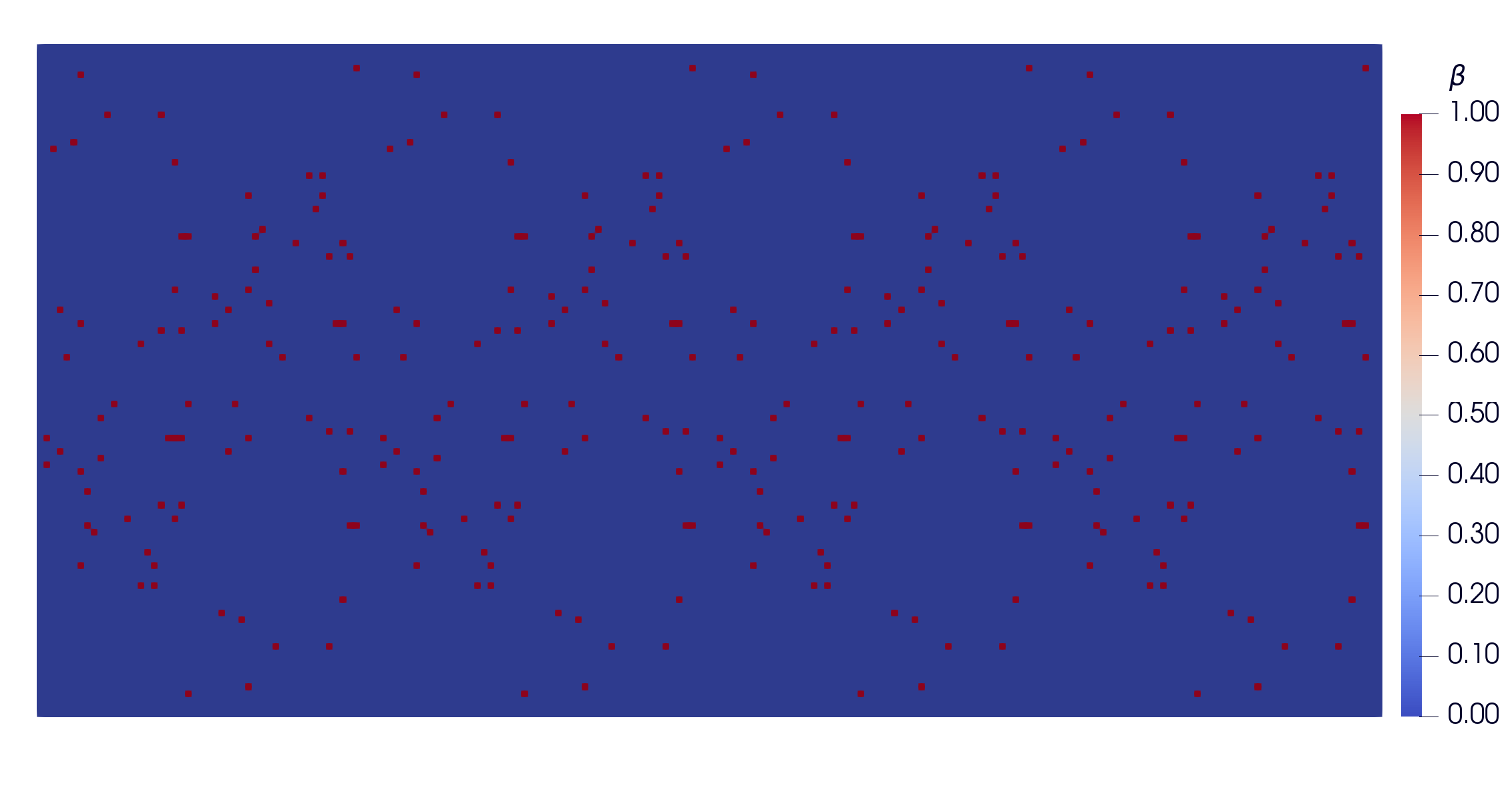}
\caption{\label{fig:khi_2d_20}%
Numerical solution of the two-dimensional Kelvin-Helmholtz instability problem (a detailed statement of the problem is presented in the text)
obtained using the ADER-DG-$\mathbb{P}_{2}$ (left) and ADER-DG-$\mathbb{P}_{5}$ (right) methods on mesh with $200 \times 100$ cells 
at the time $t = 2.0$. 
The graphs show the coordinate dependencies of the subcells finite-volume representation of density $\rho$, pressure $p$,
flow velocity magnitude $v$ and troubled cells indicator $\beta$ (bottom).
}
\end{figure*}

In this work, the problem statement is chosen in the form proposed in the work~\cite{Springel_mnras_2010} and used in the work~\cite{Springel_mnras_2015}. The coordinate domain is chosen in the form of a rectangle $\Omega = [0, 2]\times[0, 1]$ with periodic boundary conditions, unlike the works~\cite{Springel_mnras_2010, Springel_mnras_2015} where a square unit domain is chosen. The initial conditions in the two-dimensional Kelvin-Helmholtz instability problem are chosen in the form~\cite{Springel_mnras_2015}:
\begin{equation}
\begin{split}
&p(x, y, t = 0) = 2.5;\\
&\rho(x, y, t = 0) = \left\{
\begin{array}{ll}
2, & \mathrm{if}\ 0.25 < y < 0.75;\\
1, & \mathrm{if}\ y \leqslant 0.25 \lor y \geqslant 0.75;
\end{array}
\right.\\
&u(x, y, t = 0) = \left\{
\begin{array}{ll}
-0.5, & \mathrm{if}\ 0.25 < y < 0.75;\\
+0.5, & \mathrm{if}\ y \leqslant 0.25 \lor y \geqslant 0.75;
\end{array}
\right.\\
&v(x, y, t = 0) = v_{0} \sin(4 \pi x) \Bigg\{ 
	\exp\left(-\frac{(y - 0.25)^{2}}{2\sigma^{2}}\right) + \exp\left(-\frac{(y - 0.75)^{2}}{2\sigma^{2}}\right)
\Bigg\};
\end{split}
\end{equation}
where $v_{0} = 0.1$ and $\sigma = 0.05/\sqrt{2}$, the adiabatic index $\gamma = 1.4$. The Courant number $\mathtt{CFL} = 0.4$.

Numerical solutions have been obtained on a $200 \times 100$ mesh by the ADER-DG-$\mathbb{P}_{5}$ method. The ADER-WENO2 finite-volume method is used as a posteriori limiter. The main results obtained are presented in Figure~\ref{fig:khi_2d_08} at the time $t = 0.8$ and  in Figure~\ref{fig:khi_2d_20} at the time $t = 2.0$. To quantify the accuracy of the numerical solution, the results in the case of the ADER-DG-$\mathbb{P}_{2}$ method are also presented for comparison.

The presented results allowed to conclude that the ADER-DG-$\mathbb{P}_{5}$ method with a posteriori ADER-WENO2 finite-volume limitation allows us to obtain a solution of very high accuracy for the selected size of the spatial mesh $200 \times 100$. A direct comparison with the solution presented in the work~\cite{Springel_mnras_2015} seems pointless, because it, of course, has much greater accuracy, which is associated with the use of AMR and a much larger spatial mesh size. However, taking into account the differences between spatial meshes, it can be said that the solution in this work agrees well with the results presented in the work~\cite{Springel_mnras_2015}.

In the numerical solution obtained using the ADER-DG-$\mathbb{P}_{5}$ method, a high accuracy of resolution of small-scale eddies, significantly smaller than the coordinate step of the spatial mesh, is observed. In dynamics, an evolutionary process of the formation of a small-scale vortex structure is observed, followed by the coarsening of small-scale disturbances into large vortex structures. The accuracy of the resolution of the vortex structures of the shear layer is clearly visible in the coordinate dependence of the pressure $p$, where the cores of the main vortex structures correspond to local pressure minima.

The comparison with the results obtained using the ADER-DG-$\mathbb{P}_{2}$ shows that the use of $N = 2$ leads to a significantly less accurate resolution of the vortex generation process, which, in principle, was initially expected. However, it should be noted that the numerical solution in the case of $N = 2$ correctly resolves large-scale vortex structures --- a direct comparison of the coordinate dependencies of the density shows that large-scale vortex structures are located in approximately the same coordinates in the numerical solution and demonstrate similar dynamics as in the case of the solution obtained using the ADER-DG-$\mathbb{P}_{5}$ method. In numerical solution obtained using the ADER-DG-$\mathbb{P}_{2}$, premature merging of vortex structures is observed, which are resolved separately for a long time in numerical solution obtained using the ADER-DG-$\mathbb{P}_{5}$.

It should be noted that there are very few troubled cells in the solution and all of them are irregularly distributed over the spatial mesh. This is due to the fact that in solving the problem, shock waves of significant amplitude are not formed, the front of which could be distinguished against the general background of the flow. This result applies to both method ADER-DG-$\mathbb{P}_{5}$ and method ADER-DG-$\mathbb{P}_{2}$. However, in the case of method ADER-DG-$\mathbb{P}_{2}$, there are usually slightly more troubled cells than in the case of method ADER-DG-$\mathbb{P}_{5}$.

\subsection{Cylindrical and spherical explosion problems}
\label{sec:apps_cgd_problems:sod_md}

Explosion problems occupy an interesting place among two-dimensional and three-dimensional problems. On the one hand, these are conceptually one-dimensional problems, where the solution depends only on one spatial coordinate --- the distance $r$ to the center of the explosion, so these problems may seem quite simple. On the other hand, the use of two-dimensional and three-dimensional computational codes to solve these problems can reveal certain problems of numerical methods and their implementations associated with maintaining the spatial symmetry of the original problem --- axial in the two-dimensional case and spherical in the three-dimensional case. The formulation of the explosion problem has been chosen in the form of a multidimensional Sod problem:
\begin{equation}\label{eq:sod_2d_init}
(\rho, |\mathbf{v}|, p)(\mathbf{r}, t = 0) = \left\{
\begin{array}{ll}
(1.000, 0.0, 1.0), & \mathrm{if}\ r \leqslant 0.5; \\
(0.125, 0.0, 0.1), & \mathrm{if}\ r >\, 0.5; \\
\end{array}
\right.
\end{equation}
where $r$ is the distance to the center of the explosion, $|\mathbf{v}|$ is the absolute value of flow velocity: $r^{2} = x^{2} + y^{2}$ and $|\mathbf{v}|^{2} = u^{2} + v^{2}$ in two-dimensional case, $r^{2} = x^{2} + y^{2} + z^{2}$ and $|\mathbf{v}|^{2} = u^{2} + v^{2} + w^{2}$ in three-dimensional case. The computational coordinate domain is chosen in the form of a $d$-dimensional cube $\Omega = [-1, +1]^{d}$, where $d = 2$ in two-dimensional case and $d = 3$ in three-dimensional case. The boundary conditions are specified in the form of free outflow conditions. The final time is chosen $t_{\rm final} = 0.25$ for two-dimensional and three-dimensional explosion problems. The adiabatic index $\gamma = 1.4$. The Courant number $\mathtt{CFL} = 0.4$. A one-dimensional problem with a geometric source term is used to obtain a reference solution in two-dimensional and three-dimensional cases.

\begin{figure*}[h!]
\begin{minipage}{1.0\textwidth}
\centering
\includegraphics[width=0.245\textwidth]{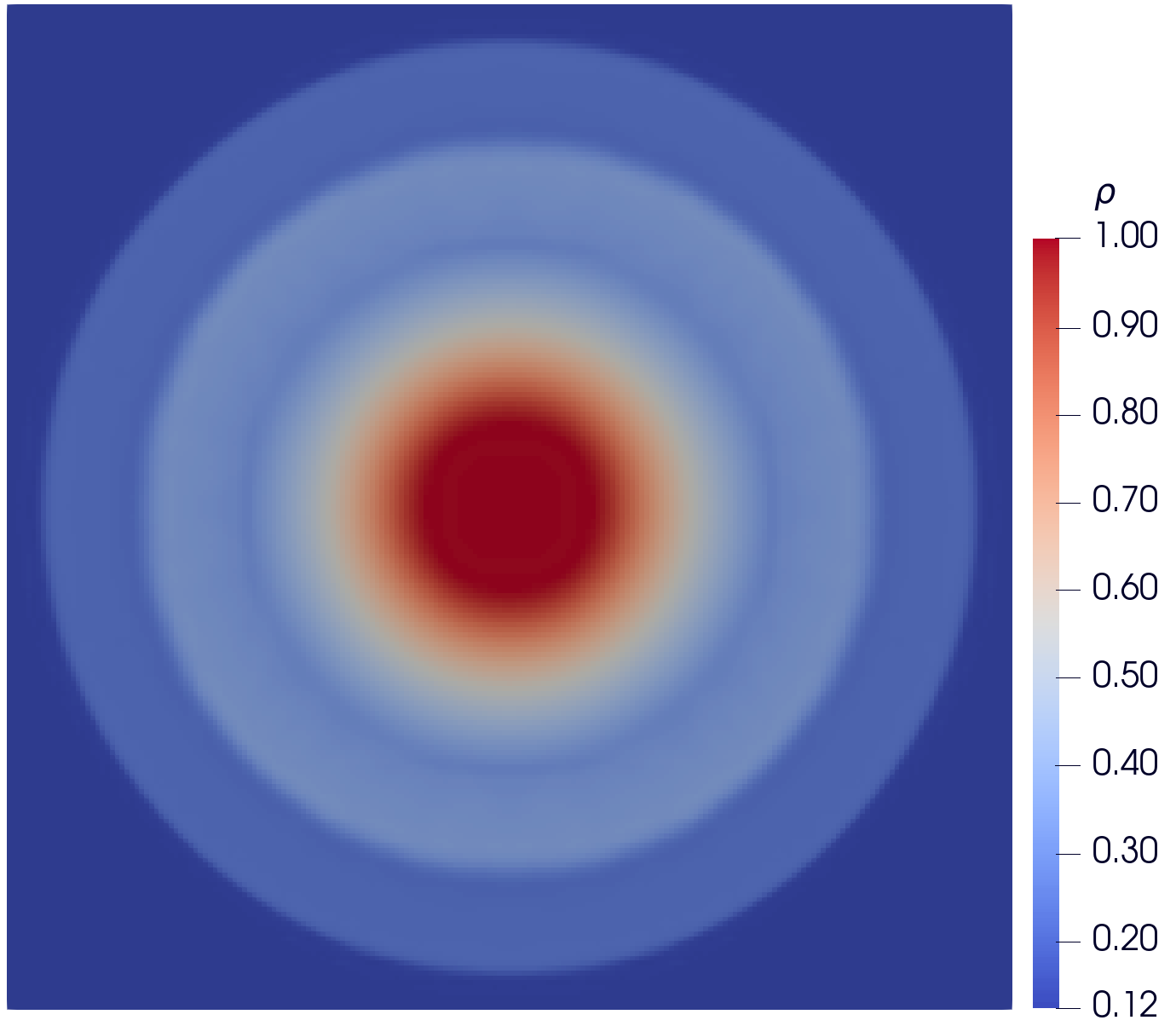}
\includegraphics[width=0.245\textwidth]{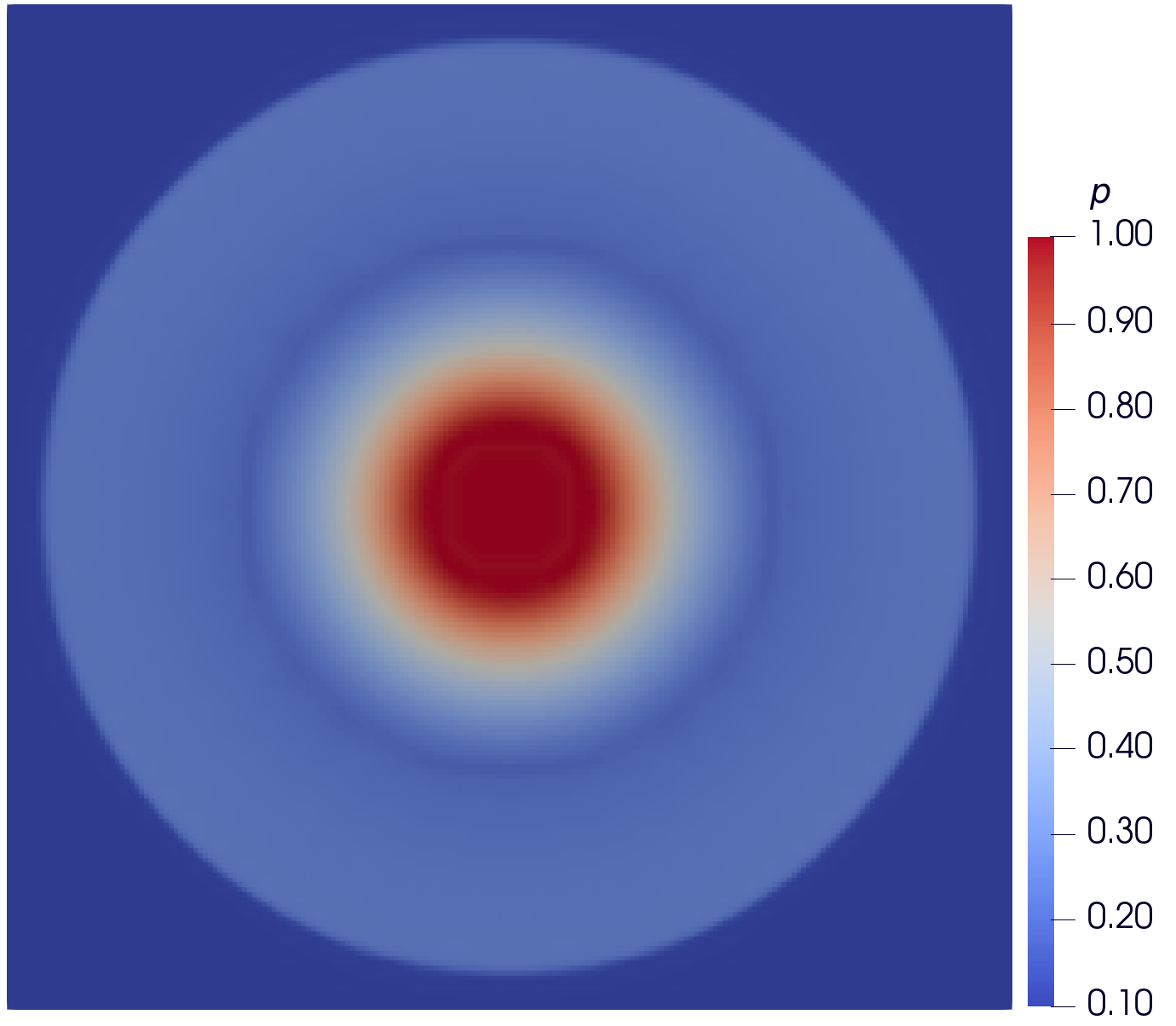}
\includegraphics[width=0.245\textwidth]{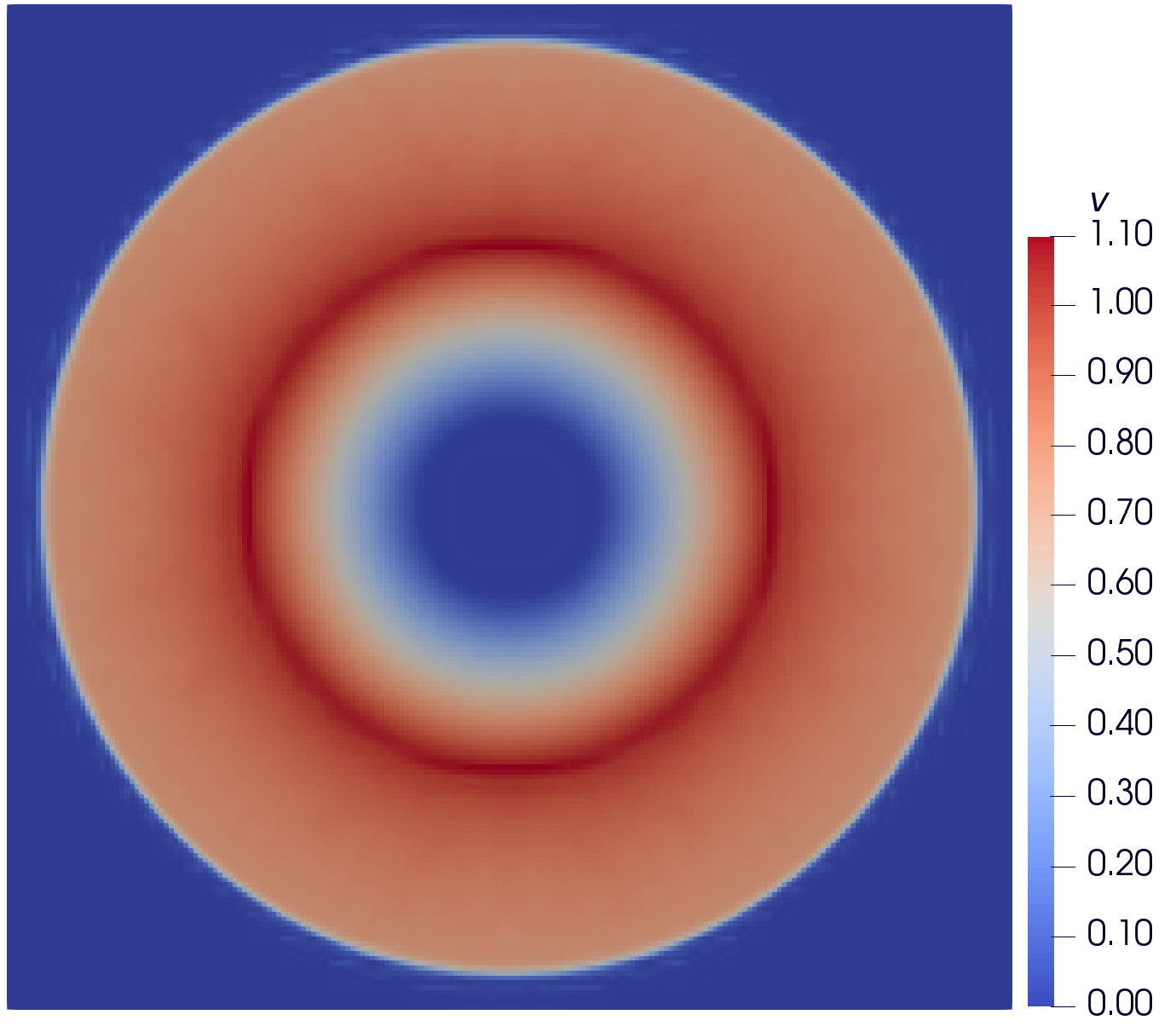}
\includegraphics[width=0.245\textwidth]{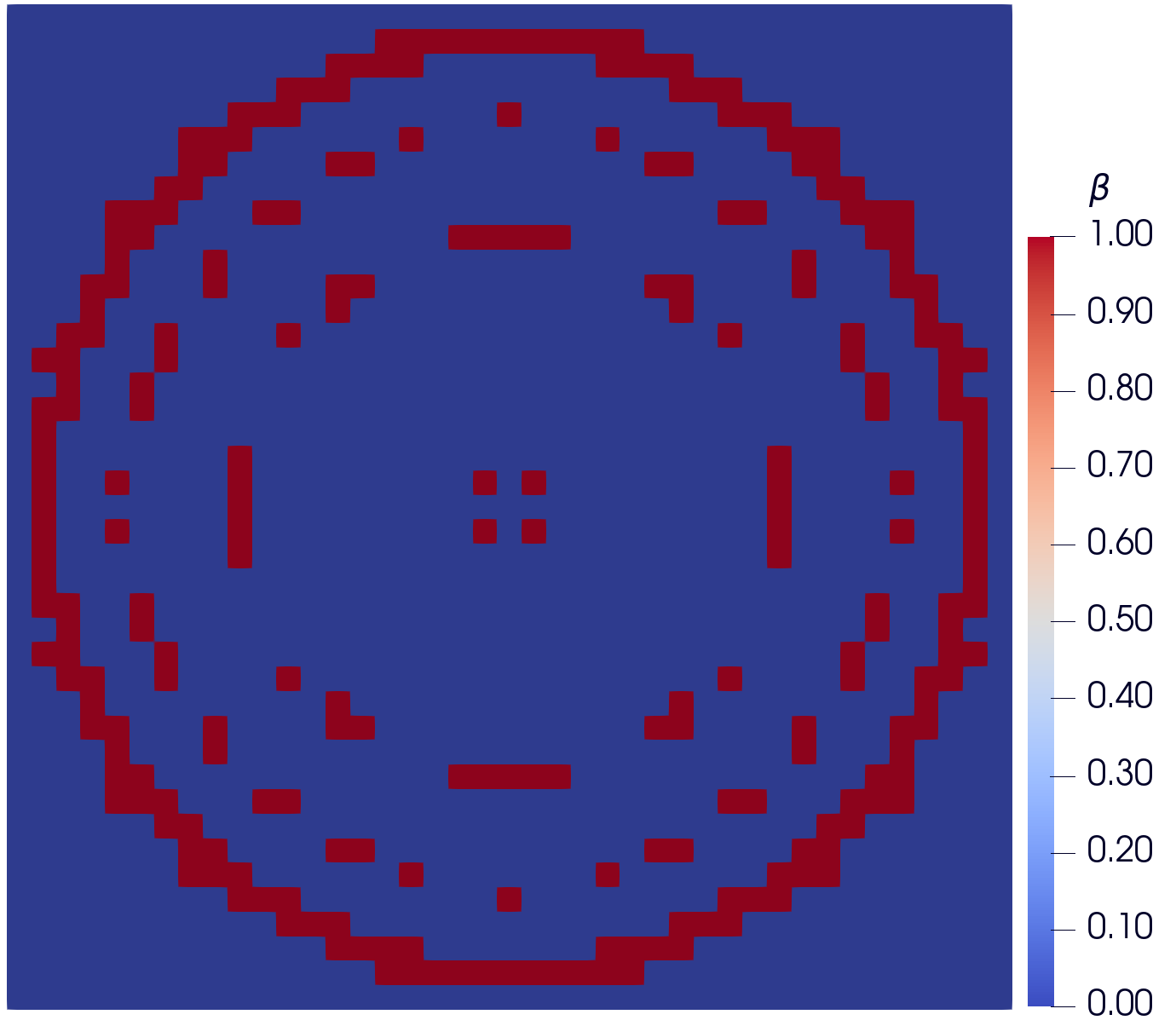}\\
\includegraphics[width=0.245\textwidth]{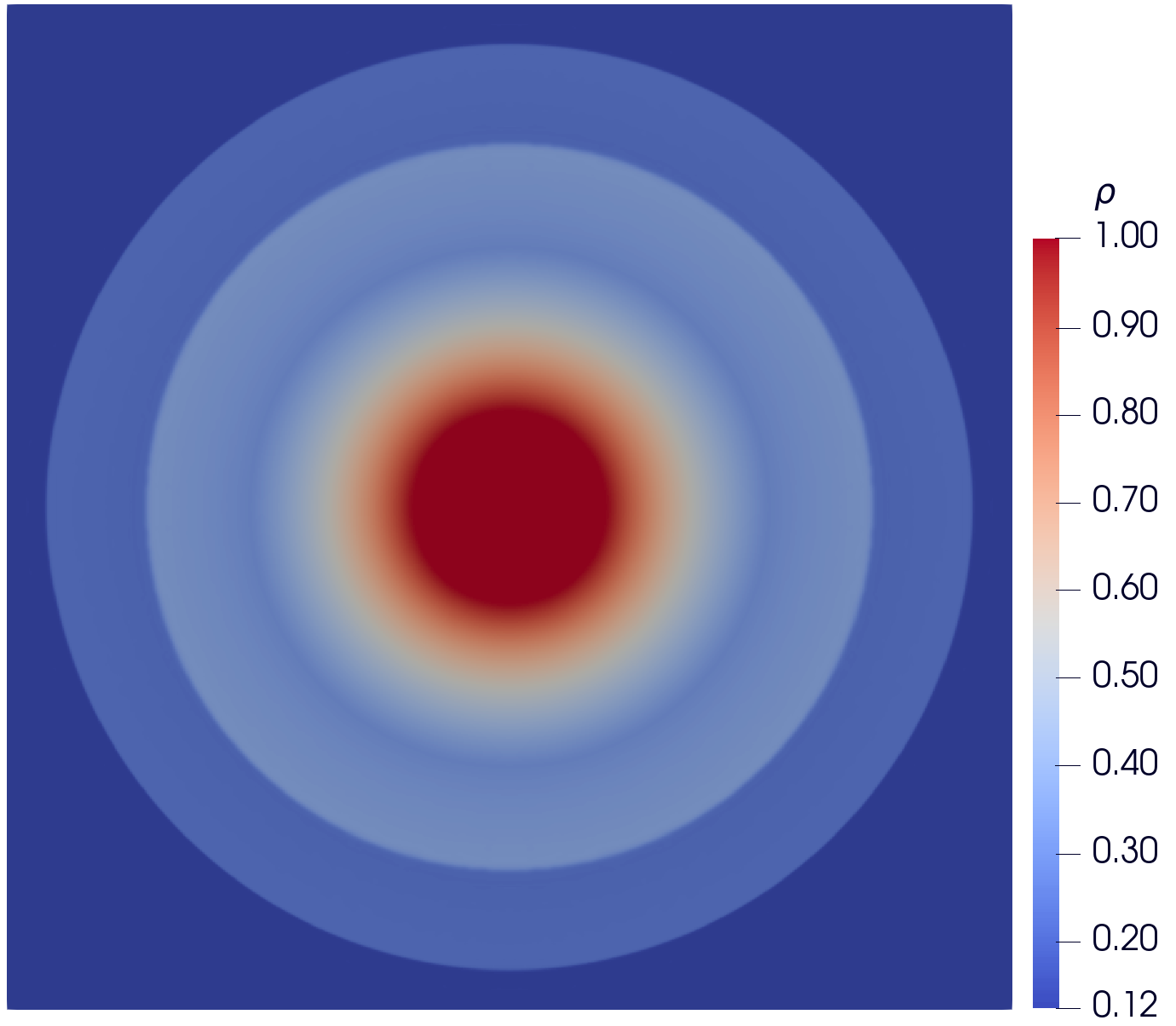}
\includegraphics[width=0.245\textwidth]{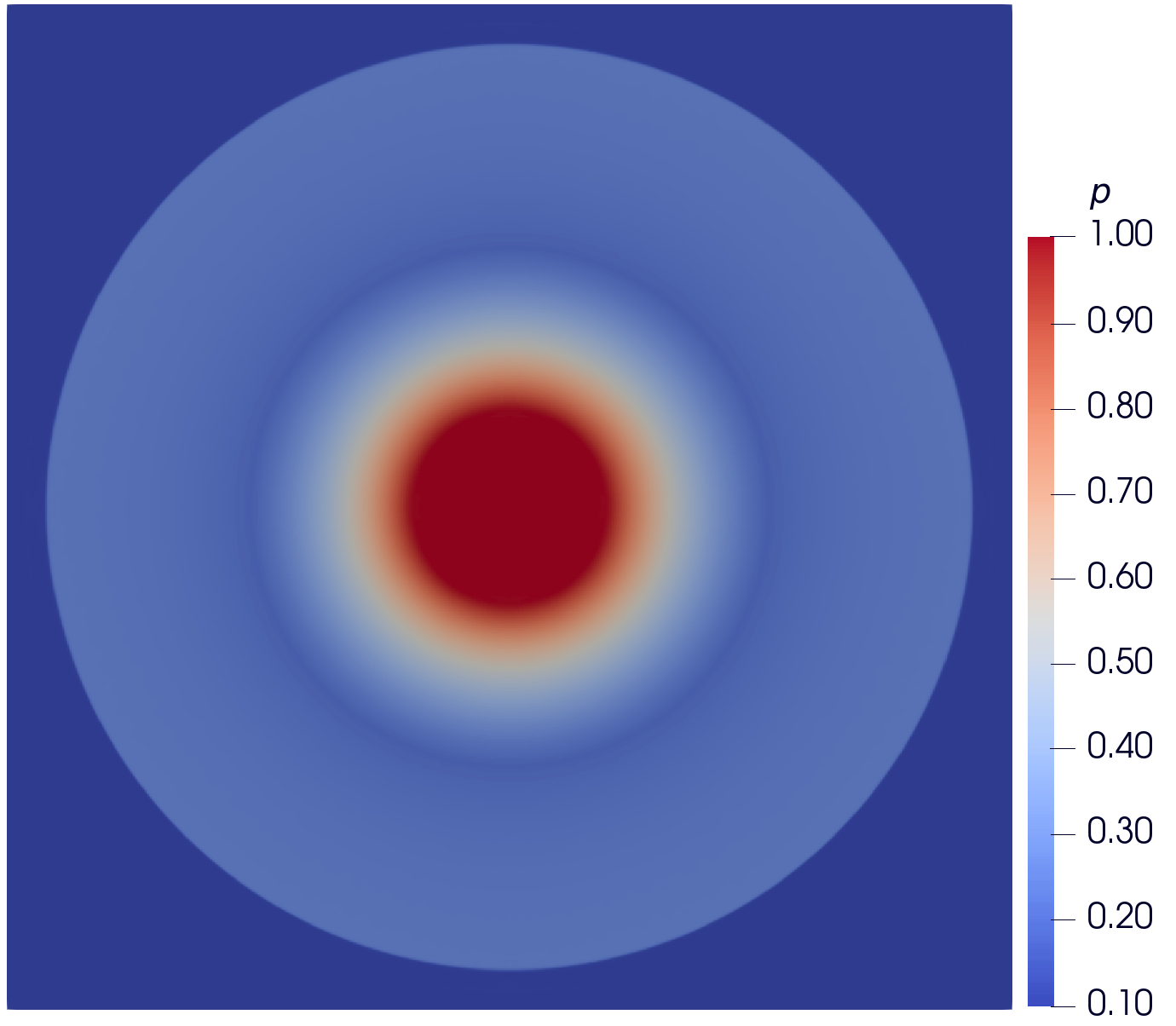}
\includegraphics[width=0.245\textwidth]{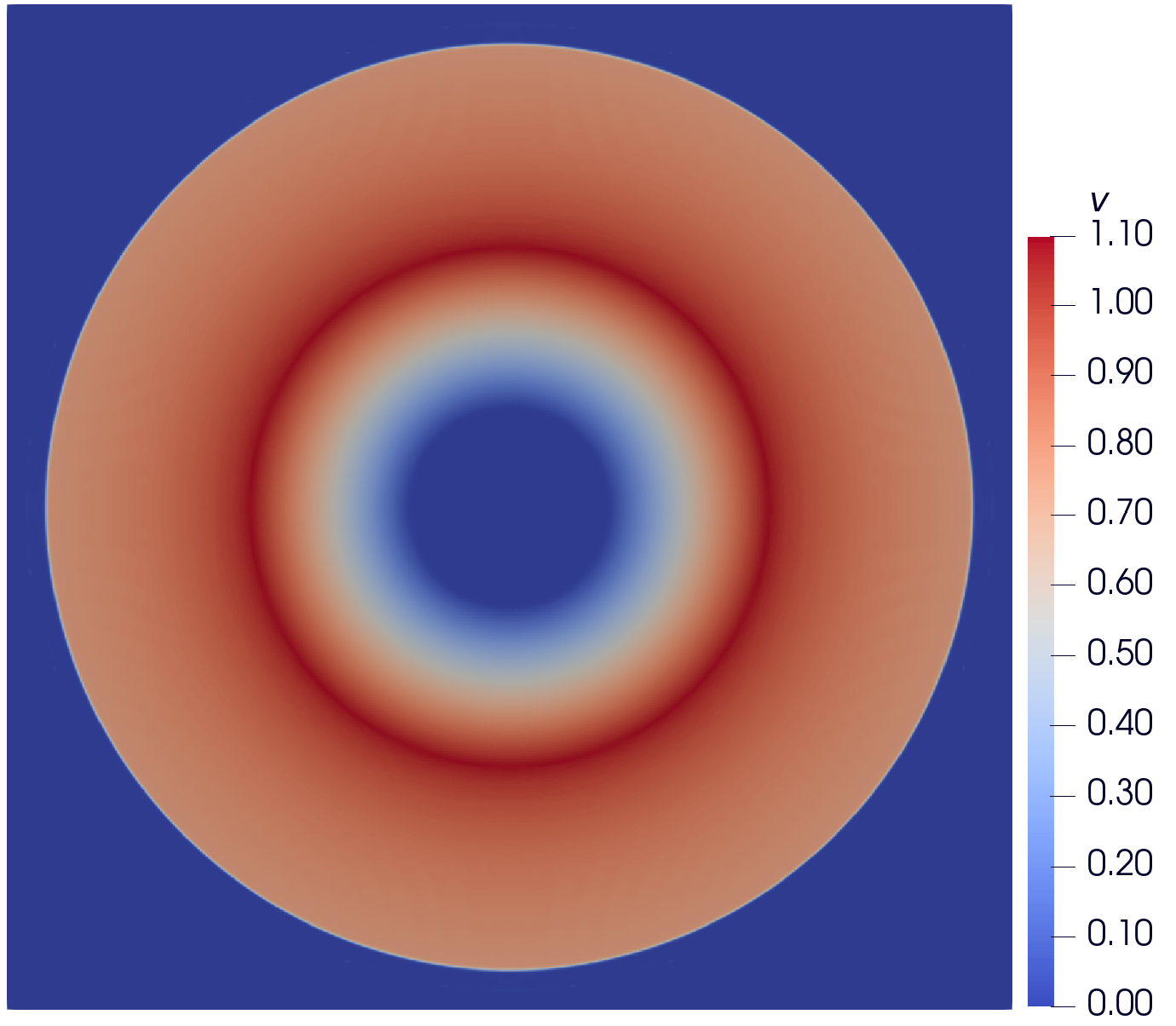}
\includegraphics[width=0.245\textwidth]{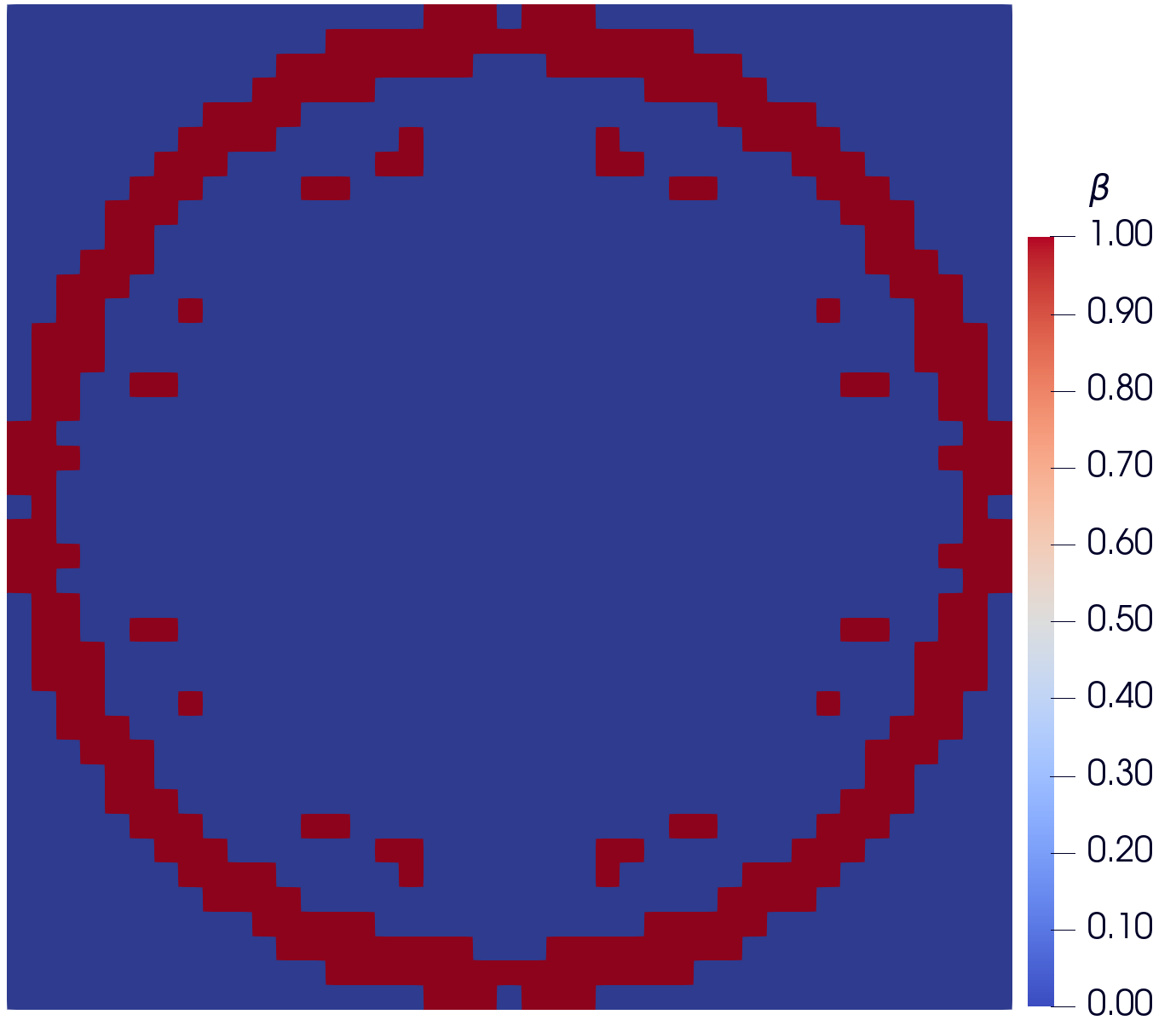}
\caption{\label{fig:sod_2d}%
Numerical solution of the two-dimensional Sod explosion problem (a detailed statement of the problem is presented in the text)
obtained using the ADER-DG-$\mathbb{P}_{2}$ (top) and ADER-DG-$\mathbb{P}_{9}$ (bottom) methods on mesh with $41 \times 41$ cells.
The graphs show the coordinate dependencies of the subcells finite-volume representation of density $\rho$, pressure $p$, flow velocity magnitude $v$ 
and troubled cells indicator $\beta$ (from left to right) at the final time $t_{\rm final} = 0.25$.
}\vspace{10mm}
\includegraphics[width=0.32\textwidth]{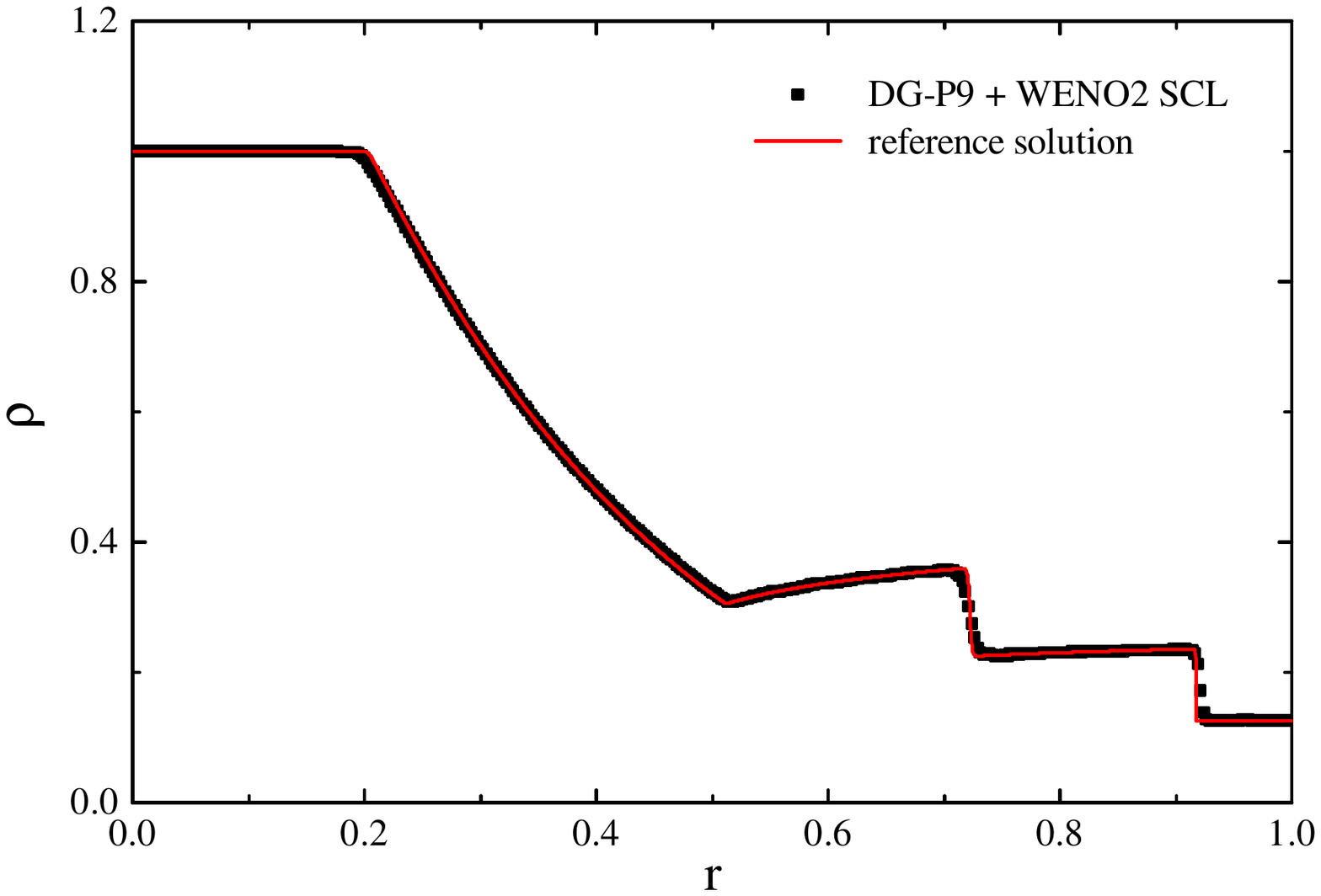}
\includegraphics[width=0.32\textwidth]{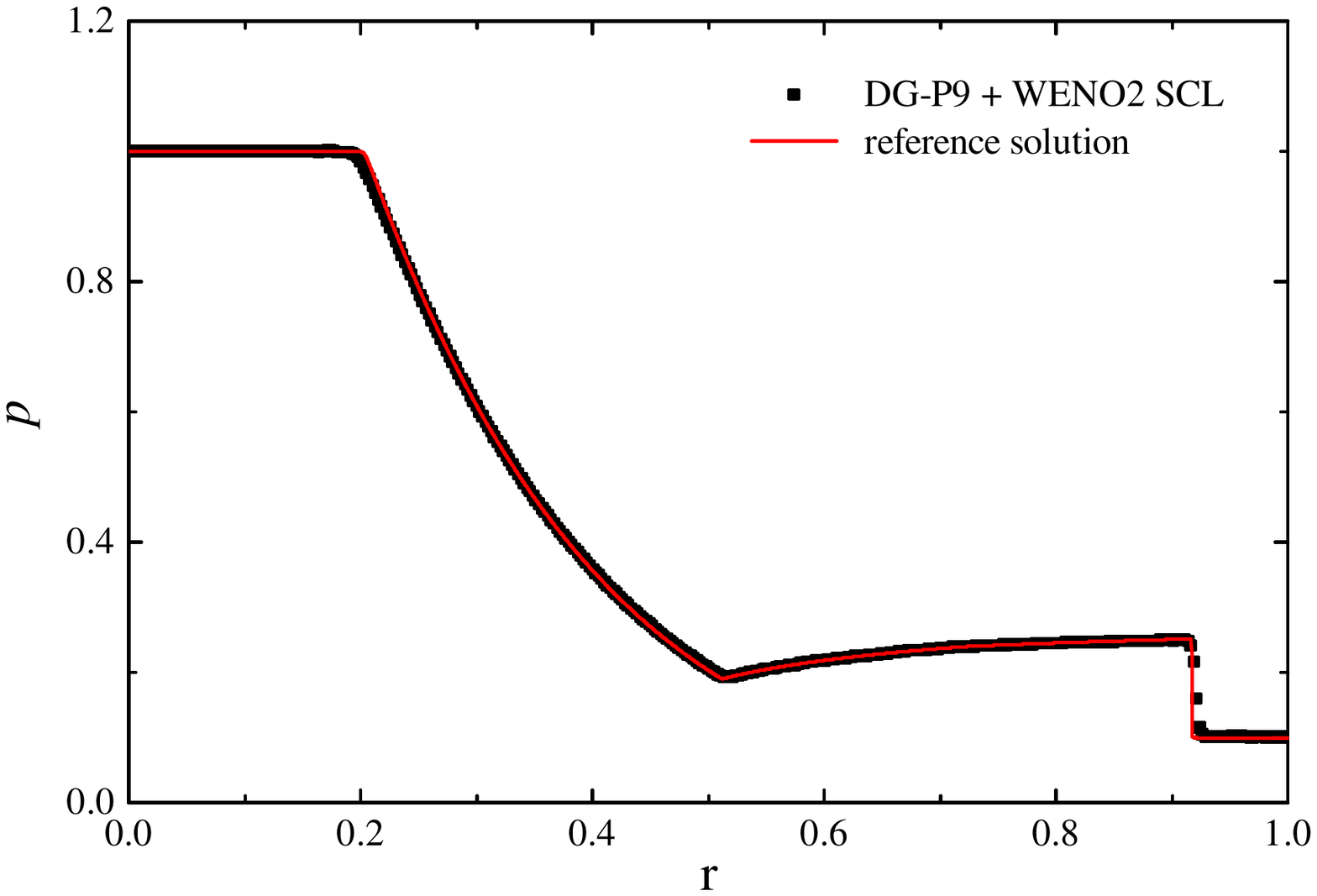}
\includegraphics[width=0.32\textwidth]{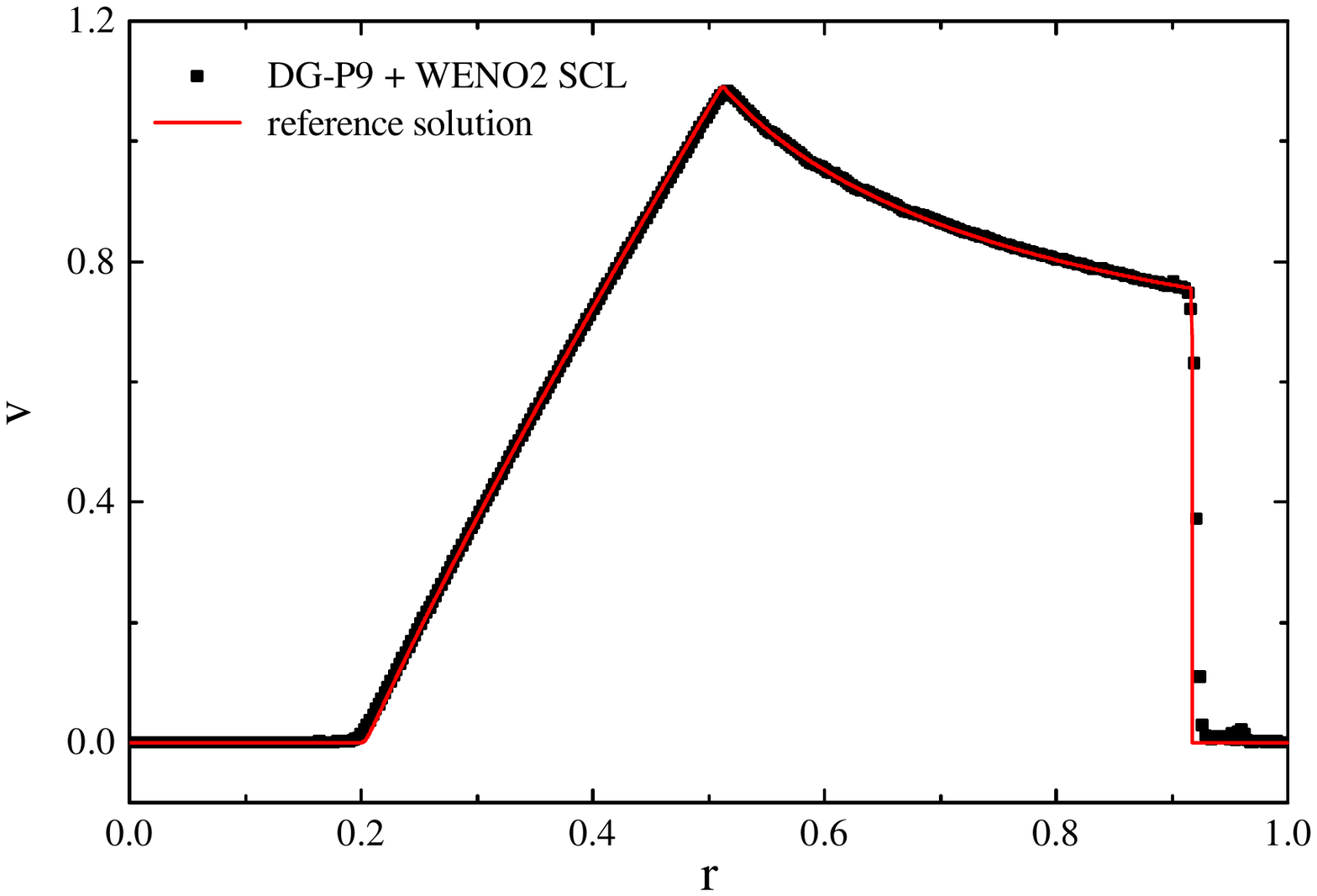}
\caption{\label{fig:sod_2d_slice}%
One dimensional cuts of the numerical solution for two-dimensional Sod explosion problem in Figure~\ref{fig:sod_2d},
obtained using the ADER-DG-$\mathbb{P}_{9}$ method on mesh with $41 \times 41$ cells.
The graphs show the coordinate dependence of density $\rho$, pressure $p$ and flow velocity magnitude $v$ (from left to right) 
on the distance $r$ to the point $(0, 0)$ along the direction $(0, 1)$.
The black square symbols represent the subcells finite-volume representation of the numerical solution; 
the red solid lines represents the reference solution of the problem.
}
\end{minipage}
\end{figure*}

\begin{figure*}[h!]
\begin{minipage}{1.0\textwidth}
\centering
\includegraphics[width=0.32\textwidth]{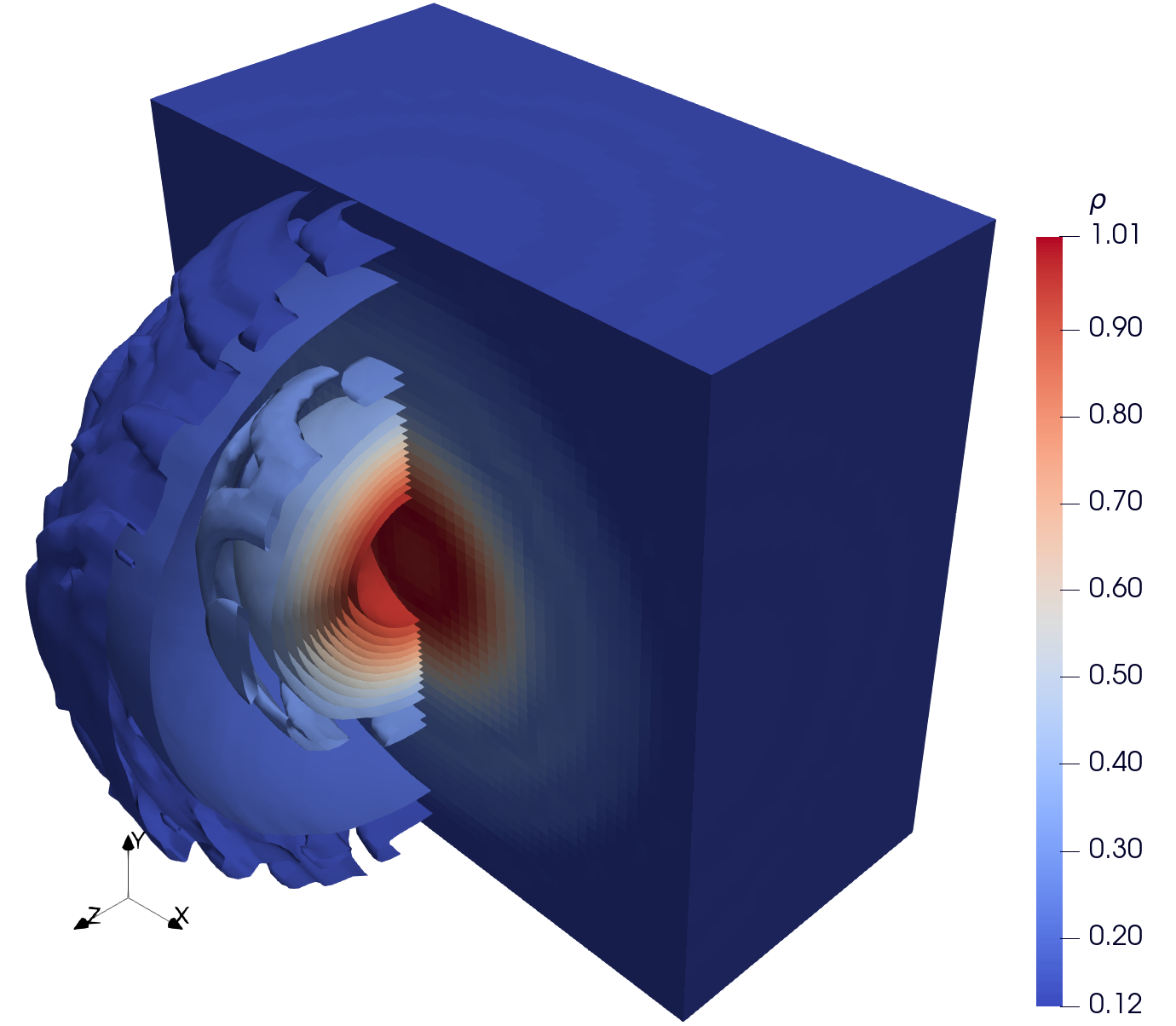}
\includegraphics[width=0.32\textwidth]{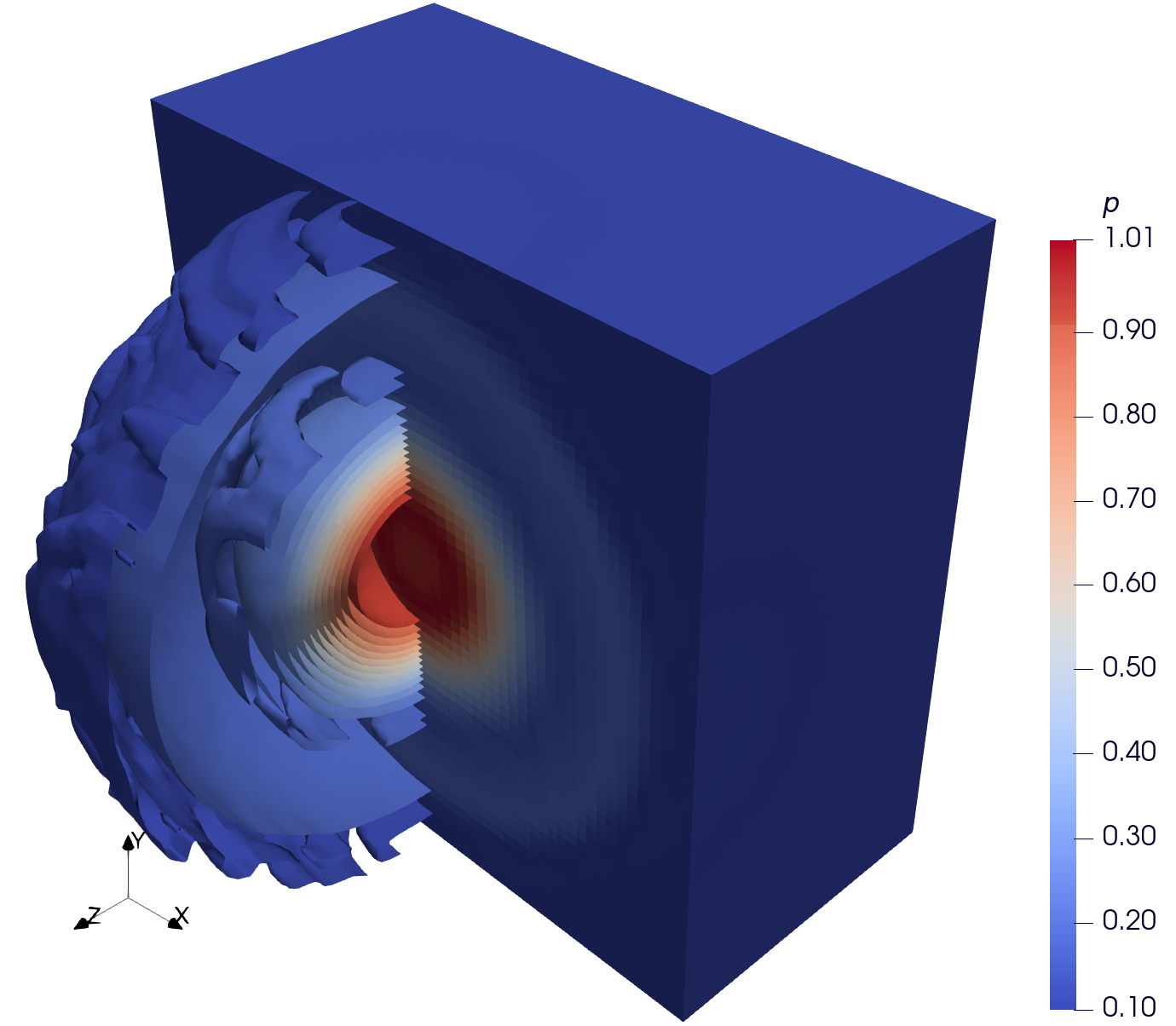}
\includegraphics[width=0.32\textwidth]{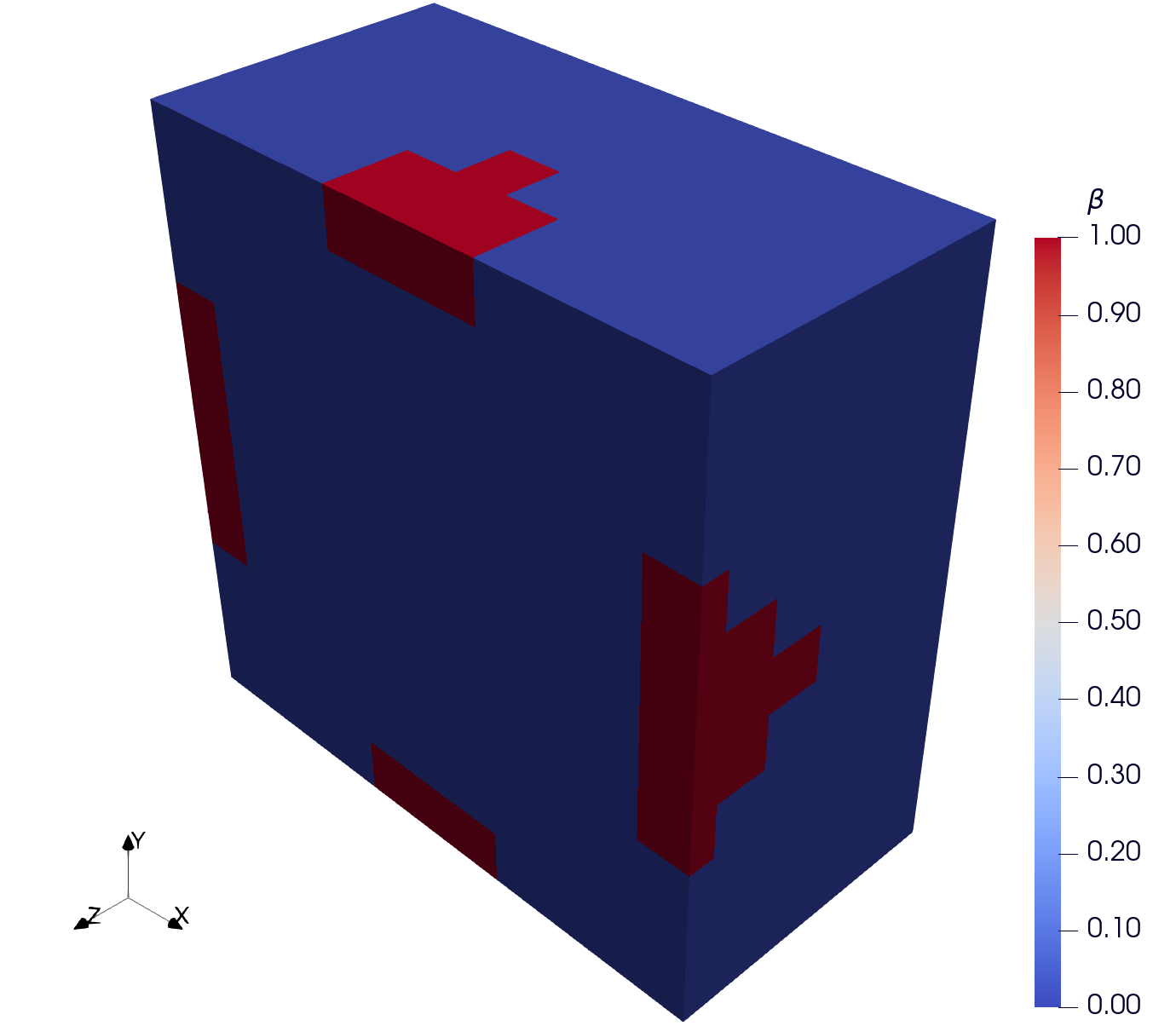}\\
\includegraphics[width=0.32\textwidth]{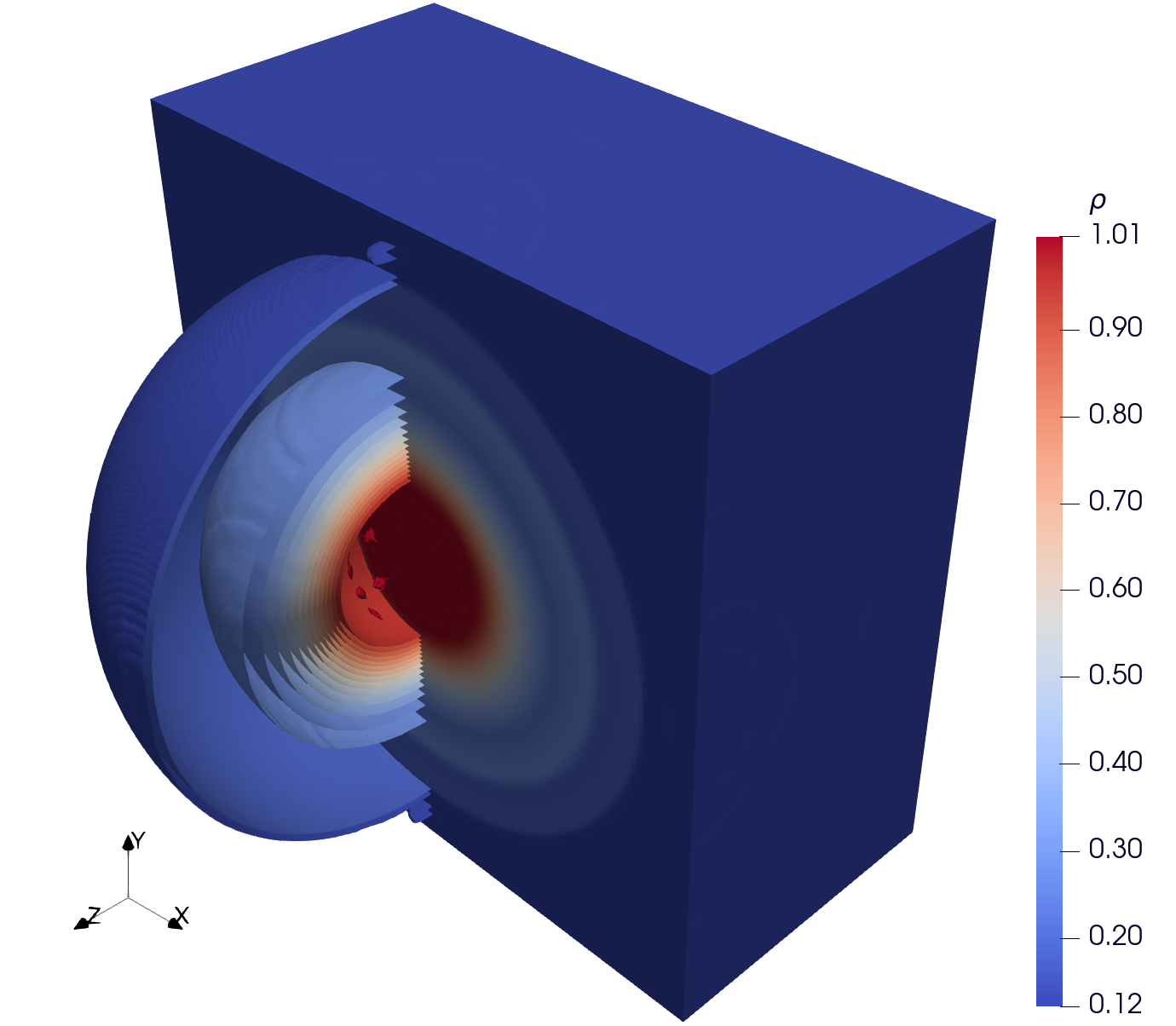}
\includegraphics[width=0.32\textwidth]{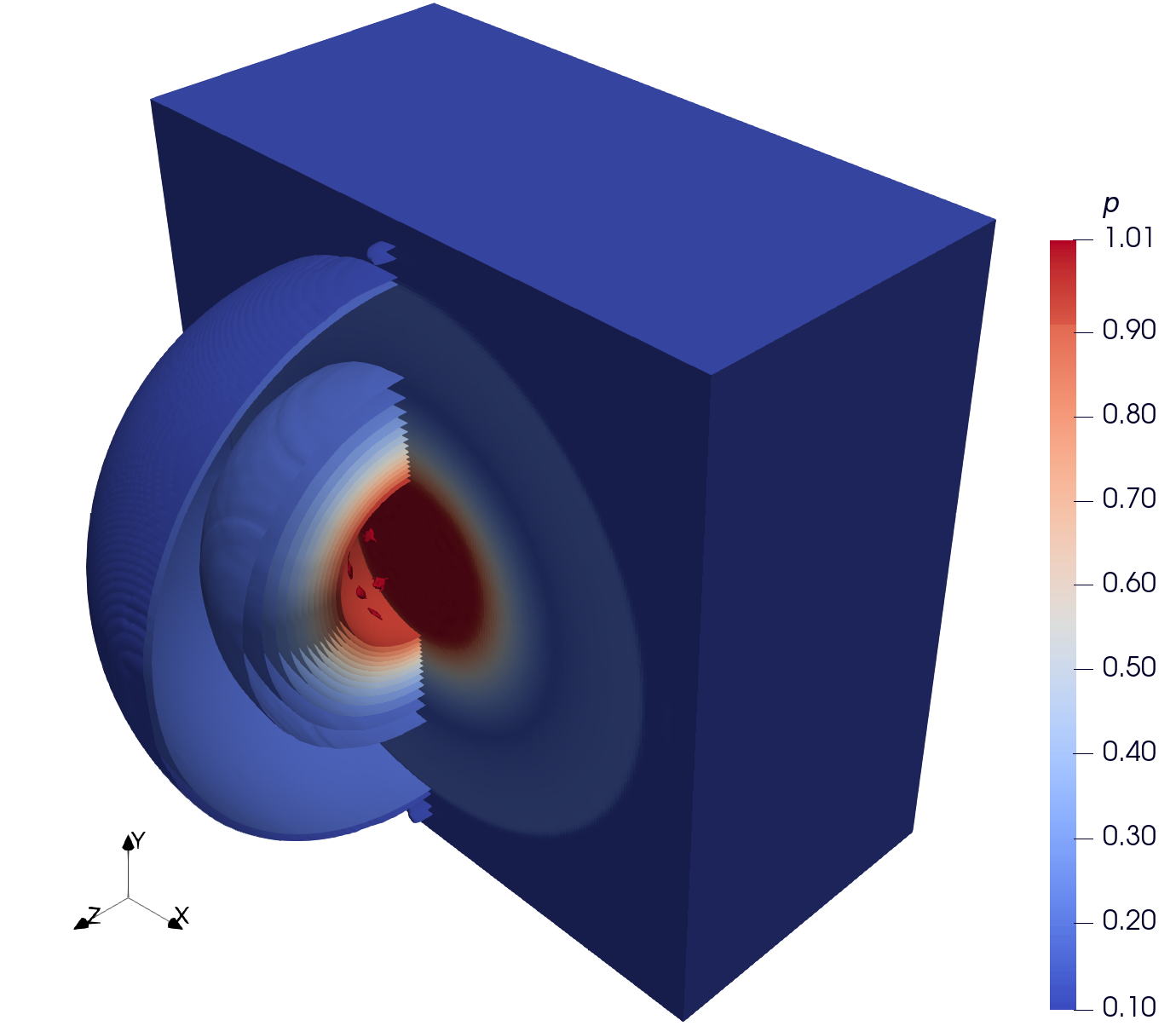}
\includegraphics[width=0.32\textwidth]{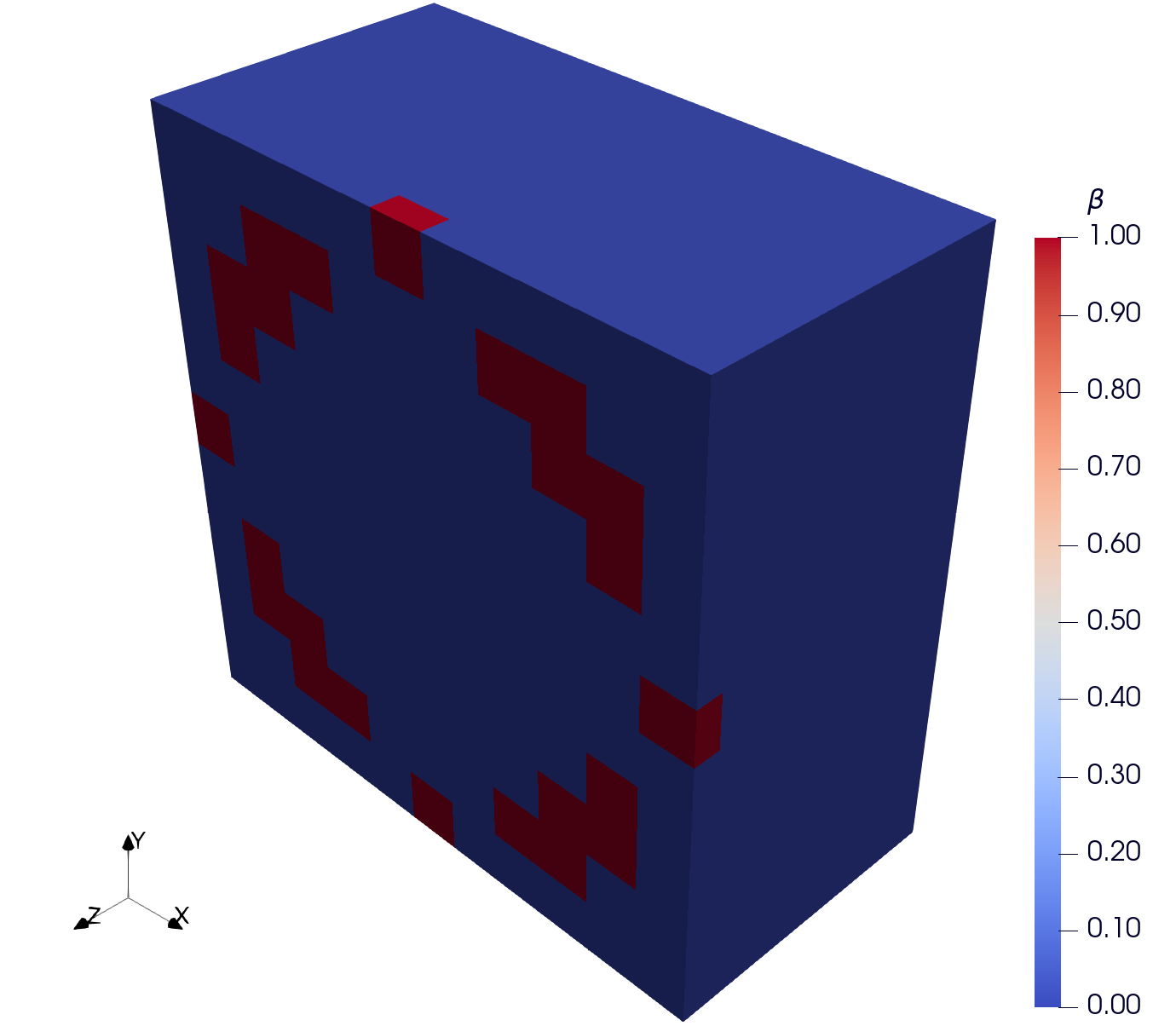}
\caption{\label{fig:sod_3d}%
Numerical solution of the three-dimensional Sod explosion problem (a detailed statement of the problem is presented in the text)
obtained using the ADER-DG-$\mathbb{P}_{2}$ (top) and ADER-DG-$\mathbb{P}_{9}$ (bottom) methods on mesh with $11 \times 11 \times 11$ cells.
The graphs show the coordinate dependencies of the subcells finite-volume representation of density $\rho$, pressure $p$
and troubled cells indicator $\beta$ (from left to right) at the final time $t_{\rm final} = 0.25$.
The opaque fill represents the coordinate domain clip $\Omega_{\rm clip} = \left\{\mathbf{r}\, |\, \mathbf{r} \in [-1, +1]^{3} \land z \geqslant 0 \right\}$; 
the left and center columns also represent density and pressure isosurfaces, uniformly distributed between the boundary values.
}\vspace{10mm}
\includegraphics[width=0.32\textwidth]{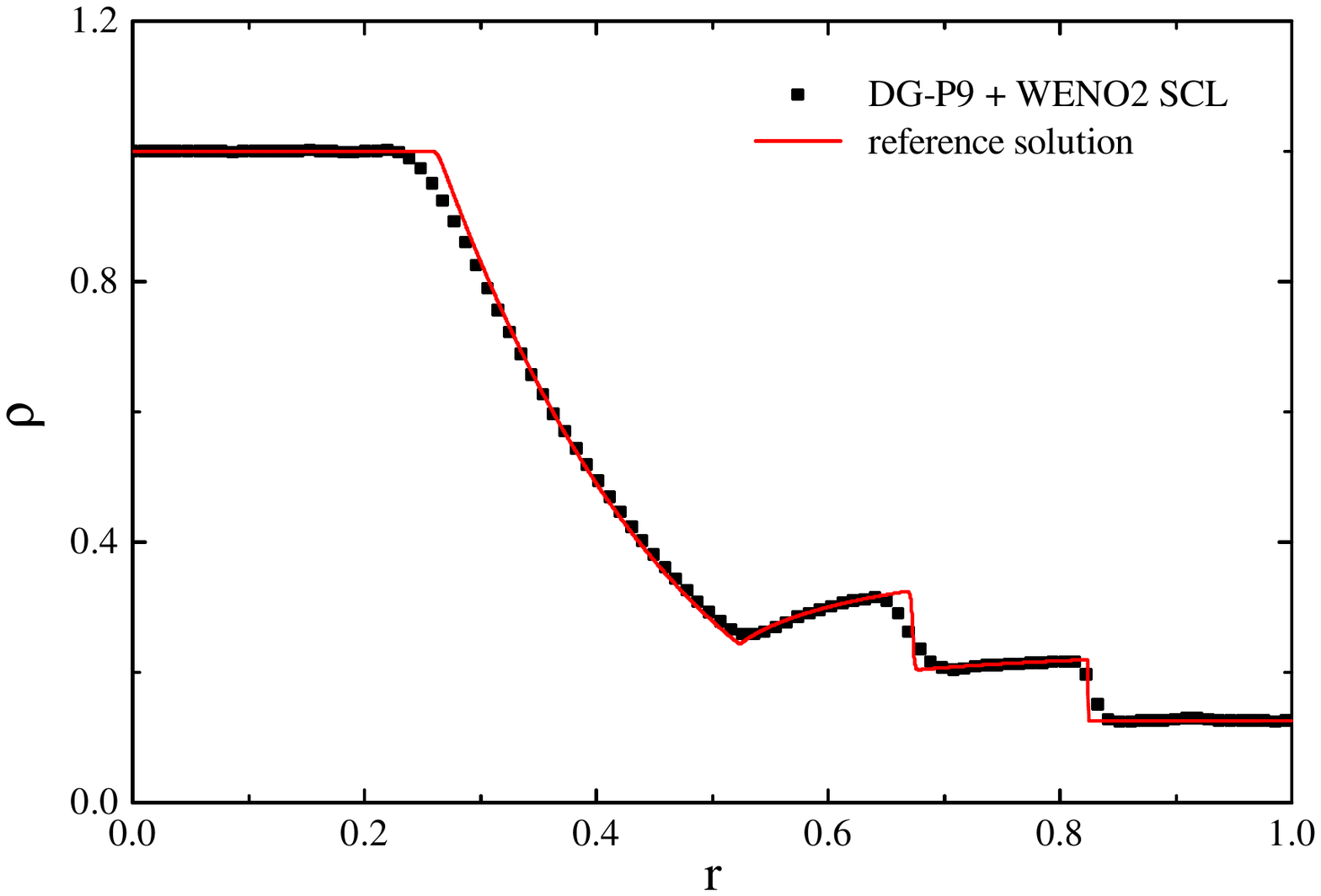}
\includegraphics[width=0.32\textwidth]{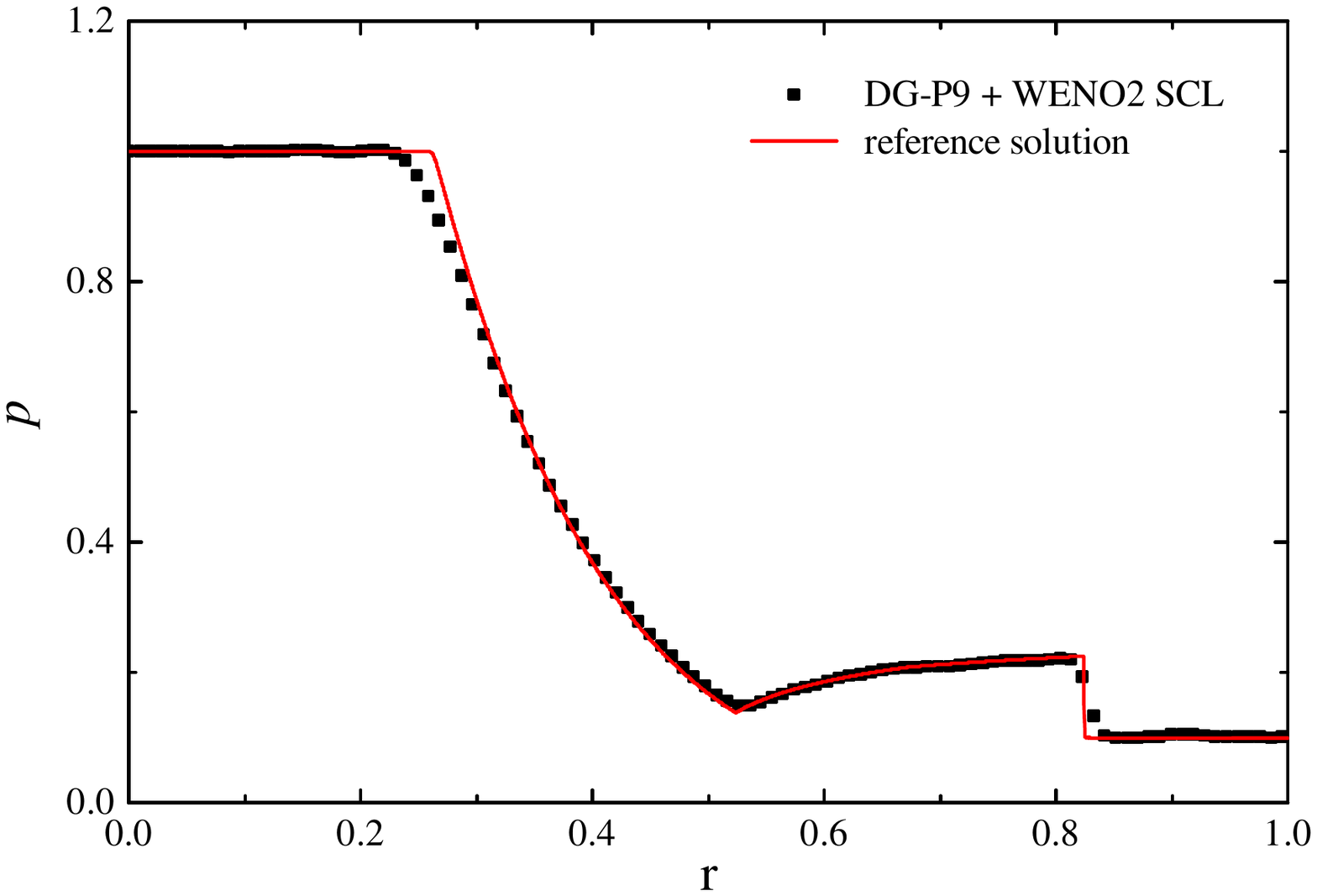}
\includegraphics[width=0.32\textwidth]{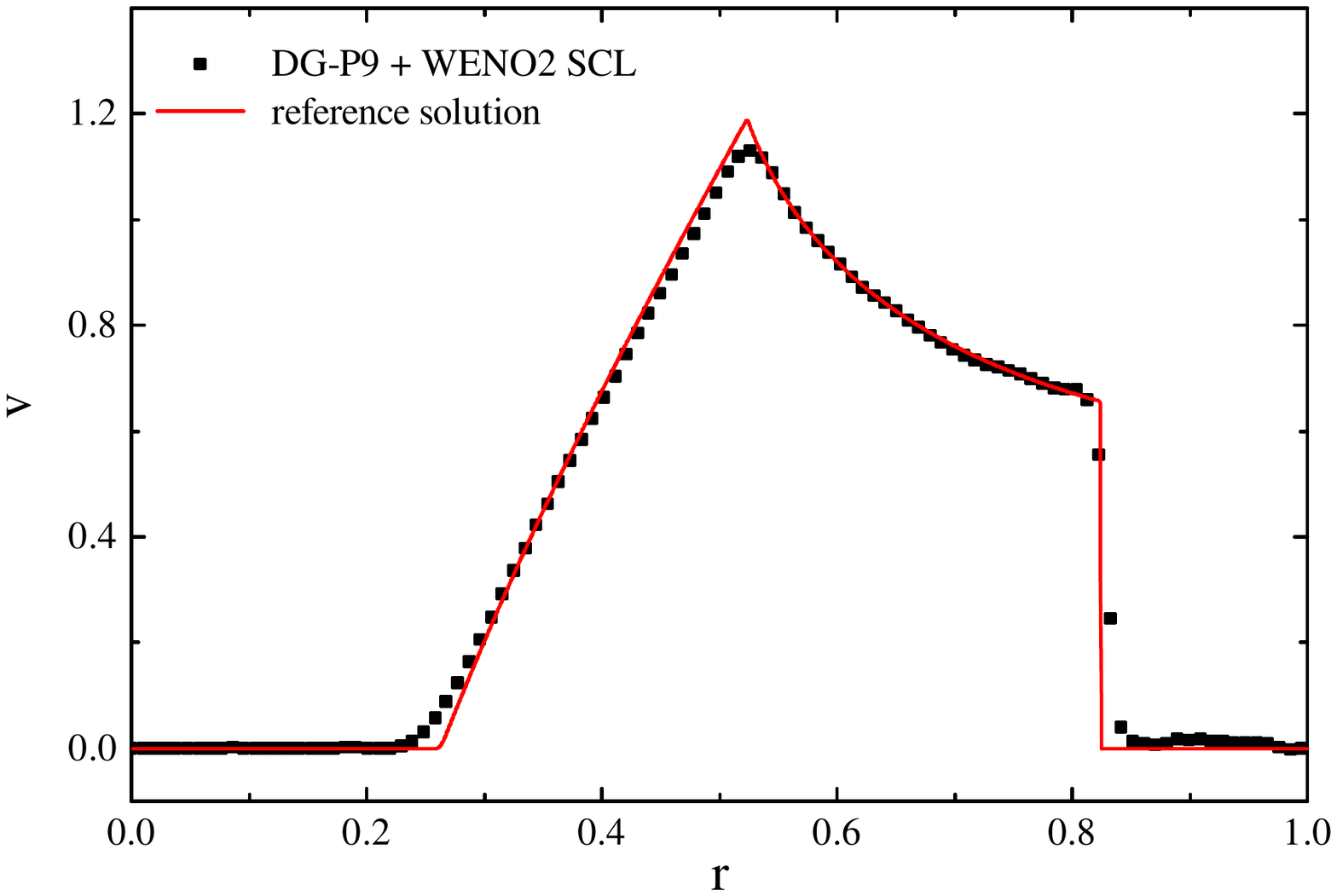}
\caption{\label{fig:sod_3d_slice}%
One dimensional cuts of the numerical solution for three-dimensional Sod explosion problem in Figure~\ref{fig:sod_3d},
obtained using the ADER-DG-$\mathbb{P}_{9}$ method on mesh with $11 \times 11 \times 11$ cells.
The graphs show the coordinate dependence of density $\rho$, pressure $p$ and flow velocity magnitude $v$ (from left to right)
on the distance $r$ to the point $(0, 0, 0)$ along the direction $(0, 0, 1)$.
The black square symbols represent the subcells finite-volume representation of the numerical solution; 
the red solid lines represents the reference solution of the problem.
}
\end{minipage}
\end{figure*}

Numerical solution to the cylindrical explosion problem has been obtained on a spatial mesh with size $41 \times 41$. This size, in this case and further in this work, \newtext{is used to obtain} the one-dimensional cut of a numerical solution along one of the coordinate directions, in order to compare it with the reference solution. The main calculations are performed using the ADER-DG-$\mathbb{P}_{9}$ method with ADER-WENO2 finite volume method used as a posteriori limiter. The main results obtained for cylindrical explosion problem are presented in Figure~\ref{fig:sod_2d}. To quantify the accuracy of the numerical solution, the results in the case of using $N = 2$ are also presented for comparison. The presented results show that the ADER-DG-$\mathbb{P}_{9}$ method with ADER-WENO2 finite volume a posteriori limiter correctly resolves the expanding shock wave and contact discontinuity and the converging rarefaction wave in a cylindrical explosive flow. Two-dimensional coordinate dependencies of pressure $p$, density $\rho$ and absolute velocity $\mathrm{v}$ are axially symmetric --- the numerical method correctly preserves the axial symmetry properties of the flow. The comparison of one dimensional cuts of the numerical solution for two-dimensional Sod explosion problem, obtained using the ADER-DG-$\mathbb{P}_{9}$ method, and reference solution, obtained using one-dimensional problem with a geometric source term, is presented in Figure~\ref{fig:sod_2d_slice}. From the presented comparison it is clear that strong discontinuous components of the flow --- the shock wave and the contact discontinuity, are resolved by a numerical method with subgrid resolution --- the width of the shock wave front is $6$-$8$ subcells, the width of the contact discontinuity front does not exceed $9$-$11$ cells, while the cell size in one direction is $2N+1 = 19$ subcells for degree $N = 9$ polynomials in DG representation. The existence of a contact discontinuity does not manifest itself in any way in the coordinate dependencies of pressure $p$ and velocity $\mathrm{v}$. The rarefaction wave compares very well with the reference solution. It can be concluded that the ADER-DG-$\mathbb{P}_{N}$ method with ADER-WENO finite volume a posteriori limiter and its software implementation make it possible to obtain a numerical solution to the cylindrical explosion problem with very high accuracy, which corresponds to the results of the basic work~\cite{ader_dg_dev_1}.

Numerical solution to the spherical explosion problem has been obtained on a spatial mesh with size $19 \times 19 \times 19$. The main calculations are performed using the ADER-DG-$\mathbb{P}_{9}$ method with ADER-WENO2 finite volume method used as a posteriori limiter. The main results obtained for spherical explosion problem are presented in Figure~\ref{fig:sod_3d}. To quantify the accuracy of the numerical solution, the results in the case of using $N = 2$ are also presented for comparison. The presented results show that the ADER-DG-$\mathbb{P}_{9}$ method with ADER-WENO2 finite volume a posteriori limiter correctly resolves the expanding shock wave and contact discontinuity and the converging rarefaction wave in a spherical explosive flow. Three-dimensional coordinate dependencies of pressure $p$, density $\rho$ and absolute velocity $\mathrm{v}$ are axially symmetric --- the numerical method correctly preserves the spherical symmetry properties of the flow. The density $\rho$ and pressure $p$ isosurfaces have well-defined spherical symmetry, despite the rather coarse mesh. The comparison of one dimensional cuts of the numerical solution for three-dimensional Sod explosion problem, obtained using the ADER-DG-$\mathbb{P}_{9}$ method, and reference solution, obtained using one-dimensional problem with a geometric source term, is presented in Figure~\ref{fig:sod_3d_slice}. From the presented comparison it is clear that strong discontinuous components of the flow --- the shock wave and the contact discontinuity, are resolved by a numerical method with subgrid resolution --- the width of the shock wave front is $5$-$8$ subcells, the width of the contact discontinuity front does not exceed $9$-$11$ cells, while the cell size in one direction is $2N+1 = 19$ subcells for degree $N = 9$ polynomials in DG representation. The existence of a contact discontinuity does not manifest itself in any way in the coordinate dependencies of pressure $p$ and velocity $\mathrm{v}$. The rarefaction wave compares very well with the reference solution. It can be concluded that ADER-DG-$\mathbb{P}_{9}$ method with ADER-WENO2 finite volume a posteriori limiter and its software implementation make it possible to obtain a numerical solution to the cylindrical explosion problem with very high accuracy, which corresponds to the results of the basic work~\cite{ader_dg_dev_1}.

\section{Computational costs}
\label{sec:comp_costs}

The main objective of this work \newtext{is} to integrate the interface of high-performance computing of the BLAS into the implementation of the finite-element numerical methods ADER-DR and finite-volume numerical methods ADER-WENO. Therefore, it \newtext{is} of interest to determine the computational costs and compare them with the computational costs of other implementations, which allowed us to determine the computational performance of the developed efficient software implementation. This \newtext{section} presents a comparison of the computational costs of the ADER-DG-$\mathbb{P}_{N}$ method with a posteriori limitation of the solution by finite-volume ADER-WENO limiter obtained in four main cases:
\begin{enumerate}
\item An implementation without using the BLAS interface (completely vanilla implementation, CV implementation), which used the formula apparatus of the numerical methods of the ADER family, described in detail in the text above. The formula apparatus \newtext{is} explicitly implemented in the computational code. The explicit optimizations of the software implementation (represented in the computational code), including loop reordering and memory access pattern to improve memory locality~\cite{Robey_Zamora_hpc_2021, Drepper_2007, CS_PP_3ed}, 
\newtext{are} not used.

\item Simple optimized vanilla implementation (OV implementation). Unlike the CV implementation, in this case a simple implementation of the \texttt{gemm} function 
\newtext{is} used (taking into account the leading dimensions of matrices in BLAS), which has locality in memory and uses a simple classical memory access pattern~\cite{Robey_Zamora_hpc_2021, Drepper_2007, CS_PP_3ed}.

\item Implementation, using general BLAS \texttt{gemm} function of the \texttt{Intel MKL} software library (GEN implementation), included in the \texttt{Intel oneAPI} toolkit, while the optimizations available using the \texttt{MKL\_DIRECT\_CALL\_JIT} option, as well as the batch and strided \texttt{gemm} functions, \newtext{are} not used. It should be noted right away that the computational costs obtained using \texttt{OpenBLAS} in this case are similar to the computational costs obtained using the general BLAS functions of \texttt{Intel MKL}.

\item Implementation, using the JIT \texttt{gemm} functions of the \texttt{Intel MKL} software library (JIT implementation), included in the \texttt{Intel oneAPI} toolkit, optimized for operations with small matrices. The jitters 
\newtext{are} created in advance due to the known parameters of all necessary calls of the \texttt{gemm} functions. The choice between \texttt{MKL} JIT and \texttt{libxsmm} in favor of \texttt{MKL} JIT is related to that the use of \texttt{libxsmm} has already been presented in detail in the works~\cite{ader_dg_eff_impl, exahype}.
\end{enumerate}
The software implementation of the method \newtext{has been} developed using the \texttt{C++} programming language. Multithreading and multiprocessing execution was organized using the \texttt{OpenMP} and \texttt{MPI} interfaces. Software implementations were carried out separately for one-dimensional, two-dimensional and three-dimensional problems. All calculations were carried out on the HEDT class workstation with Intel i9-10980xe processor and 256 GB RAM. The computational costs were determined for calculations with single MPI process and 4 OpenMP threads.

The compilation and linking of the project's program code was performed using compiler \texttt{icpx}, included in the \texttt{Intel oneAPI} toolkit. The options for strict compliance with the C++23 standard were selected; it was from this standard that the possibility of effective use of \texttt{goto} to the label at end of compound statement for breaking a nested loop when checking the admissibility criteria was taken, which is recommended by ES.76 C++ Core Guidelines recommendation. The \texttt{O3}, \texttt{funroll-loops}, \texttt{xHOST} and \texttt{qopt-zmm-usage=high} optimization options were selected. The optimization report, generated by using \texttt{qopt-report} option, showed that for the selected optimization options, in the case of an optimized vanilla implementation (OV), the compiler managed to vectorized the loops associated with stream triad, included in the inner loop of the \texttt{gemm} function using a simple memory access pattern. In case of completely vanilla implementation (CV), the compiler optimization report showed unrolling and vectorization of only some loops.

The computational costs of finite-element ADER-DG method and finite-volume ADER-WENO method were determined using the problem of advection of a periodic wave in two-dimensional and three-dimensional formulations of the problem. The initial conditions and complete problem statement were chosen the same to the problem of sine wave advection considered in Subsection~\ref{sec:apps:acc_conv}. Therefore, this problem allowed us to carry out a separate study both for the finite-element ADER-DG method, the use of which in the case of discontinuous solutions would necessarily require the use of a limiter, and for the finite-volume ADER-WENO method. The presented problem has a smooth solution, therefore in the numerical ADER-DG methods trouble cells do not occur and the limiter was not activated, however, the numerical solution of the problem was obtained using the full implementation of the ADER-DG method with a posteriori correction of the solution in subcells by the finite-volume ADER-WENO limiter, therefore, despite the inactivity of the limiter, the admissibility criteria were fully verified at each time step. In the case of obtaining a numerical solution using the finite-volume ADER-WENO method, the same software implementation of the ADER-DG method with a posteriori correction of the solution in subcells by the finite-volume ADER-WENO limiter was used, however, the calculation of the candidate solution by the ADER-DG method was disabled, and the admissibility criteria explicitly designated all cells as trouble. In this case, the ``basic'' polynomial degree $N = 2$ was chosen for the ADER-DG method, for which the subgrid size was $N_{s} = 2N+1 = 5$ in each cell, so the choice of the mesh with sizes $1$, $2$, $3$, $4$ in each coordinate direction allowed us to perform calculations by the finite-volume ADER-WENO method on a mesh of $5$, $10$, $15$, $20$, respectively. The subgrid size of $N_{s} = 5$ in each direction was sufficient to perform dimension-by-dimension WENO reconstruction for ADER-WENO2, ADER-WENO3, ADER-WENO4 (and also for ADER-WENO5) methods. Therefore, in the case of determining the computational costs of the finite-volume ADER-WENO method, the number of mesh cells is understood as the number of mesh subcells.

\begin{table*}[h!]
\begin{center}
\caption{\label{tab:comp_costs_dg_2d}
The absolute computational costs (CPU time measured in seconds) $T_\mathrm{JIT}$, $T_\mathrm{GEN}$, $T_\mathrm{OV}$, $T_\mathrm{CV}$ for ADER-DG-$\mathbb{P}_{N}$ method for two-dimensional sine wave advection problem. The relative computational costs $\Upsilon_\mathrm{JIT}$, $\Upsilon_\mathrm{GEN}$, $\Upsilon_\mathrm{OV}$, $\Upsilon_\mathrm{CV}$ are the absolute computational costs $T$ normalized to $T_\mathrm{JIT}$.
}
\begin{tabular}{|r|r|llll|cccc|}
\hline
& cells 
& $T_\mathrm{JIT}$ & $T_\mathrm{GEN}$ & $T_\mathrm{OV}$ & $T_\mathrm{CV}$ 
& $\Upsilon_\mathrm{JIT} \quad$ & $\quad \Upsilon_\mathrm{GEN} \quad$ & $\quad \Upsilon_\mathrm{OV} \quad$ & $\quad \Upsilon_\mathrm{CV}$ \\
\hline
DG-$\mathbb{P}_{1}$ &	$4^{2}$ &		1.16E$-$01 &	2.45E$-$01 &	1.22E$-$01 &	1.79E$-$01 &		$1.00$ &	$2.12$ &	$1.05$ &	$1.55$\\ 
	&	$8^{2}$ &		2.65E$-$01 &	5.59E$-$01 &	3.01E$-$01 &	4.50E$-$01 &		$1.00$ &	$2.11$ &	$1.14$ &	$1.70$\\ 
	&	$12^{2}$ &		5.54E$-$01 &	9.92E$-$01 &	6.26E$-$01 &	9.18E$-$01 &		$1.00$ &	$1.79$ &	$1.13$ &	$1.66$\\ 
	&	$16^{2}$ &		9.31E$-$01 &	1.61E$+$00 &	1.06E$+$00 &	1.50E$+$00 &		$1.00$ &	$1.73$ &	$1.14$ &	$1.61$\\ \hline
DG-$\mathbb{P}_{2}$ &	$4^{2}$ &		2.54E$-$01 &	6.11E$-$01 &	3.34E$-$01 &	5.18E$-$01 &		$1.00$ &	$2.41$ &	$1.31$ &	$2.04$\\ 
	&	$8^{2}$ &		4.93E$-$01 &	1.13E$+$00 &	6.06E$-$01 &	9.98E$-$01 &		$1.00$ &	$2.30$ &	$1.23$ &	$2.02$\\ 
	&	$12^{2}$ &		1.04E$+$00 &	2.40E$+$00 &	1.32E$+$00 &	2.22E$+$00 &		$1.00$ &	$2.31$ &	$1.27$ &	$2.13$\\ 
	&	$16^{2}$ &		2.10E$+$00 &	4.72E$+$00 &	2.52E$+$00 &	4.40E$+$00 &		$1.00$ &	$2.24$ &	$1.20$ &	$2.09$\\ \hline
DG-$\mathbb{P}_{3}$ &	$4^{2}$ &		4.34E$-$01 &	1.21E$+$00 &	6.60E$-$01 &	1.04E$+$00 &		$1.00$ &	$2.79$ &	$1.52$ &	$2.41$\\ 
	&	$8^{2}$ &		8.98E$-$01 &	2.22E$+$00 &	1.36E$+$00 &	2.50E$+$00 &		$1.00$ &	$2.47$ &	$1.52$ &	$2.78$\\ 
	&	$12^{2}$ &		2.02E$+$00 &	5.15E$+$00 &	3.65E$+$00 &	6.21E$+$00 &		$1.00$ &	$2.55$ &	$1.81$ &	$3.08$\\ 
	&	$16^{2}$ &		3.72E$+$00 &	1.00E$+$01 &	7.15E$+$00 &	1.34E$+$01 &		$1.00$ &	$2.69$ &	$1.92$ &	$3.61$\\ \hline
DG-$\mathbb{P}_{4}$ &	$4^{2}$ &		6.07E$-$01 &	1.66E$+$00 &	9.56E$-$01 &	1.62E$+$00 &		$1.00$ &	$2.73$ &	$1.57$ &	$2.66$\\ 
	&	$8^{2}$ &		1.44E$+$00 &	4.07E$+$00 &	2.90E$+$00 &	5.55E$+$00 &		$1.00$ &	$2.82$ &	$2.01$ &	$3.85$\\ 
	&	$12^{2}$ &		3.24E$+$00 &	9.23E$+$00 &	8.66E$+$00 &	1.47E$+$01 &		$1.00$ &	$2.85$ &	$2.67$ &	$4.55$\\ 
	&	$16^{2}$ &		6.30E$+$00 &	1.76E$+$01 &	1.82E$+$01 &	3.21E$+$01 &		$1.00$ &	$2.80$ &	$2.88$ &	$5.10$\\ \hline
DG-$\mathbb{P}_{5}$ &	$4^{2}$ &		1.03E$+$00 &	2.85E$+$00 &	1.84E$+$00 &	3.26E$+$00 &		$1.00$ &	$2.78$ &	$1.80$ &	$3.17$\\ 
	&	$8^{2}$ &		6.09E$+$00 &	1.93E$+$01 &	1.28E$+$01 &	2.53E$+$01 &		$1.00$ &	$3.16$ &	$2.10$ &	$4.15$\\ 
	&	$12^{2}$ &		1.40E$+$01 &	3.90E$+$01 &	3.75E$+$01 &	6.22E$+$01 &		$1.00$ &	$2.79$ &	$2.68$ &	$4.44$\\ 
	&	$16^{2}$ &		2.95E$+$01 &	7.90E$+$01 &	8.03E$+$01 &	1.31E$+$02 &		$1.00$ &	$2.68$ &	$2.72$ &	$4.44$\\ \hline
DG-$\mathbb{P}_{6}$ &	$4^{2}$ &		4.07E$+$00 &	1.02E$+$01 &	5.93E$+$00 &	1.16E$+$01 &		$1.00$ &	$2.50$ &	$1.46$ &	$2.85$\\ 
	&	$8^{2}$ &		1.22E$+$01 &	3.26E$+$01 &	2.62E$+$01 &	4.81E$+$01 &		$1.00$ &	$2.67$ &	$2.14$ &	$3.94$\\ 
	&	$12^{2}$ &		3.52E$+$01 &	8.43E$+$01 &	9.17E$+$01 &	1.31E$+$02 &		$1.00$ &	$2.40$ &	$2.61$ &	$3.73$\\ 
	&	$16^{2}$ &		7.83E$+$01 &	1.86E$+$02 &	2.04E$+$02 &	3.03E$+$02 &		$1.00$ &	$2.37$ &	$2.60$ &	$3.87$\\ \hline
DG-$\mathbb{P}_{7}$ &	$4^{2}$ &		5.50E$+$00 &	1.71E$+$01 &	9.53E$+$00 &	2.04E$+$01 &		$1.00$ &	$3.11$ &	$1.73$ &	$3.71$\\ 
	&	$8^{2}$ &		2.07E$+$01 &	5.79E$+$01 &	3.55E$+$01 &	8.39E$+$01 &		$1.00$ &	$2.79$ &	$1.71$ &	$4.05$\\ 
	&	$12^{2}$ &		7.00E$+$01 &	1.79E$+$02 &	1.14E$+$02 &	2.61E$+$02 &		$1.00$ &	$2.56$ &	$1.62$ &	$3.73$\\ 
	&	$16^{2}$ &		1.68E$+$02 &	4.14E$+$02 &	2.77E$+$02 &	6.42E$+$02 &		$1.00$ &	$2.46$ &	$1.65$ &	$3.82$\\ \hline
DG-$\mathbb{P}_{8}$ &	$4^{2}$ &		8.52E$+$00 &	2.58E$+$01 &	1.50E$+$01 &	3.80E$+$01 &		$1.00$ &	$3.03$ &	$1.76$ &	$4.46$\\ 
	&	$8^{2}$ &		4.25E$+$01 &	1.10E$+$02 &	7.54E$+$01 &	2.26E$+$02 &		$1.00$ &	$2.59$ &	$1.77$ &	$5.31$\\ 
	&	$12^{2}$ &		1.44E$+$02 &	3.35E$+$02 &	2.44E$+$02 &	6.67E$+$02 &		$1.00$ &	$2.33$ &	$1.69$ &	$4.64$\\ 
	&	$16^{2}$ &		3.22E$+$02 &	7.72E$+$02 &	5.72E$+$02 &	1.64E$+$03 &		$1.00$ &	$2.39$ &	$1.78$ &	$5.09$\\ \hline
DG-$\mathbb{P}_{9}$ &	$4^{2}$ &		1.19E$+$01 &	3.59E$+$01 &	2.18E$+$01 &	5.87E$+$01 &		$1.00$ &	$3.02$ &	$1.83$ &	$4.94$\\ 
	&	$8^{2}$ &		7.42E$+$01 &	1.85E$+$02 &	1.28E$+$02 &	3.51E$+$02 &		$1.00$ &	$2.49$ &	$1.72$ &	$4.73$\\ 
	&	$12^{2}$ &		2.45E$+$02 &	5.76E$+$02 &	4.44E$+$02 &	1.13E$+$03 &		$1.00$ &	$2.35$ &	$1.81$ &	$4.61$\\ 
	&	$16^{2}$ &		6.25E$+$02 &	1.34E$+$03 &	1.07E$+$03 &	2.74E$+$03 &		$1.00$ &	$2.14$ &	$1.71$ &	$4.39$\\ 
\hline
\end{tabular}
\end{center}
\end{table*}

\begin{table*}[h!]
\begin{center}
\caption{\label{tab:comp_costs_dg_3d}
The absolute computational costs (CPU time measured in seconds) $T_\mathrm{JIT}$, $T_\mathrm{GEN}$, $T_\mathrm{OV}$, $T_\mathrm{CV}$ for ADER-DG-$\mathbb{P}_{N}$ method for three-dimensional sine wave advection problem. The relative computational costs $\Upsilon_\mathrm{JIT}$, $\Upsilon_\mathrm{GEN}$, $\Upsilon_\mathrm{OV}$, $\Upsilon_\mathrm{CV}$ are the absolute computational costs $T$ normalized to $T_\mathrm{JIT}$.
}
\begin{tabular}{|r|r|llll|cccc|}
\hline
& cells 
& $T_\mathrm{JIT}$ & $T_\mathrm{GEN}$ & $T_\mathrm{OV}$ & $T_\mathrm{CV}$ 
& $\Upsilon_\mathrm{JIT} \quad$ & $\quad \Upsilon_\mathrm{GEN} \quad$ & $\quad \Upsilon_\mathrm{OV} \quad$ & $\quad \Upsilon_\mathrm{CV}$ \\
\hline
DG-$\mathbb{P}_{1}$ &	$4^{3}$ &		2.46E$+$00 &	8.12E$+$00 &	4.47E$+$00 &	6.73E$+$00 &		$1.00$ &	$3.30$ &	$1.82$ &	$2.73$\\ 
	&	$5^{3}$ &		3.28E$+$00 &	1.17E$+$01 &	6.14E$+$00 &	9.81E$+$00 &		$1.00$ &	$3.57$ &	$1.87$ &	$2.99$\\ 
	&	$6^{3}$ &		5.89E$+$00 &	1.74E$+$01 &	9.33E$+$00 &	1.42E$+$01 &		$1.00$ &	$2.96$ &	$1.59$ &	$2.41$\\ 
	&	$7^{3}$ &		5.20E$+$00 &	1.81E$+$01 &	8.78E$+$00 &	1.49E$+$01 &		$1.00$ &	$3.49$ &	$1.69$ &	$2.88$\\ \hline
DG-$\mathbb{P}_{2}$ &	$4^{3}$ &		6.77E$+$00 &	2.38E$+$01 &	1.26E$+$01 &	2.03E$+$01 &		$1.00$ &	$3.52$ &	$1.86$ &	$3.00$\\ 
	&	$5^{3}$ &		8.93E$+$00 &	3.36E$+$01 &	1.65E$+$01 &	2.75E$+$01 &		$1.00$ &	$3.76$ &	$1.85$ &	$3.09$\\ 
	&	$6^{3}$ &		1.10E$+$01 &	4.85E$+$01 &	2.69E$+$01 &	4.45E$+$01 &		$1.00$ &	$4.39$ &	$2.44$ &	$4.03$\\ 
	&	$7^{3}$ &		1.23E$+$01 &	5.55E$+$01 &	2.85E$+$01 &	5.25E$+$01 &		$1.00$ &	$4.52$ &	$2.32$ &	$4.28$\\ \hline
DG-$\mathbb{P}_{3}$ &	$4^{3}$ &		1.99E$+$01 &	5.29E$+$01 &	2.74E$+$01 &	4.66E$+$01 &		$1.00$ &	$2.66$ &	$1.37$ &	$2.34$\\ 
	&	$5^{3}$ &		1.68E$+$01 &	5.30E$+$01 &	2.90E$+$01 &	5.38E$+$01 &		$1.00$ &	$3.16$ &	$1.73$ &	$3.21$\\ 
	&	$6^{3}$ &		3.57E$+$01 &	1.29E$+$02 &	6.81E$+$01 &	1.29E$+$02 &		$1.00$ &	$3.61$ &	$1.91$ &	$3.61$\\ 
	&	$7^{3}$ &		4.65E$+$01 &	1.48E$+$02 &	8.46E$+$01 &	1.61E$+$02 &		$1.00$ &	$3.19$ &	$1.82$ &	$3.47$\\ \hline
DG-$\mathbb{P}_{4}$ &	$4^{3}$ &		2.94E$+$01 &	1.04E$+$02 &	5.76E$+$01 &	1.01E$+$02 &		$1.00$ &	$3.56$ &	$1.96$ &	$3.43$\\ 
	&	$5^{3}$ &		4.76E$+$01 &	1.50E$+$02 &	8.88E$+$01 &	1.64E$+$02 &		$1.00$ &	$3.15$ &	$1.87$ &	$3.46$\\ 
	&	$6^{3}$ &		6.34E$+$01 &	2.09E$+$02 &	1.36E$+$02 &	2.47E$+$02 &		$1.00$ &	$3.30$ &	$2.14$ &	$3.89$\\ 
	&	$7^{3}$ &		8.74E$+$01 &	2.81E$+$02 &	1.97E$+$02 &	3.82E$+$02 &		$1.00$ &	$3.21$ &	$2.25$ &	$4.37$\\ \hline
DG-$\mathbb{P}_{5}$ &	$4^{3}$ &		5.42E$+$01 &	1.89E$+$02 &	1.17E$+$02 &	2.04E$+$02 &		$1.00$ &	$3.48$ &	$2.16$ &	$3.77$\\ 
	&	$5^{3}$ &		8.41E$+$01 &	2.68E$+$02 &	1.86E$+$02 &	3.38E$+$02 &		$1.00$ &	$3.19$ &	$2.21$ &	$4.02$\\ 
	&	$6^{3}$ &		1.14E$+$02 &	3.52E$+$02 &	2.88E$+$02 &	5.41E$+$02 &		$1.00$ &	$3.10$ &	$2.54$ &	$4.76$\\ 
	&	$7^{3}$ &		1.70E$+$02 &	5.66E$+$02 &	4.73E$+$02 &	9.41E$+$02 &		$1.00$ &	$3.32$ &	$2.78$ &	$5.52$\\ \hline
DG-$\mathbb{P}_{6}$ &	$4^{3}$ &		8.18E$+$01 &	2.54E$+$02 &	1.83E$+$02 &	3.13E$+$02 &		$1.00$ &	$3.10$ &	$2.24$ &	$3.82$\\ 
	&	$5^{3}$ &		1.31E$+$02 &	4.00E$+$02 &	3.40E$+$02 &	5.63E$+$02 &		$1.00$ &	$3.06$ &	$2.60$ &	$4.30$\\ 
	&	$6^{3}$ &		2.16E$+$02 &	6.55E$+$02 &	6.12E$+$02 &	1.08E$+$03 &		$1.00$ &	$3.04$ &	$2.84$ &	$4.99$\\ 
	&	$7^{3}$ &		3.33E$+$02 &	9.87E$+$02 &	1.01E$+$03 &	1.82E$+$03 &		$1.00$ &	$2.96$ &	$3.04$ &	$5.46$\\ \hline
DG-$\mathbb{P}_{7}$ &	$4^{3}$ &		1.39E$+$02 &	3.99E$+$02 &	2.74E$+$02 &	5.48E$+$02 &		$1.00$ &	$2.88$ &	$1.98$ &	$3.95$\\ 
	&	$5^{3}$ &		2.29E$+$02 &	6.58E$+$02 &	5.02E$+$02 &	1.02E$+$03 &		$1.00$ &	$2.88$ &	$2.20$ &	$4.46$\\ 
	&	$6^{3}$ &		3.90E$+$02 &	1.12E$+$03 &	8.62E$+$02 &	1.94E$+$03 &		$1.00$ &	$2.88$ &	$2.21$ &	$4.98$\\ 
	&	$7^{3}$ &		6.38E$+$02 &	1.79E$+$03 &	1.43E$+$03 &	3.39E$+$03 &		$1.00$ &	$2.81$ &	$2.25$ &	$5.31$\\ \hline
DG-$\mathbb{P}_{8}$ &	$4^{3}$ &		3.00E$+$02 &	7.14E$+$02 &	4.85E$+$02 &	1.46E$+$03 &		$1.00$ &	$2.38$ &	$1.62$ &	$4.87$\\ 
	&	$5^{3}$ &		5.02E$+$02 &	1.16E$+$03 &	8.83E$+$02 &	2.98E$+$03 &		$1.00$ &	$2.31$ &	$1.76$ &	$5.93$\\ 
	&	$6^{3}$ &		8.19E$+$02 &	1.72E$+$03 &	1.69E$+$03 &	5.59E$+$03 &		$1.00$ &	$2.10$ &	$2.06$ &	$6.83$\\ 
	&	$7^{3}$ &		1.22E$+$03 &	2.91E$+$03 &	2.86E$+$03 &	1.03E$+$04 &		$1.00$ &	$2.39$ &	$2.35$ &	$8.46$\\ \hline
DG-$\mathbb{P}_{9}$ &	$4^{3}$ &		5.49E$+$02 &	1.09E$+$03 &	8.20E$+$02 &	2.36E$+$03 &		$1.00$ &	$1.99$ &	$1.49$ &	$4.30$\\ 
	&	$5^{3}$ &		1.05E$+$03 &	1.91E$+$03 &	1.76E$+$03 &	5.10E$+$03 &		$1.00$ &	$1.82$ &	$1.67$ &	$4.84$\\ 
	&	$6^{3}$ &		2.23E$+$03 &	3.47E$+$03 &	3.20E$+$03 &	1.04E$+$04 &		$1.00$ &	$1.55$ &	$1.43$ &	$4.65$\\ 
	&	$7^{3}$ &		3.69E$+$03 &	5.43E$+$03 &	5.65E$+$03 &	1.83E$+$04 &		$1.00$ &	$1.47$ &	$1.53$ &	$4.96$\\ 
\hline
\end{tabular}
\end{center}
\end{table*}

\begin{table*}[h!]
\begin{center}
\caption{\label{tab:comp_costs_fv_2d}
The absolute computational costs (CPU time measured in seconds) $T_\mathrm{JIT}$, $T_\mathrm{GEN}$, $T_\mathrm{OV}$, $T_\mathrm{CV}$ for finite-volume ADER-WENO($N+1$) method for two-dimensional sine wave advection problem. The relative computational costs $\Upsilon_\mathrm{JIT}$, $\Upsilon_\mathrm{GEN}$, $\Upsilon_\mathrm{OV}$, $\Upsilon_\mathrm{CV}$ are the absolute computational costs $T$ normalized to $T_\mathrm{JIT}$.
}
\begin{tabular}{|r|r|llll|cccc|}
\hline
& cells 
& $T_\mathrm{JIT}$ & $T_\mathrm{GEN}$ & $T_\mathrm{OV}$ & $T_\mathrm{CV}$ 
& $\Upsilon_\mathrm{JIT} \quad$ & $\quad \Upsilon_\mathrm{GEN} \quad$ & $\quad \Upsilon_\mathrm{OV} \quad$ & $\quad \Upsilon_\mathrm{CV}$ \\
\hline
WENO$2$ &	$5^{2}$ &		4.31E$-$02 &	1.20E$-$01 &	6.45E$-$02 &	9.68E$-$02 &		$1.00$ &	$2.78$ &	$1.50$ &	$2.25$\\ 
	&	$10^{2}$ &		1.33E$-$01 &	2.83E$-$01 &	1.52E$-$01 &	2.41E$-$01 &		$1.00$ &	$2.13$ &	$1.14$ &	$1.82$\\ 
	&	$15^{2}$ &		2.22E$-$01 &	5.17E$-$01 &	3.02E$-$01 &	4.39E$-$01 &		$1.00$ &	$2.33$ &	$1.36$ &	$1.97$\\ 
	&	$20^{2}$ &		4.40E$-$01 &	8.27E$-$01 &	4.87E$-$01 &	7.33E$-$01 &		$1.00$ &	$1.88$ &	$1.11$ &	$1.66$\\ \hline
WENO$3$ &	$5^{2}$ &		8.50E$-$02 &	2.92E$-$01 &	1.91E$-$01 &	3.21E$-$01 &		$1.00$ &	$3.43$ &	$2.25$ &	$3.77$\\ 
	&	$10^{2}$ &		2.27E$-$01 &	7.38E$-$01 &	4.64E$-$01 &	8.07E$-$01 &		$1.00$ &	$3.25$ &	$2.04$ &	$3.55$\\ 
	&	$15^{2}$ &		5.89E$-$01 &	1.66E$+$00 &	1.13E$+$00 &	1.73E$+$00 &		$1.00$ &	$2.82$ &	$1.91$ &	$2.94$\\ 
	&	$20^{2}$ &		1.35E$+$00 &	2.79E$+$00 &	1.80E$+$00 &	3.35E$+$00 &		$1.00$ &	$2.06$ &	$1.33$ &	$2.48$\\ \hline
WENO$4$ &	$5^{2}$ &		1.65E$-$01 &	6.72E$-$01 &	5.58E$-$01 &	1.01E$+$00 &		$1.00$ &	$4.07$ &	$3.38$ &	$6.11$\\ 
	&	$10^{2}$ &		4.64E$-$01 &	1.75E$+$00 &	1.51E$+$00 &	2.64E$+$00 &		$1.00$ &	$3.77$ &	$3.27$ &	$5.69$\\ 
	&	$15^{2}$ &		8.52E$-$01 &	3.19E$+$00 &	2.51E$+$00 &	4.95E$+$00 &		$1.00$ &	$3.75$ &	$2.95$ &	$5.81$\\ 
	&	$20^{2}$ &		1.84E$+$00 &	5.68E$+$00 &	4.96E$+$00 &	9.72E$+$00 &		$1.00$ &	$3.08$ &	$2.69$ &	$5.27$\\ 
\hline
\end{tabular}
\end{center}
\end{table*}

\begin{table*}[h!]
\begin{center}
\caption{\label{tab:comp_costs_fv_3d}
The absolute computational costs (CPU time measured in seconds) $T_\mathrm{JIT}$, $T_\mathrm{GEN}$, $T_\mathrm{OV}$, $T_\mathrm{CV}$ for finite-volume ADER-WENO($N+1$) method for three-dimensional sine wave advection problem. The relative computational costs $\Upsilon_\mathrm{JIT}$, $\Upsilon_\mathrm{GEN}$, $\Upsilon_\mathrm{OV}$, $\Upsilon_\mathrm{CV}$ are the absolute computational costs $T$ normalized to $T_\mathrm{JIT}$.
}
\begin{tabular}{|r|r|llll|cccc|}
\hline
& cells 
& $T_\mathrm{JIT}$ & $T_\mathrm{GEN}$ & $T_\mathrm{OV}$ & $T_\mathrm{CV}$ 
& $\Upsilon_\mathrm{JIT} \quad$ & $\quad \Upsilon_\mathrm{GEN} \quad$ & $\quad \Upsilon_\mathrm{OV} \quad$ & $\quad \Upsilon_\mathrm{CV}$ \\
\hline
WENO$2$ &	$5^{3}$ &		1.45E$+$00 &	5.51E$+$00 &	2.72E$+$00 &	4.37E$+$00 &		$1.00$ &	$3.81$ &	$1.88$ &	$3.02$\\ 
	&	$10^{3}$ &		4.71E$+$00 &	1.48E$+$01 &	7.64E$+$00 &	1.17E$+$01 &		$1.00$ &	$3.15$ &	$1.62$ &	$2.49$\\ 
	&	$15^{3}$ &		2.00E$+$01 &	5.37E$+$01 &	2.90E$+$01 &	4.37E$+$01 &		$1.00$ &	$2.68$ &	$1.45$ &	$2.18$\\ 
	&	$20^{3}$ &		5.87E$+$01 &	1.78E$+$02 &	8.99E$+$01 &	1.28E$+$02 &		$1.00$ &	$3.04$ &	$1.53$ &	$2.18$\\ \hline
WENO$3$ &	$5^{3}$ &		4.78E$+$00 &	2.16E$+$01 &	1.25E$+$01 &	2.09E$+$01 &		$1.00$ &	$4.52$ &	$2.61$ &	$4.37$\\ 
	&	$10^{3}$ &		1.34E$+$01 &	5.33E$+$01 &	3.21E$+$01 &	5.57E$+$01 &		$1.00$ &	$3.97$ &	$2.39$ &	$4.15$\\ 
	&	$15^{3}$ &		6.18E$+$01 &	2.07E$+$02 &	1.30E$+$02 &	2.07E$+$02 &		$1.00$ &	$3.36$ &	$2.10$ &	$3.35$\\ 
	&	$20^{3}$ &		2.23E$+$02 &	7.74E$+$02 &	4.59E$+$02 &	8.12E$+$02 &		$1.00$ &	$3.46$ &	$2.05$ &	$3.64$\\ \hline
WENO$4$ &	$5^{3}$ &		1.33E$+$01 &	5.92E$+$01 &	4.52E$+$01 &	7.52E$+$01 &		$1.00$ &	$4.45$ &	$3.40$ &	$5.66$\\ 
	&	$10^{3}$ &		4.39E$+$01 &	1.53E$+$02 &	1.20E$+$02 &	1.97E$+$02 &		$1.00$ &	$3.49$ &	$2.74$ &	$4.49$\\ 
	&	$15^{3}$ &		1.96E$+$02 &	5.93E$+$02 &	4.81E$+$02 &	7.44E$+$02 &		$1.00$ &	$3.02$ &	$2.45$ &	$3.79$\\ 
	&	$20^{3}$ &		6.65E$+$02 &	2.03E$+$03 &	1.66E$+$03 &	2.81E$+$03 &		$1.00$ &	$3.06$ &	$2.49$ &	$4.22$\\ 
\hline
\end{tabular}
\end{center}
\end{table*}

\begin{table*}[h!]
\begin{center}
\caption{\label{tab:comp_costs_dg_scl_2d}
The absolute computational costs (CPU time measured in seconds) $T_\mathrm{JIT}$, $T_\mathrm{GEN}$, $T_\mathrm{OV}$, $T_\mathrm{CV}$ for finite-element ADER-DG-$\mathbb{P}_{N}$ method with a posteriori correction of the solution in subcells by a finite-volume ADER-WENO2 limiter for cylindrical explosion problem. The relative computational costs $\Upsilon_\mathrm{JIT}$, $\Upsilon_\mathrm{GEN}$, $\Upsilon_\mathrm{OV}$, $\Upsilon_\mathrm{CV}$ are the absolute computational costs $T$ normalized to $T_\mathrm{JIT}$.
}
\begin{tabular}{|r|r|llll|cccc|}
\hline
& cells 
& $T_\mathrm{JIT}$ & $T_\mathrm{GEN}$ & $T_\mathrm{OV}$ & $T_\mathrm{CV}$ 
& $\Upsilon_\mathrm{JIT} \quad$ & $\quad \Upsilon_\mathrm{GEN} \quad$ & $\quad \Upsilon_\mathrm{OV} \quad$ & $\quad \Upsilon_\mathrm{CV}$ \\
\hline
DG-$\mathbb{P}_{1}$ &	$11^{2}$ &		2.66E$-$01 &	6.29E$-$01 &	3.72E$-$01 &	5.60E$-$01 &		$1.00$ &	$2.37$ &	$1.40$ &	$2.11$\\ 
	&	$21^{2}$ &		1.03E$+$00 &	2.73E$+$00 &	1.62E$+$00 &	2.64E$+$00 &		$1.00$ &	$2.66$ &	$1.57$ &	$2.57$\\ 
	&	$31^{2}$ &		2.48E$+$00 &	6.65E$+$00 &	3.69E$+$00 &	6.42E$+$00 &		$1.00$ &	$2.68$ &	$1.49$ &	$2.58$\\ 
	&	$41^{2}$ &		5.16E$+$00 &	1.32E$+$01 &	7.92E$+$00 &	1.34E$+$01 &		$1.00$ &	$2.56$ &	$1.54$ &	$2.60$\\ \hline
DG-$\mathbb{P}_{2}$ &	$11^{2}$ &		6.54E$-$01 &	1.94E$+$00 &	1.04E$+$00 &	1.86E$+$00 &		$1.00$ &	$2.97$ &	$1.59$ &	$2.84$\\ 
	&	$21^{2}$ &		2.93E$+$00 &	7.74E$+$00 &	4.23E$+$00 &	7.63E$+$00 &		$1.00$ &	$2.64$ &	$1.44$ &	$2.61$\\ 
	&	$31^{2}$ &		6.94E$+$00 &	2.00E$+$01 &	1.15E$+$01 &	1.94E$+$01 &		$1.00$ &	$2.89$ &	$1.65$ &	$2.80$\\ 
	&	$41^{2}$ &		1.37E$+$01 &	4.02E$+$01 &	2.38E$+$01 &	4.02E$+$01 &		$1.00$ &	$2.94$ &	$1.74$ &	$2.94$\\ \hline
DG-$\mathbb{P}_{3}$ &	$11^{2}$ &		1.71E$+$00 &	5.00E$+$00 &	2.99E$+$00 &	5.24E$+$00 &		$1.00$ &	$2.92$ &	$1.75$ &	$3.07$\\ 
	&	$21^{2}$ &		6.50E$+$00 &	2.01E$+$01 &	1.22E$+$01 &	2.08E$+$01 &		$1.00$ &	$3.10$ &	$1.88$ &	$3.20$\\ 
	&	$31^{2}$ &		1.76E$+$01 &	5.17E$+$01 &	3.35E$+$01 &	5.52E$+$01 &		$1.00$ &	$2.95$ &	$1.91$ &	$3.14$\\ 
	&	$41^{2}$ &		3.52E$+$01 &	9.92E$+$01 &	6.60E$+$01 &	1.10E$+$02 &		$1.00$ &	$2.82$ &	$1.87$ &	$3.14$\\ \hline
DG-$\mathbb{P}_{4}$ &	$11^{2}$ &		2.95E$+$00 &	8.84E$+$00 &	5.58E$+$00 &	9.28E$+$00 &		$1.00$ &	$3.00$ &	$1.89$ &	$3.15$\\ 
	&	$21^{2}$ &		1.18E$+$01 &	3.55E$+$01 &	2.55E$+$01 &	4.10E$+$01 &		$1.00$ &	$3.01$ &	$2.16$ &	$3.48$\\ 
	&	$31^{2}$ &		2.93E$+$01 &	9.36E$+$01 &	6.51E$+$01 &	1.09E$+$02 &		$1.00$ &	$3.20$ &	$2.23$ &	$3.74$\\ 
	&	$41^{2}$ &		5.97E$+$01 &	1.80E$+$02 &	1.39E$+$02 &	2.25E$+$02 &		$1.00$ &	$3.01$ &	$2.33$ &	$3.78$\\ \hline
DG-$\mathbb{P}_{5}$ &	$11^{2}$ &		5.71E$+$00 &	1.64E$+$01 &	1.14E$+$01 &	1.87E$+$01 &		$1.00$ &	$2.88$ &	$1.99$ &	$3.27$\\ 
	&	$21^{2}$ &		2.13E$+$01 &	6.40E$+$01 &	5.10E$+$01 &	7.93E$+$01 &		$1.00$ &	$3.01$ &	$2.40$ &	$3.73$\\ 
	&	$31^{2}$ &		5.46E$+$01 &	1.63E$+$02 &	1.38E$+$02 &	2.20E$+$02 &		$1.00$ &	$2.99$ &	$2.52$ &	$4.02$\\ 
	&	$41^{2}$ &		1.06E$+$02 &	3.23E$+$02 &	2.75E$+$02 &	4.40E$+$02 &		$1.00$ &	$3.06$ &	$2.61$ &	$4.17$\\ \hline
DG-$\mathbb{P}_{6}$ &	$11^{2}$ &		9.17E$+$00 &	2.48E$+$01 &	1.86E$+$01 &	3.02E$+$01 &		$1.00$ &	$2.71$ &	$2.03$ &	$3.29$\\ 
	&	$21^{2}$ &		3.53E$+$01 &	1.00E$+$02 &	8.63E$+$01 &	1.33E$+$02 &		$1.00$ &	$2.84$ &	$2.44$ &	$3.78$\\ 
	&	$31^{2}$ &		8.67E$+$01 &	2.57E$+$02 &	2.36E$+$02 &	3.67E$+$02 &		$1.00$ &	$2.97$ &	$2.72$ &	$4.23$\\ 
	&	$41^{2}$ &		1.73E$+$02 &	5.05E$+$02 &	5.02E$+$02 &	7.53E$+$02 &		$1.00$ &	$2.91$ &	$2.89$ &	$4.34$\\ \hline
DG-$\mathbb{P}_{7}$ &	$11^{2}$ &		1.42E$+$01 &	3.87E$+$01 &	2.78E$+$01 &	4.91E$+$01 &		$1.00$ &	$2.73$ &	$1.96$ &	$3.46$\\ 
	&	$21^{2}$ &		5.57E$+$01 &	1.56E$+$02 &	1.19E$+$02 &	2.20E$+$02 &		$1.00$ &	$2.80$ &	$2.13$ &	$3.94$\\ 
	&	$31^{2}$ &		1.37E$+$02 &	3.96E$+$02 &	3.00E$+$02 &	6.21E$+$02 &		$1.00$ &	$2.88$ &	$2.18$ &	$4.52$\\ 
	&	$41^{2}$ &		2.78E$+$02 &	7.68E$+$02 &	6.31E$+$02 &	1.26E$+$03 &		$1.00$ &	$2.76$ &	$2.27$ &	$4.53$\\ \hline
DG-$\mathbb{P}_{8}$ &	$11^{2}$ &		2.09E$+$01 &	5.66E$+$01 &	4.08E$+$01 &	9.31E$+$01 &		$1.00$ &	$2.71$ &	$1.95$ &	$4.45$\\ 
	&	$21^{2}$ &		8.15E$+$01 &	2.26E$+$02 &	1.82E$+$02 &	4.77E$+$02 &		$1.00$ &	$2.77$ &	$2.23$ &	$5.85$\\ 
	&	$31^{2}$ &		2.18E$+$02 &	5.92E$+$02 &	4.84E$+$02 &	1.37E$+$03 &		$1.00$ &	$2.71$ &	$2.22$ &	$6.29$\\ 
	&	$41^{2}$ &		4.29E$+$02 &	1.14E$+$03 &	8.66E$+$02 &	3.24E$+$03 &		$1.00$ &	$2.67$ &	$2.02$ &	$7.55$\\ \hline
DG-$\mathbb{P}_{9}$ &	$11^{2}$ &		3.04E$+$01 &	8.01E$+$01 &	6.91E$+$01 &	1.47E$+$02 &		$1.00$ &	$2.64$ &	$2.28$ &	$4.84$\\ 
	&	$21^{2}$ &		1.21E$+$02 &	3.22E$+$02 &	2.19E$+$02 &	7.33E$+$02 &		$1.00$ &	$2.66$ &	$1.81$ &	$6.07$\\ 
	&	$31^{2}$ &		3.30E$+$02 &	8.18E$+$02 &	6.06E$+$02 &	1.98E$+$03 &		$1.00$ &	$2.48$ &	$1.84$ &	$6.00$\\ 
	&	$41^{2}$ &		6.57E$+$02 &	1.65E$+$03 &	1.27E$+$03 &	4.15E$+$03 &		$1.00$ &	$2.52$ &	$1.93$ &	$6.32$\\ 
\hline
\end{tabular}
\end{center}
\end{table*}

\begin{table*}[h!]
\begin{center}
\caption{\label{tab:comp_costs_dg_scl_3d}
The absolute computational costs (CPU time measured in seconds) $T_\mathrm{JIT}$, $T_\mathrm{GEN}$, $T_\mathrm{OV}$, $T_\mathrm{CV}$ for finite-element ADER-DG-$\mathbb{P}_{N}$ method with a posteriori correction of the solution in subcells by a finite-volume ADER-WENO2 limiter for spherical explosion problem. The relative computational costs $\Upsilon_\mathrm{JIT}$, $\Upsilon_\mathrm{GEN}$, $\Upsilon_\mathrm{OV}$, $\Upsilon_\mathrm{CV}$ are the absolute computational costs $T$ normalized to $T_\mathrm{JIT}$.
}
\begin{tabular}{|r|r|llll|cccc|}
\hline
& cells 
& $T_\mathrm{JIT}$ & $T_\mathrm{GEN}$ & $T_\mathrm{OV}$ & $T_\mathrm{CV}$ 
& $\Upsilon_\mathrm{JIT} \quad$ & $\quad \Upsilon_\mathrm{GEN} \quad$ & $\quad \Upsilon_\mathrm{OV} \quad$ & $\quad \Upsilon_\mathrm{CV}$ \\
\hline
DG-$\mathbb{P}_{1}$ &	$5^{3}$ &		7.41E$-$01 &	2.40E$+$00 &	1.34E$+$00 &	2.35E$+$00 &		$1.00$ &	$3.23$ &	$1.81$ &	$3.17$\\ 
	&	$7^{3}$ &		4.37E$+$00 &	6.13E$+$00 &	7.43E$+$00 &	6.34E$+$00 &		$1.00$ &	$1.40$ &	$1.70$ &	$1.45$\\ 
	&	$9^{3}$ &		5.12E$+$00 &	1.67E$+$01 &	9.28E$+$00 &	1.61E$+$01 &		$1.00$ &	$3.25$ &	$1.81$ &	$3.14$\\ 
	&	$11^{3}$ &		7.68E$+$00 &	2.63E$+$01 &	1.48E$+$01 &	2.60E$+$01 &		$1.00$ &	$3.43$ &	$1.92$ &	$3.39$\\ \hline
DG-$\mathbb{P}_{2}$ &	$5^{3}$ &		3.22E$+$00 &	1.26E$+$01 &	6.85E$+$00 &	1.25E$+$01 &		$1.00$ &	$3.91$ &	$2.13$ &	$3.88$\\ 
	&	$7^{3}$ &		1.22E$+$01 &	3.46E$+$01 &	1.93E$+$01 &	3.39E$+$01 &		$1.00$ &	$2.84$ &	$1.58$ &	$2.78$\\ 
	&	$9^{3}$ &		1.83E$+$01 &	5.38E$+$01 &	3.10E$+$01 &	5.40E$+$01 &		$1.00$ &	$2.94$ &	$1.69$ &	$2.95$\\ 
	&	$11^{3}$ &		4.77E$+$01 &	1.44E$+$02 &	8.49E$+$01 &	1.47E$+$02 &		$1.00$ &	$3.02$ &	$1.78$ &	$3.09$\\ \hline
DG-$\mathbb{P}_{3}$ &	$5^{3}$ &		1.10E$+$01 &	2.82E$+$01 &	1.65E$+$01 &	2.87E$+$01 &		$1.00$ &	$2.57$ &	$1.50$ &	$2.62$\\ 
	&	$7^{3}$ &		5.14E$+$01 &	1.47E$+$02 &	8.69E$+$01 &	1.46E$+$02 &		$1.00$ &	$2.86$ &	$1.69$ &	$2.84$\\ 
	&	$9^{3}$ &		1.33E$+$02 &	2.76E$+$02 &	1.64E$+$02 &	2.79E$+$02 &		$1.00$ &	$2.07$ &	$1.23$ &	$2.09$\\ 
	&	$11^{3}$ &		1.99E$+$02 &	5.39E$+$02 &	3.29E$+$02 &	5.58E$+$02 &		$1.00$ &	$2.70$ &	$1.65$ &	$2.80$\\ \hline
DG-$\mathbb{P}_{4}$ &	$5^{3}$ &		3.10E$+$01 &	6.91E$+$01 &	4.58E$+$01 &	7.20E$+$01 &		$1.00$ &	$2.23$ &	$1.48$ &	$2.32$\\ 
	&	$7^{3}$ &		1.81E$+$02 &	4.81E$+$02 &	2.93E$+$02 &	4.86E$+$02 &		$1.00$ &	$2.65$ &	$1.61$ &	$2.68$\\ 
	&	$9^{3}$ &		4.02E$+$02 &	1.04E$+$03 &	6.66E$+$02 &	1.08E$+$03 &		$1.00$ &	$2.58$ &	$1.66$ &	$2.70$\\ 
	&	$11^{3}$ &		6.11E$+$02 &	1.61E$+$03 &	1.02E$+$03 &	1.66E$+$03 &		$1.00$ &	$2.63$ &	$1.67$ &	$2.71$\\ \hline
DG-$\mathbb{P}_{5}$ &	$5^{3}$ &		1.25E$+$02 &	2.35E$+$02 &	1.70E$+$02 &	2.37E$+$02 &		$1.00$ &	$1.88$ &	$1.36$ &	$1.90$\\ 
	&	$7^{3}$ &		3.84E$+$02 &	8.53E$+$02 &	5.59E$+$02 &	8.57E$+$02 &		$1.00$ &	$2.22$ &	$1.46$ &	$2.23$\\ 
	&	$9^{3}$ &		9.54E$+$02 &	2.31E$+$03 &	1.53E$+$03 &	2.33E$+$03 &		$1.00$ &	$2.42$ &	$1.60$ &	$2.45$\\ 
	&	$11^{3}$ &		1.52E$+$03 &	3.71E$+$03 &	2.44E$+$03 &	3.88E$+$03 &		$1.00$ &	$2.44$ &	$1.60$ &	$2.55$\\ \hline
DG-$\mathbb{P}_{6}$ &	$5^{3}$ &		4.00E$+$02 &	7.77E$+$02 &	5.55E$+$02 &	7.80E$+$02 &		$1.00$ &	$1.94$ &	$1.39$ &	$1.95$\\ 
	&	$7^{3}$ &		1.09E$+$03 &	2.29E$+$03 &	1.62E$+$03 &	2.34E$+$03 &		$1.00$ &	$2.10$ &	$1.49$ &	$2.15$\\ 
	&	$9^{3}$ &		2.20E$+$03 &	4.82E$+$03 &	3.21E$+$03 &	4.92E$+$03 &		$1.00$ &	$2.19$ &	$1.46$ &	$2.24$\\ 
	&	$11^{3}$ &		3.85E$+$03 &	8.28E$+$03 &	5.85E$+$03 &	1.01E$+$04 &		$1.00$ &	$2.15$ &	$1.52$ &	$2.61$\\ \hline
DG-$\mathbb{P}_{7}$ &	$5^{3}$ &		8.89E$+$02 &	1.50E$+$03 &	1.72E$+$03 &	3.59E$+$03 &		$1.00$ &	$1.69$ &	$1.94$ &	$4.04$\\ 
	&	$7^{3}$ &		2.11E$+$03 &	4.31E$+$03 &	3.97E$+$03 &	9.29E$+$03 &		$1.00$ &	$2.04$ &	$1.88$ &	$4.40$\\ 
	&	$9^{3}$ &		4.37E$+$03 &	8.81E$+$03 &	8.65E$+$03 &	2.27E$+$04 &		$1.00$ &	$2.02$ &	$1.98$ &	$5.21$\\ 
	&	$11^{3}$ &		7.65E$+$03 &	1.56E$+$04 &	1.64E$+$04 &	4.32E$+$04 &		$1.00$ &	$2.04$ &	$2.14$ &	$5.64$\\ \hline
DG-$\mathbb{P}_{8}$ &	$5^{3}$ &		1.74E$+$03 &	2.66E$+$03 &	2.67E$+$03 &	8.43E$+$03 &		$1.00$ &	$1.53$ &	$1.53$ &	$4.83$\\ 
	&	$7^{3}$ &		4.61E$+$03 &	7.50E$+$03 &	7.37E$+$03 &	2.43E$+$04 &		$1.00$ &	$1.63$ &	$1.60$ &	$5.27$\\ 
	&	$9^{3}$ &		8.99E$+$03 &	1.57E$+$04 &	1.58E$+$04 &	5.95E$+$04 &		$1.00$ &	$1.75$ &	$1.76$ &	$6.62$\\ 
	&	$11^{3}$ &		1.49E$+$04 &	2.76E$+$04 &	2.90E$+$04 &	1.20E$+$05 &		$1.00$ &	$1.86$ &	$1.95$ &	$8.07$\\ \hline
DG-$\mathbb{P}_{9}$ &	$5^{3}$ &		3.32E$+$03 &	5.20E$+$03 &	4.03E$+$03 &	1.32E$+$04 &		$1.00$ &	$1.56$ &	$1.21$ &	$3.97$\\ 
	&	$7^{3}$ &		8.81E$+$03 &	1.35E$+$04 &	1.29E$+$04 &	3.55E$+$04 &		$1.00$ &	$1.54$ &	$1.47$ &	$4.03$\\ 
	&	$9^{3}$ &		1.70E$+$04 &	2.69E$+$04 &	2.60E$+$04 &	7.49E$+$04 &		$1.00$ &	$1.58$ &	$1.53$ &	$4.41$\\ 
	&	$11^{3}$ &		2.88E$+$04 &	3.20E$+$04 &	5.31E$+$04 &	1.36E$+$05 &		$1.00$ &	$1.11$ &	$1.84$ &	$4.73$\\ 
\hline
\end{tabular}
\end{center}
\end{table*}

The computational costs of the method \newtext{have been} investigated using cylindrical and spherical explosion problems --- cylindrical in the two-dimensional case and spherical in the three-dimensional case. The initial conditions and the complete problem statement are chosen in the form presented in Subsection~\ref{sec:apps_cgd_problems:sod_md}. The explosion problem is characterized by a discontinuous solution, which allowed us to investigate not only the finite-element ADER-DG method and the finite-volume ADER-WENO method separately, but also the transformation of solution representations from piecewise polynomial in cells to finite-volume in subcells and the inverse transformation.

The computational costs of the finite-element method ADER-DR for polynomial degrees $N = 1, \ldots, 9$ and the finite-volume method ADER-WENO for polynomial degrees $N = 1, \ldots, 3$ \newtext{are} separately investigated. In the context of using numerical methods of ADER, the finite-volume method ADER-WENO is usually used as a limiter, and the ARER-WENO3 method with $N = 2$ is mainly used~\cite{ader_dg_ideal_flows, ader_dg_diss_flows}, therefore, the computational costs of polynomial degrees $N \geqslant 4$ \newtext{are} not investigated. In the basic works~\cite{ader_weno_lstdg_ideal, ader_weno_lstdg_diss} in which the numerical method ADER-WENO was developed, methods with degrees $N \leqslant 3$ were used. The computational costs of two-dimensional and three-dimensional implementation were also investigated separately. Computational costs were determined not only for different polynomial degrees $N$, but also for different mesh sizes.

Absolute computational costs $T$ \newtext{are} determined as the CPU time (measured in seconds) to perform the calculations in the four main software implementations: $T_\mathrm{JIT}$, $T_\mathrm{GEN}$, $T_\mathrm{OV}$, $T_\mathrm{CV}$, which correspond to variants JIT, GEN, OV, CV, introduced above in the text, respectively. To compare the obtained results and determine the increase of performance as a result of using the BLAS interface integration method presented in this paper, relative computational costs $\Upsilon = T/T_\mathrm{JIT}$ 
\newtext{are} also determined: $\Upsilon_\mathrm{JIT}$, $\Upsilon_\mathrm{GEN}$, $\Upsilon_\mathrm{OV}$, $\Upsilon_\mathrm{CV}$, defined as absolute computational costs $T$ normalized by the computational costs $T_\mathrm{JIT}$ of the JIT implementation, therefore $\Upsilon_\mathrm{JIT} = 1$. Calculations of absolute computational costs \newtext{are} made based on averaging the CPU time of calculations for more than 5 program runs, while the total execution time \newtext{is} taken to be at least 5 minutes, therefore samples executed for less than 1 second \newtext{are} repeated more than 300 times to average the execution time.

The obtained values of absolute $T$ and relative computational costs $\Upsilon$ for the finite-element ADER-DG method in the two-dimensional case are presented in Table~\ref{tab:comp_costs_dg_2d} and in the three-dimensional case are presented in Table~\ref{tab:comp_costs_dg_3d}. The obtained values of absolute and relative computational costs for the finite-volume ADER-WENO method in the two-dimensional case are presented in Table~\ref{tab:comp_costs_fv_2d} and in the three-dimensional case are presented in Table~\ref{tab:comp_costs_fv_3d}. The obtained values of absolute and relative computational costs for the finite-element ADER-DG methods with a posteriori correction of the solution in subcells by a finite-volume ADER-WENO limiter in the two-dimensional case are presented in Table~\ref{tab:comp_costs_dg_scl_2d} and in the three-dimensional case are presented in Table~\ref{tab:comp_costs_dg_scl_3d}.

Note that the increase of absolute computational costs $T$ is slower with increase of the polynomial degree $N$ compared with increase of the number of mesh cells. This is due to the following factors. Initial conditions are calculated only once, so the relative contribution of this procedure is decreased with increase of the total number of time steps and decrease the coordinate step $h$. The boundary has a dimension one less than the dimension of the problem, and the increase of the number of boundary cells with increase in the mesh size has a different dimensional asymptotics. Another factor is the change in the number of iterations of the LST-DG predictor --- for very coarse meshes, the number of iterations is close to the largest expected value $\sim N+1$, and when the mesh is refined, the relative variability of the smooth solution in the cell decreases, and along with it, the number of iterations decreases.

The relative performance gain depends significantly on the polynomials degree $N$, that determine the sizes of the matrices with which the solution representations are convolved, and on the size of the spatial mesh on which the calculations are performed. There is also a significant dependence on the spatial dimension of the problem and a specific numerical method --- finite-element ADER-DG or finite-volume ADER-WENO. It should be noted that the nature of these dependencies is significantly non-monotonic, however, these patterns can be explained in the context of reducing the overhead costs of function calls (especially in the case of the GEN implementation) and changing the number of \texttt{gemm} function calls. 

The obtained relative values of computational costs $\Upsilon$ \newtext{allow} to immediately conclude that the implementation presented in this work, based on the use of the JIT functions of the BLAS, outperforms both the implementation based on the general BLAS functions and the vanilla implementations using explicit multiplications and summations in simple loops. The JIT implementation outperforms the OV implementation in computation speed by $1.5$-$2.5$ times in most cases, the GEN implementation by $1.5$-$4.0$ times, and the CV implementation by 2-8 times. In this case, this advantage is observed precisely in the computational costs of executing the entire algorithm, and not just matrix-matrix multiplications. 

In almost all cases, the next smallest computational cost is the OV implementation based on a simply optimized \texttt{gemm} function. In the case of the ADER-DG-$\mathbb{P}_{1}$ method for two-dimensional problem, the OV implementation is only 6-14\% slower than the JIT implementation, which is due to the relatively small number of calls to the \texttt{gemm} function in loops and the commensurability of the performance of the functions in the case of multiplication by a $2\times2$ matrix. However, in the case of ADER-DG-$\mathbb{P}_{1}$ method for three-dimensional problem, the differences are already $1.5$-$2.0$ times. With an increase in the polynomials degree $N$, there is a relative increase in the computational costs of the OV implementation compared to JIT. The OV implementation shows results slower than GEN only in case of ADER-DG-$\mathbb{P}_{6}$ method for three-dimensional problem with mesh size $7^{3}$. It should be noted that the simplicity of the OV implementation vs JIT is not great --- it is still necessary to explicitly use the \texttt{gemm} function, only with a simple implementation instead of creating, storing and destroying jitters in JIT implementation. In this case, the relative efficiency of OV implementation compared to CV is due to the allocation of a separate \texttt{gemm} function as a bottleneck in execution, and its implementation in such a way that the compiler could effectively optimize this function.

The high computational costs $\Upsilon$ of the GEN implementation are expected --- the \texttt{cblas\_(d)gemm} functions are originally intended for calculations with relatively large matrices, unlike JIT BLAS. However, in many of the presented cases, GEN has lower computational costs compared to CV. The high computational costs $\Upsilon$ of the CV implementation are also expected and are related to the inefficiency of compiler optimization of operations with low memory locality and very deeply nested loops. However, the CV implementation is the simplest of those proposed --- it is a simple coding of the formula apparatus of the numerical method.

\section{Performance analysis}
\label{sec:perf_anal}

The results of the computational cost evaluation presented in the previous Section~\ref{sec:comp_costs} showed that in terms of computational costs, the JIT implementation outperforms the optimized vanilla implementation in most cases by $1.5$-$2.5$ times, the implementation with common BLAS functions by $1.5$-$4.0$ times, and the completely vanilla implementation without using the BLAS interface by 2-8 times. It is clear that the total number of floating-point operations when solving the same problems is the same, up to possible compiler optimizations associated with the reordering of arithmetic expressions. Therefore, it is expected that such significant differences in computational costs for optimized implementations are mainly due to more efficient use of the computing architecture, primarily due to increased memory locality and more efficient use of the instructions related to floating-point numbers. It should be noted that reducing computational costs while maintaining the accuracy and correctness of calculations is usually a necessary and sufficient reason for adopting the selected optimization methods. 

However, the relative decreasing in computational costs alone is not sufficient to understand the reasons for the increase in computational performance associated with optimization. The performance analysis in this work is carried out using the naive roofline model. The roofline model is widely used to obtain estimates of the performance of computational codes~\cite{Robey_Zamora_hpc_2021} with respect to the main limitations inherent in computing systems, especially the bandwidth of various levels of memory organization (DRAM, processor caches) and the system's performance in floating-point operations, which differs significantly for different instructions sets associated with the use of the SIMD paradigm (from scalar operations to using SSE, AVX, AVX-512 FMA instructions sets).

The performance analysis results in this work have been obtained using the \texttt{likwid} (stands for ``Like I Knew What I’m Doing'') application~\cite{likwid_wiki, likwid_git}. The roofline model for the used computing system was obtained using the \texttt{likwid-bench} application~\cite{likwid_bench_git}. All data on the performance of computing implementations have been obtained using the \texttt{likwid-perfctr} application~\cite{likwid_perfctr_git}. The \texttt{likwid} toolkit is one of the recommended tools~\cite{Robey_Zamora_hpc_2021, doe_hpc_recommend} for performance analysis in the field of high-performance and parallel computing. The peak roofline plot has been constructed based on the application output using the formula for performance $P(I) = \min(P_{\rm max},\ I \cdot B)$, where $P_{\rm max}$ is the peak performance for the instruction set involved (separately for scalar operations and SSE, AVX, AVX-512 FMA instructions sets), $B$ is the peak bandwidth when the same instruction set is involved, $I$ is the arithmetic intensity --- the number of floating point operations per transferred byte.

It is interesting to note~\cite{doe_ai_code} that typical finite-volume methods based on the use of reconstruction on mesh stencils, such as ENO and WENO, are bandwidth-bound problems (limited by bandwidth of memory) and are characterized by an arithmetic intensity of $I \sim 0.1$-$1.0$, spectral methods are intermediate between bandwidth-bound and compute-bound (limited by performance in floating-point operations) problems and are characterized by an arithmetic intensity of $I \sim 1.0$-$2.0$, while level 3 BLAS operations for dense matrices are classic examples of compute-bound problems and can be characterized by an arithmetic intensity $I$ of up to $10$ or more. 

\begin{figure*}[h!]
\centering
\includegraphics[width=0.32\textwidth]{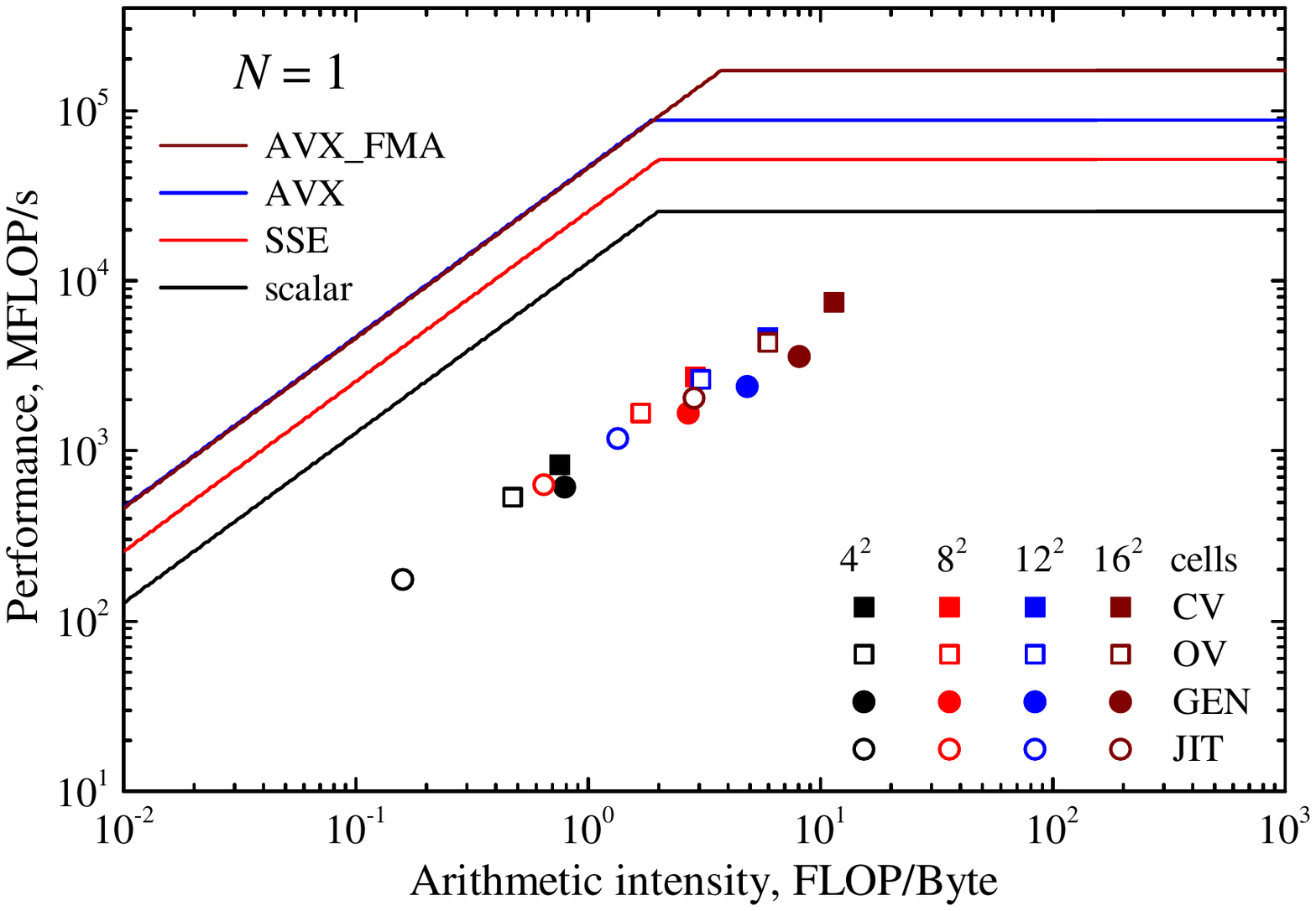}
\includegraphics[width=0.32\textwidth]{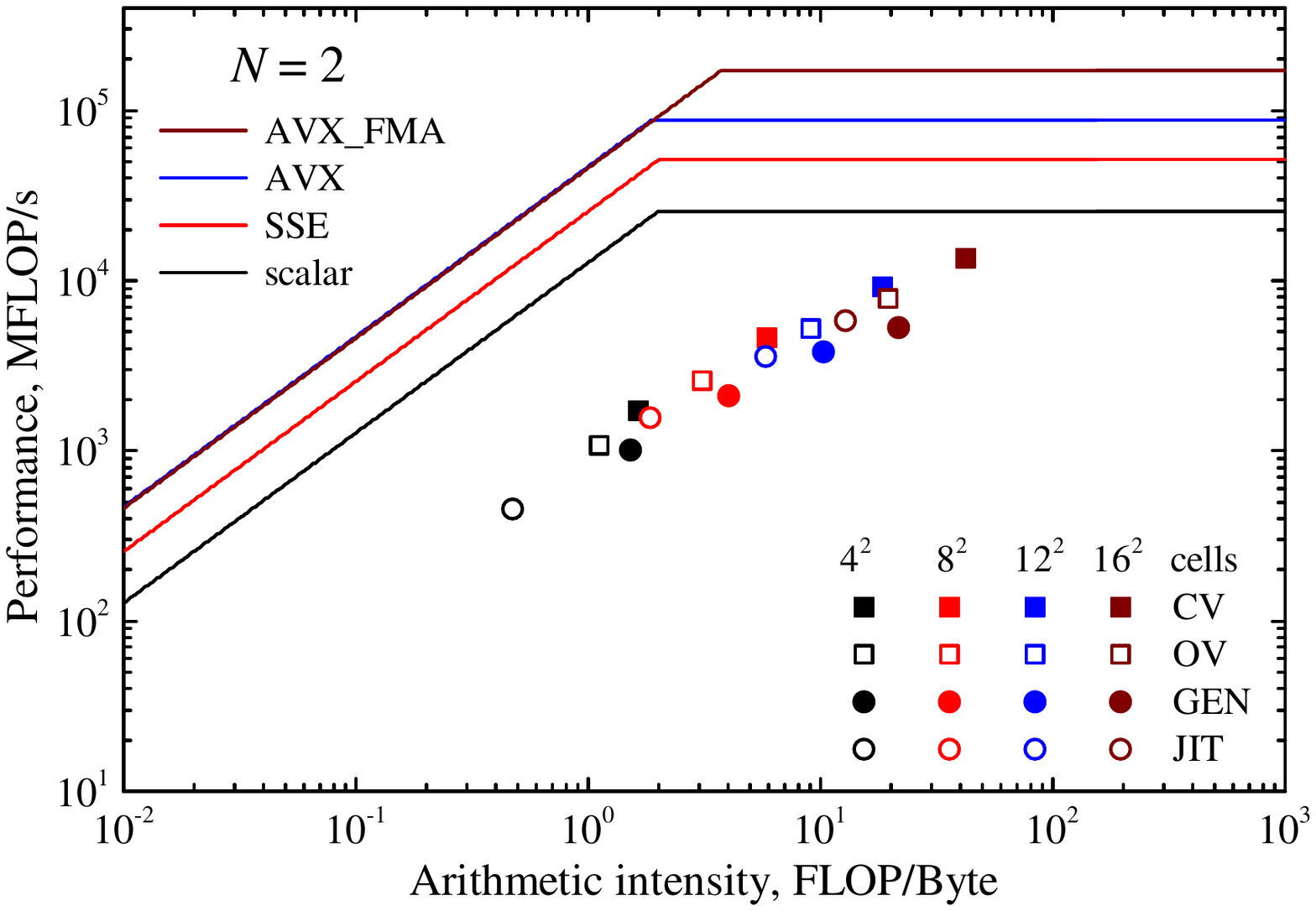}
\includegraphics[width=0.32\textwidth]{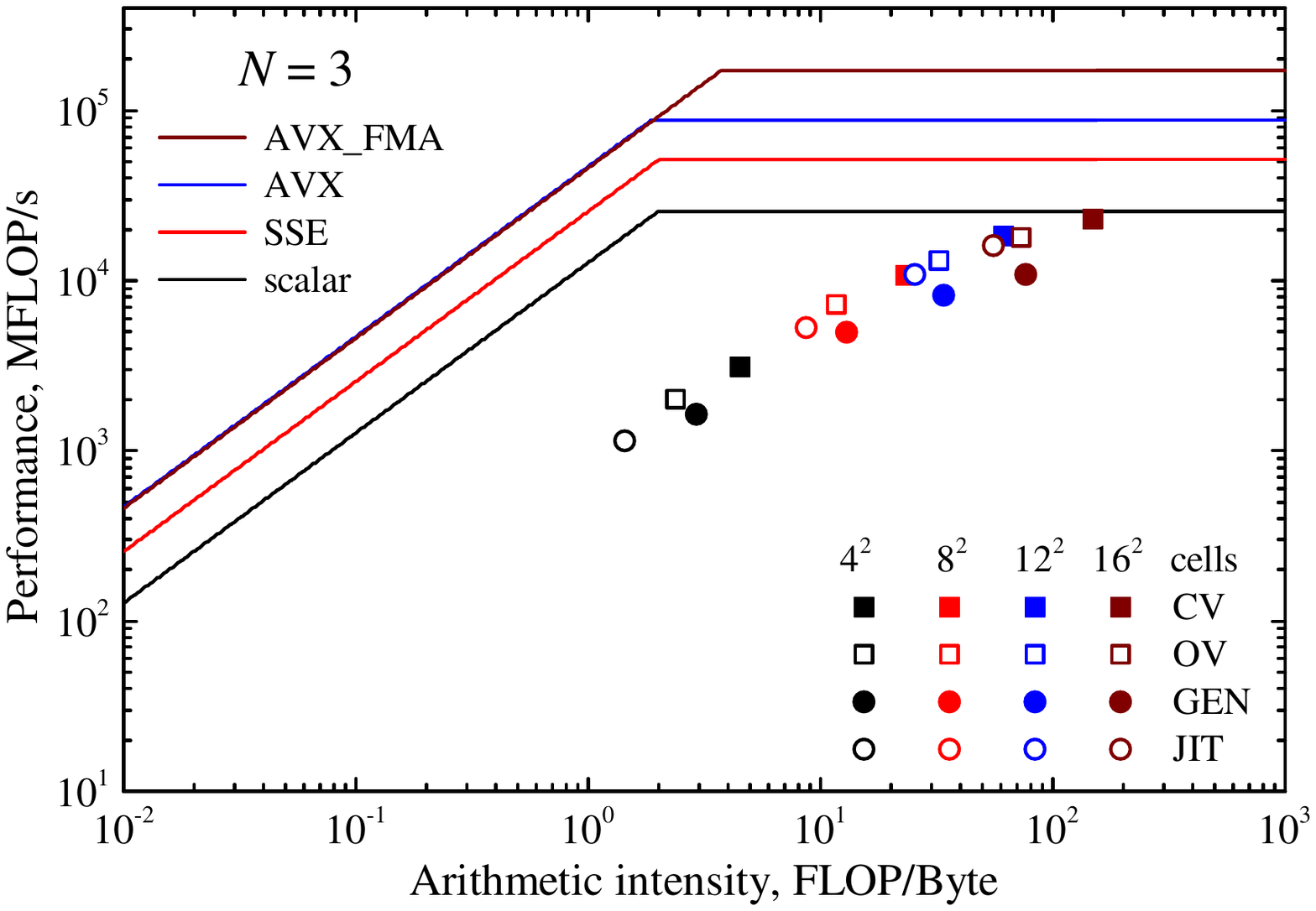}\\
\includegraphics[width=0.32\textwidth]{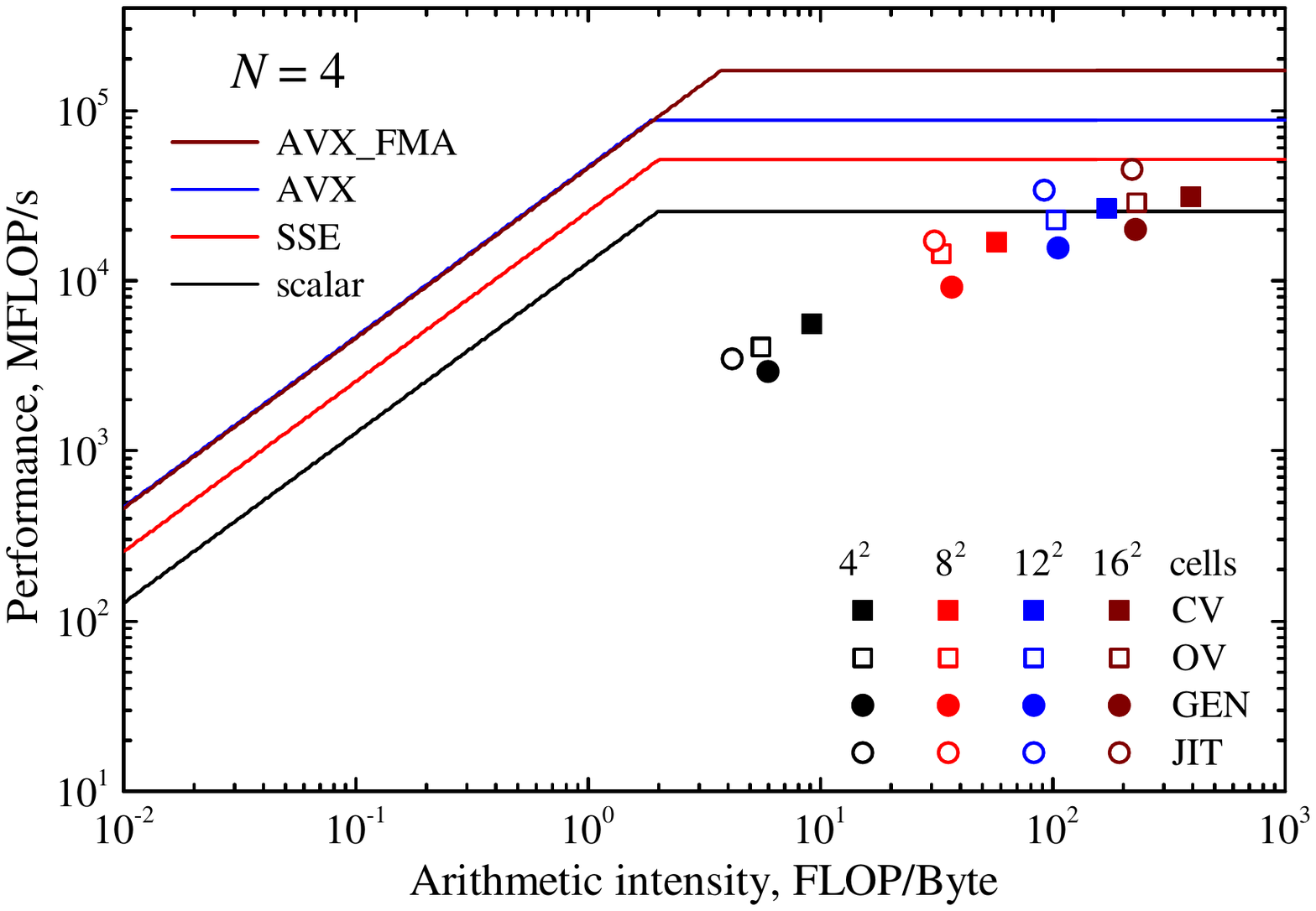}
\includegraphics[width=0.32\textwidth]{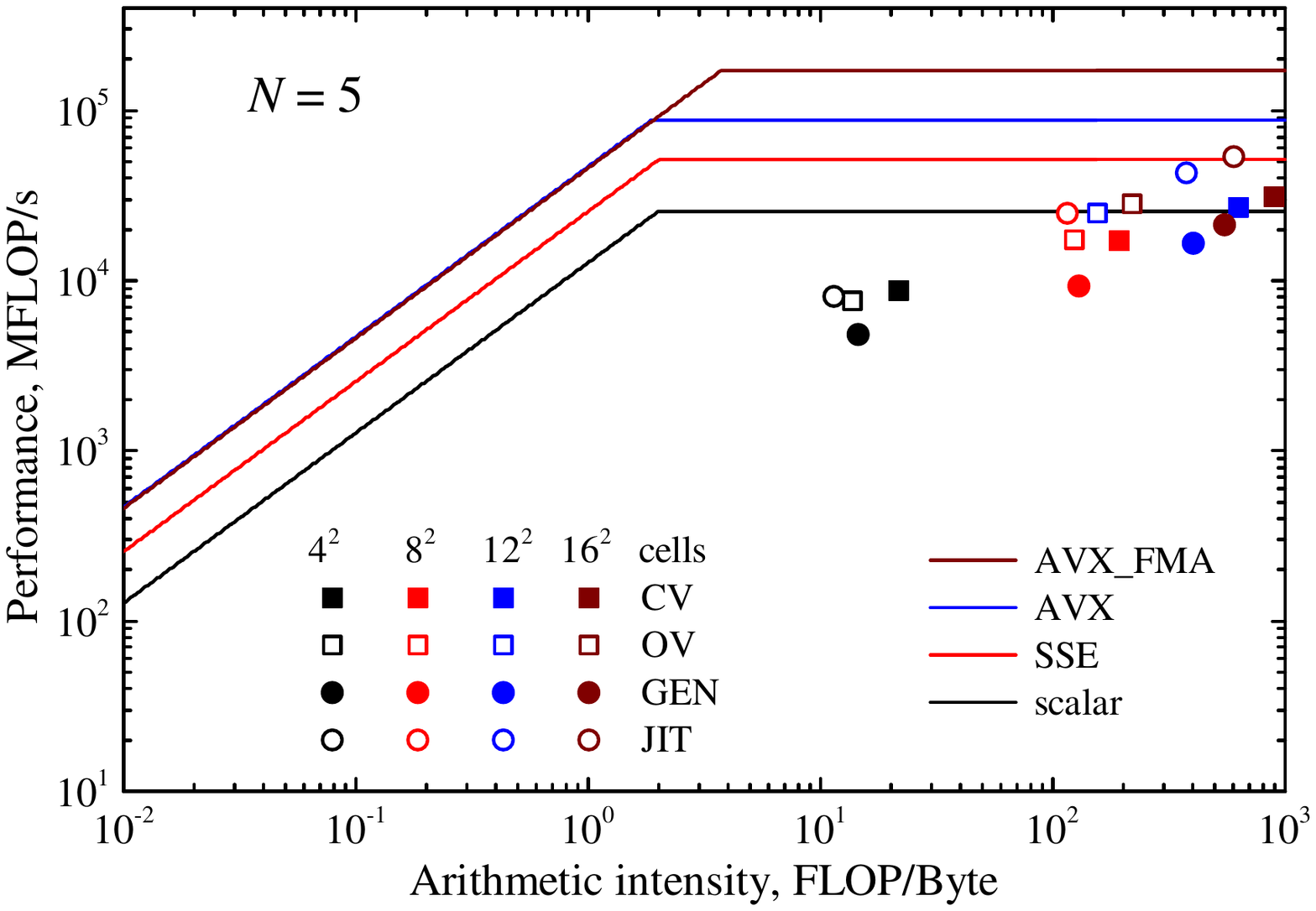}
\includegraphics[width=0.32\textwidth]{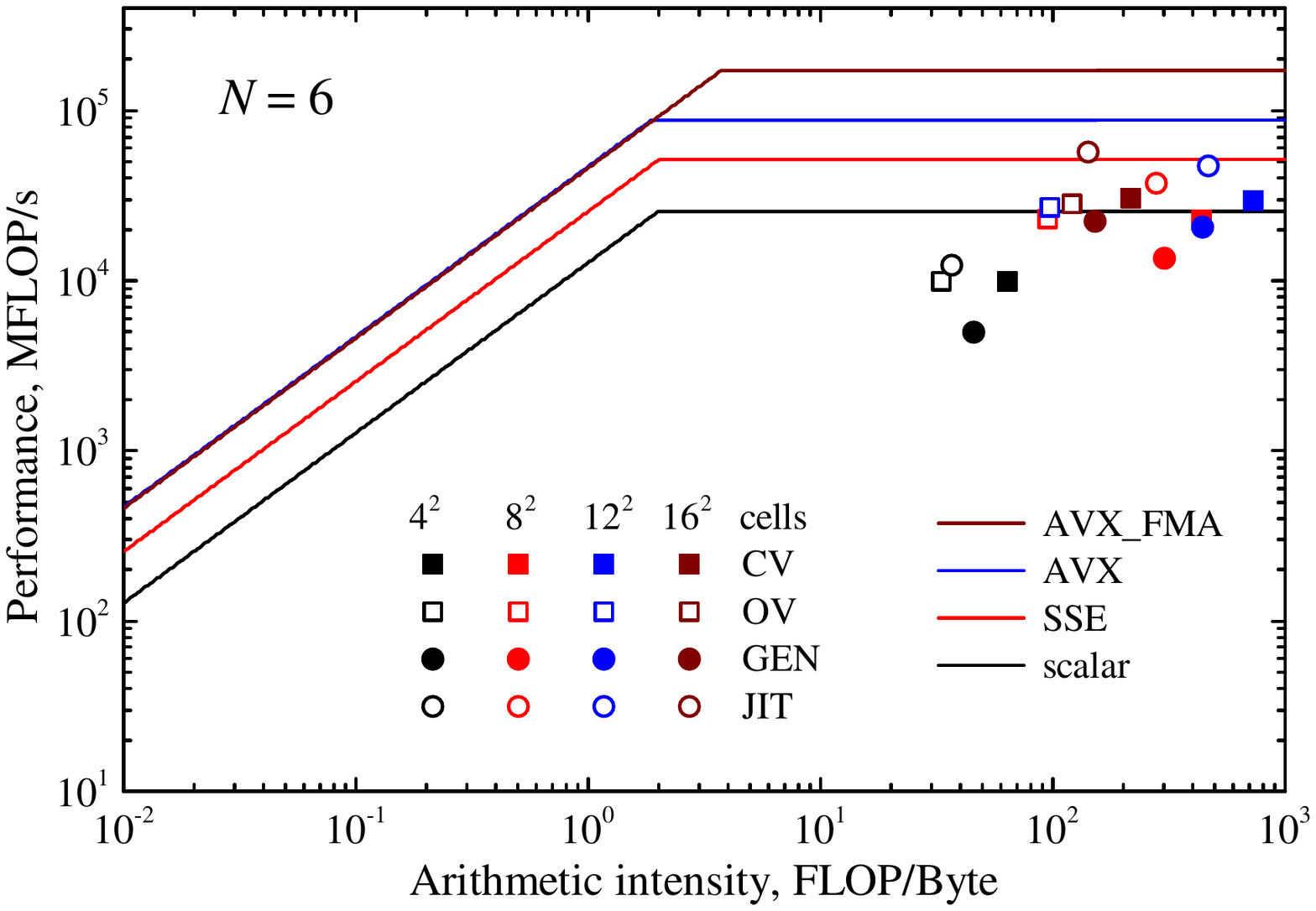}\\
\includegraphics[width=0.32\textwidth]{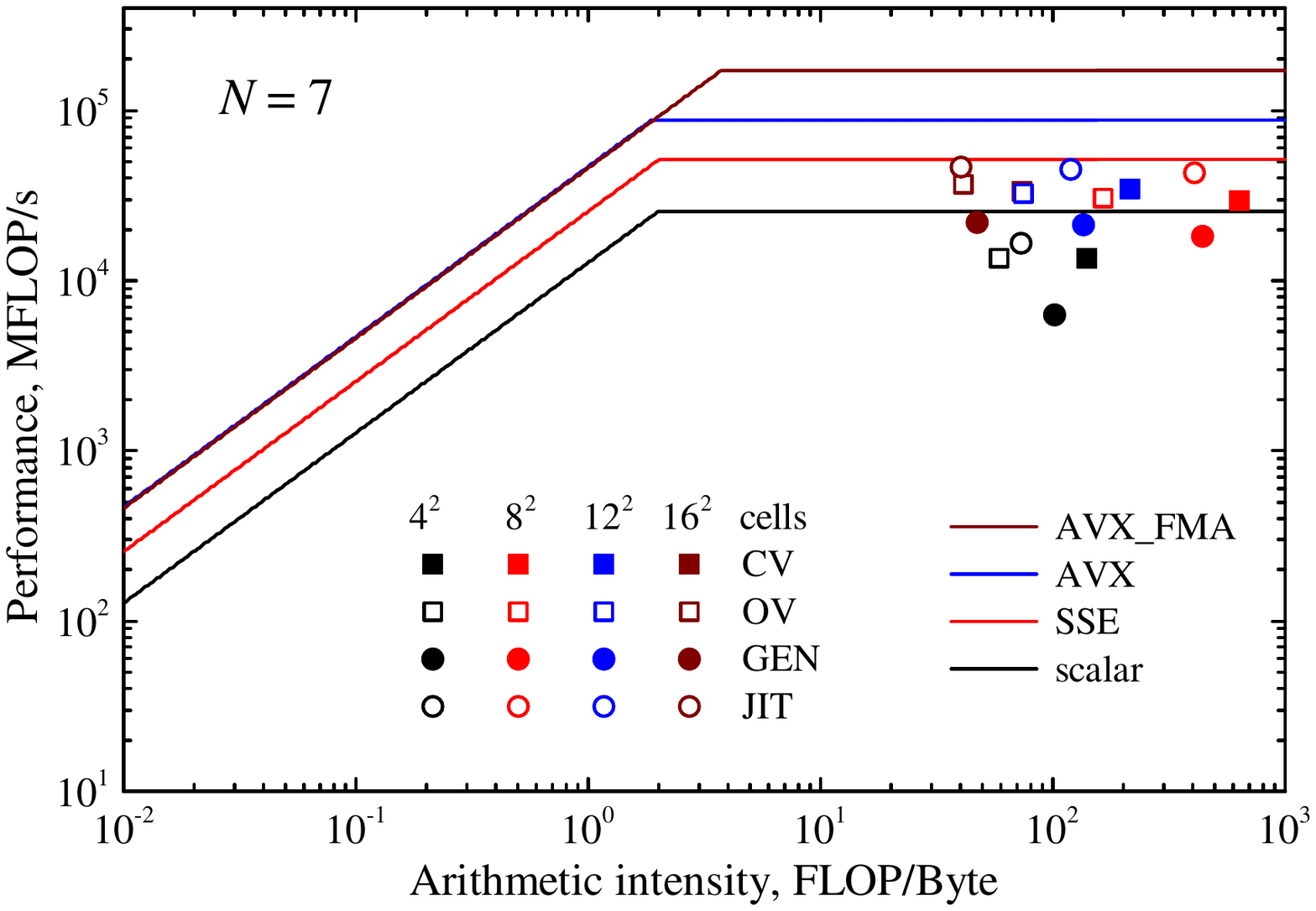}
\includegraphics[width=0.32\textwidth]{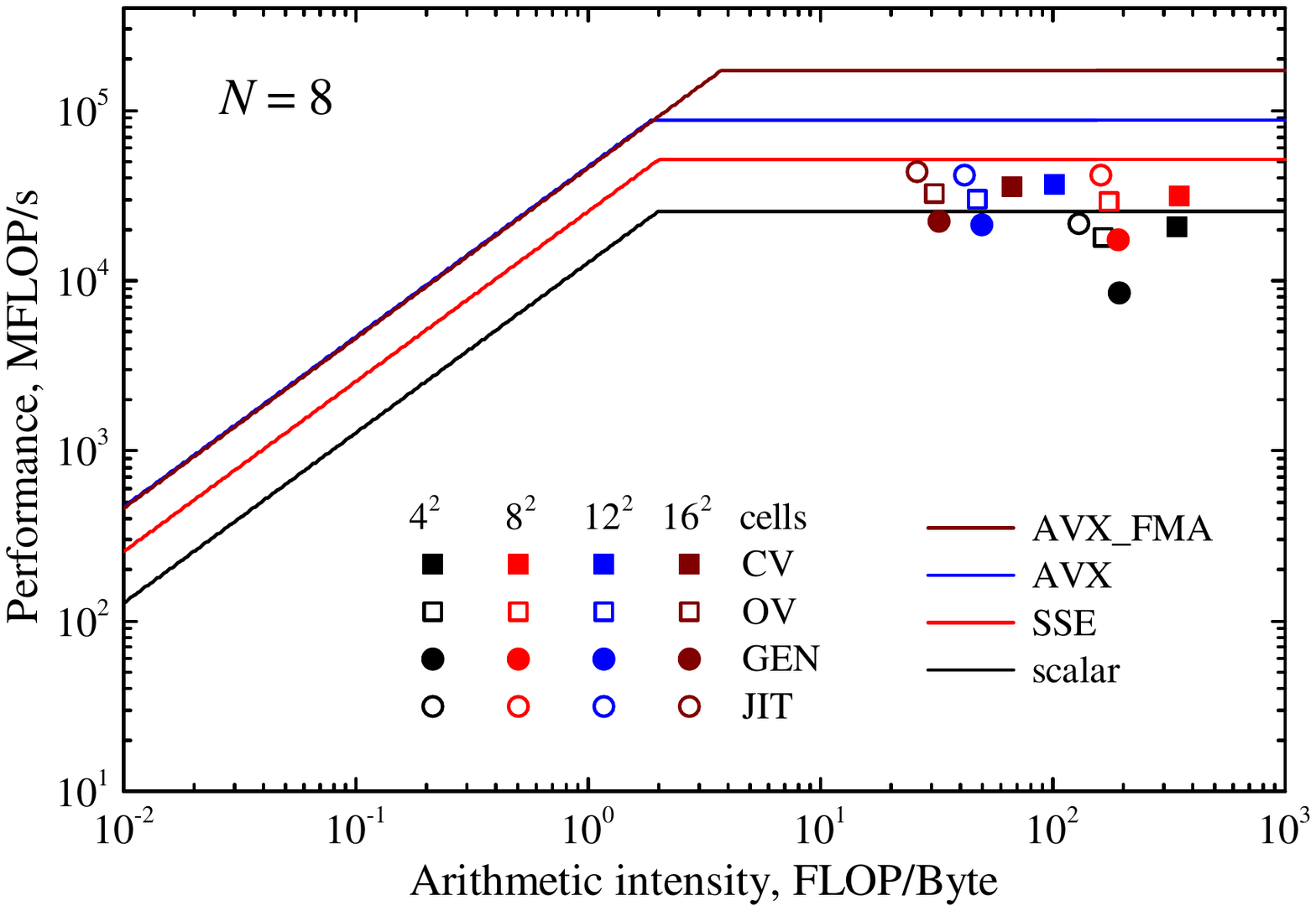}
\includegraphics[width=0.32\textwidth]{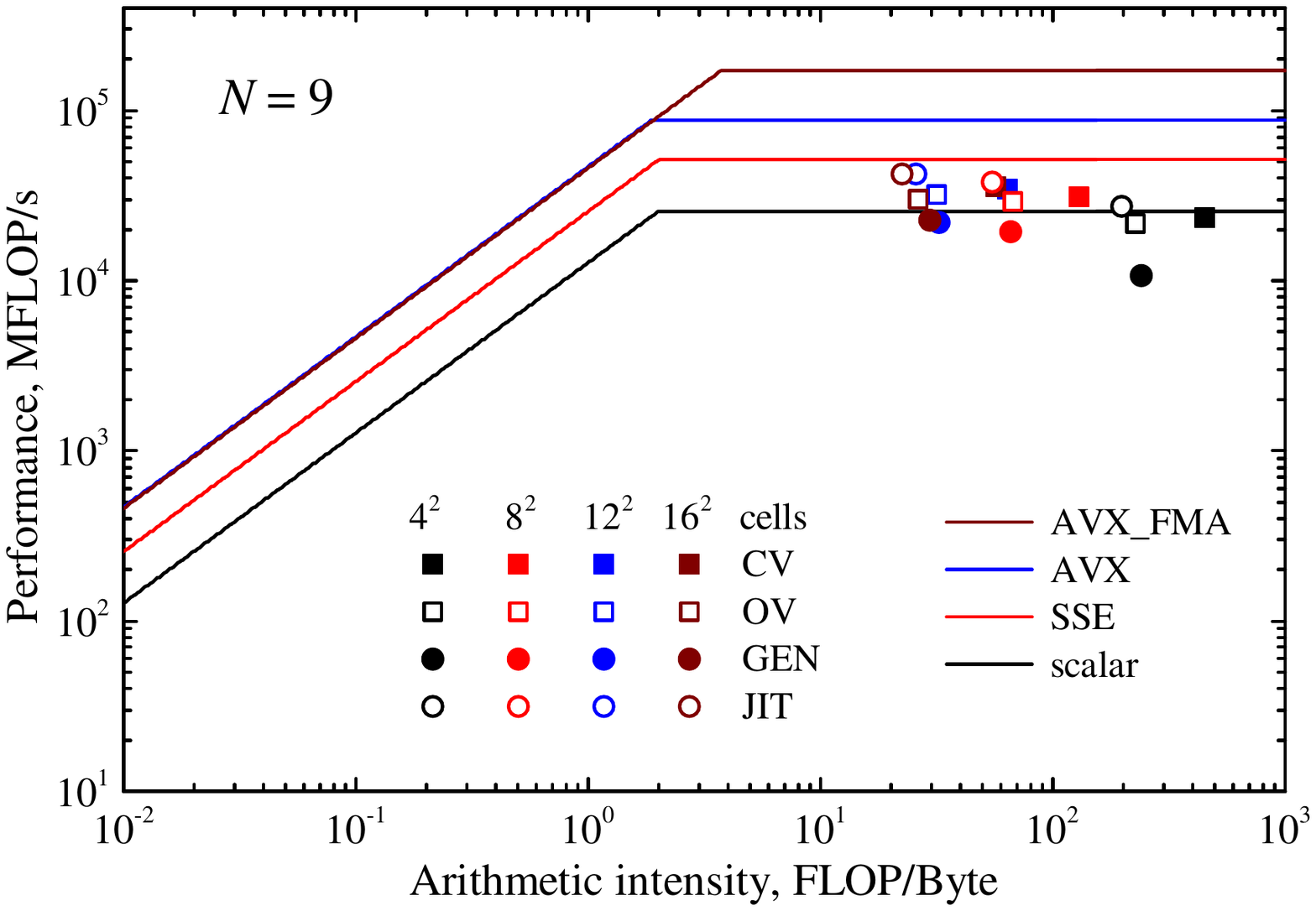}
\caption{\label{fig:roofline_DG_2d_degrees}%
Roofline models for ADER-DG-$\mathbb{P}_{N}$ method for two-dimensional sine wave advection problem --- performance (measured in MFLOP/s) versus arithmetic intensity (measured in FLOP/Byte) on a double logarithmic scale. Peak performance values are presented for scalar operations, for operations using the SSE instruction set, using the AVX instruction set, and using the AVX-512 FMA instruction set. Data points are presented for CV, OV, general BLAS and JIT BLAS implementations. A comparison is presented for different mesh sizes --- $4^{2}$, $8^{2}$, $12^{2}$, $16^{2}$, at different polynomial degrees $N = 1, \ldots, 9$.
}
\end{figure*}

\begin{figure*}[h!]
\centering
\includegraphics[width=0.245\textwidth]{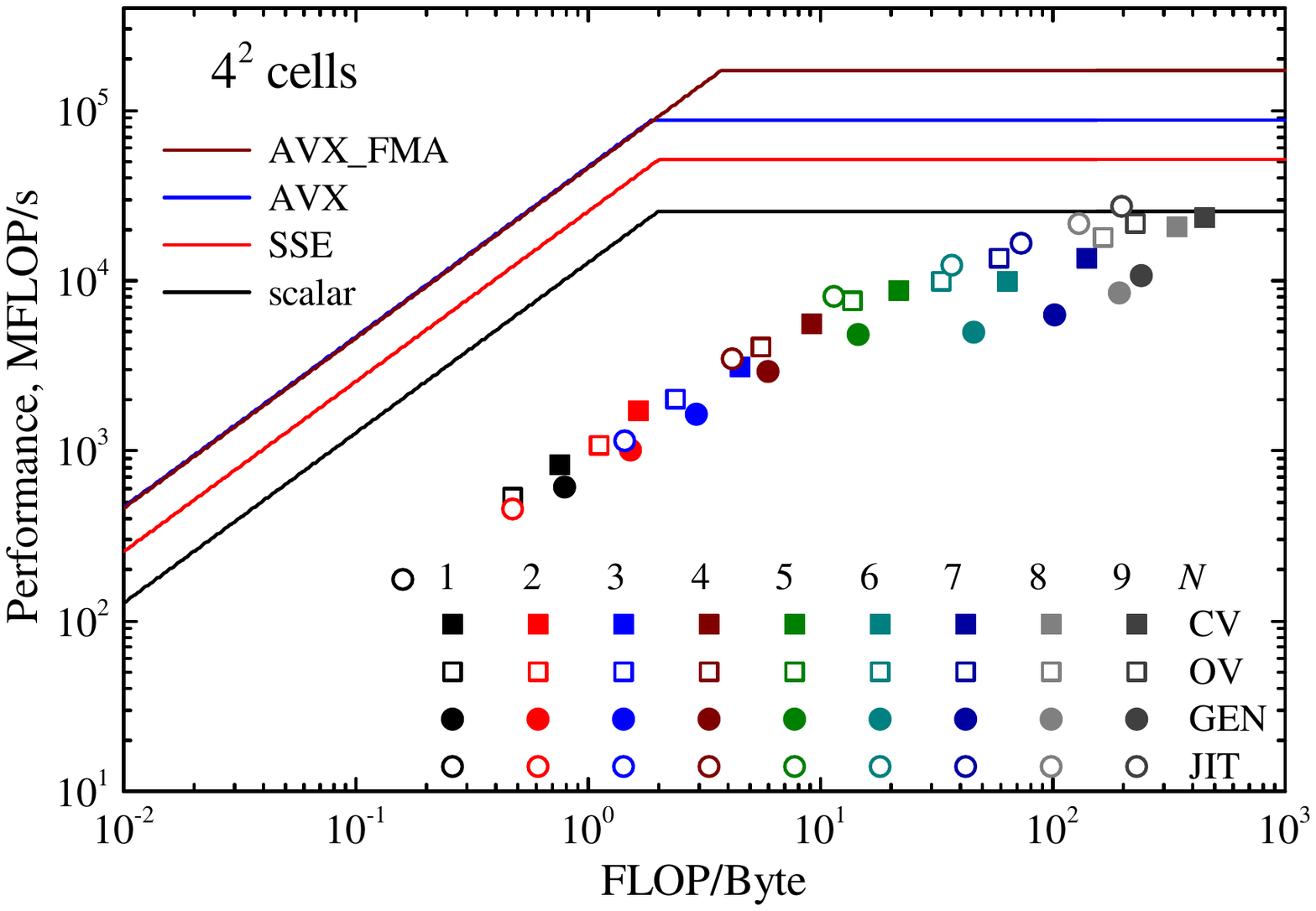}
\includegraphics[width=0.245\textwidth]{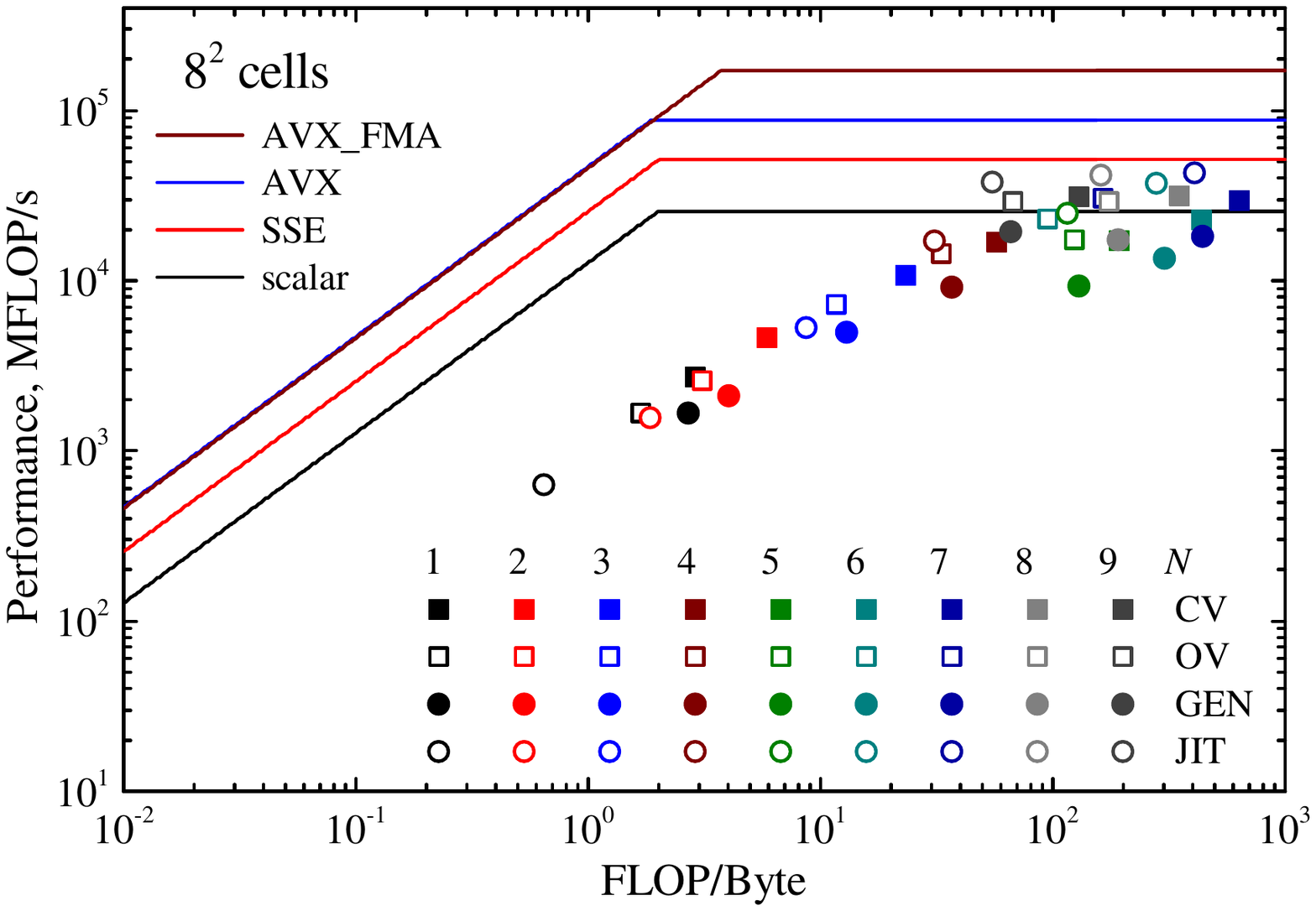}
\includegraphics[width=0.245\textwidth]{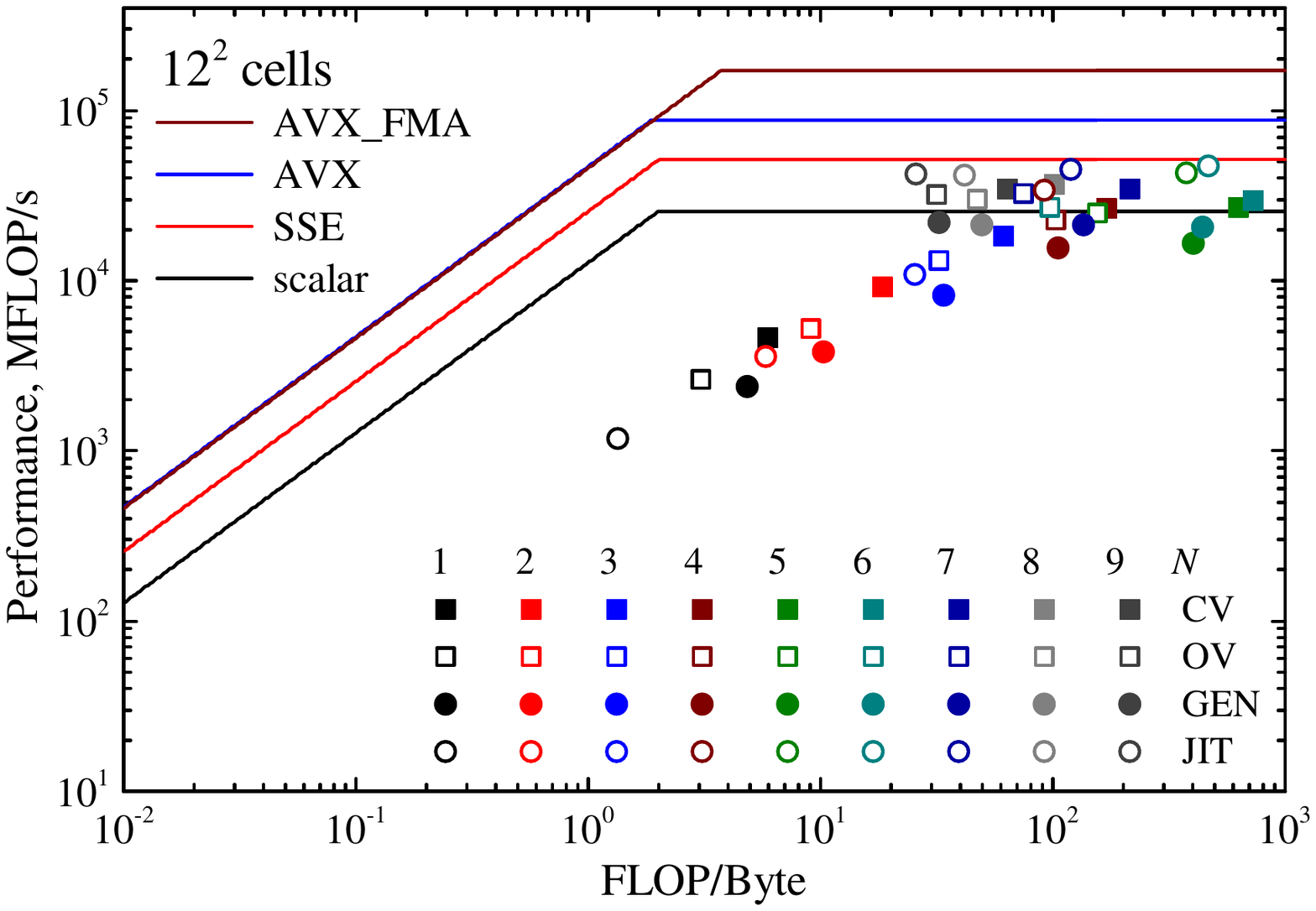}
\includegraphics[width=0.245\textwidth]{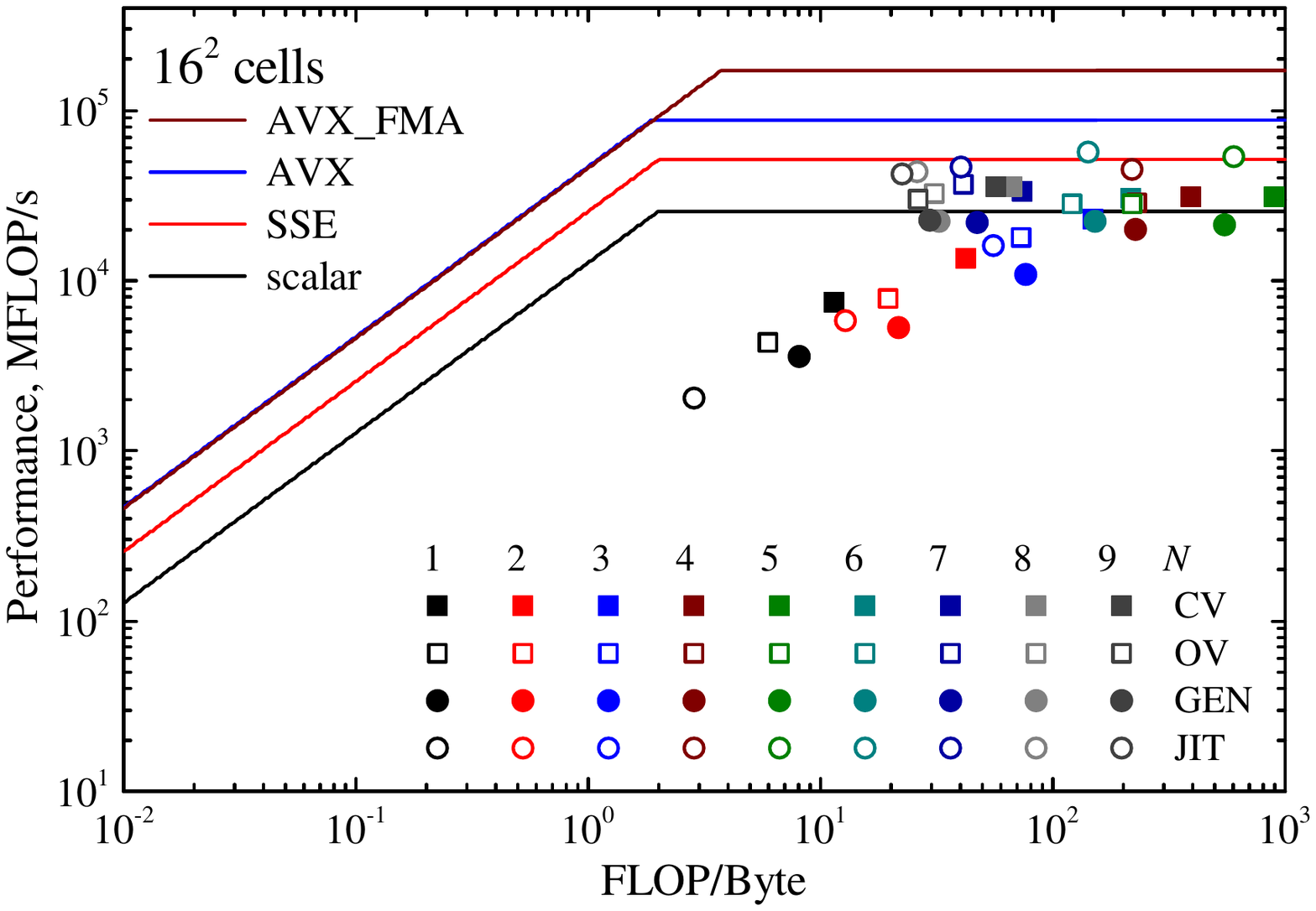}
\caption{\label{fig:roofline_DG_2d_mesh_sizes}%
Roofline models for ADER-DG-$\mathbb{P}_{N}$ method for two-dimensional sine wave advection problem --- performance (measured in MFLOP/s) versus arithmetic intensity (measured in FLOP/Byte) on a double logarithmic scale. Peak performance values are presented for scalar operations, for operations using the SSE instruction set, using the AVX instruction set, and using the AVX-512 FMA instruction set. Data points are presented for CV, OV, general BLAS and JIT BLAS implementations. A comparison is presented for different polynomial degrees $N = 1, \ldots, 9$, at mesh sizes $4^{2}$, $8^{2}$, $12^{2}$, $16^{2}$.
}
\end{figure*}

The finite-element ADER-DG and finite-volume ADER-WENO methods separately, and the finite-element ADER-DG methods with a posteriori correction of the solution in subcells by a finite-volume ADER-WENO limiter, which combines the ADER-DG and ADER-WENO methods and some additional procedures for analyzing and rebuilding the solution, presented in this work purely fit into all these classes of typical methods. The finite element ADER-DG-$\mathbb{P}_{N}$ method does not use the reconstruction of the solution on mesh stencils and has a very compact stencil, while it is expected that the main bottleneck in this case is the matrix-matrix multiplication operations, as well as the calculation of mutual transformations of conservative values, flow, non-conservative and source terms and primitive values. The ADER-WENO($N+1$) method uses the procedure of reconstruction of the solution on stencils, but it is expected that the use of the LST-DG-predictor and convolutions in the calculation of flow, non-conservative and source terms will also lead to a shift in the roofline data to the right towards compute-bound problems. In the case of the finite-element ADER-DG-$\mathbb{P}_{N}$ method with a posteriori correction of the solution in subcells by a finite-volume ADER-WENO limiter, the occurrence of features inherent in both finite-element ADER-DG-$\mathbb{P}_{N}$ method and finite-volume ADER-WENO method is expected, and the specifics will depend on the proportion of trouble cells on the mesh.

\begin{figure*}[h!]
\centering
\includegraphics[width=0.32\textwidth]{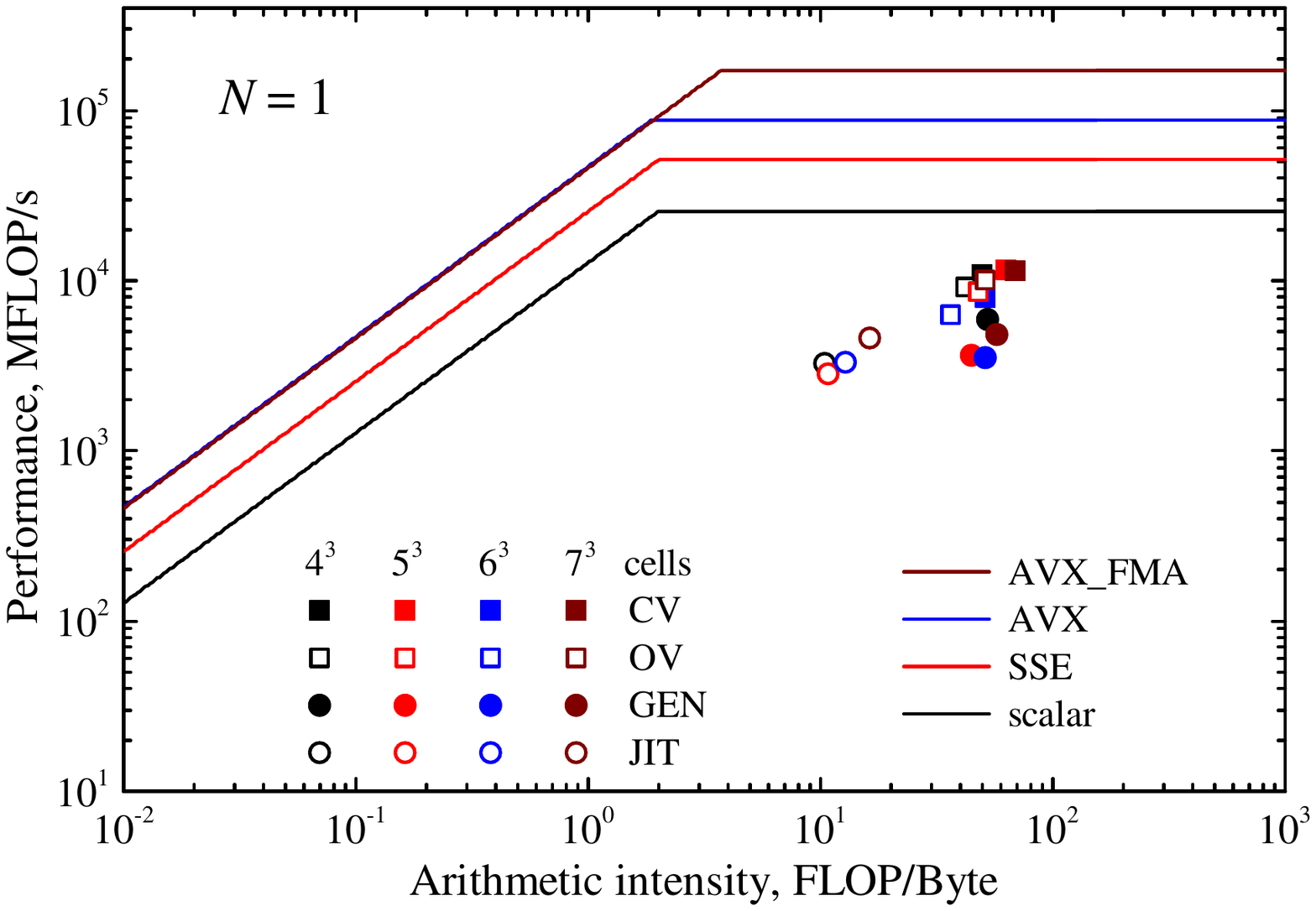}
\includegraphics[width=0.32\textwidth]{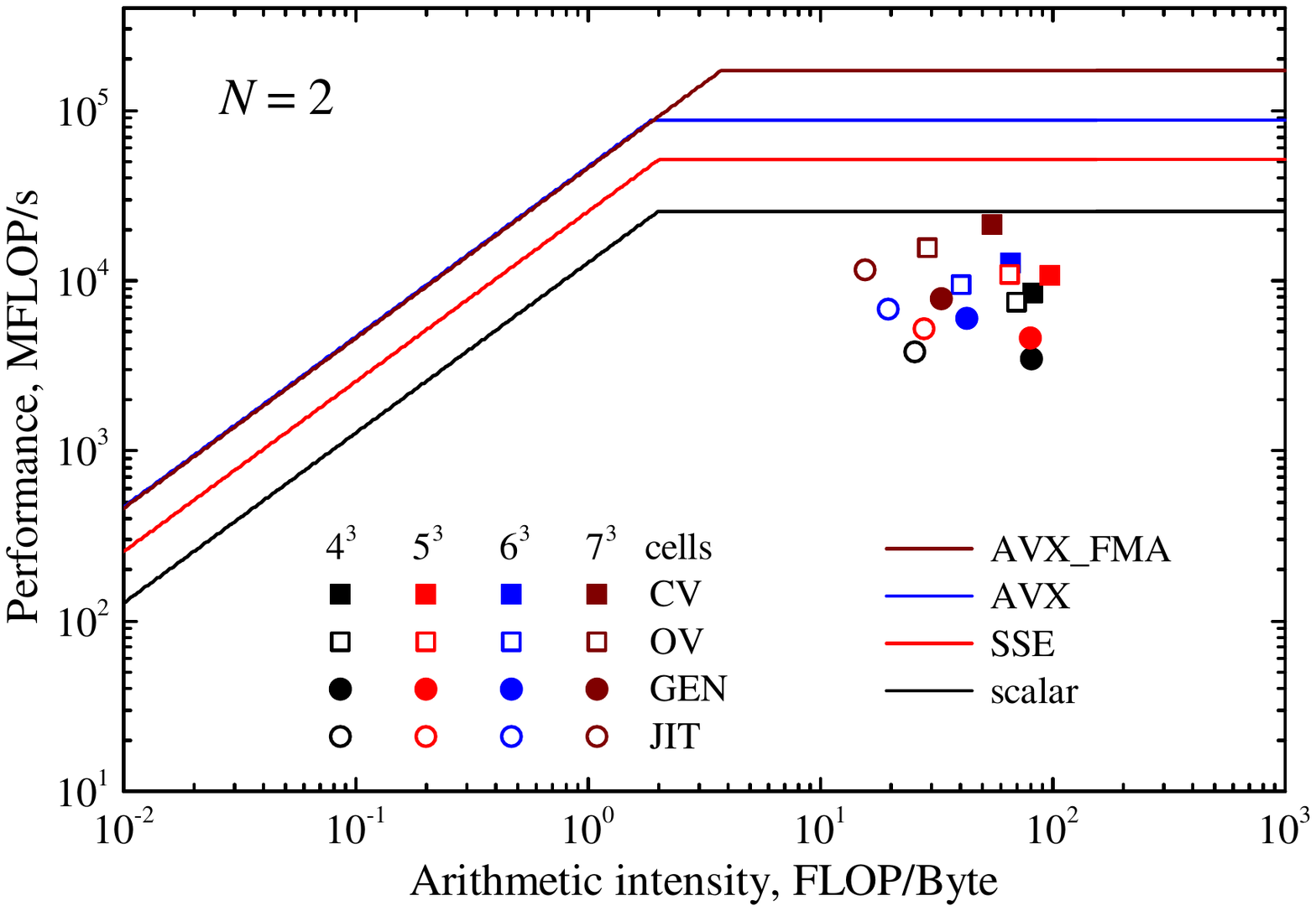}
\includegraphics[width=0.32\textwidth]{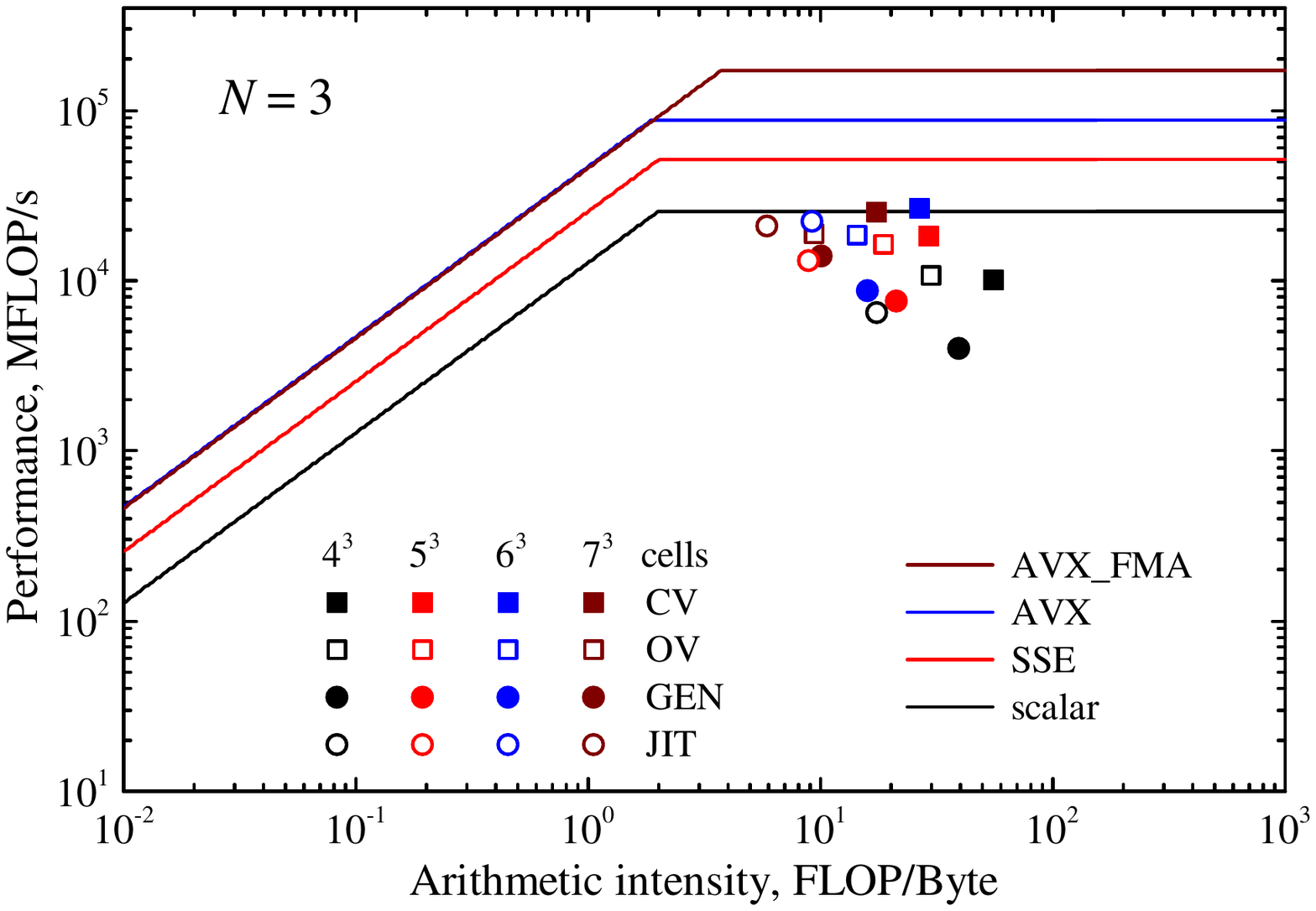}\\
\includegraphics[width=0.32\textwidth]{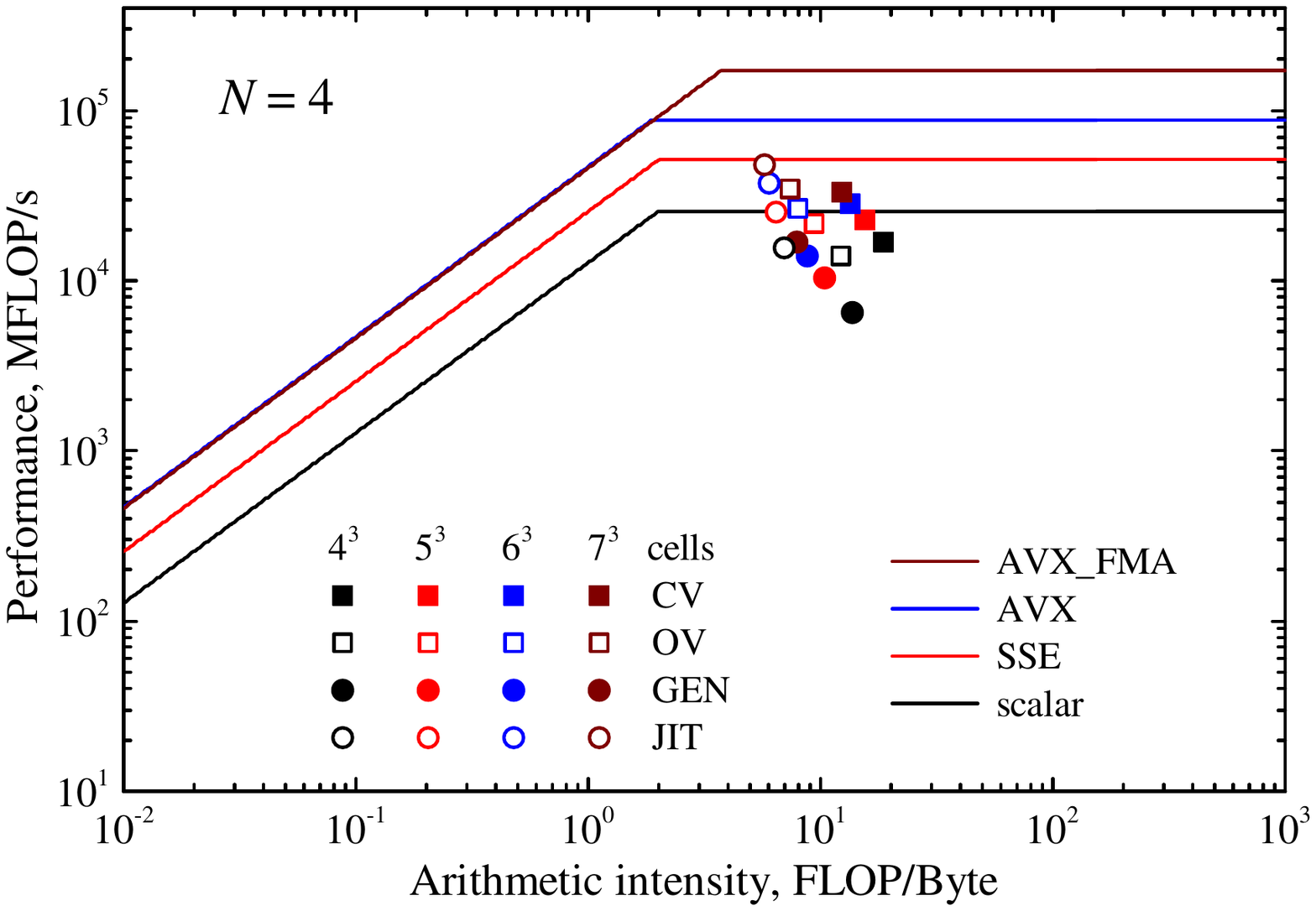}
\includegraphics[width=0.32\textwidth]{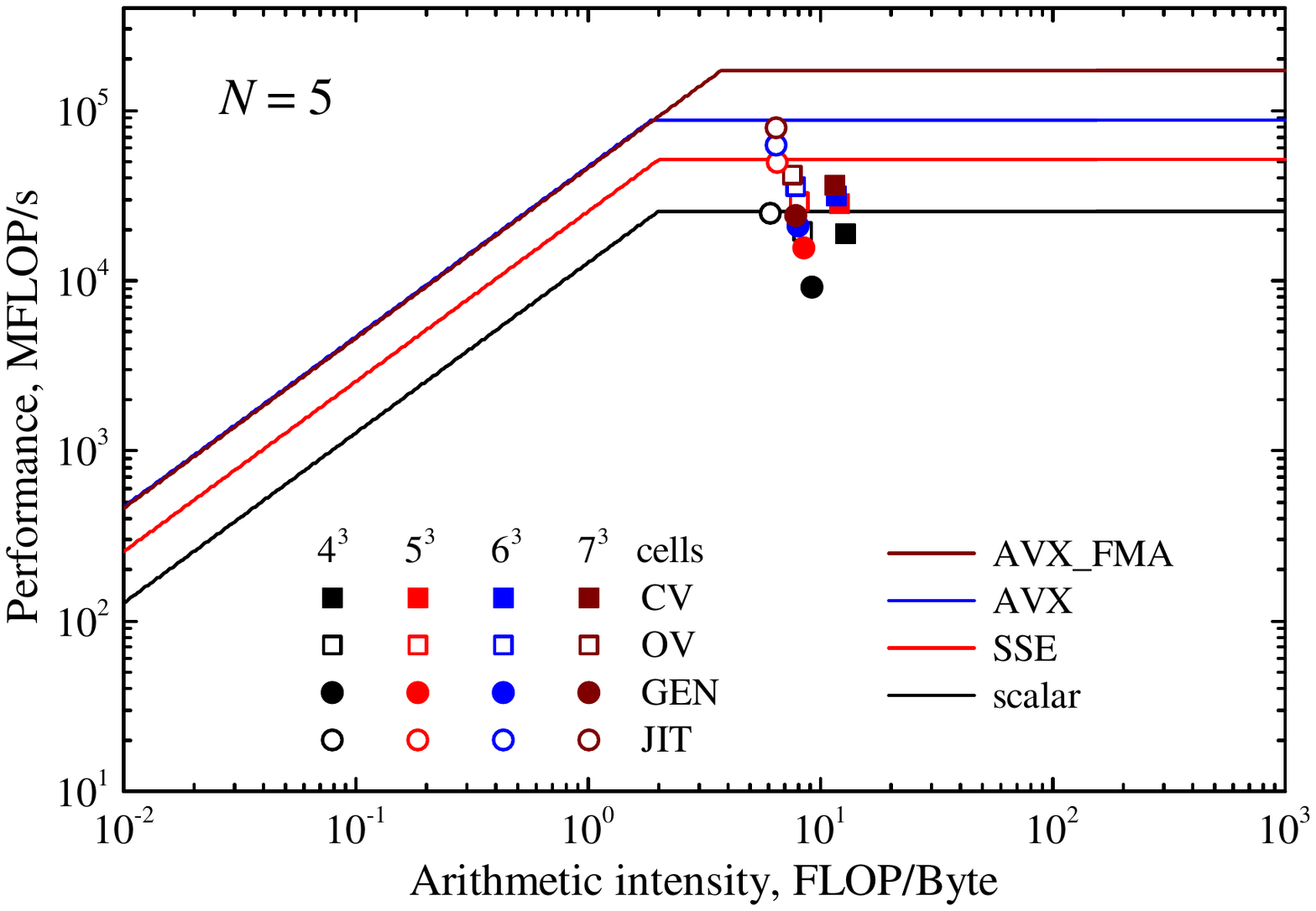}
\includegraphics[width=0.32\textwidth]{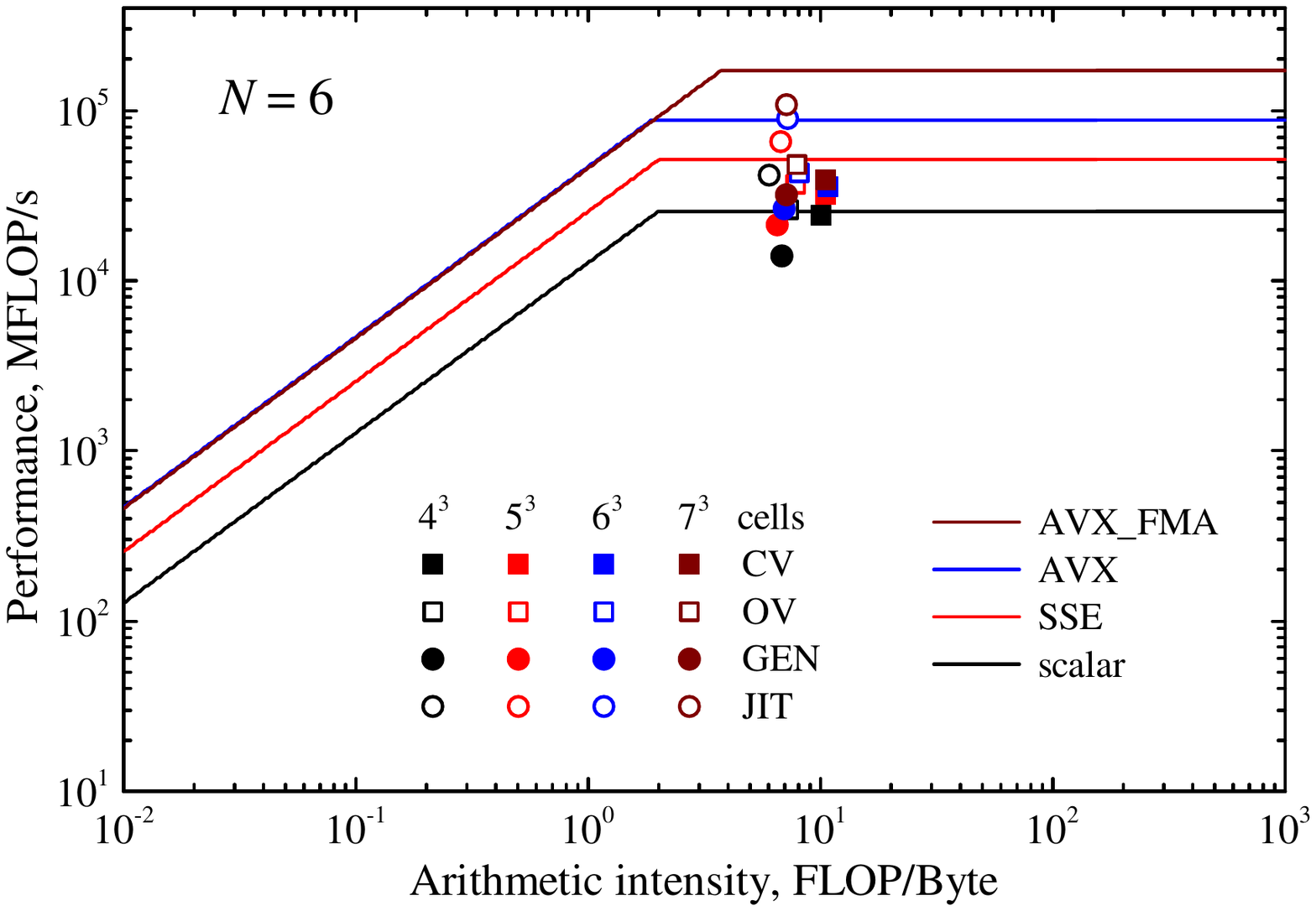}\\
\includegraphics[width=0.32\textwidth]{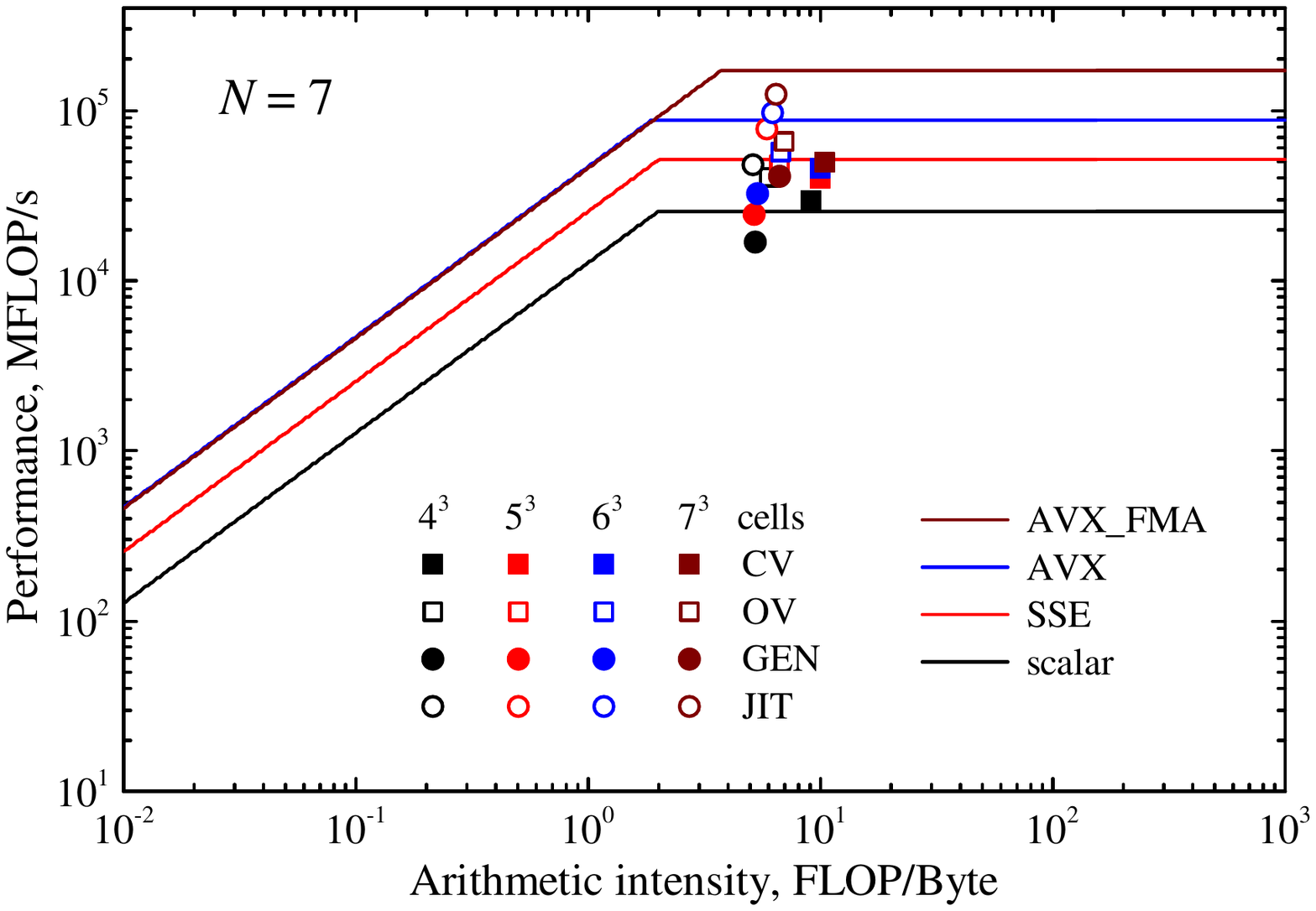}
\includegraphics[width=0.32\textwidth]{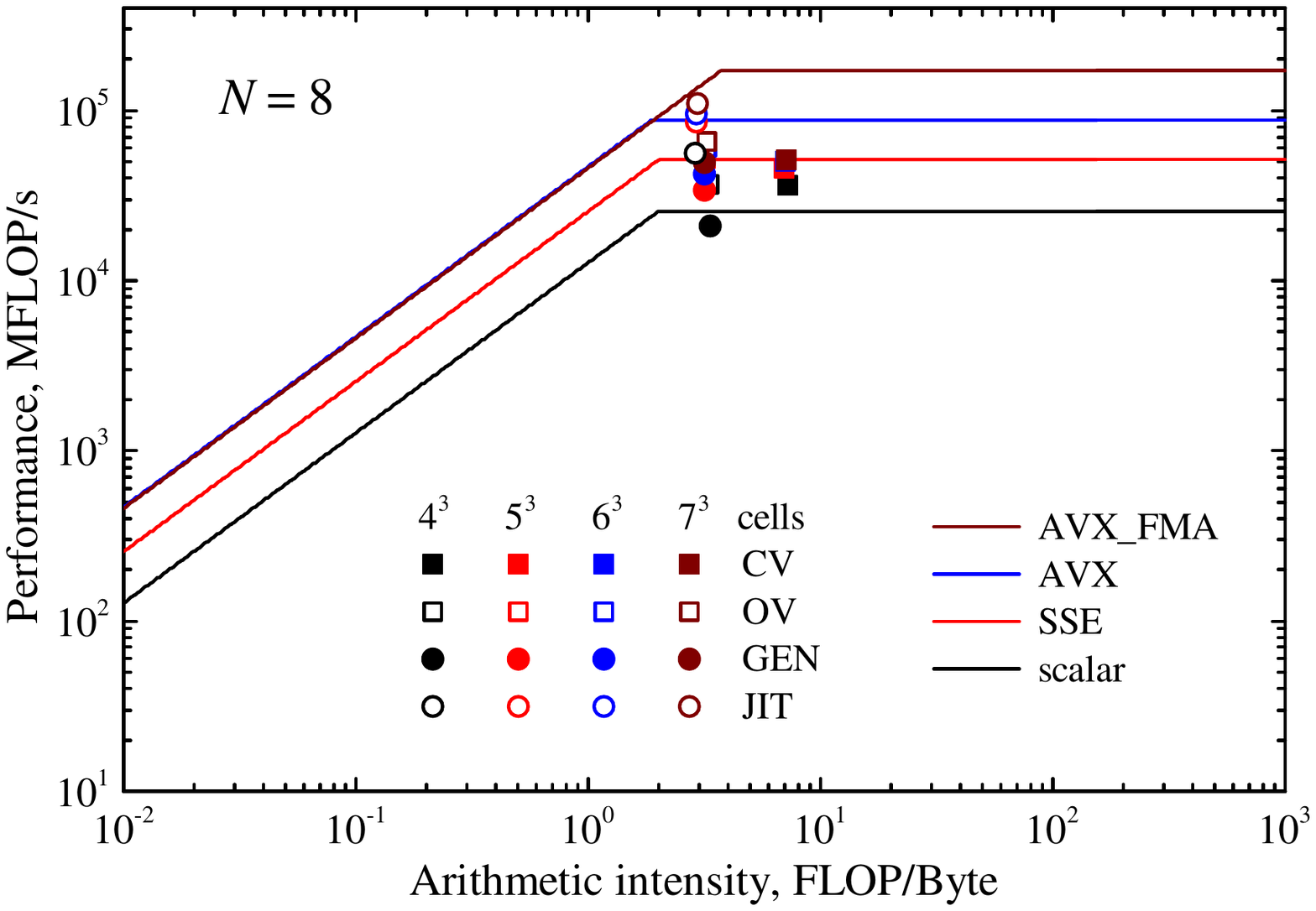}
\includegraphics[width=0.32\textwidth]{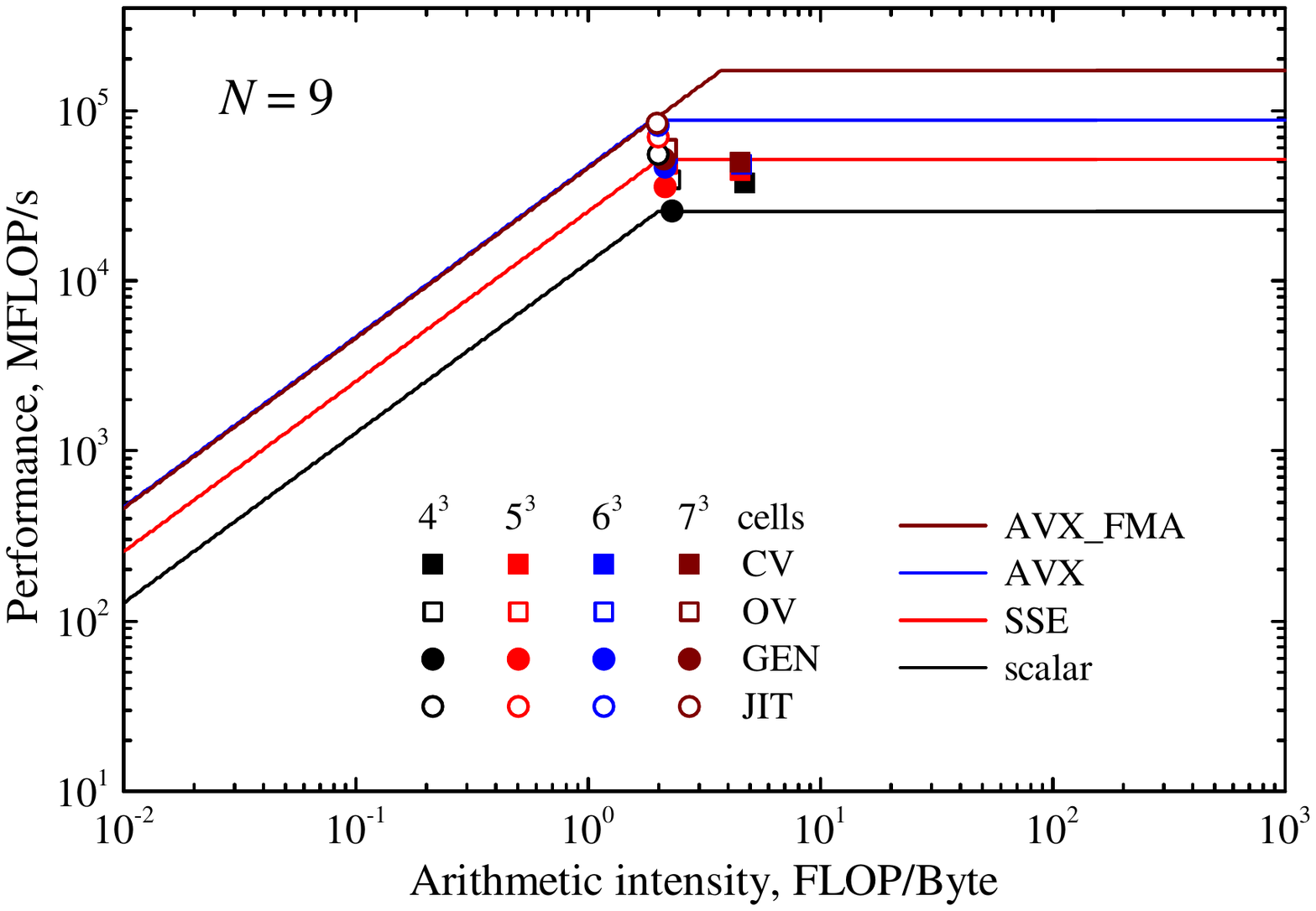}
\caption{\label{fig:roofline_DG_3d_degrees}%
Roofline models for ADER-DG-$\mathbb{P}_{N}$ method for three-dimensional sine wave advection problem --- performance (measured in MFLOP/s) versus arithmetic intensity (measured in FLOP/Byte) on a double logarithmic scale. Peak performance values are presented for scalar operations, for operations using the SSE instruction set, using the AVX instruction set, and using the AVX-512 FMA instruction set. Data points are presented for CV, OV, general BLAS and JIT BLAS implementations. A comparison is presented for different mesh sizes --- $4^{3}$, $5^{3}$, $6^{3}$, $7^{3}$, at different polynomial degrees $N = 1, \ldots, 9$.
}
\end{figure*}

\begin{figure*}[h!]
\centering
\includegraphics[width=0.245\textwidth]{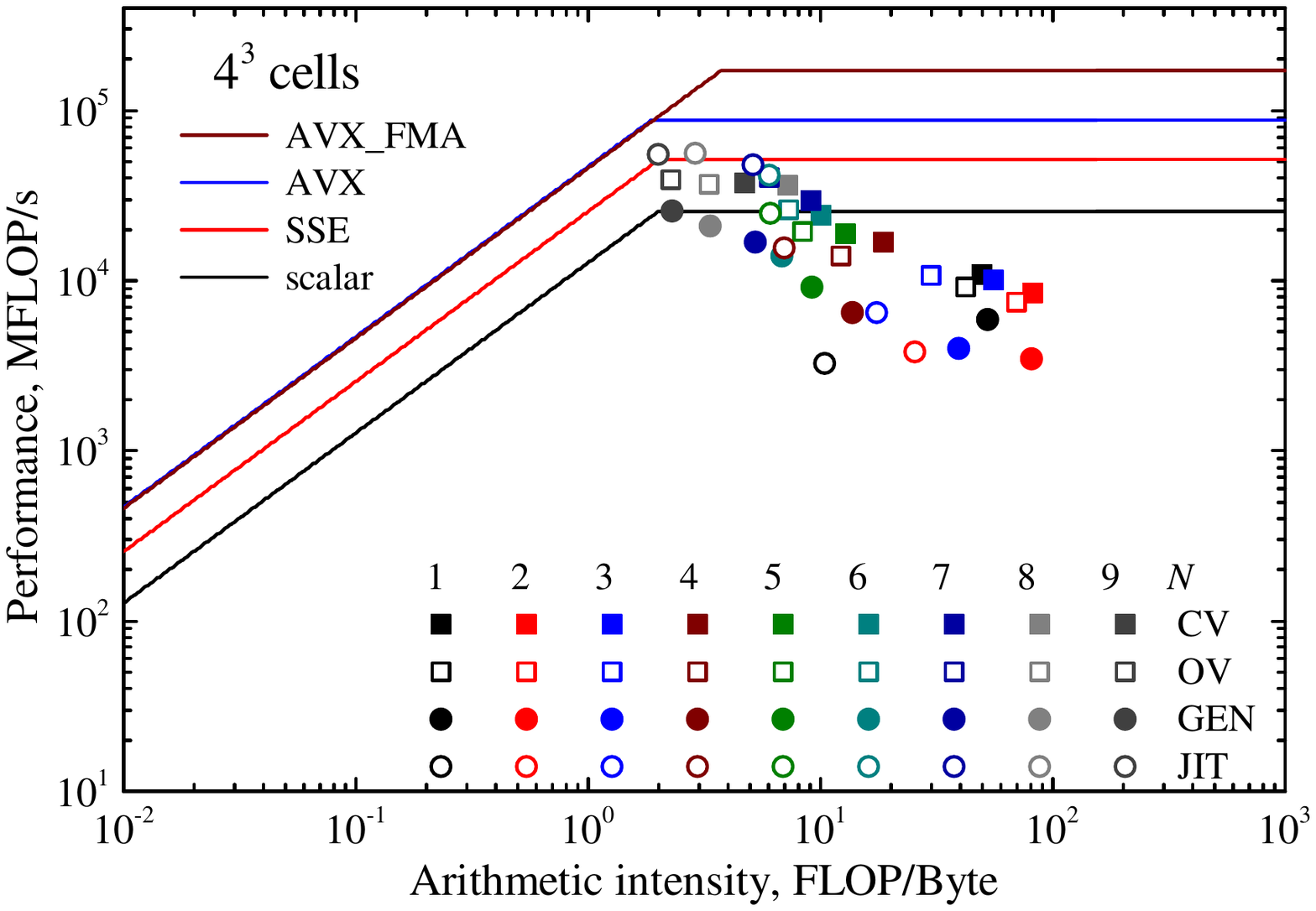}
\includegraphics[width=0.245\textwidth]{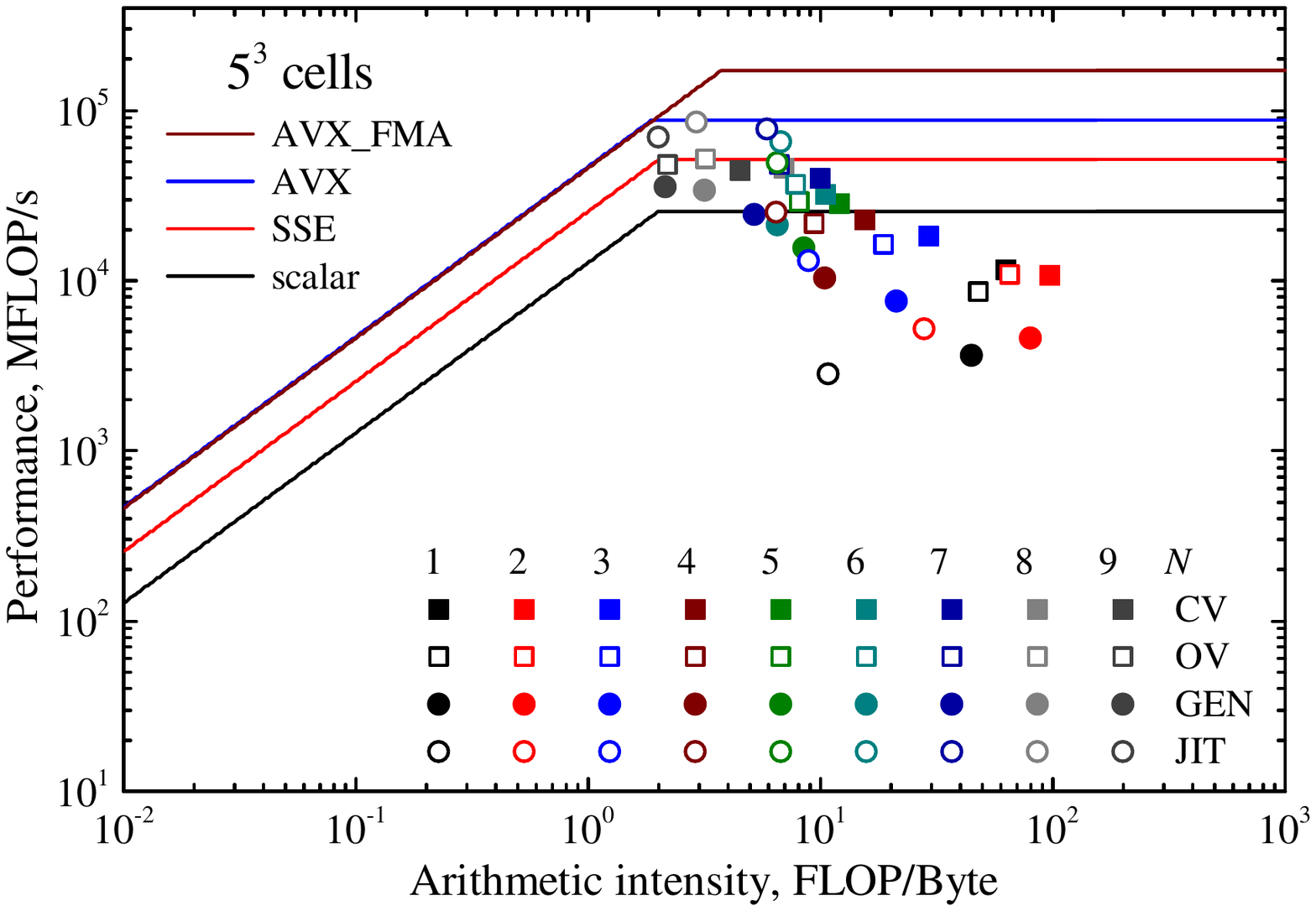}
\includegraphics[width=0.245\textwidth]{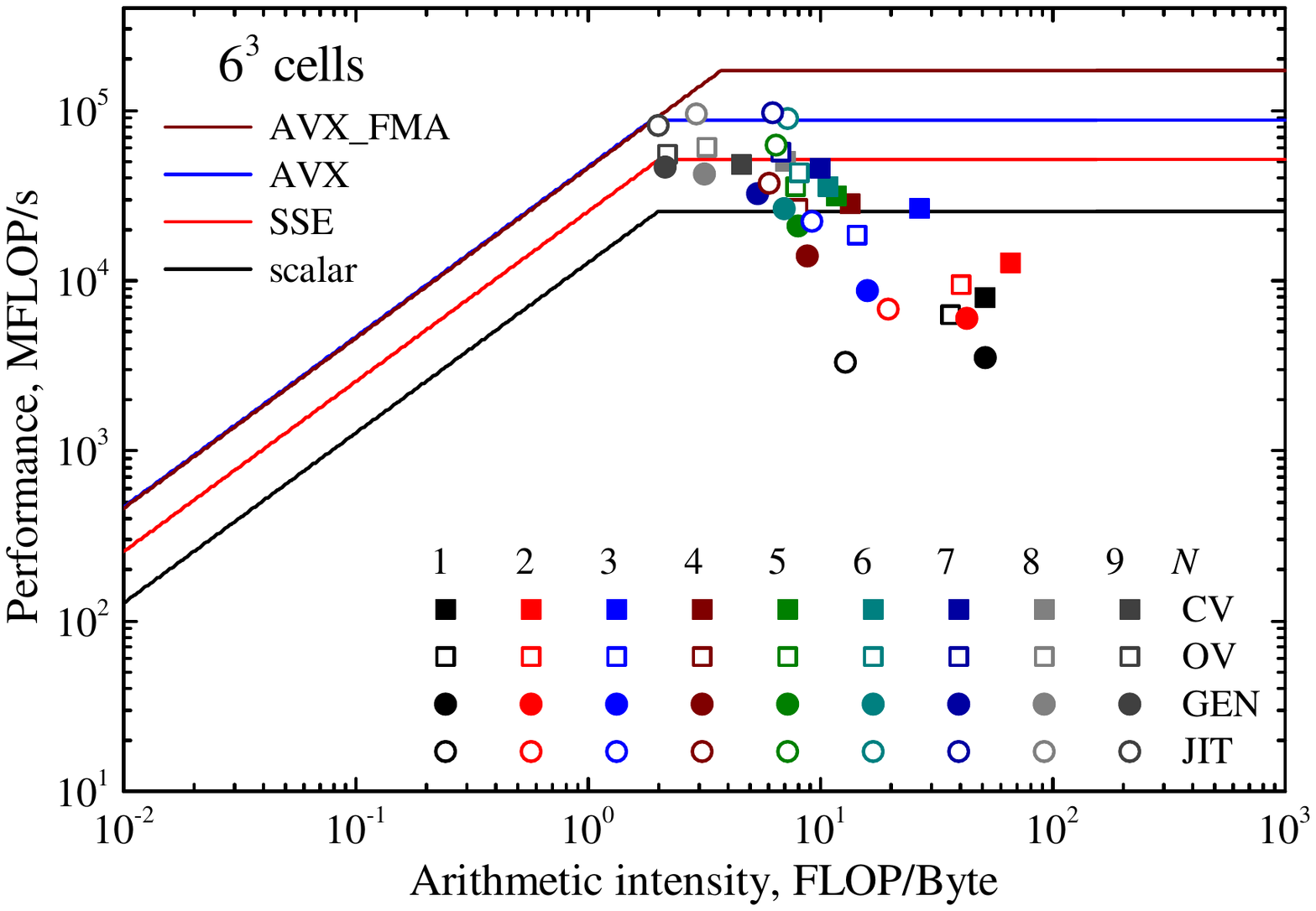}
\includegraphics[width=0.245\textwidth]{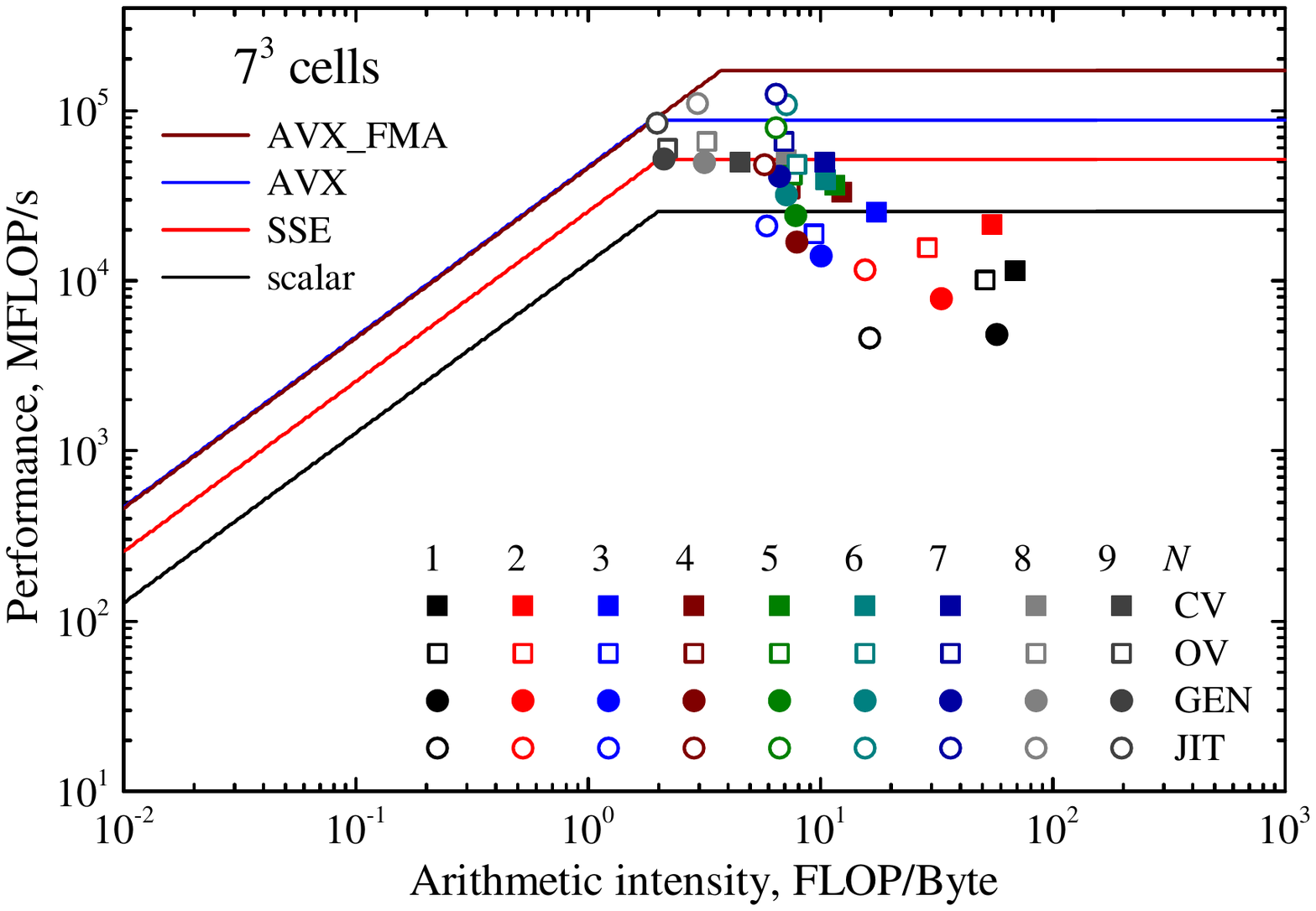}
\caption{\label{fig:roofline_DG_3d_mesh_sizes}%
Roofline models for ADER-DG-$\mathbb{P}_{N}$ method for three-dimensional sine wave advection problem --- performance (measured in MFLOP/s) versus arithmetic intensity (measured in FLOP/Byte) on a double logarithmic scale. Peak performance values are presented for scalar operations, for operations using the SSE instruction set, using the AVX instruction set, and using the AVX-512 FMA instruction set. Data points are presented for CV, OV, general BLAS and JIT BLAS implementations. A comparison is presented for different polynomial degrees $N = 1, \ldots, 9$, at mesh sizes $4^{3}$, $5^{3}$, $6^{3}$, $7^{3}$.
}
\end{figure*}

\begin{figure*}[h!]
\centering
\includegraphics[width=0.32\textwidth]{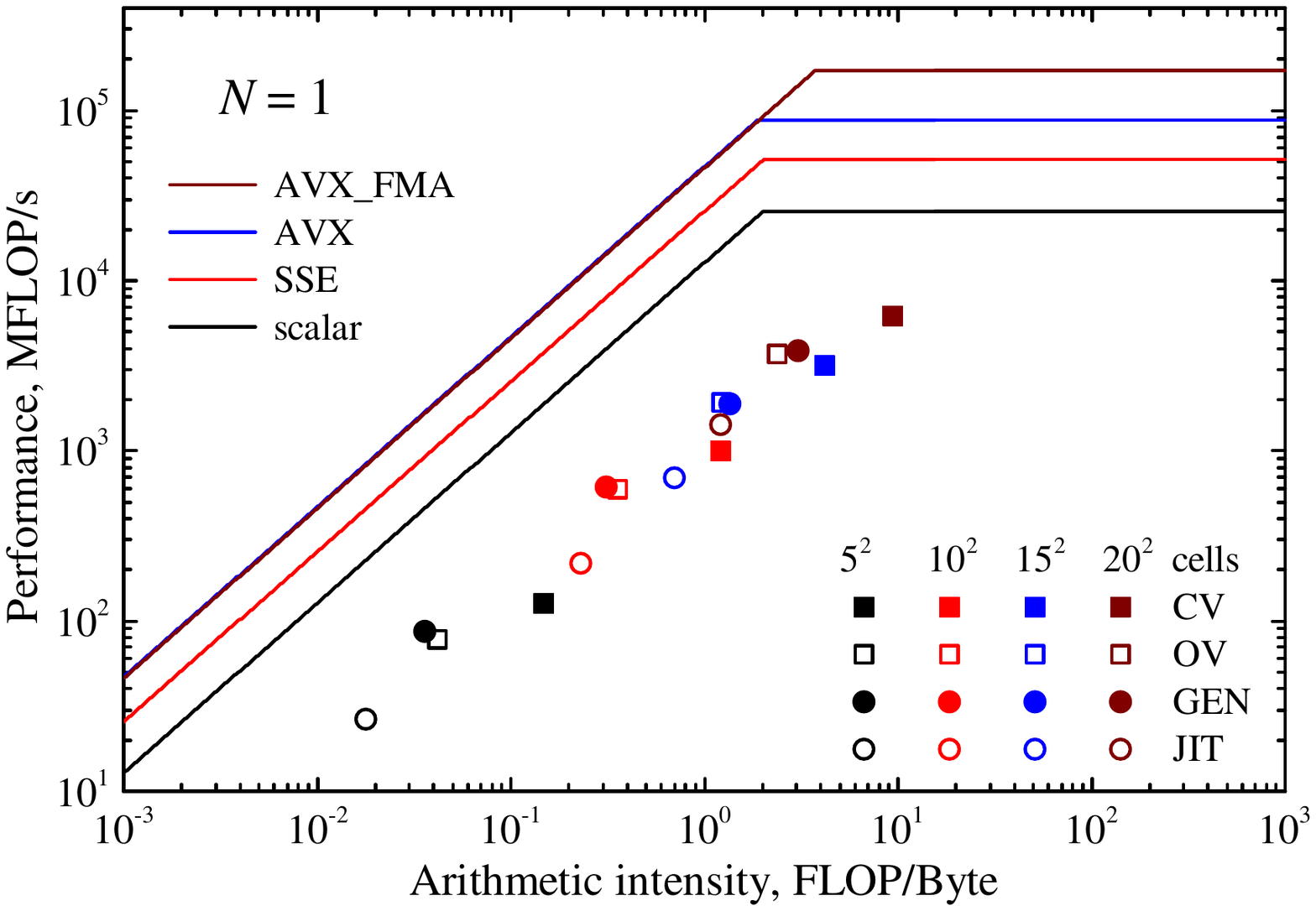}
\includegraphics[width=0.32\textwidth]{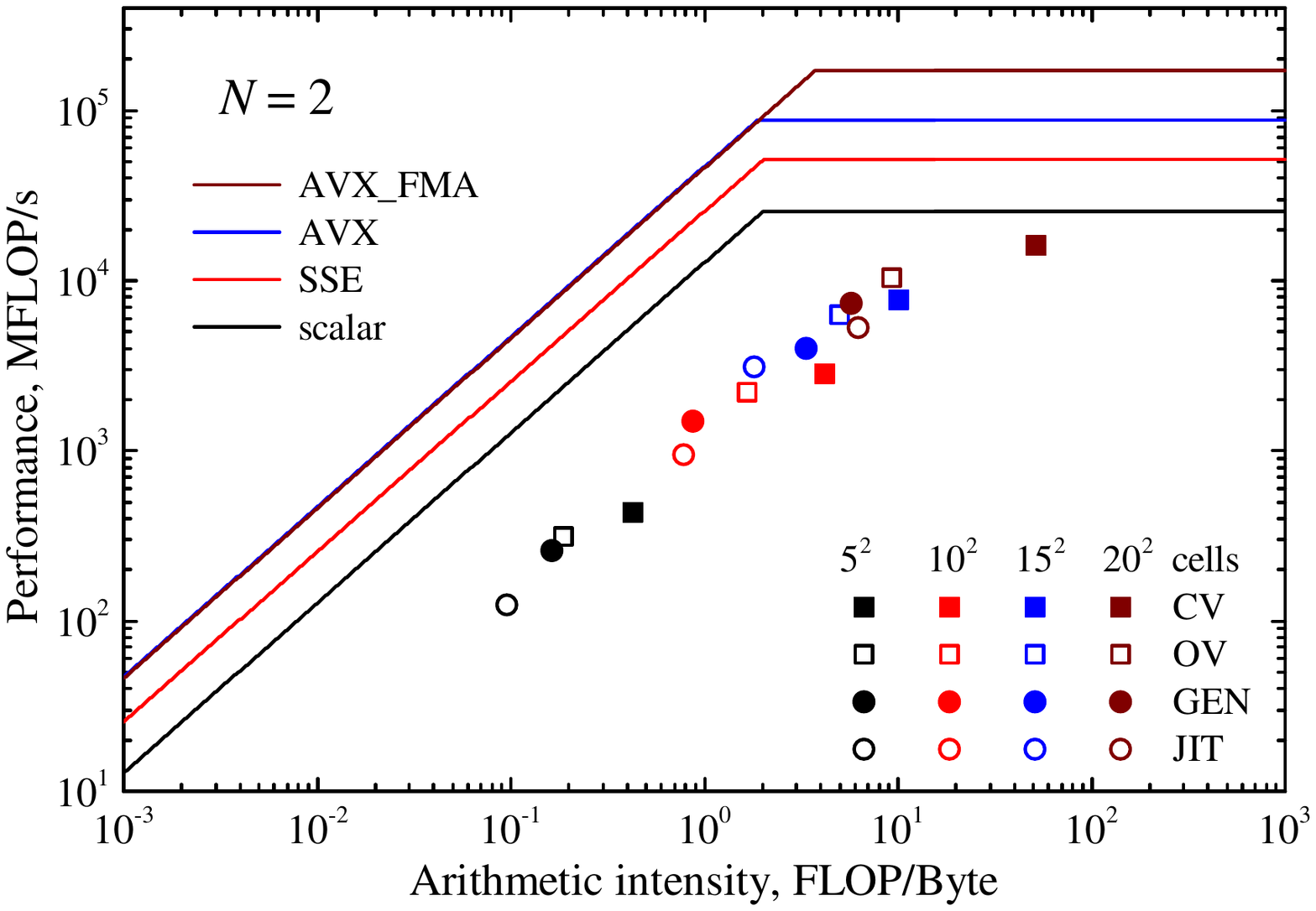}
\includegraphics[width=0.32\textwidth]{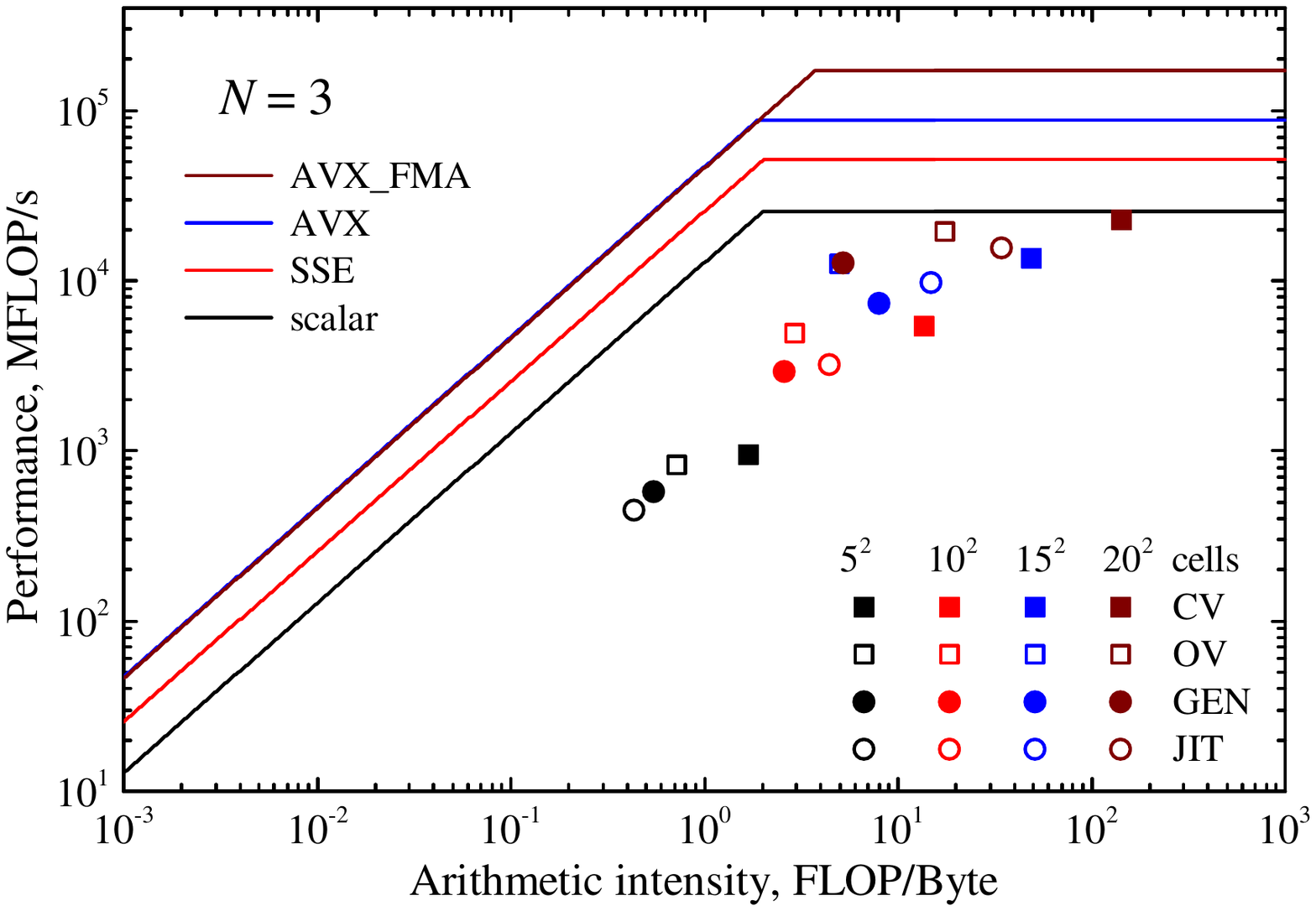}
\caption{\label{fig:roofline_WENO_2d_degrees}%
Roofline models for finite-volume ADER-WENO($N+1$) method for two-dimensional sine wave advection problem --- performance (measured in MFLOP/s) versus arithmetic intensity (measured in FLOP/Byte) on a double logarithmic scale. Peak performance values are presented for scalar operations, for operations using the SSE instruction set, using the AVX instruction set, and using the AVX-512 FMA instruction set. Data points are presented for CV, OV, general BLAS and JIT BLAS implementations. A comparison is presented for different mesh sizes --- $5^{2}$, $10^{2}$, $15^{2}$, $20^{2}$, at different polynomial degrees $N = 1, 2, 3$.
}
\end{figure*}

\begin{figure*}[h!]
\centering
\includegraphics[width=0.245\textwidth]{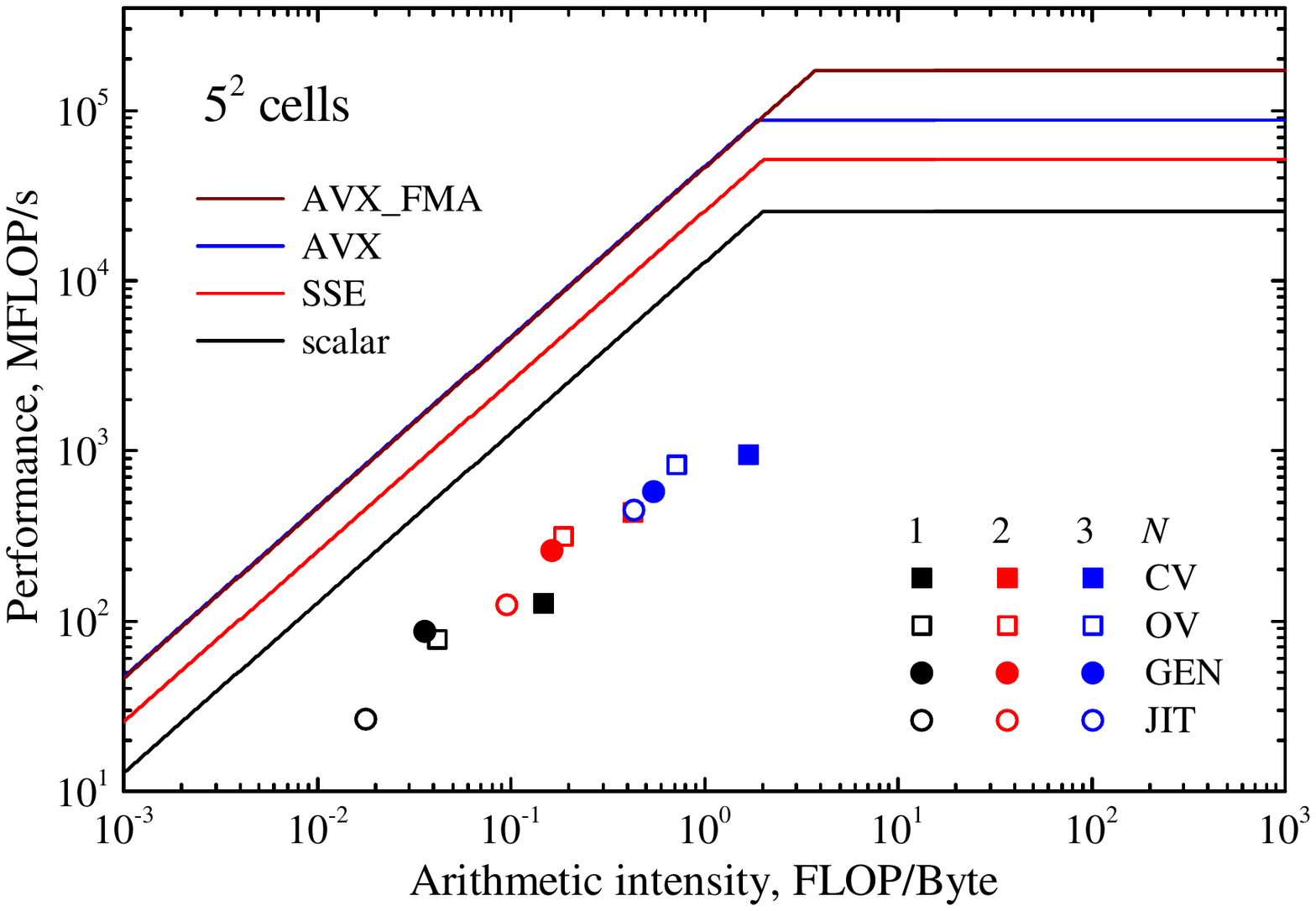}
\includegraphics[width=0.245\textwidth]{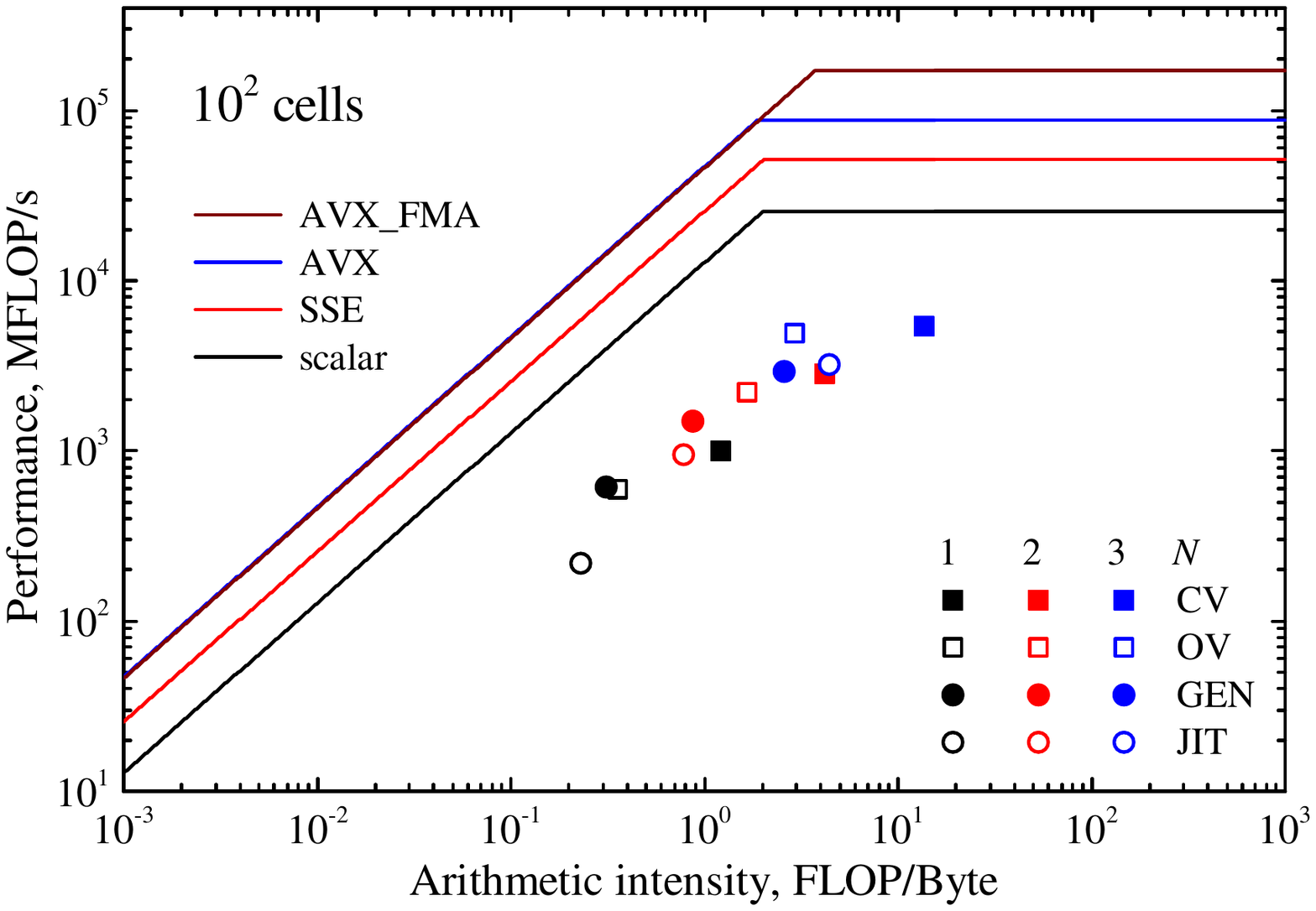}
\includegraphics[width=0.245\textwidth]{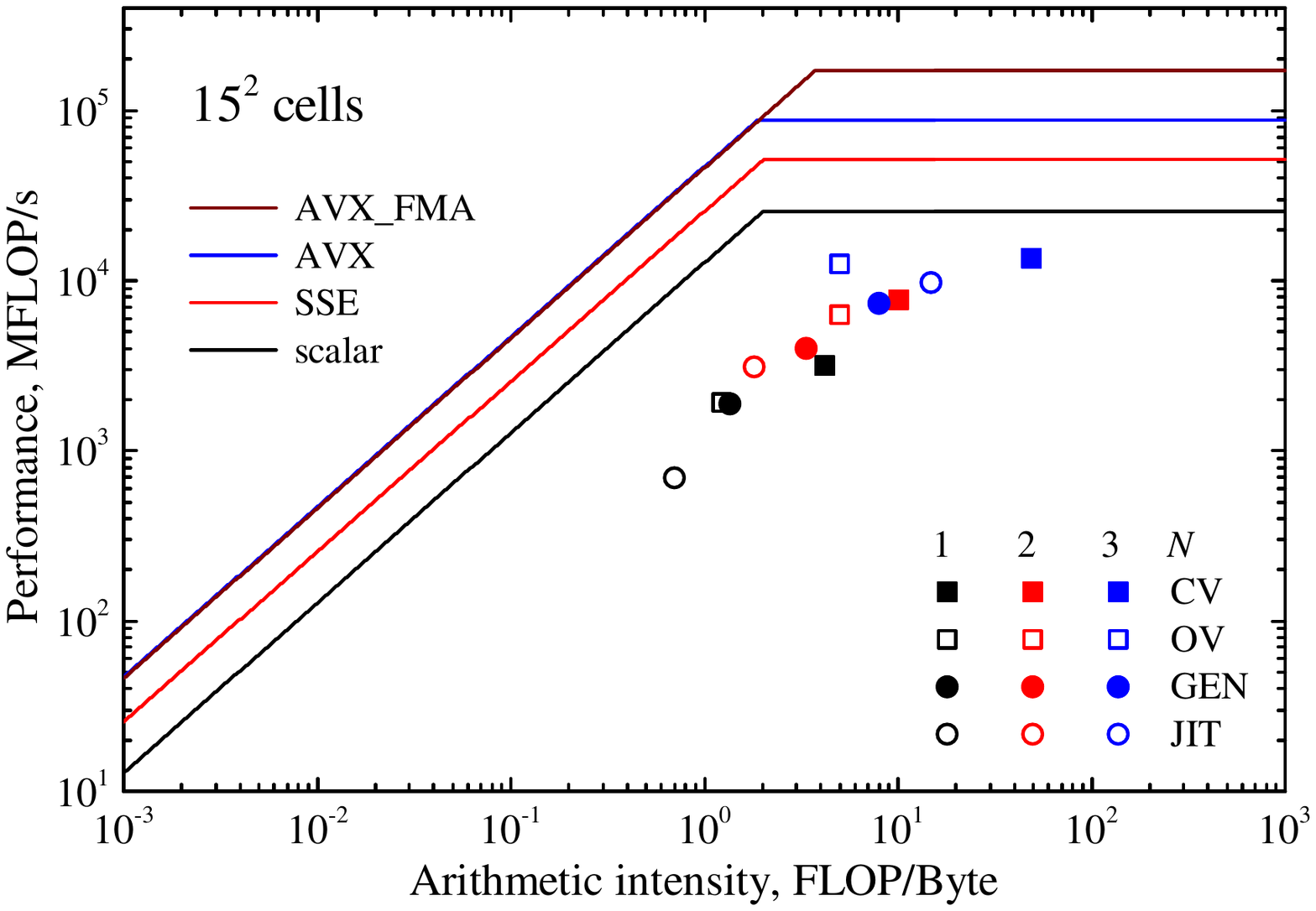}
\includegraphics[width=0.245\textwidth]{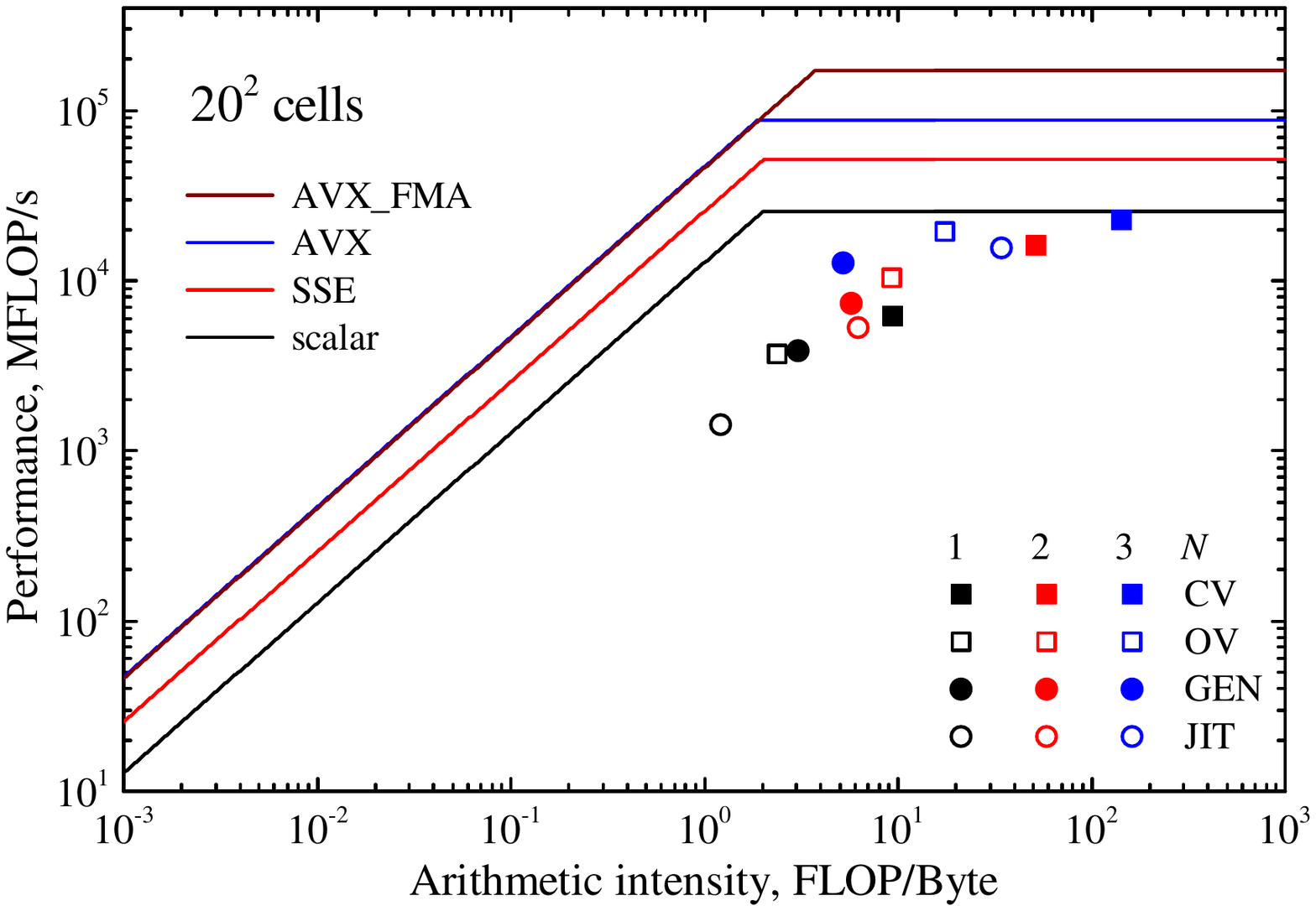}
\caption{\label{fig:roofline_WENO_2d_mesh_sizes}%
Roofline models for finite-volume ADER-WENO($N+1$) method for two-dimensional sine wave advection problem --- performance (measured in MFLOP/s) versus arithmetic intensity (measured in FLOP/Byte) on a double logarithmic scale. Peak performance values are presented for scalar operations, for operations using the SSE instruction set, using the AVX instruction set, and using the AVX-512 FMA instruction set. Data points are presented for CV, OV, general BLAS and JIT BLAS implementations. A comparison is presented for different polynomial degrees $N = 1, 2, 3$, at different mesh sizes $5^{2}$, $10^{2}$, $15^{2}$, $20^{2}$.
}
\end{figure*}

\begin{figure*}[h!]
\centering
\includegraphics[width=0.32\textwidth]{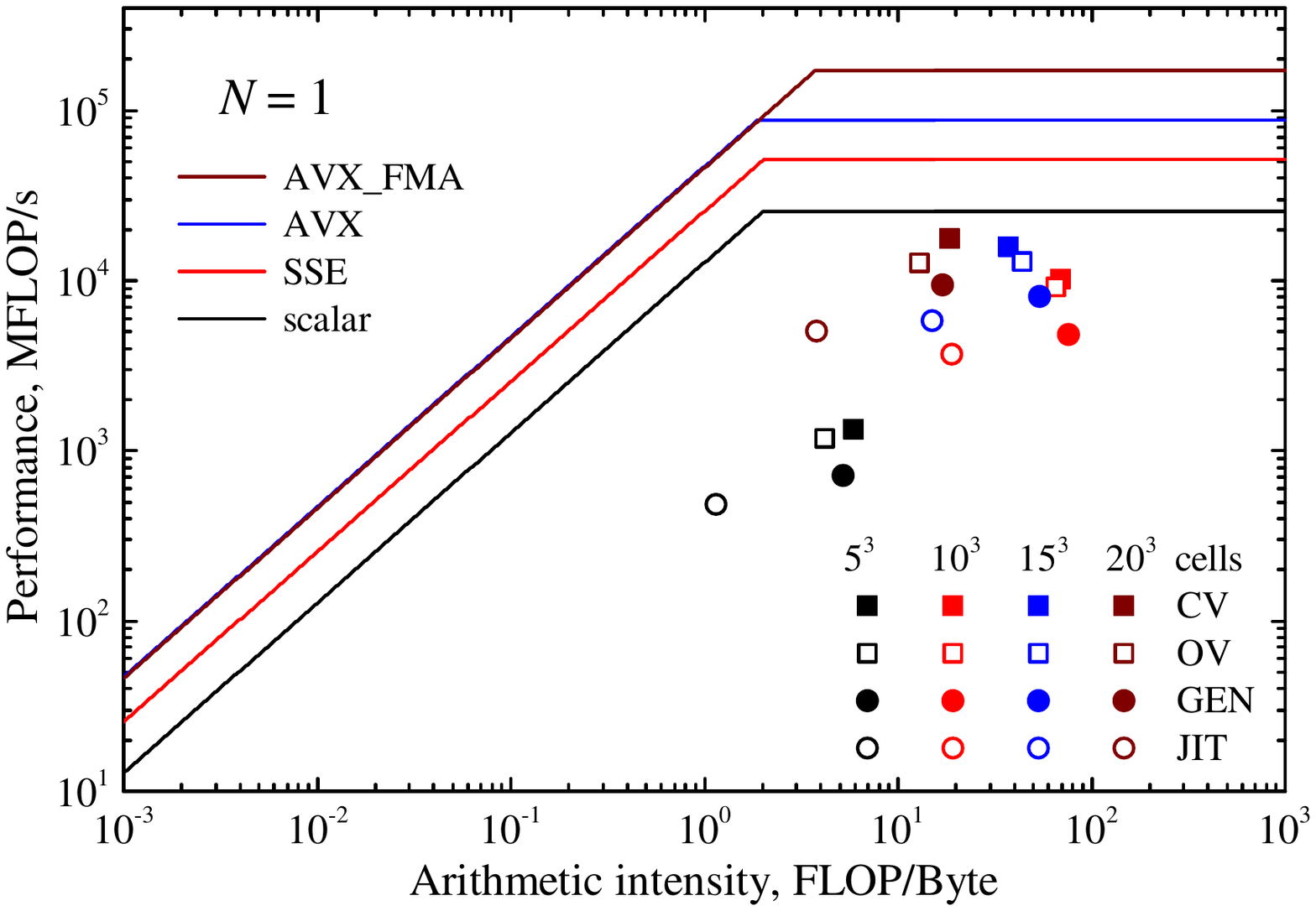}
\includegraphics[width=0.32\textwidth]{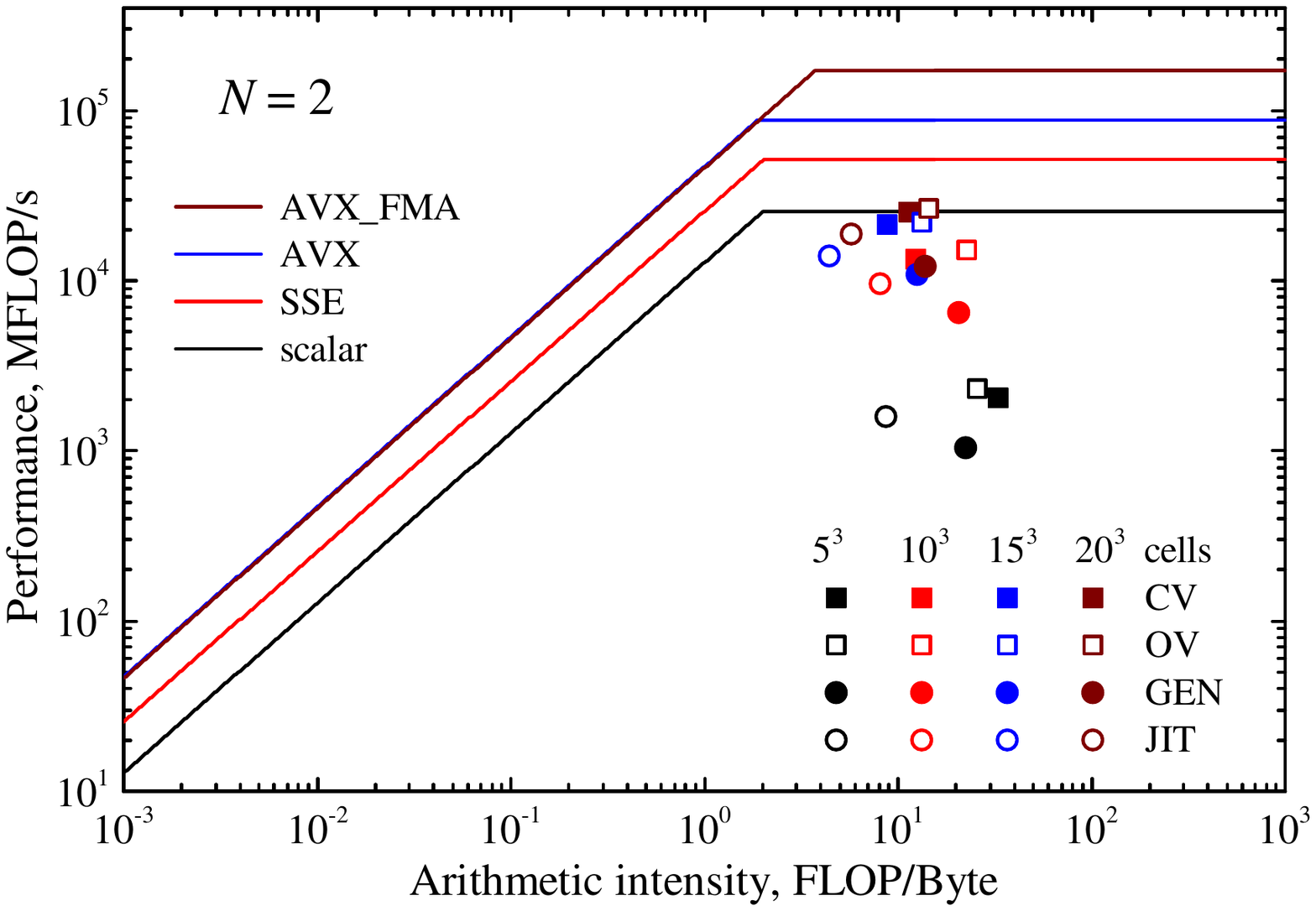}
\includegraphics[width=0.32\textwidth]{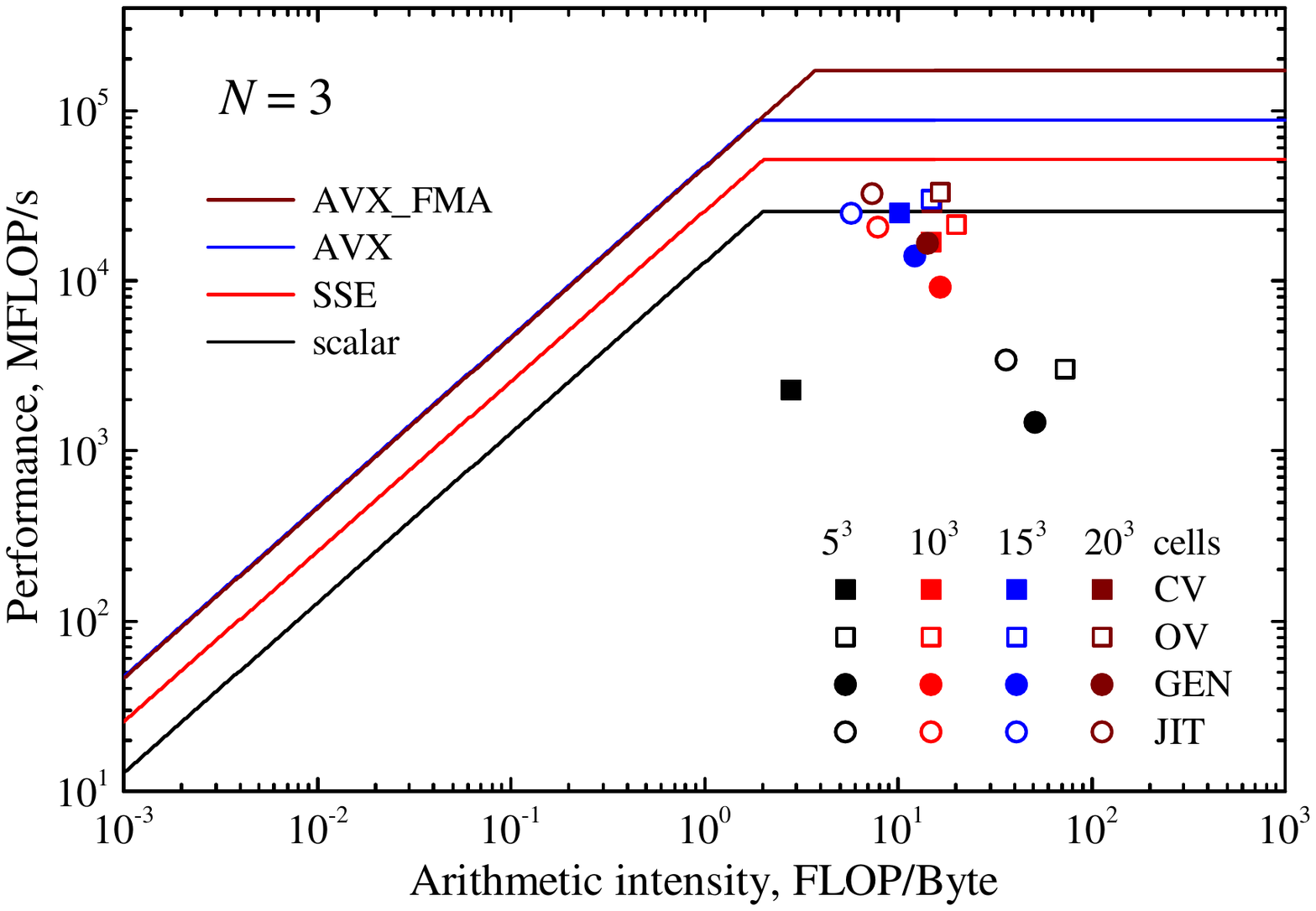}
\caption{\label{fig:roofline_WENO_3d_degrees}%
Roofline models for finite-volume ADER-WENO($N+1$) method for three-dimensional sine wave advection problem --- performance (measured in MFLOP/s) versus arithmetic intensity (measured in FLOP/Byte) on a double logarithmic scale. Peak performance values are presented for scalar operations, for operations using the SSE instruction set, using the AVX instruction set, and using the AVX-512 FMA instruction set. Data points are presented for CV, OV, general BLAS and JIT BLAS implementations. A comparison is presented for different mesh sizes --- $5^{3}$, $10^{3}$, $15^{3}$, $20^{3}$, at different polynomial degrees $N = 1, 2, 3$.
}
\end{figure*}

\begin{figure*}[h!]
\centering
\includegraphics[width=0.245\textwidth]{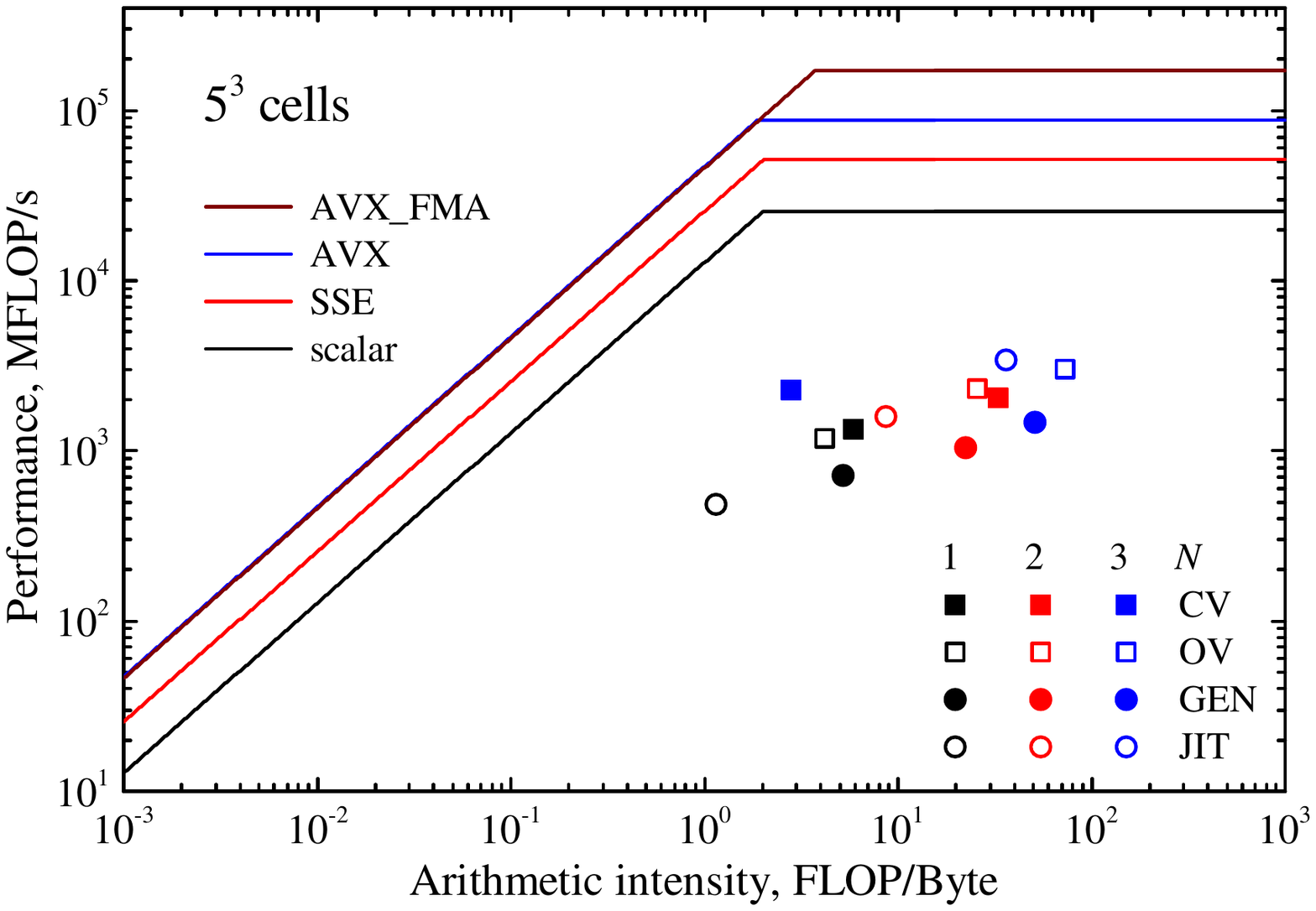}
\includegraphics[width=0.245\textwidth]{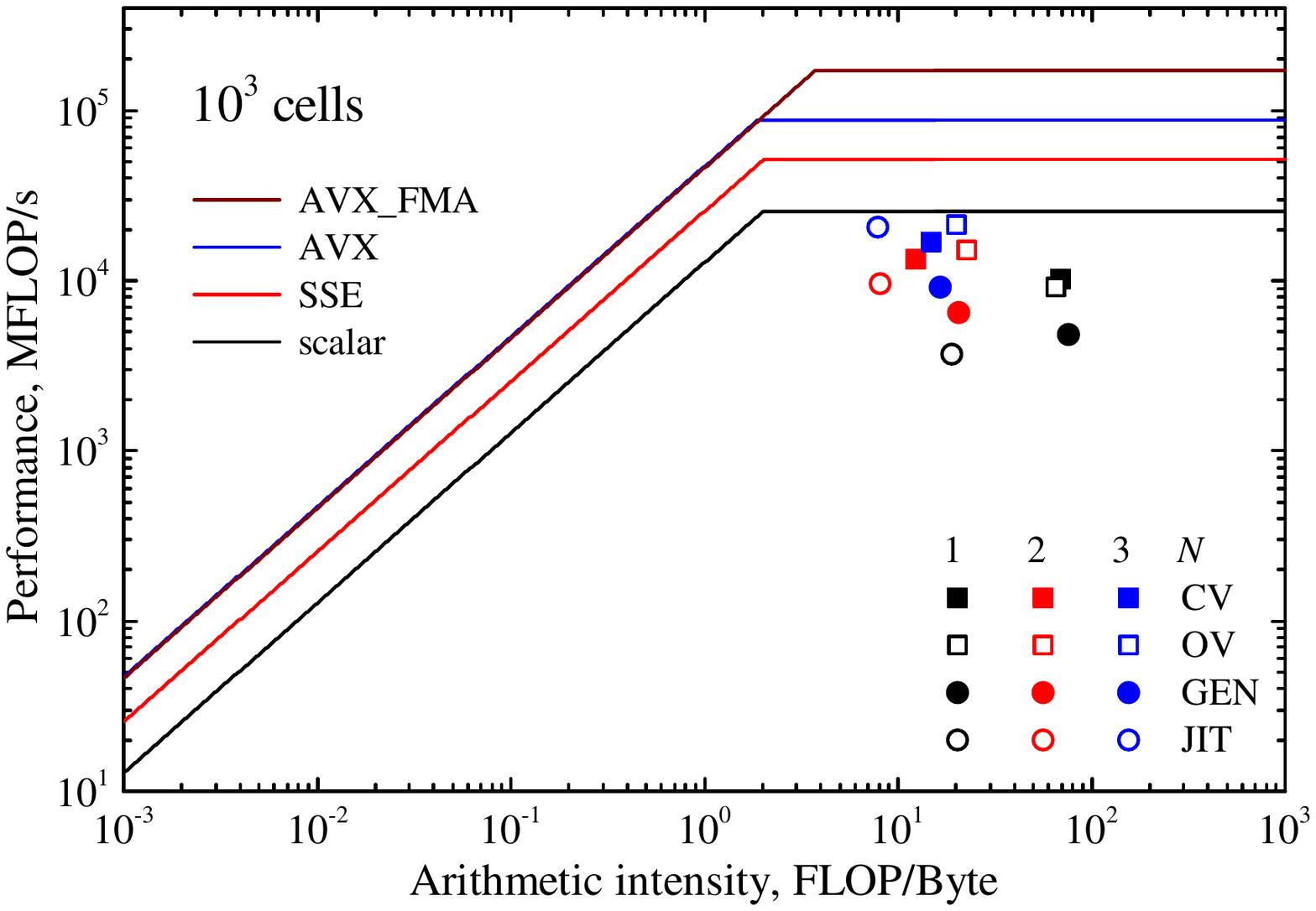}
\includegraphics[width=0.245\textwidth]{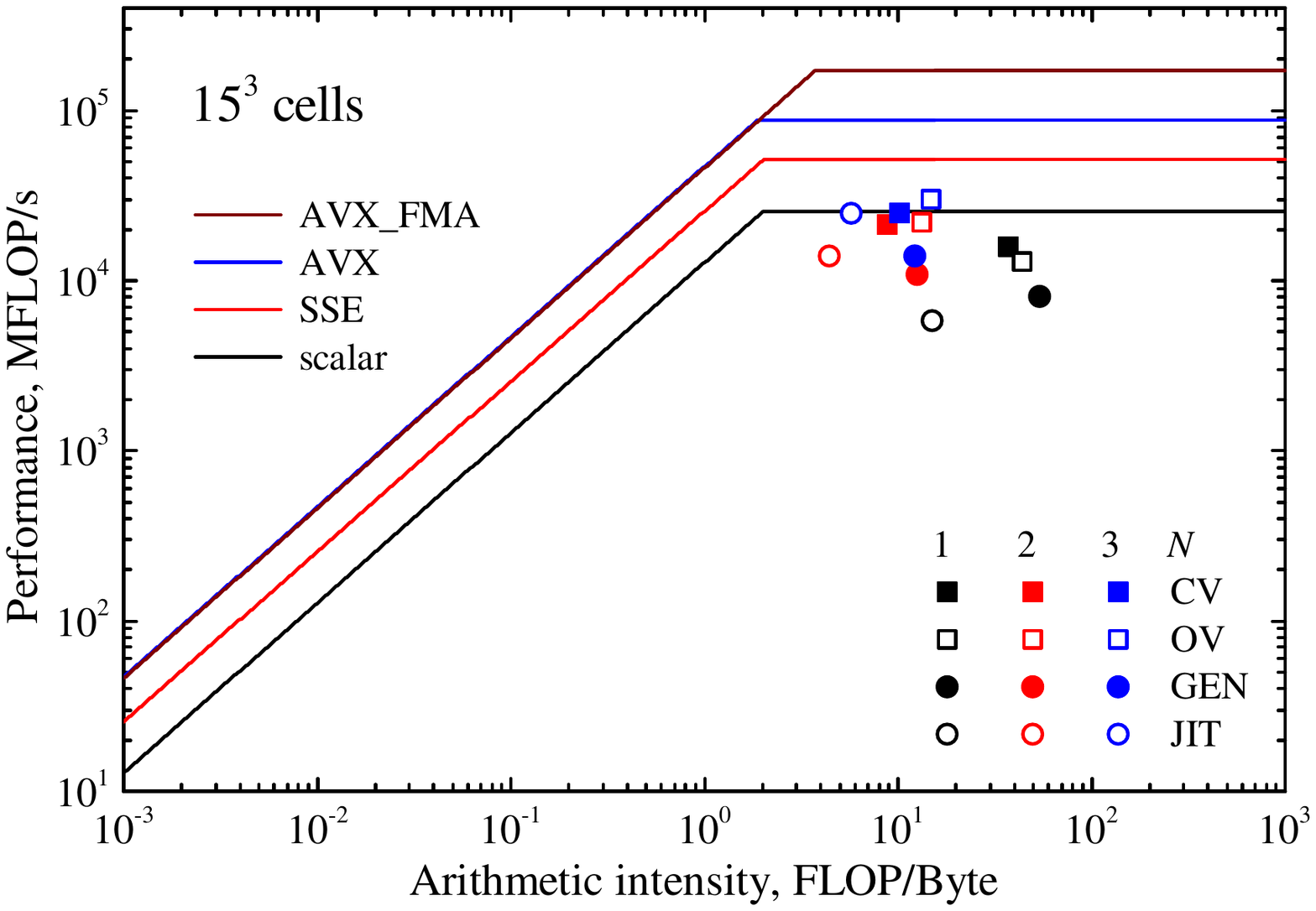}
\includegraphics[width=0.245\textwidth]{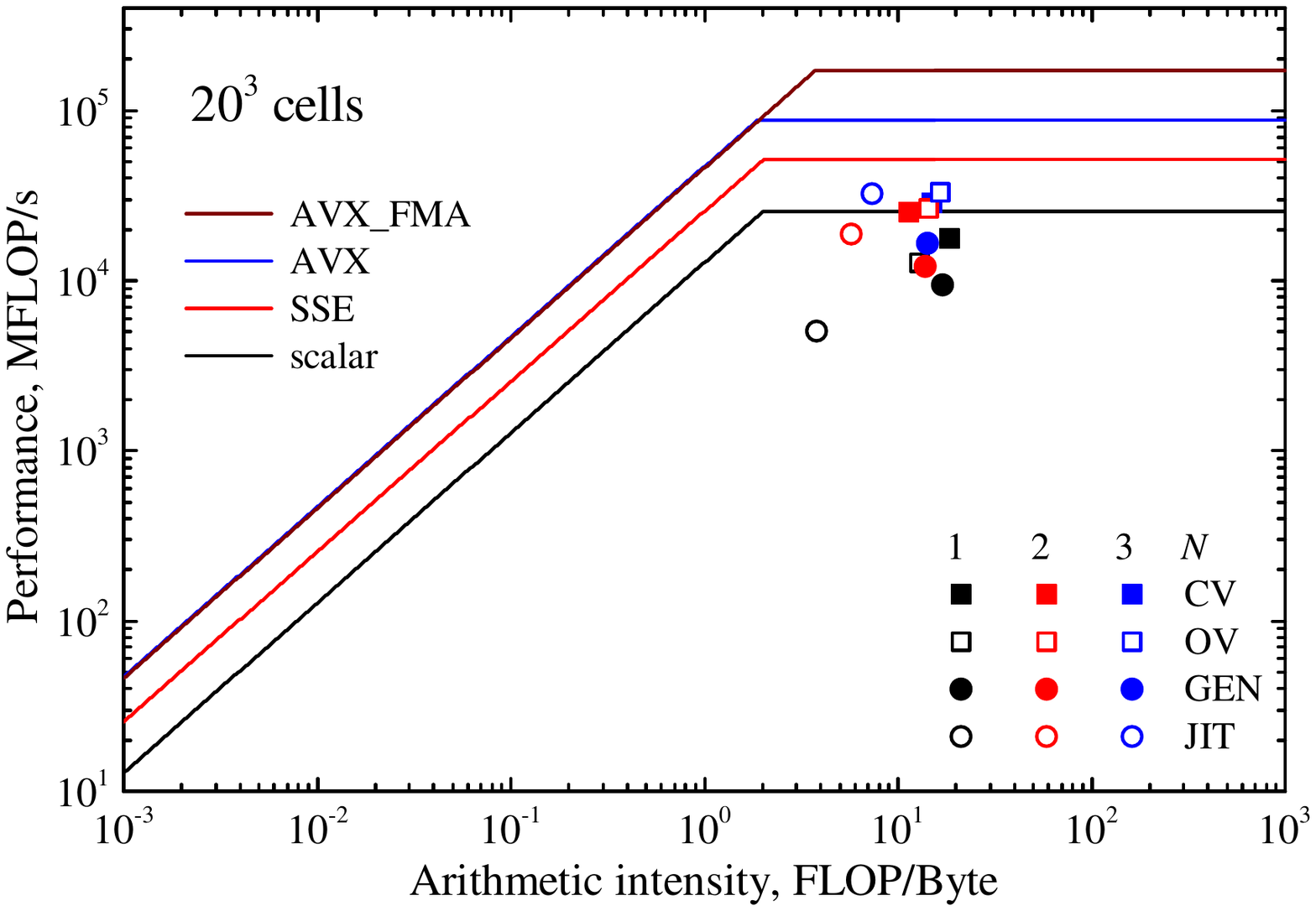}
\caption{\label{fig:roofline_WENO_3d_mesh_sizes}%
Roofline models for finite-volume ADER-WENO($N+1$) method for three-dimensional sine wave advection problem --- performance (measured in MFLOP/s) versus arithmetic intensity (measured in FLOP/Byte) on a double logarithmic scale. Peak performance values are presented for scalar operations, for operations using the SSE instruction set, using the AVX instruction set, and using the AVX-512 FMA instruction set. Data points are presented for CV, OV, general BLAS and JIT BLAS implementations. A comparison is presented for different polynomial degrees $N = 1, 2, 3$, at different mesh sizes $5^{3}$, $10^{3}$, $15^{3}$, $20^{3}$.
}
\end{figure*}

\begin{figure*}[h!]
\centering
\includegraphics[width=0.32\textwidth]{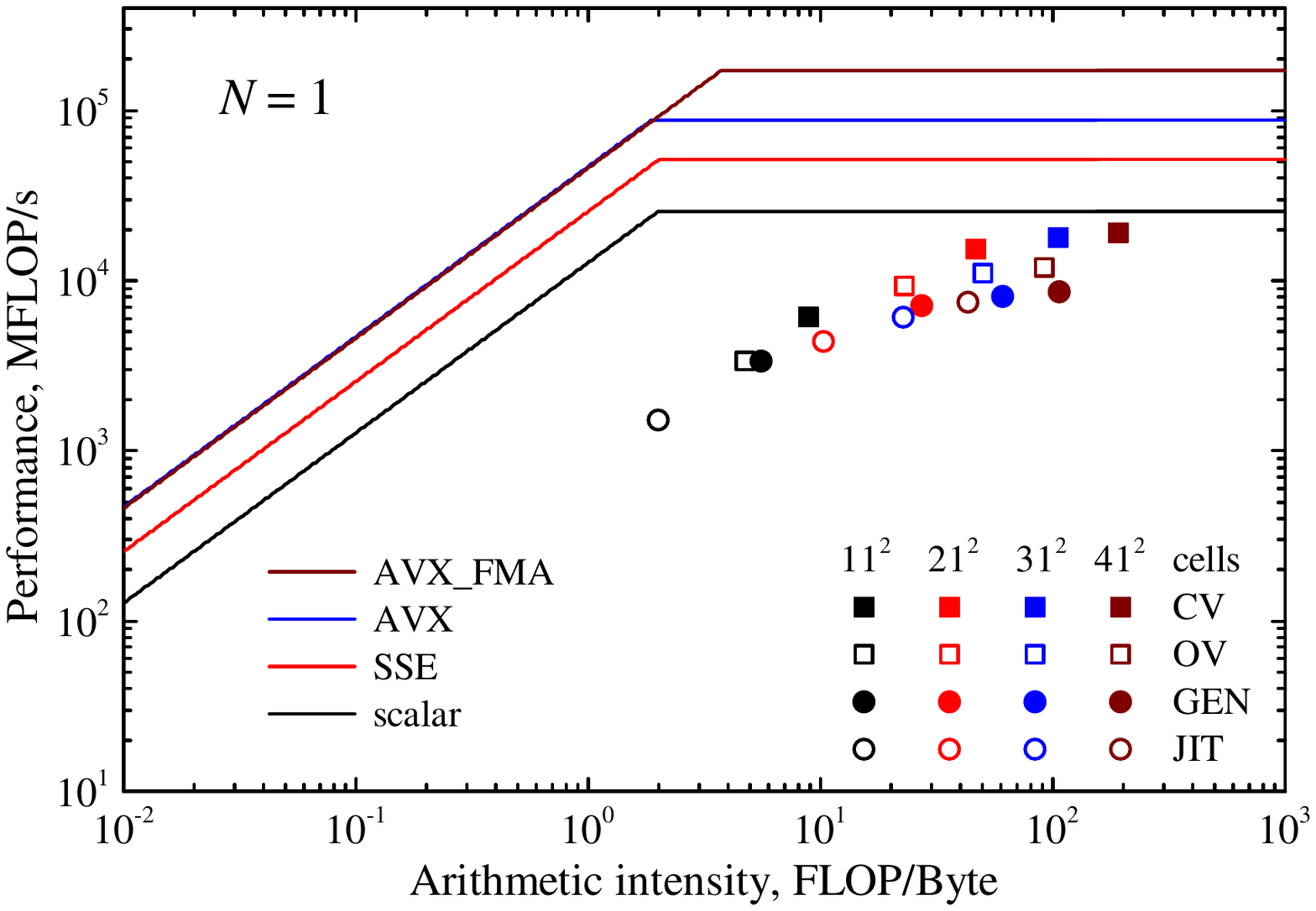}
\includegraphics[width=0.32\textwidth]{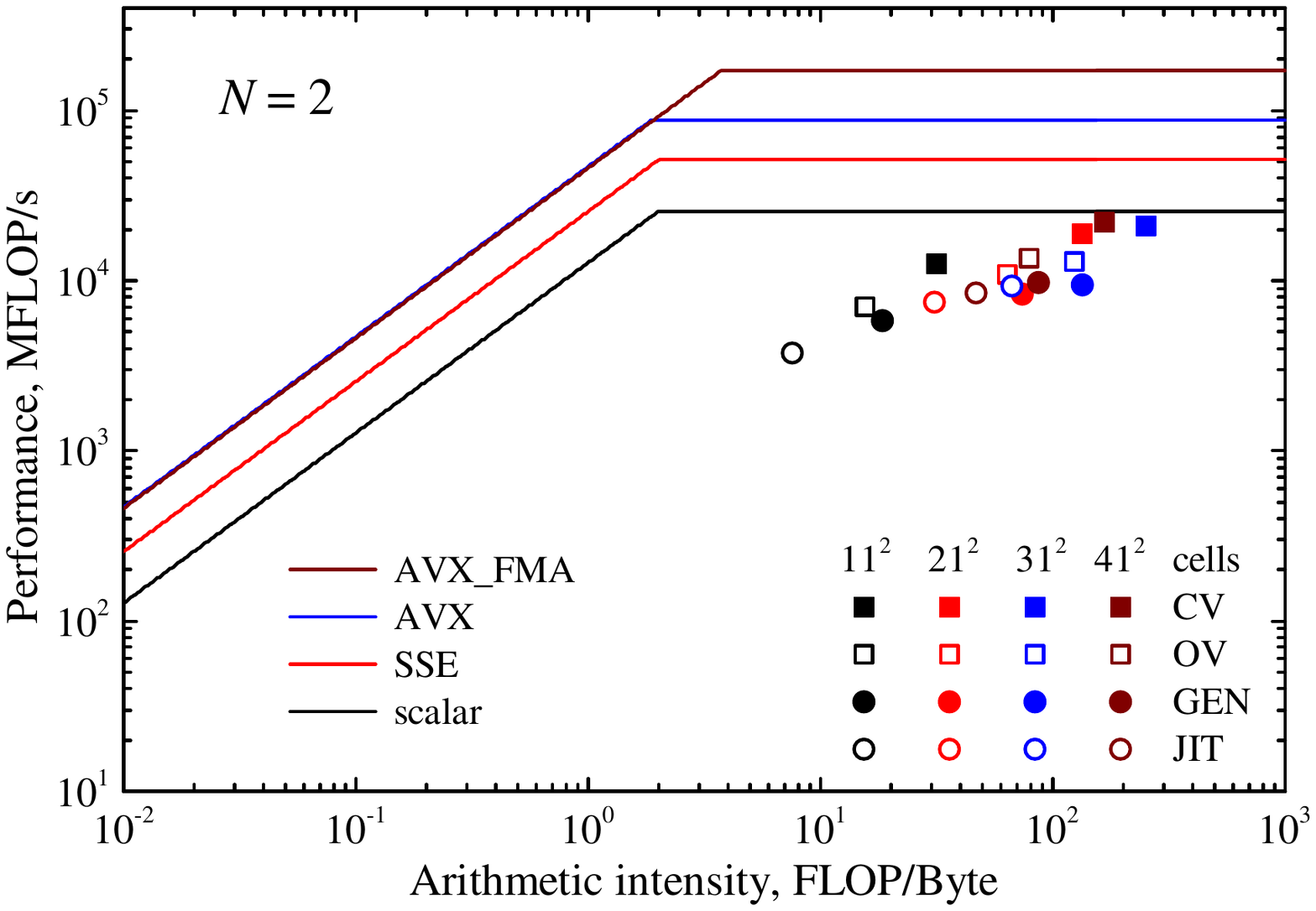}
\includegraphics[width=0.32\textwidth]{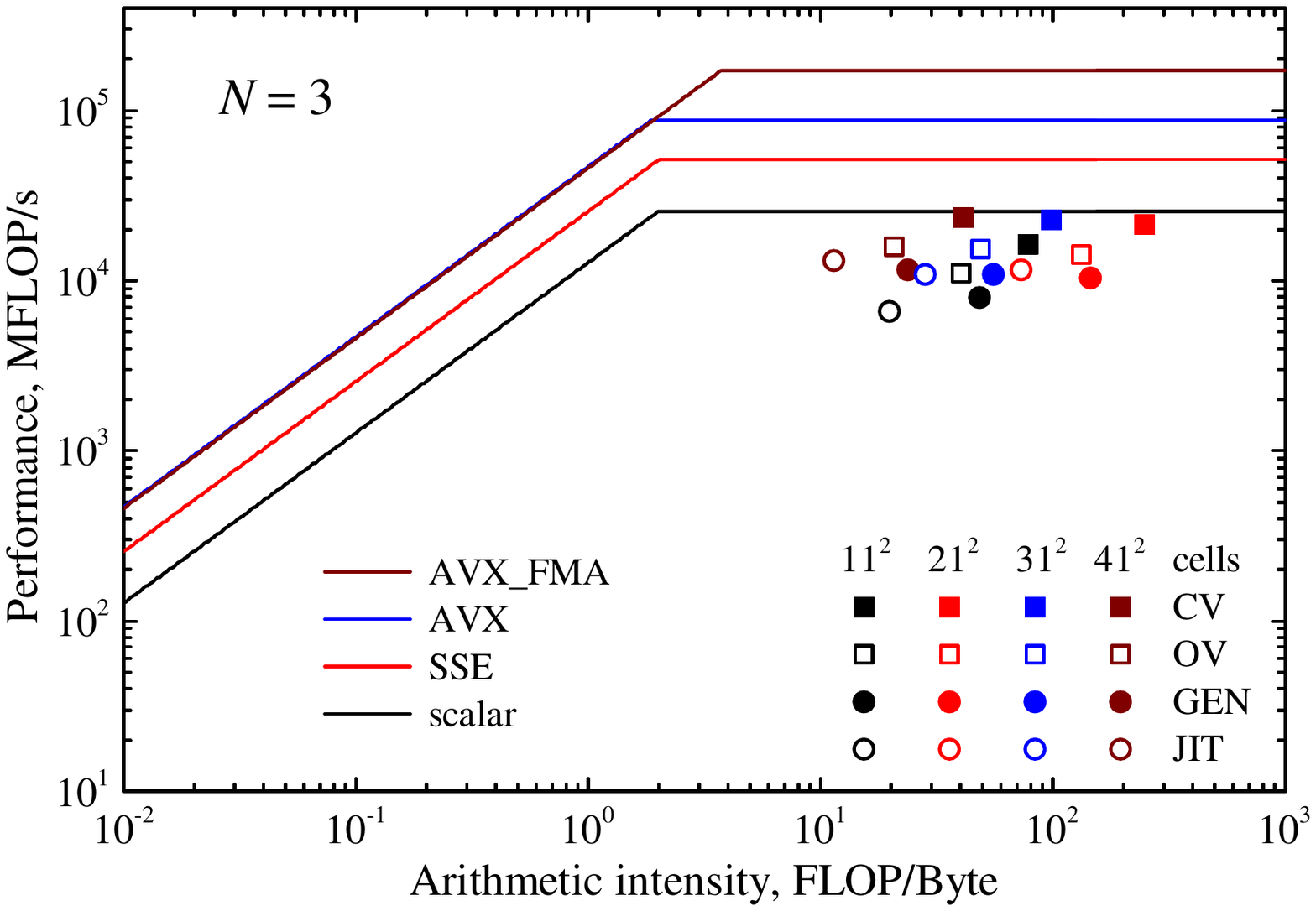}\\
\includegraphics[width=0.32\textwidth]{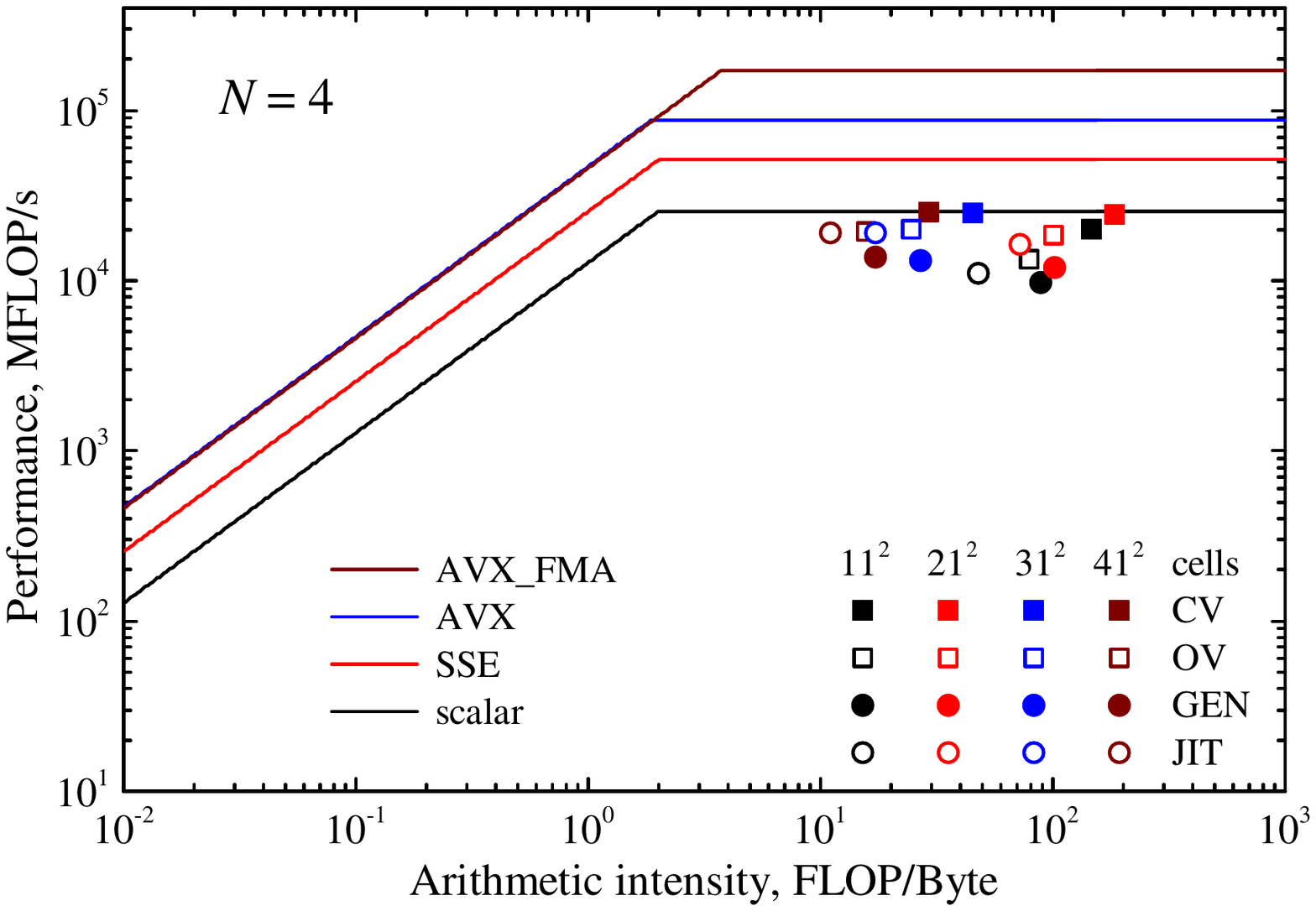}
\includegraphics[width=0.32\textwidth]{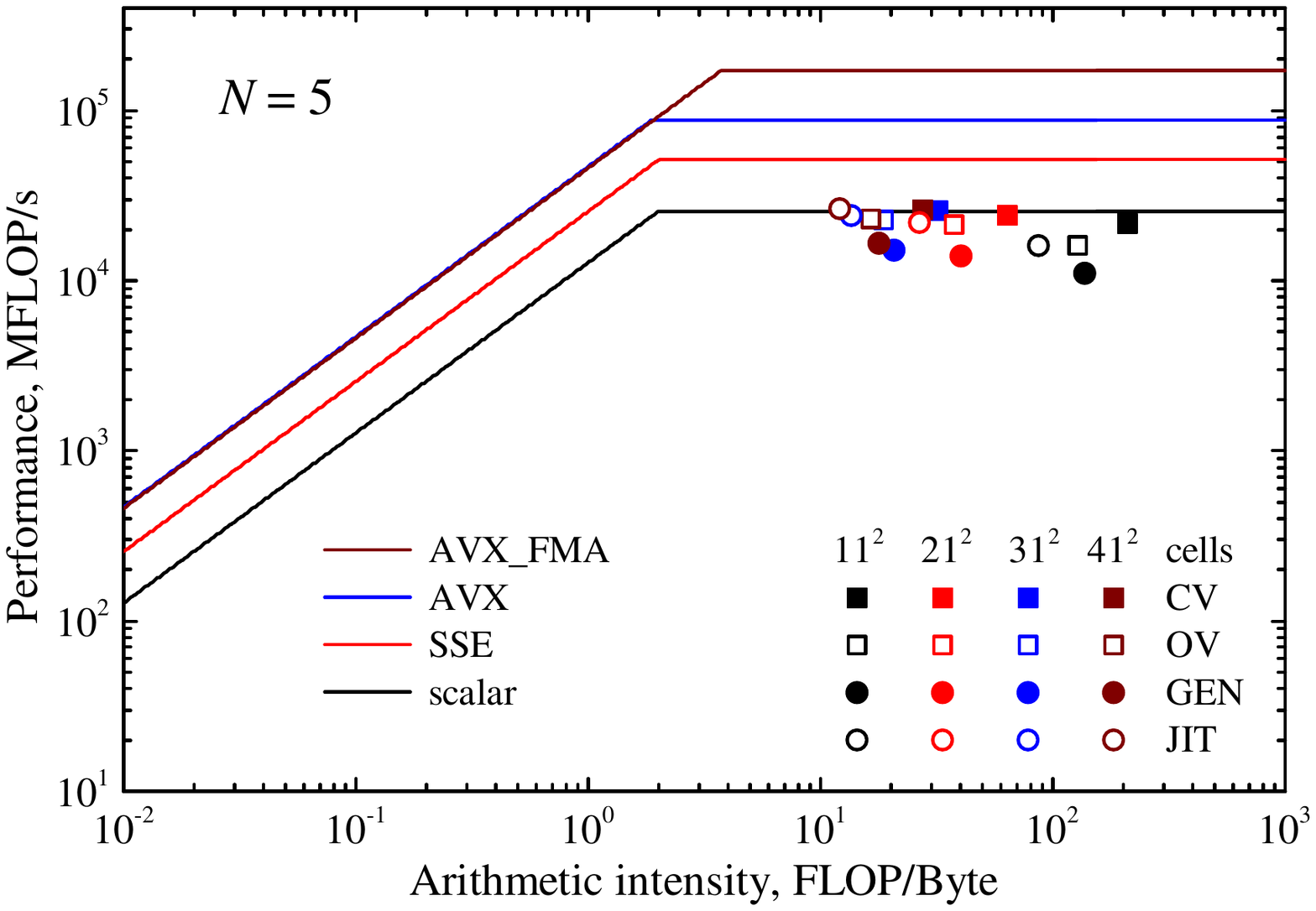}
\includegraphics[width=0.32\textwidth]{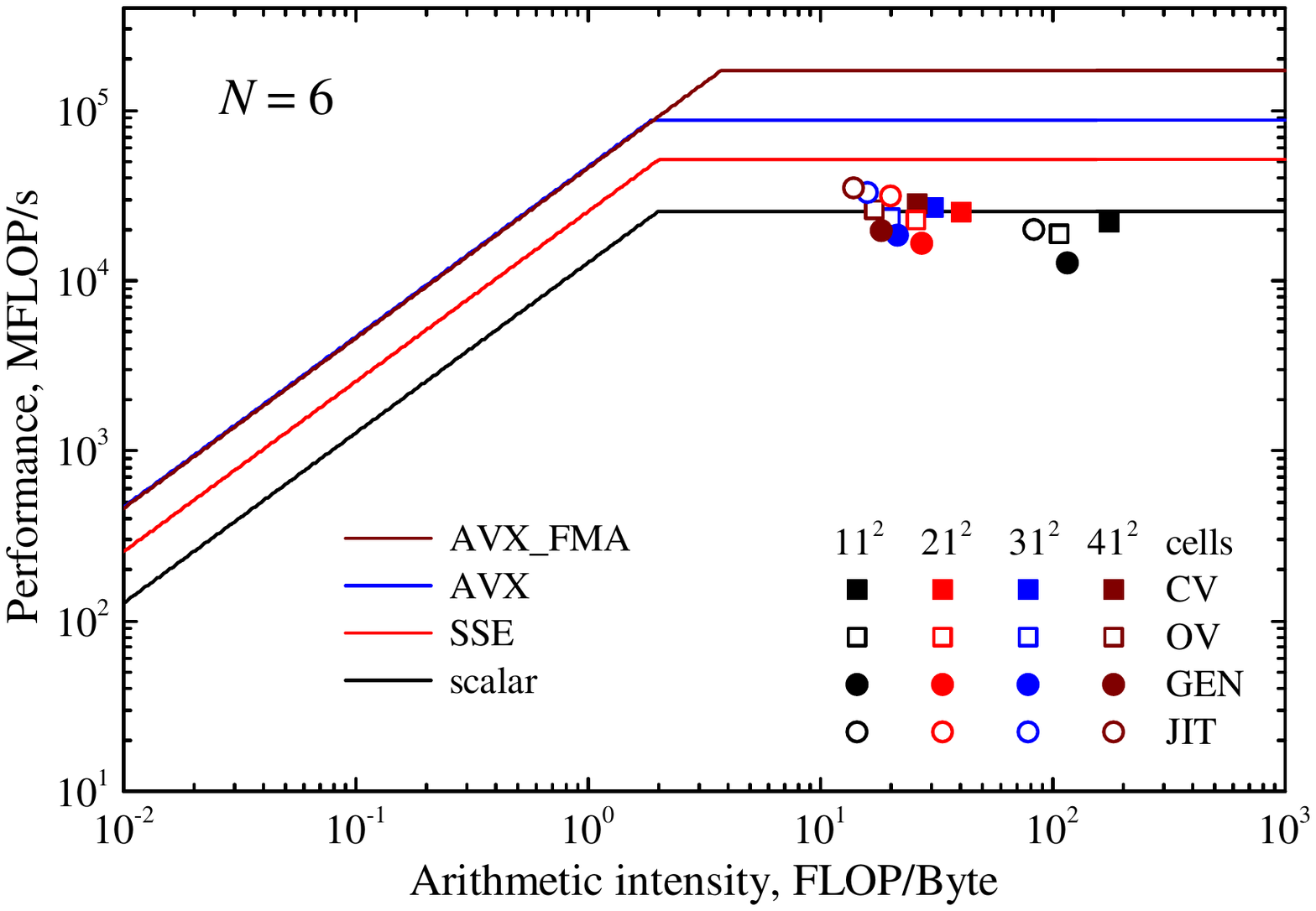}\\
\includegraphics[width=0.32\textwidth]{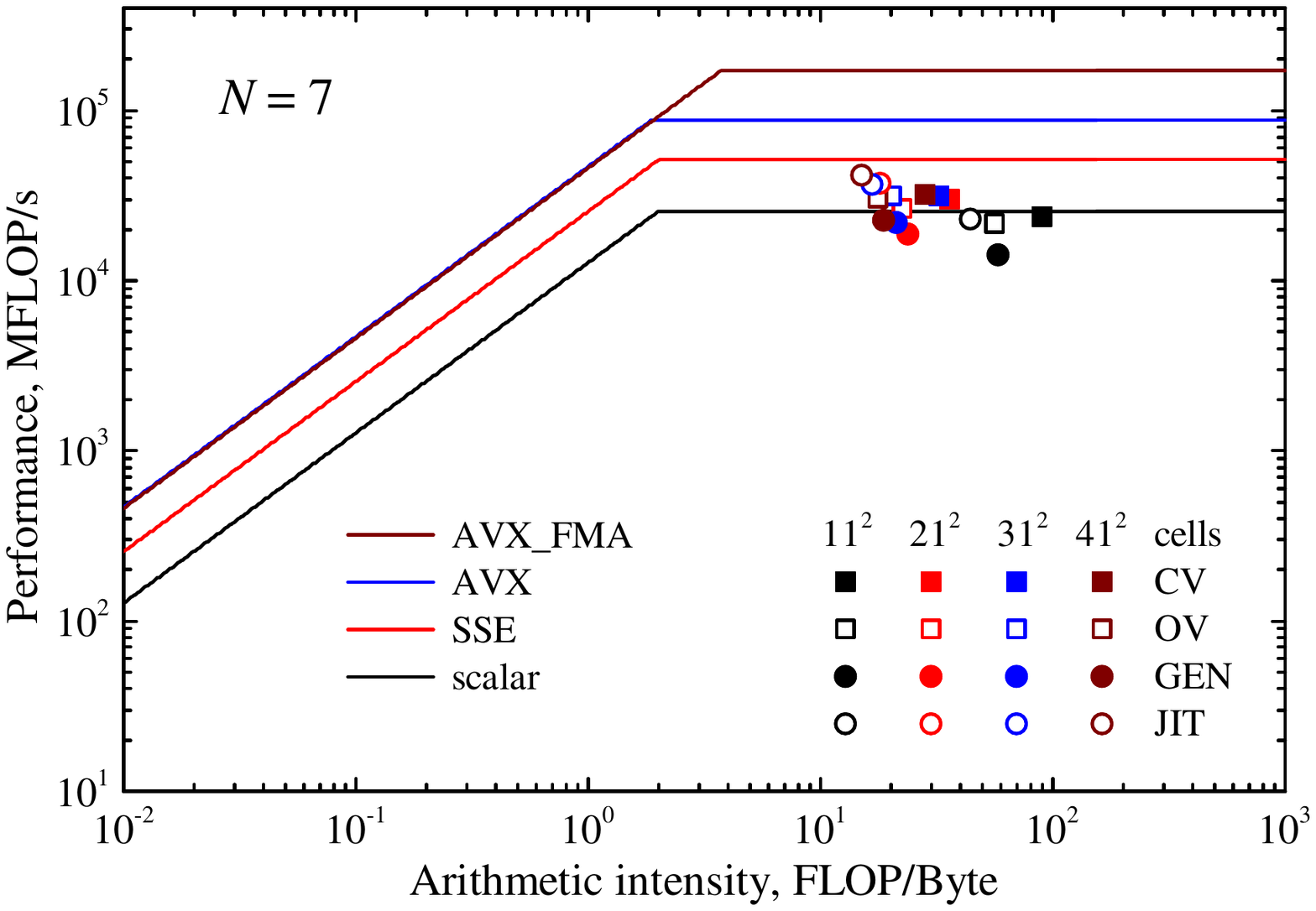}
\includegraphics[width=0.32\textwidth]{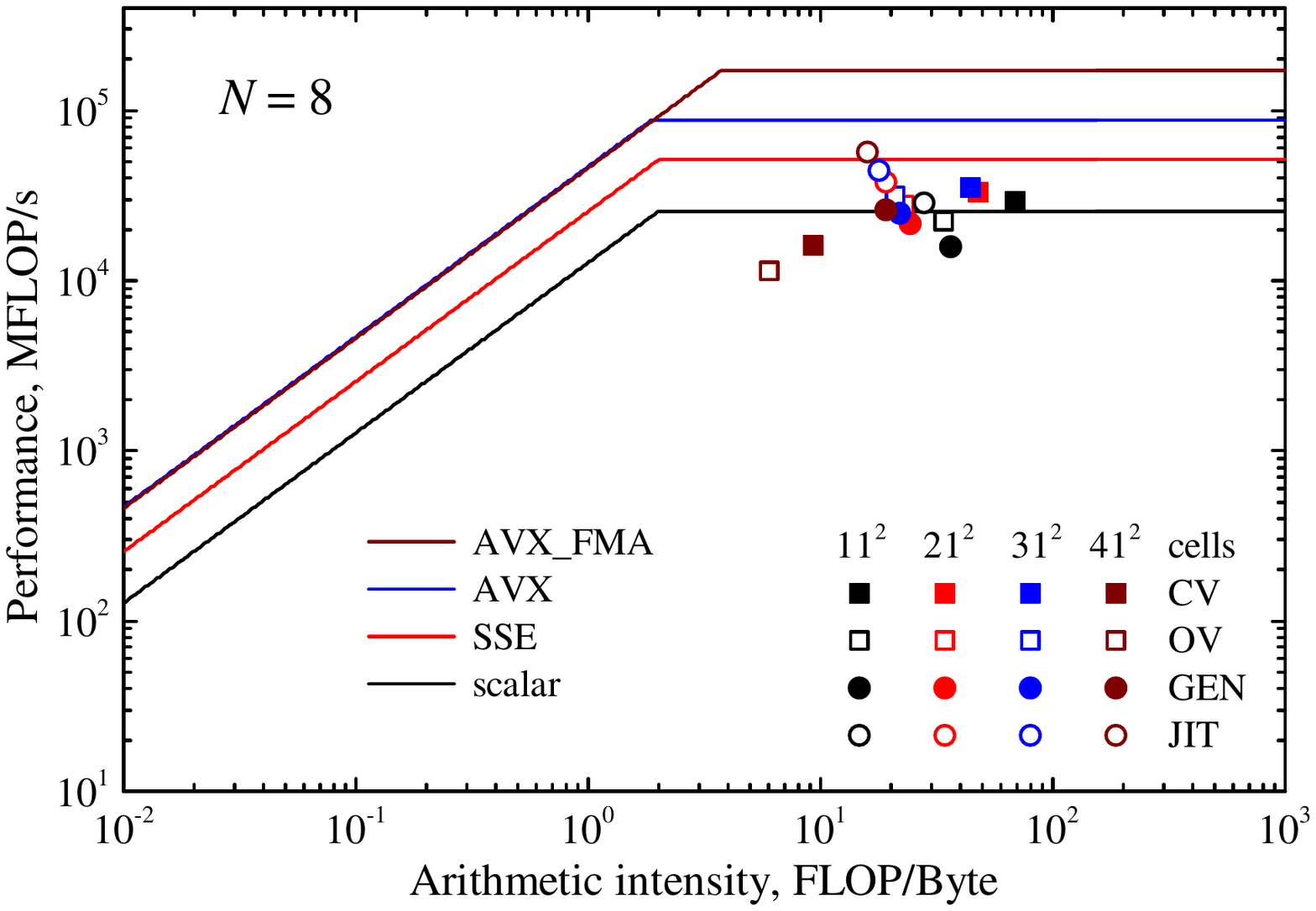}
\includegraphics[width=0.32\textwidth]{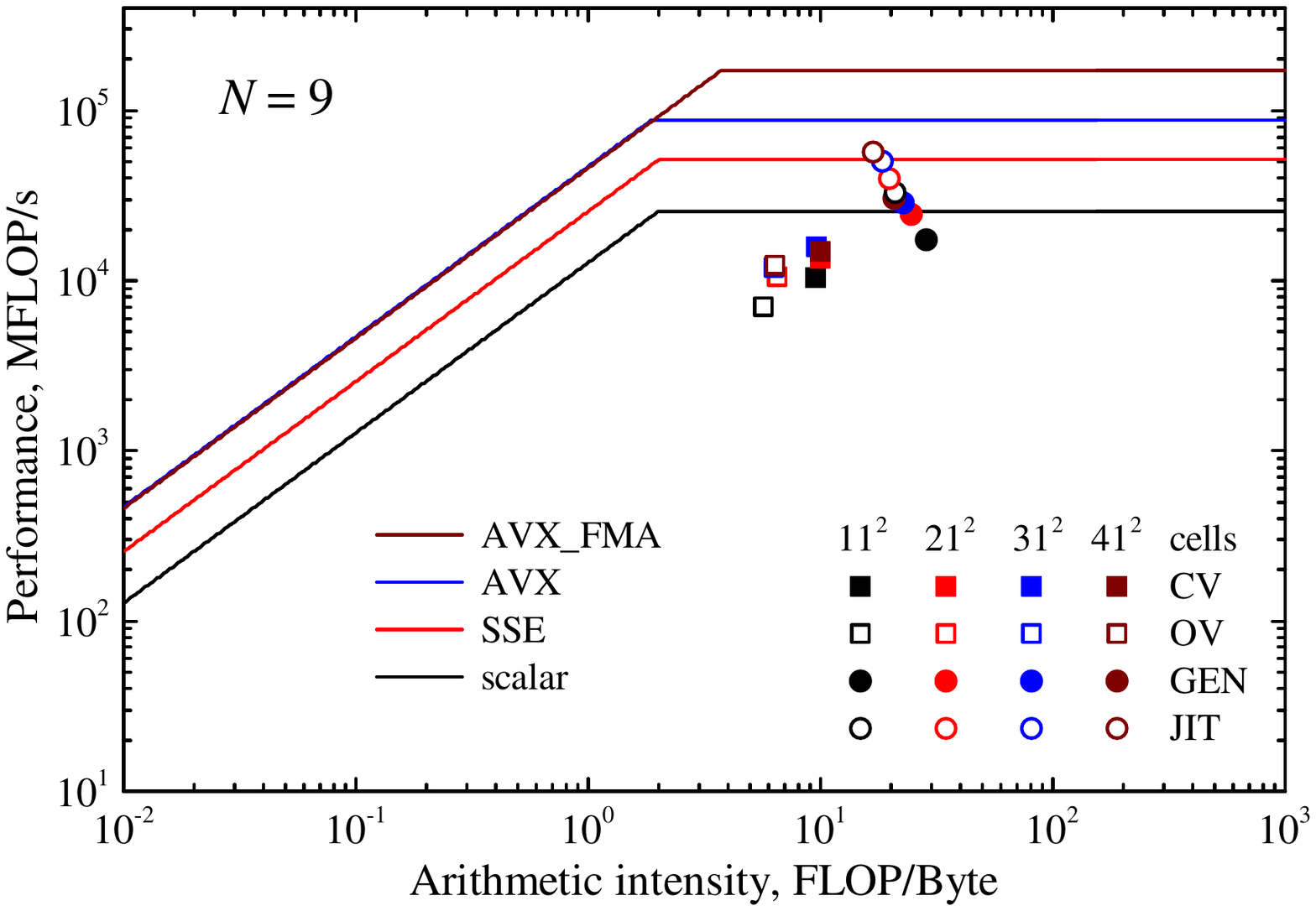}
\caption{\label{fig:roofline_DG_SCL_2d_degrees}%
Roofline models for finite-element ADER-DG-$\mathbb{P}_{N}$ method with a posteriori correction of the solution in subcells by a finite-volume ADER-WENO2 limiter for cylindrical explosion problem --- performance (measured in MFLOP/s) versus arithmetic intensity (measured in FLOP/Byte) on a double logarithmic scale. Peak performance values are presented for scalar operations, for operations using the SSE instruction set, using the AVX instruction set, and using the AVX-512 FMA instruction set. Data points are presented for CV, OV, general BLAS and JIT BLAS implementations. A comparison is presented for different mesh sizes --- $11^{2}$, $21^{2}$, $31^{2}$, $41^{2}$, at different polynomial degrees $N = 1, \ldots, 9$.
}
\end{figure*}

\begin{figure*}[h!]
\centering
\includegraphics[width=0.245\textwidth]{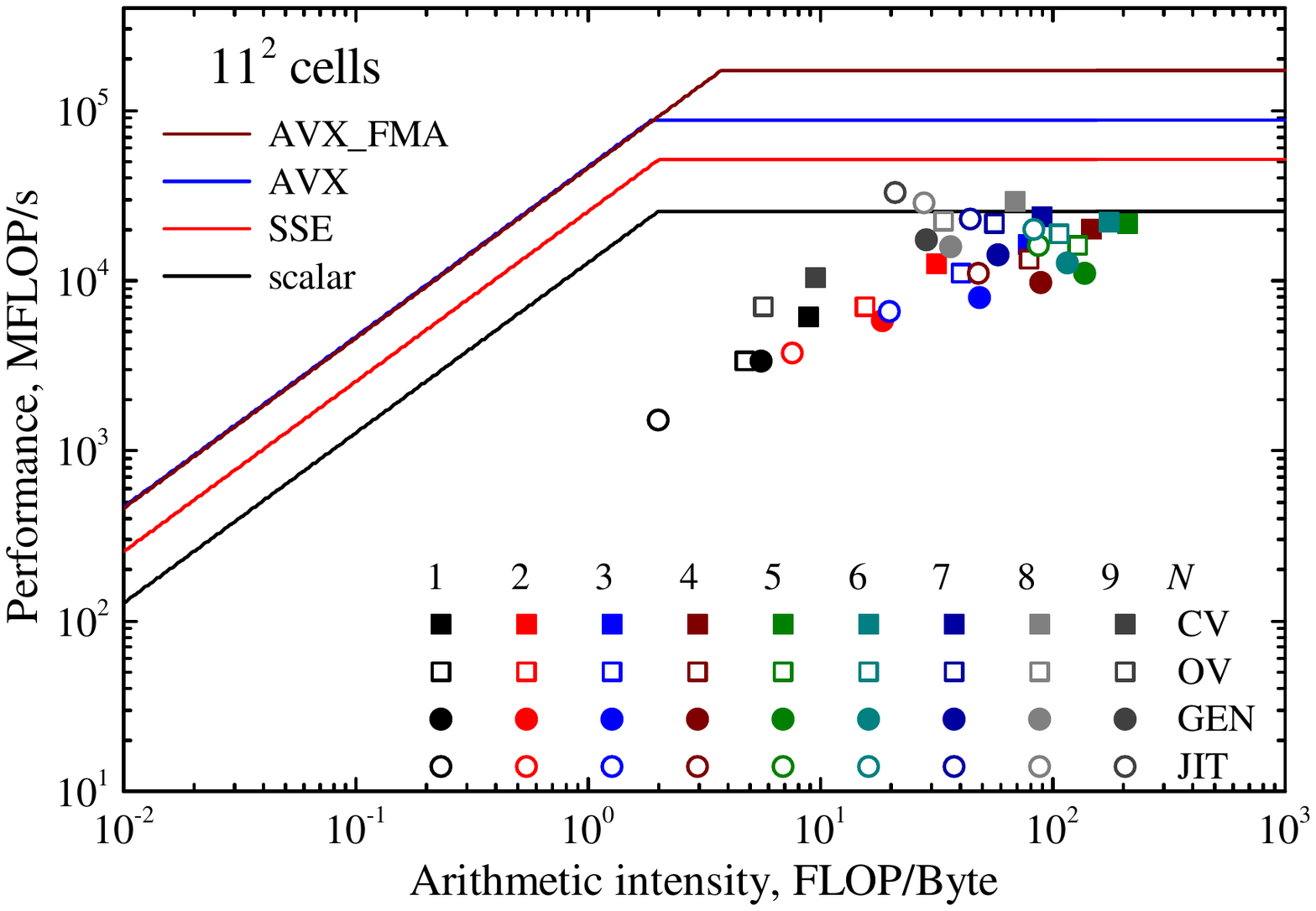}
\includegraphics[width=0.245\textwidth]{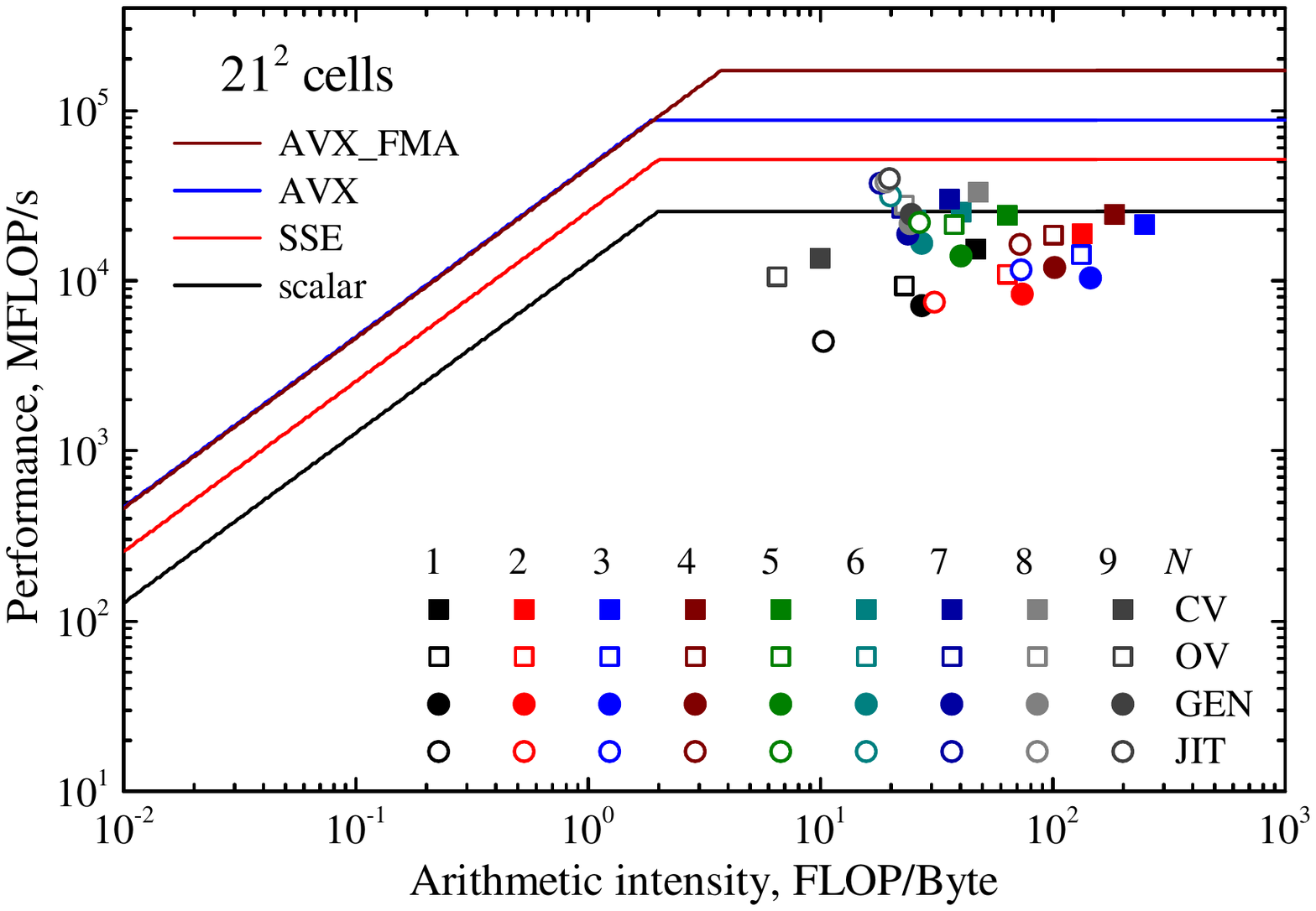}
\includegraphics[width=0.245\textwidth]{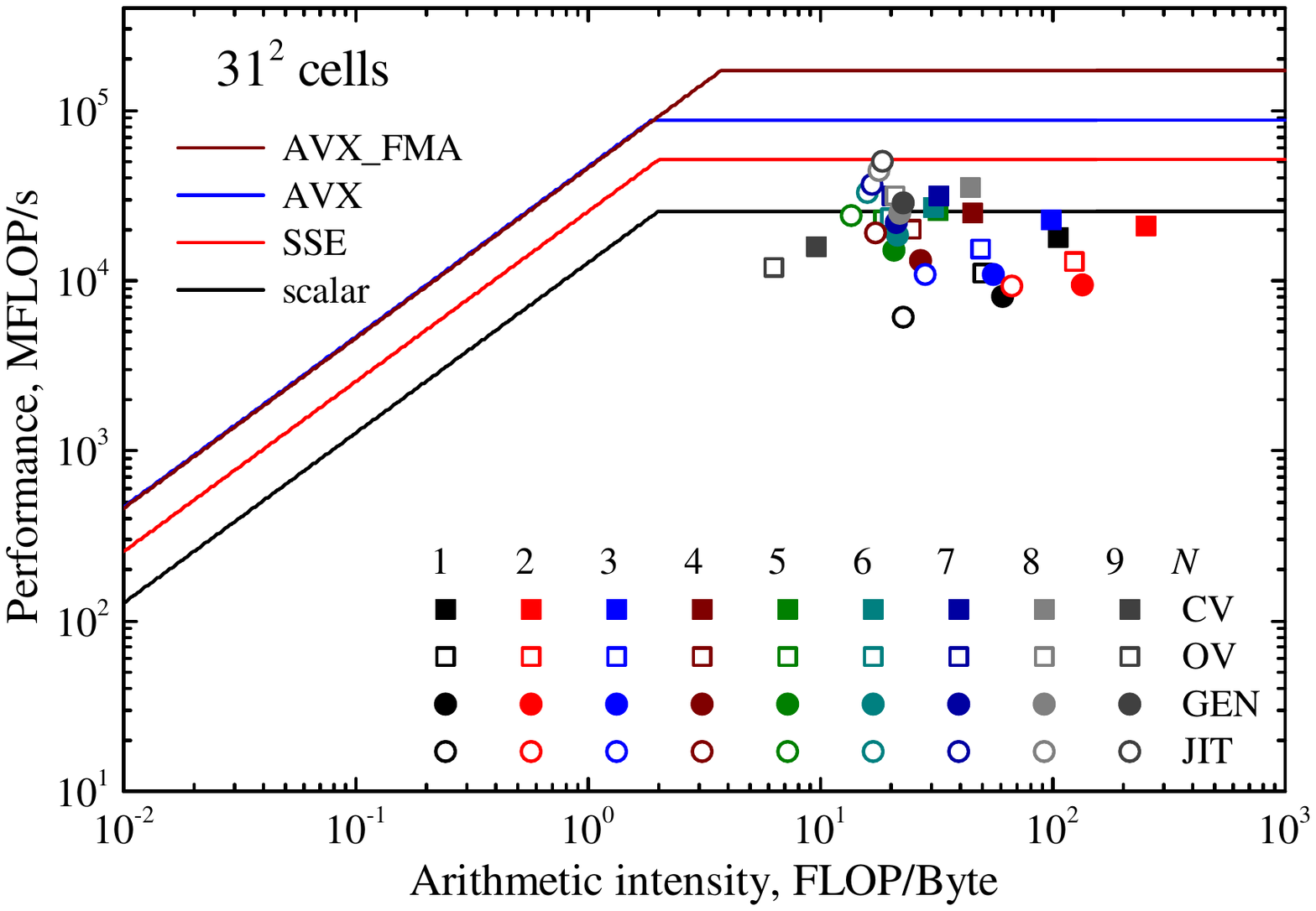}
\includegraphics[width=0.245\textwidth]{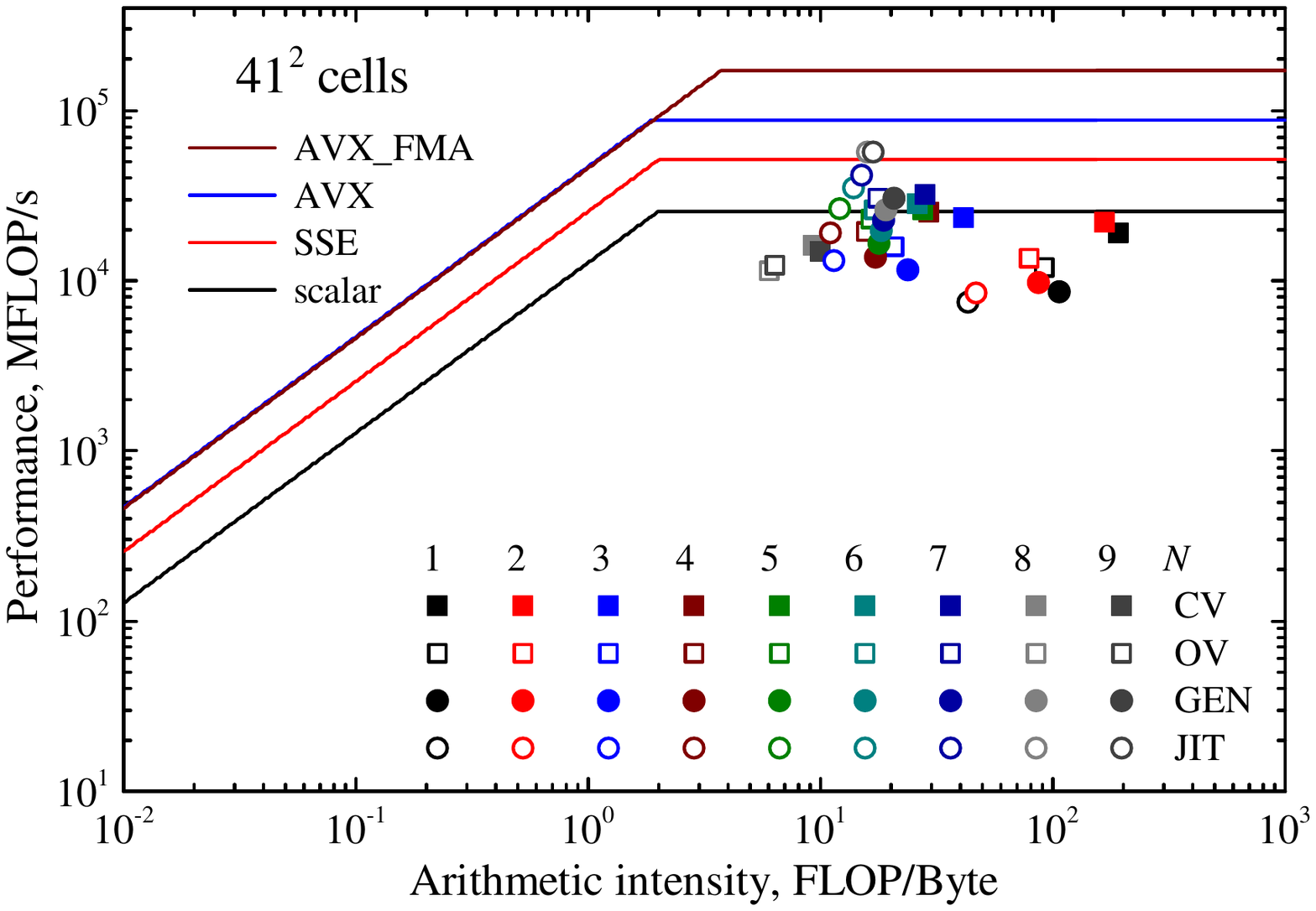}
\caption{\label{fig:roofline_DG_SCL_2d_mesh_sizes}%
Roofline models for finite-element ADER-DG-$\mathbb{P}_{N}$ method with a posteriori correction of the solution in subcells by a finite-volume ADER-WENO2 limiter for cylindrical explosion problem --- performance (measured in MFLOP/s) versus arithmetic intensity (measured in FLOP/Byte) on a double logarithmic scale. Peak performance values are presented for scalar operations, for operations using the SSE instruction set, using the AVX instruction set, and using the AVX-512 FMA instruction set. Data points are presented for CV, OV, general BLAS and JIT BLAS implementations. A comparison is presented for different polynomial degrees $N = 1, \ldots, 9$, at different mesh sizes $11^{2}$, $21^{2}$, $31^{2}$, $41^{2}$.
}
\end{figure*}

\begin{figure*}[h!]
\centering
\includegraphics[width=0.32\textwidth]{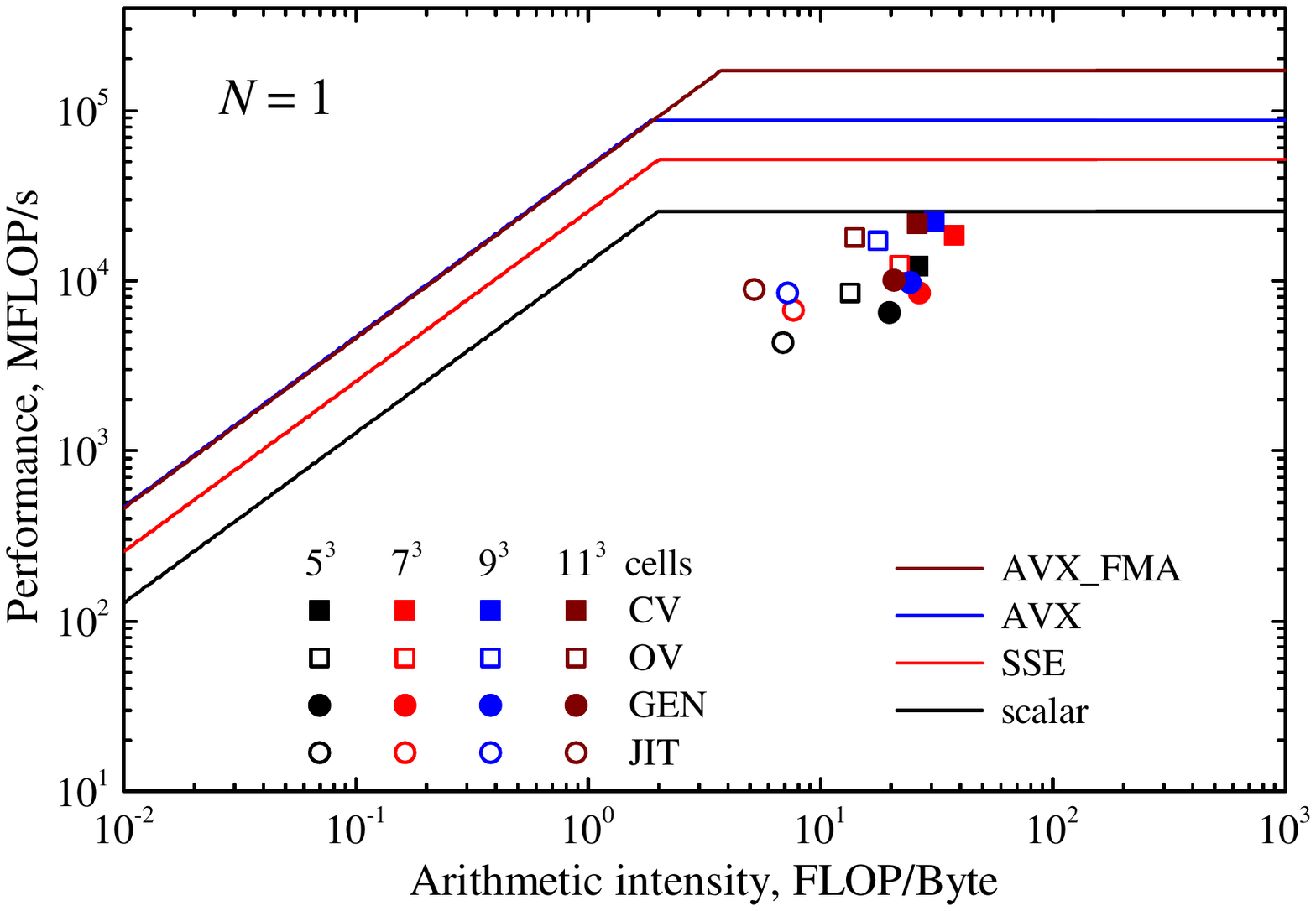}
\includegraphics[width=0.32\textwidth]{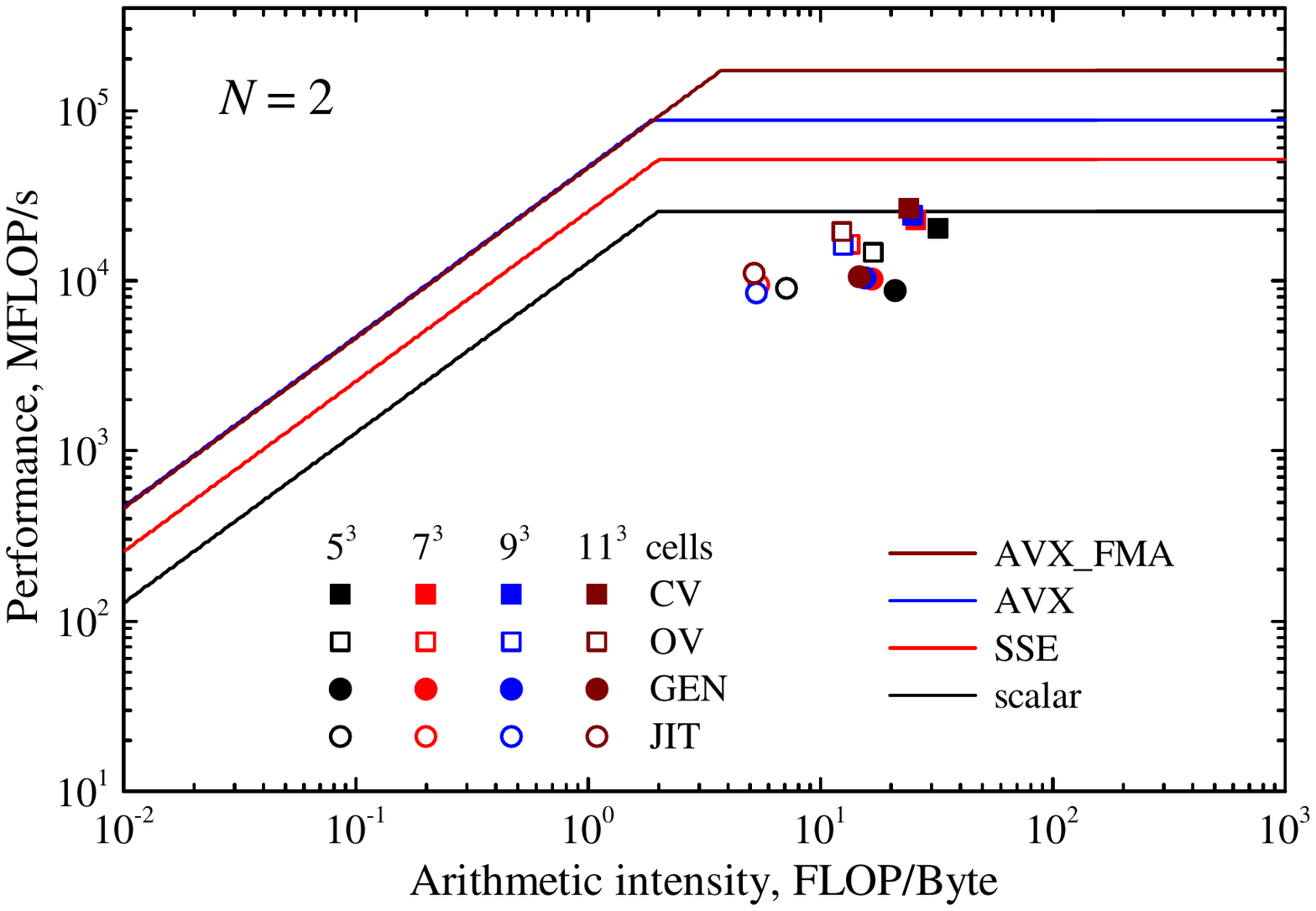}
\includegraphics[width=0.32\textwidth]{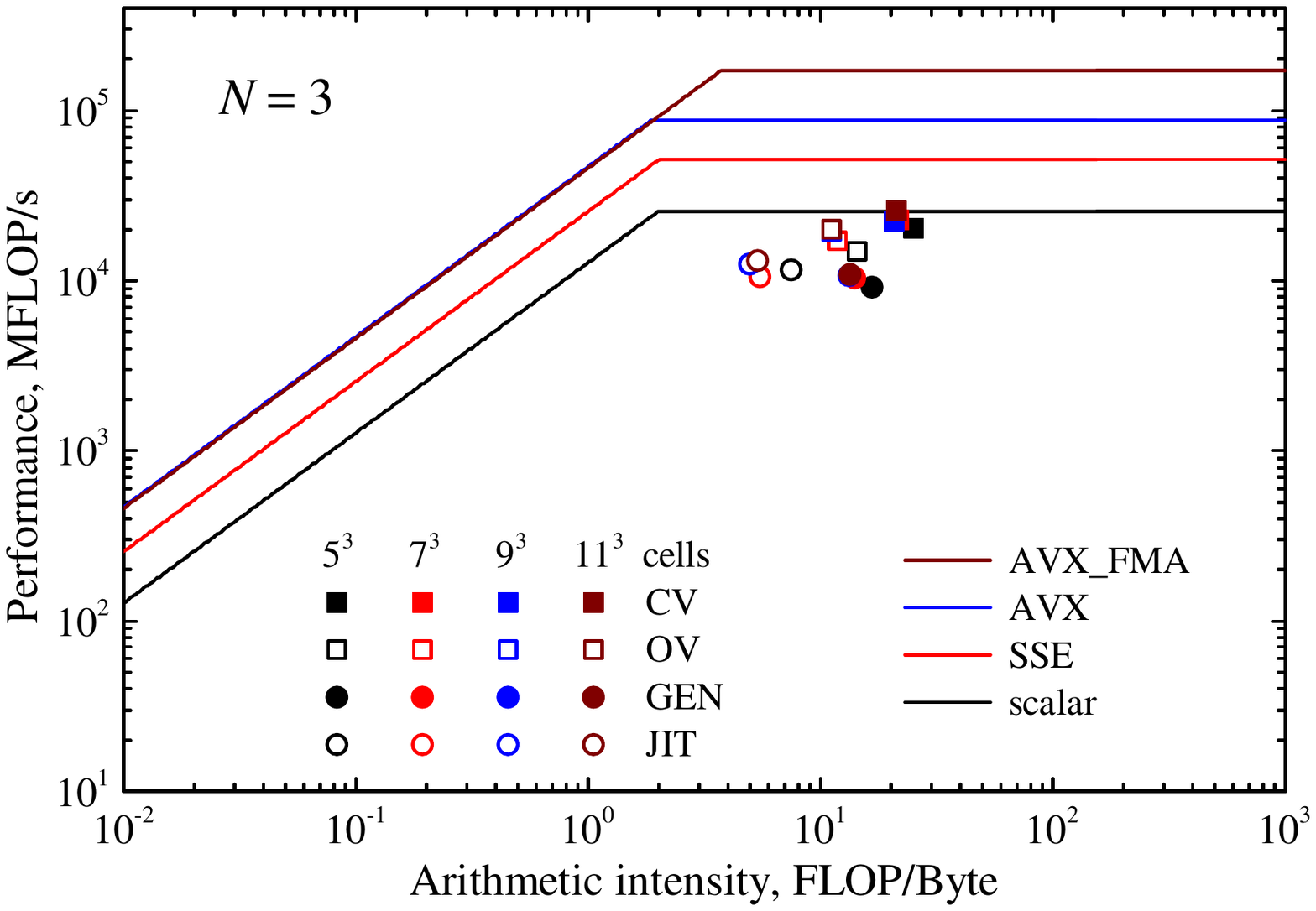}\\
\includegraphics[width=0.32\textwidth]{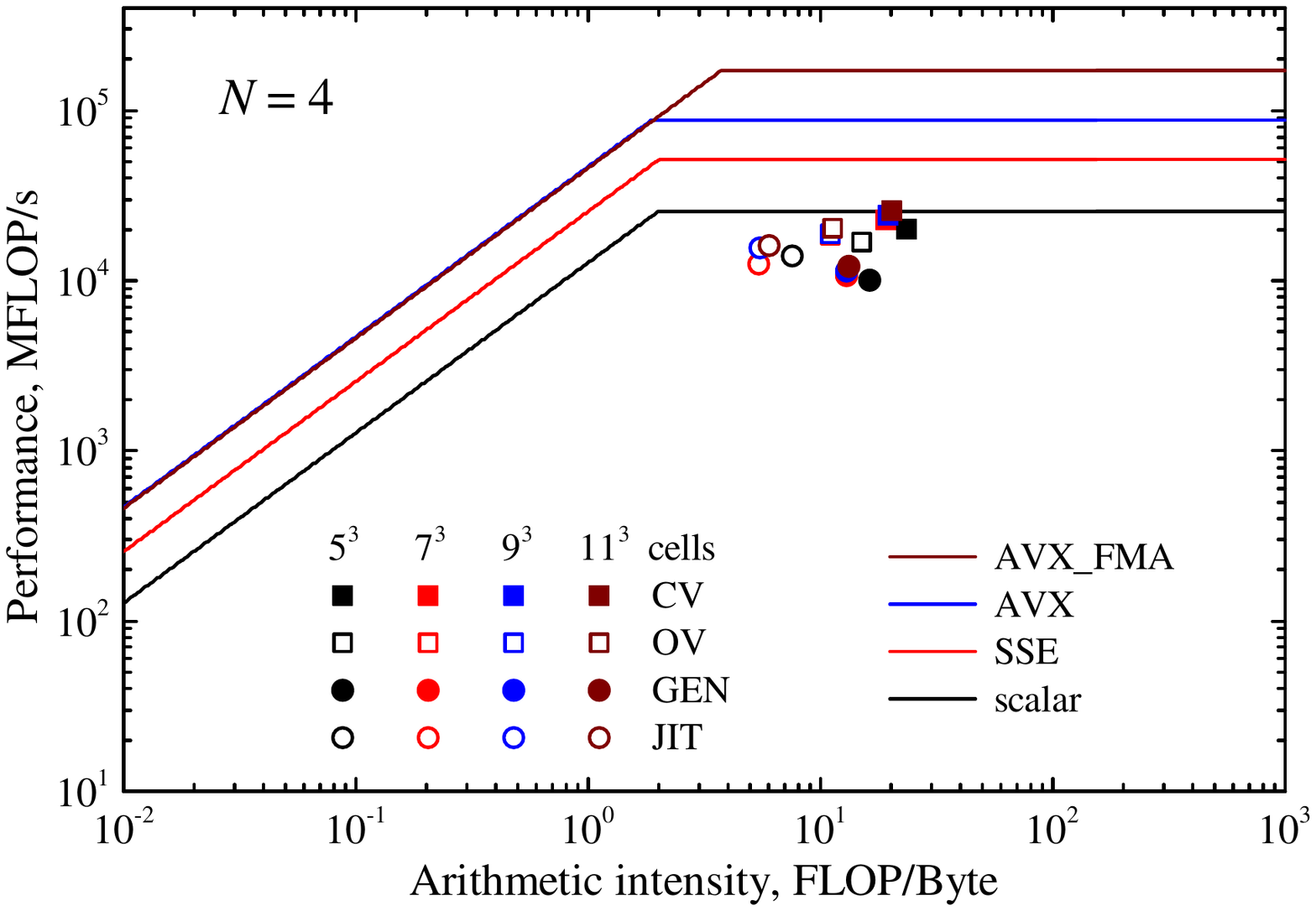}
\includegraphics[width=0.32\textwidth]{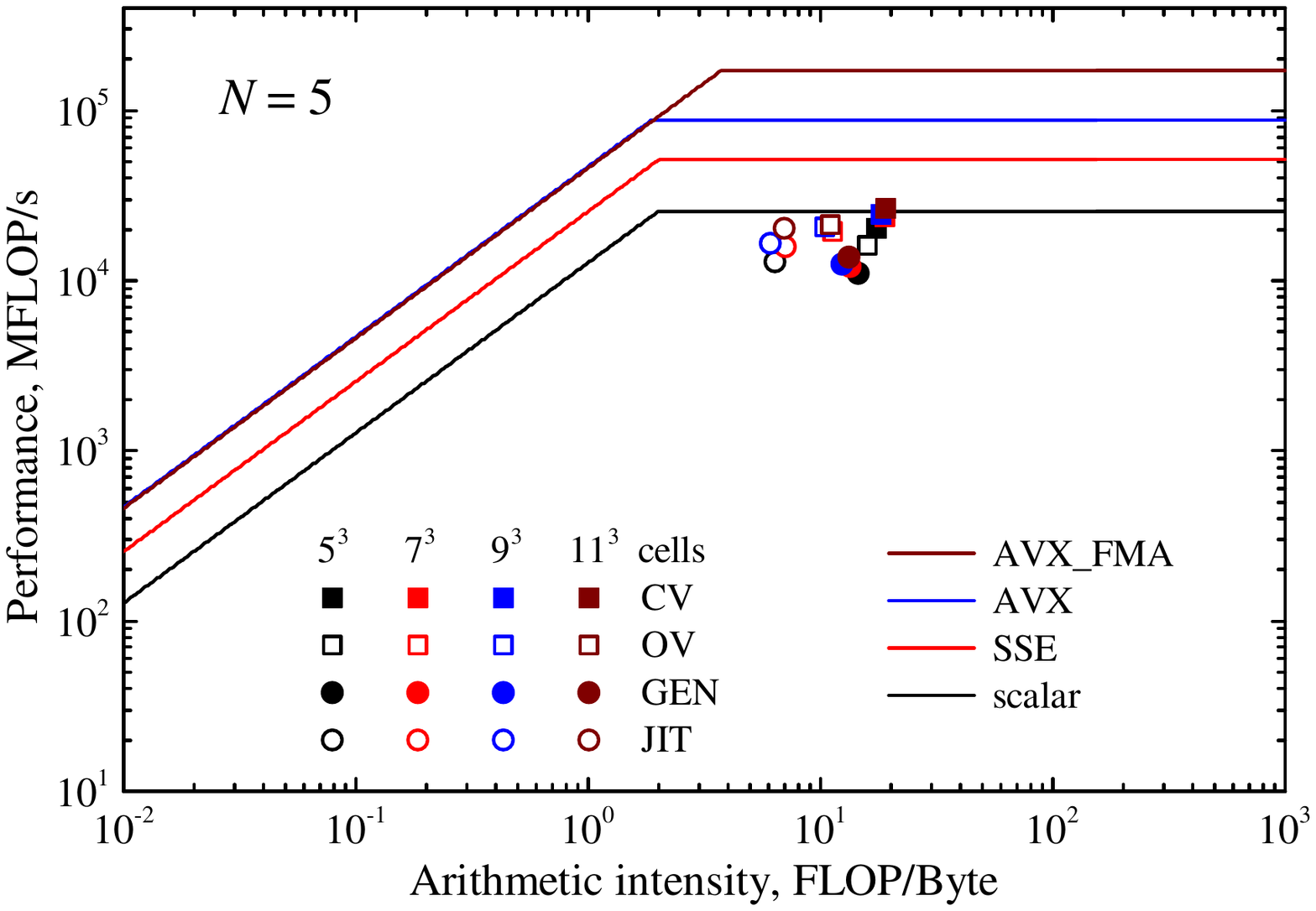}
\includegraphics[width=0.32\textwidth]{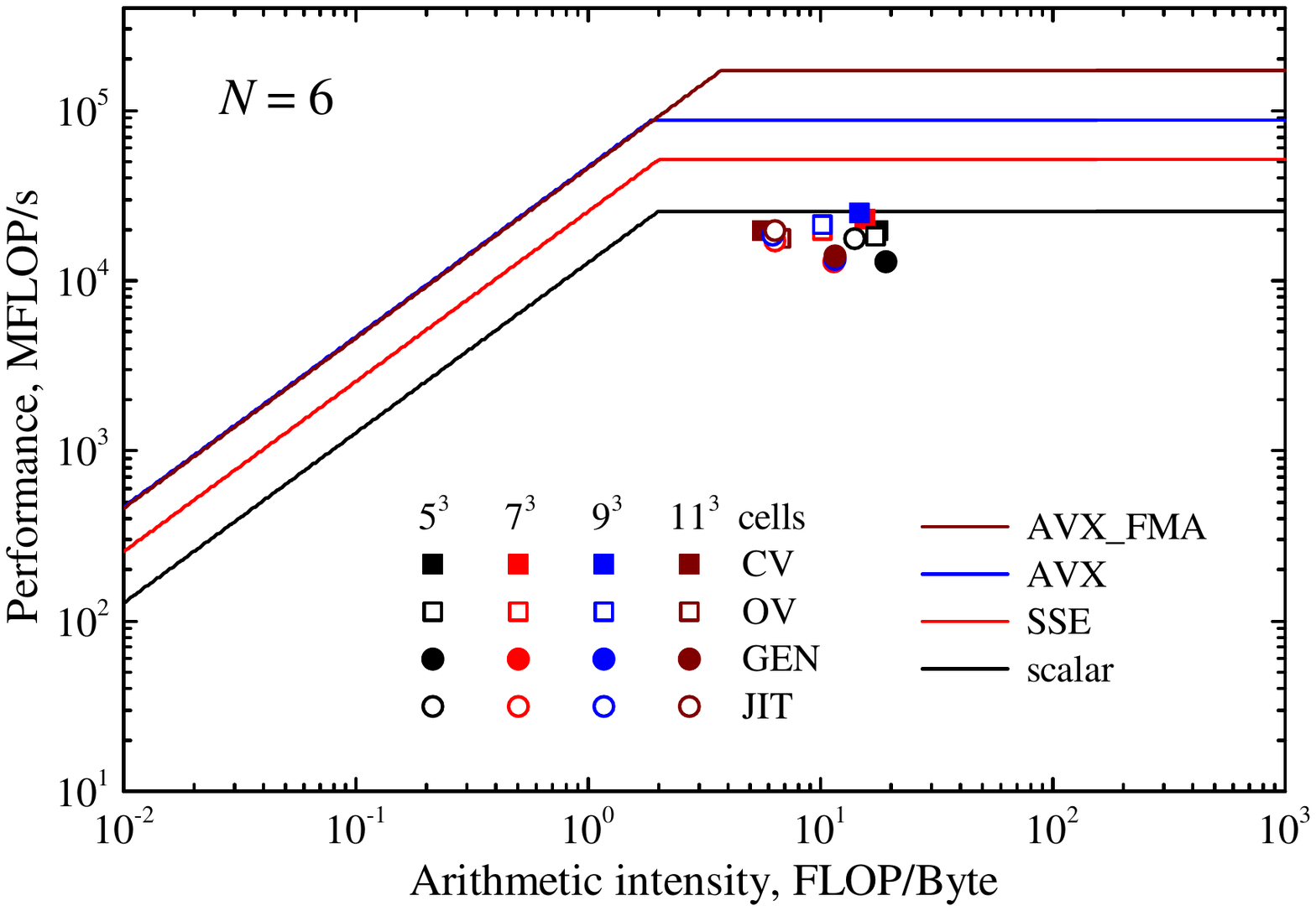}\\
\includegraphics[width=0.32\textwidth]{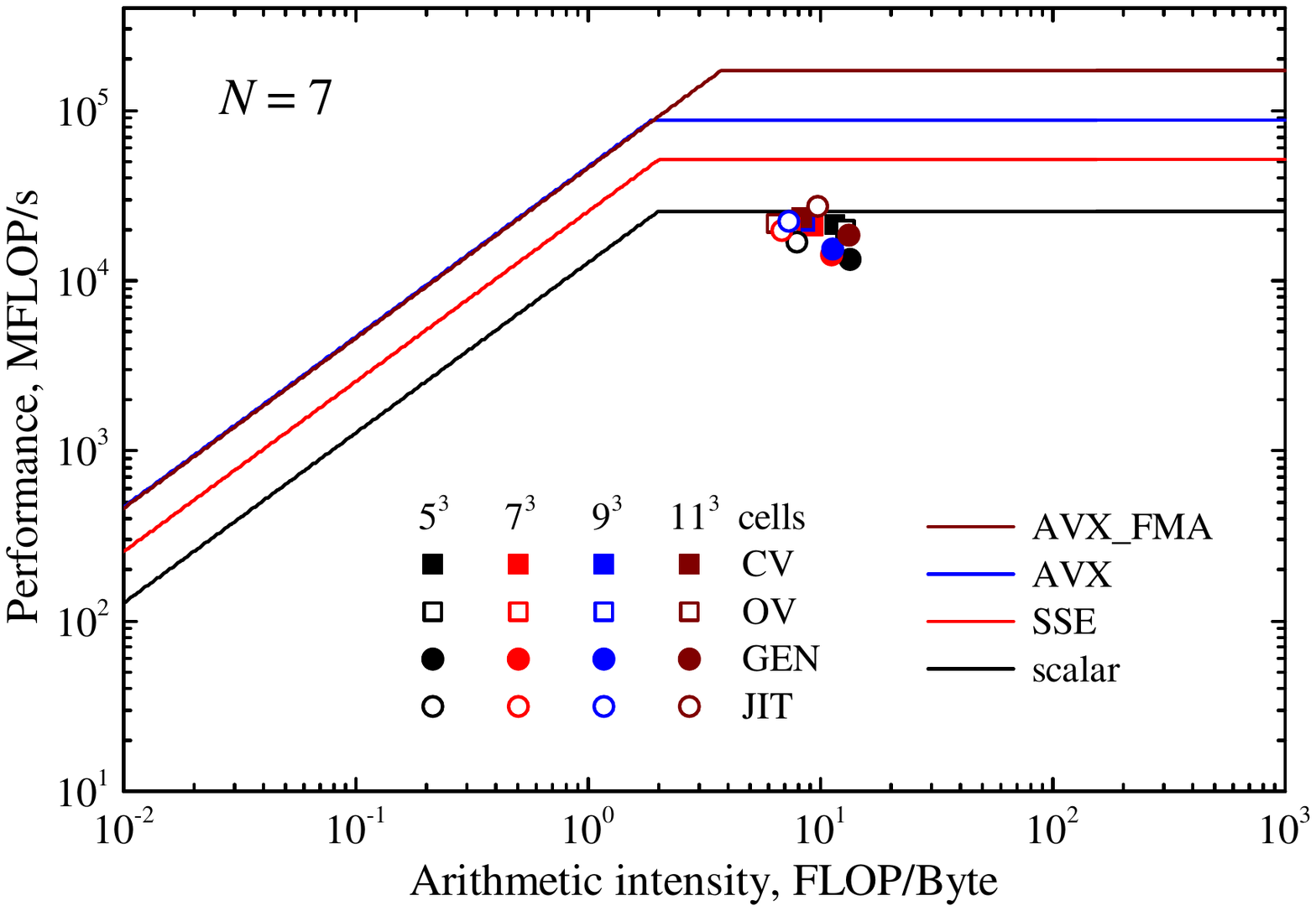}
\includegraphics[width=0.32\textwidth]{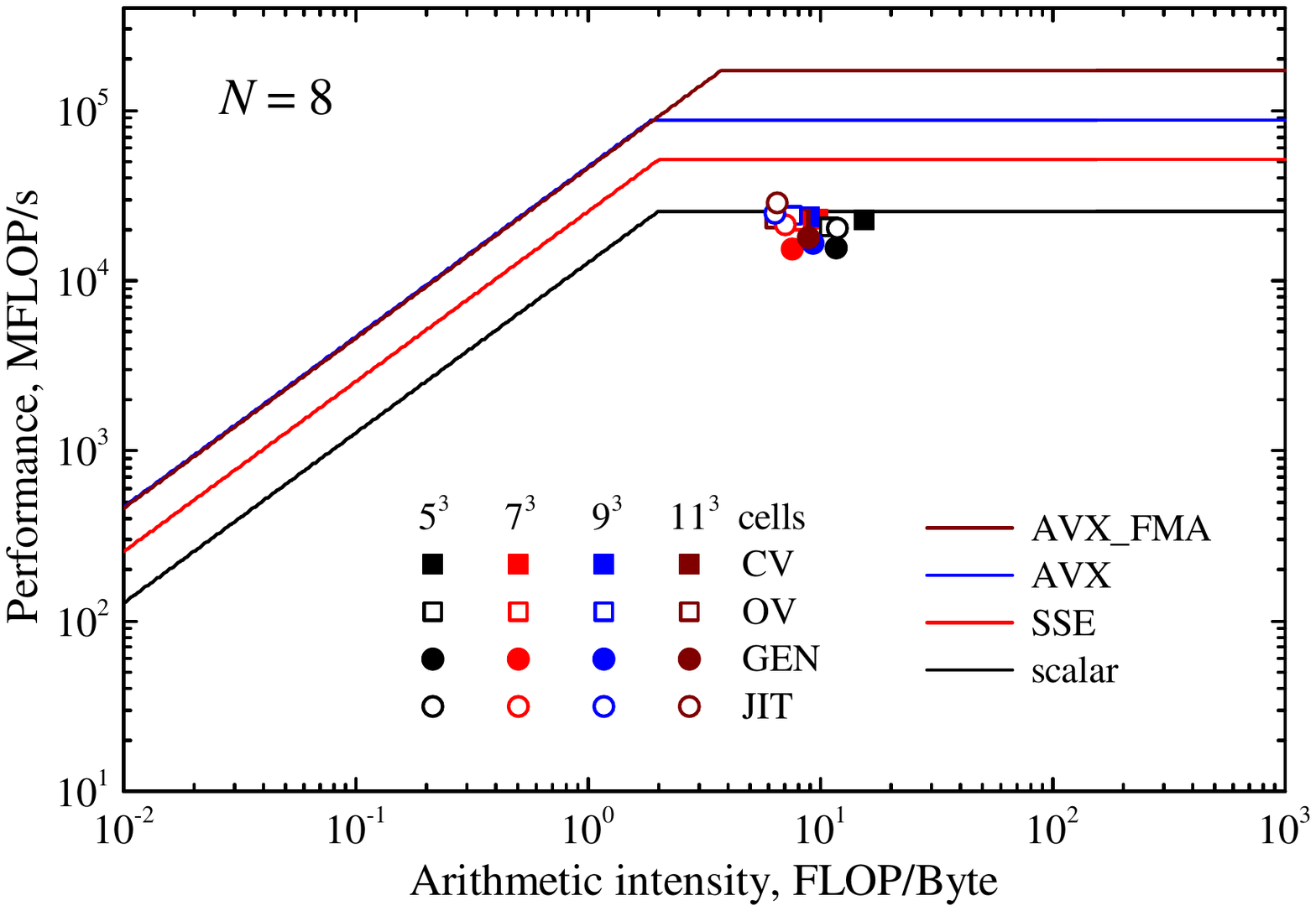}
\includegraphics[width=0.32\textwidth]{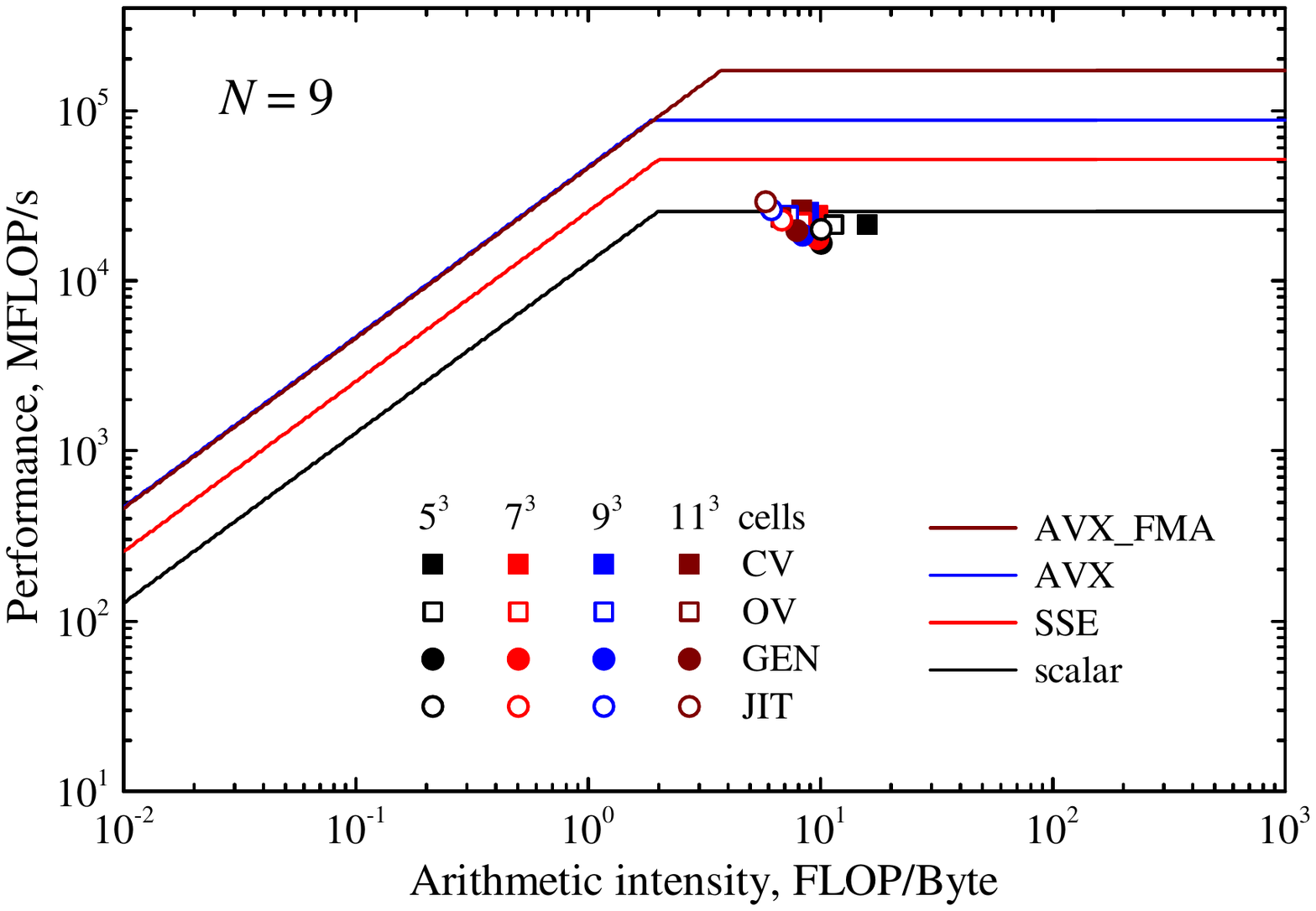}
\caption{\label{fig:roofline_DG_SCL_3d_degrees}%
Roofline models for finite-element ADER-DG-$\mathbb{P}_{N}$ method with a posteriori correction of the solution in subcells by a finite-volume ADER-WENO2 limiter for spherical explosion problem --- performance (measured in MFLOP/s) versus arithmetic intensity (measured in FLOP/Byte) on a double logarithmic scale. Peak performance values are presented for scalar operations, for operations using the SSE instruction set, using the AVX instruction set, and using the AVX-512 FMA instruction set. Data points are presented for CV, OV, general BLAS and JIT BLAS implementations. A comparison is presented for different mesh sizes --- $5^{3}$, $7^{3}$, $9^{3}$, $11^{3}$, at different polynomial degrees $N = 1, \ldots, 9$.
}
\end{figure*}

\begin{figure*}[h!]
\centering
\includegraphics[width=0.245\textwidth]{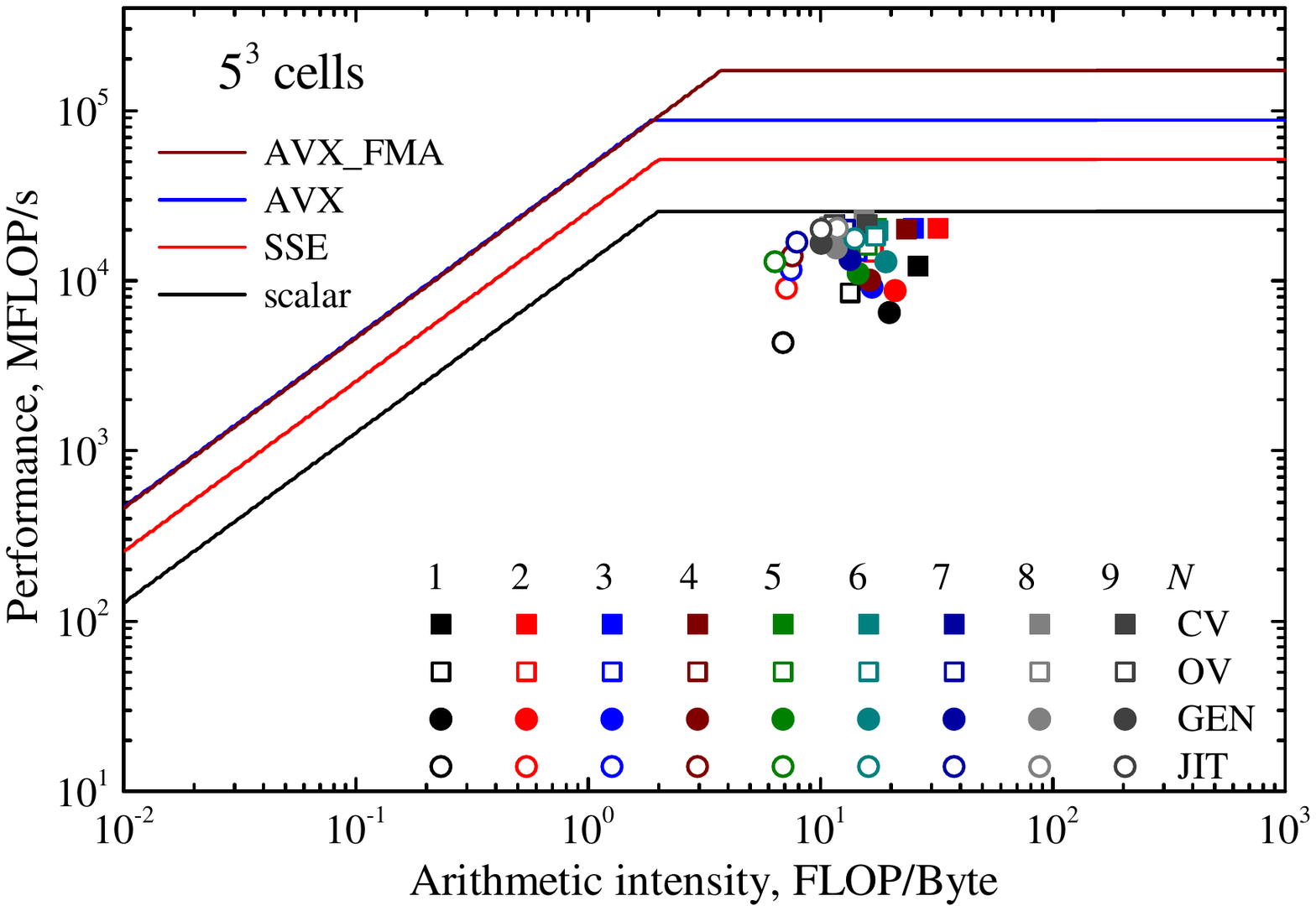}
\includegraphics[width=0.245\textwidth]{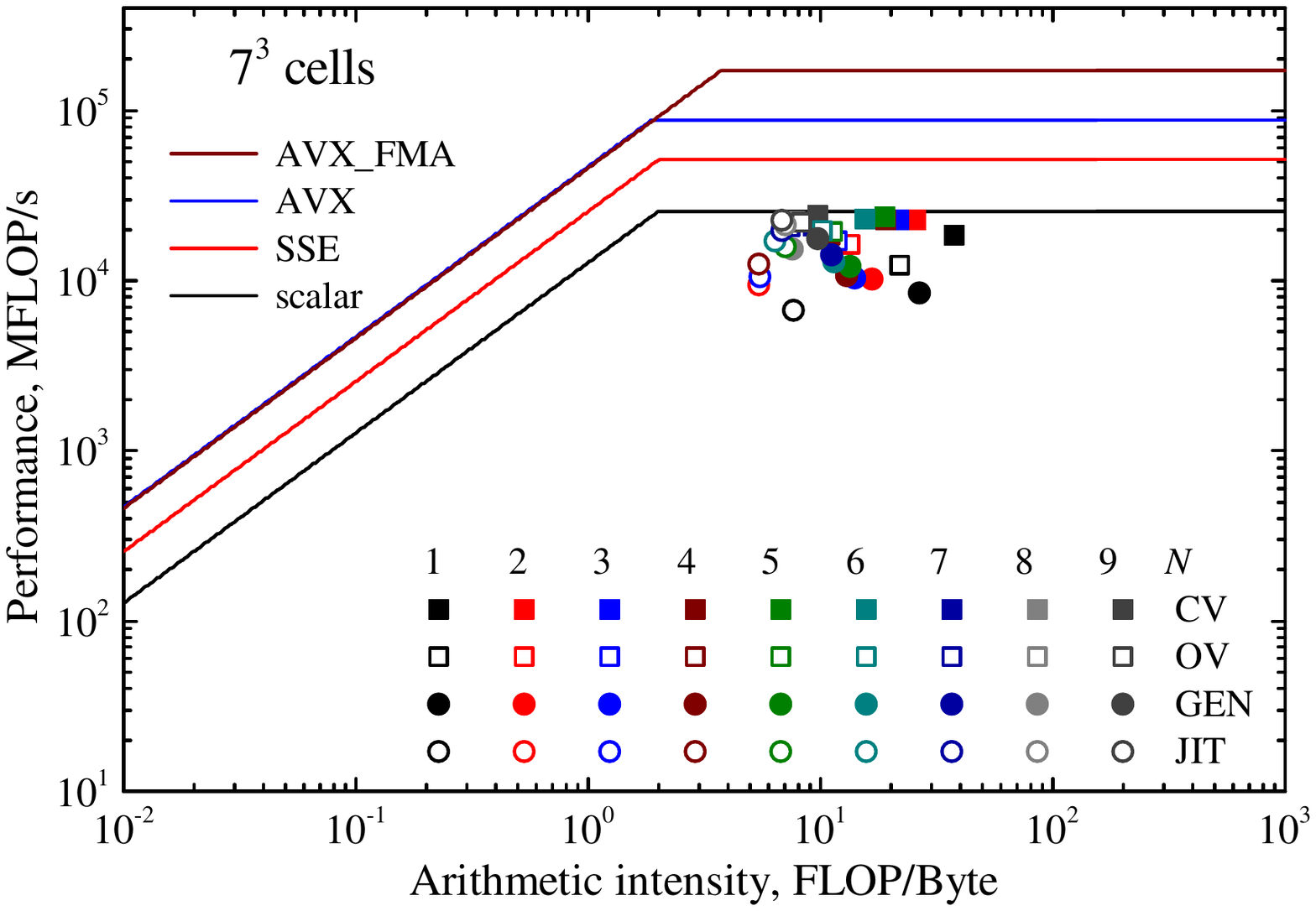}
\includegraphics[width=0.245\textwidth]{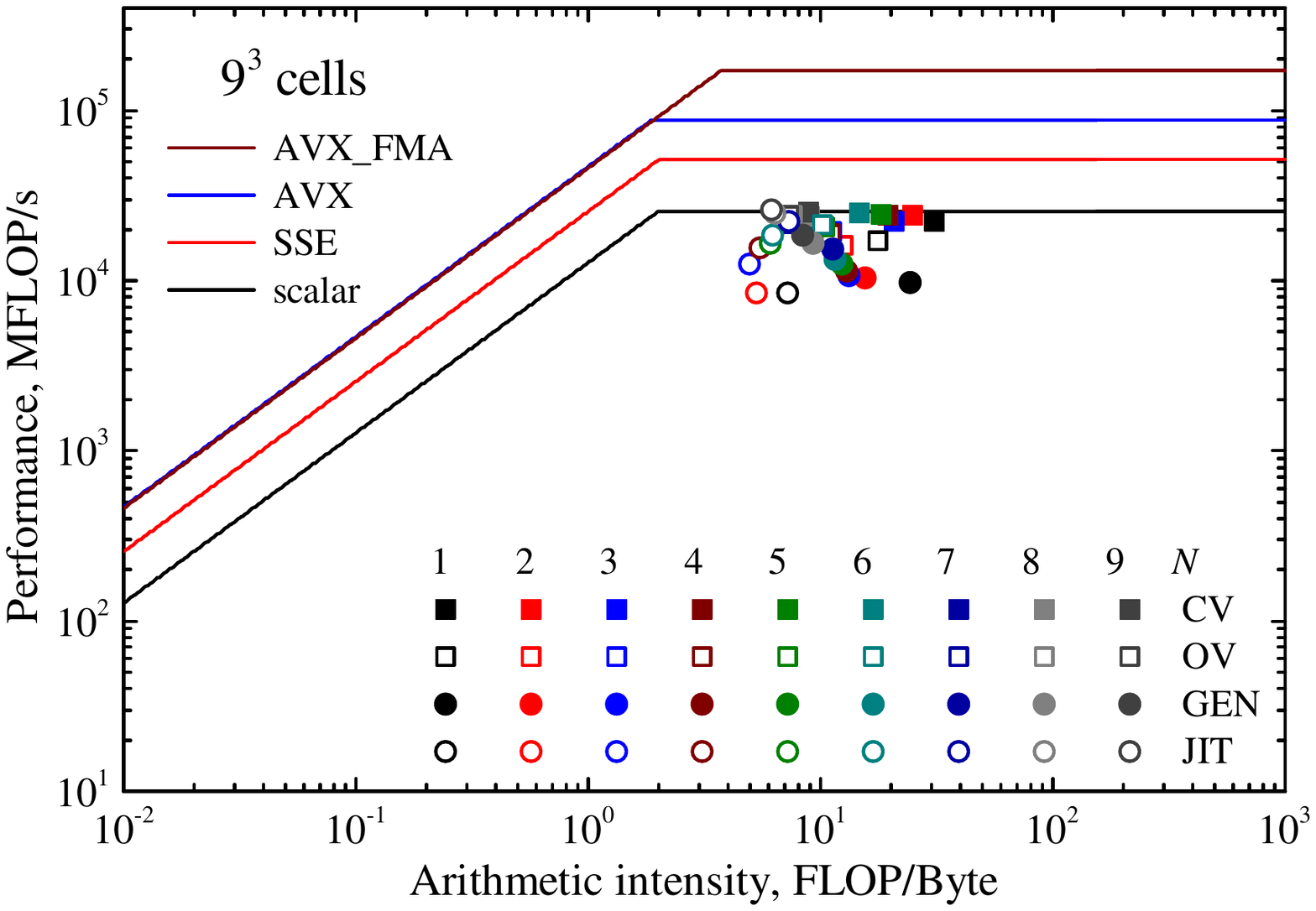}
\includegraphics[width=0.245\textwidth]{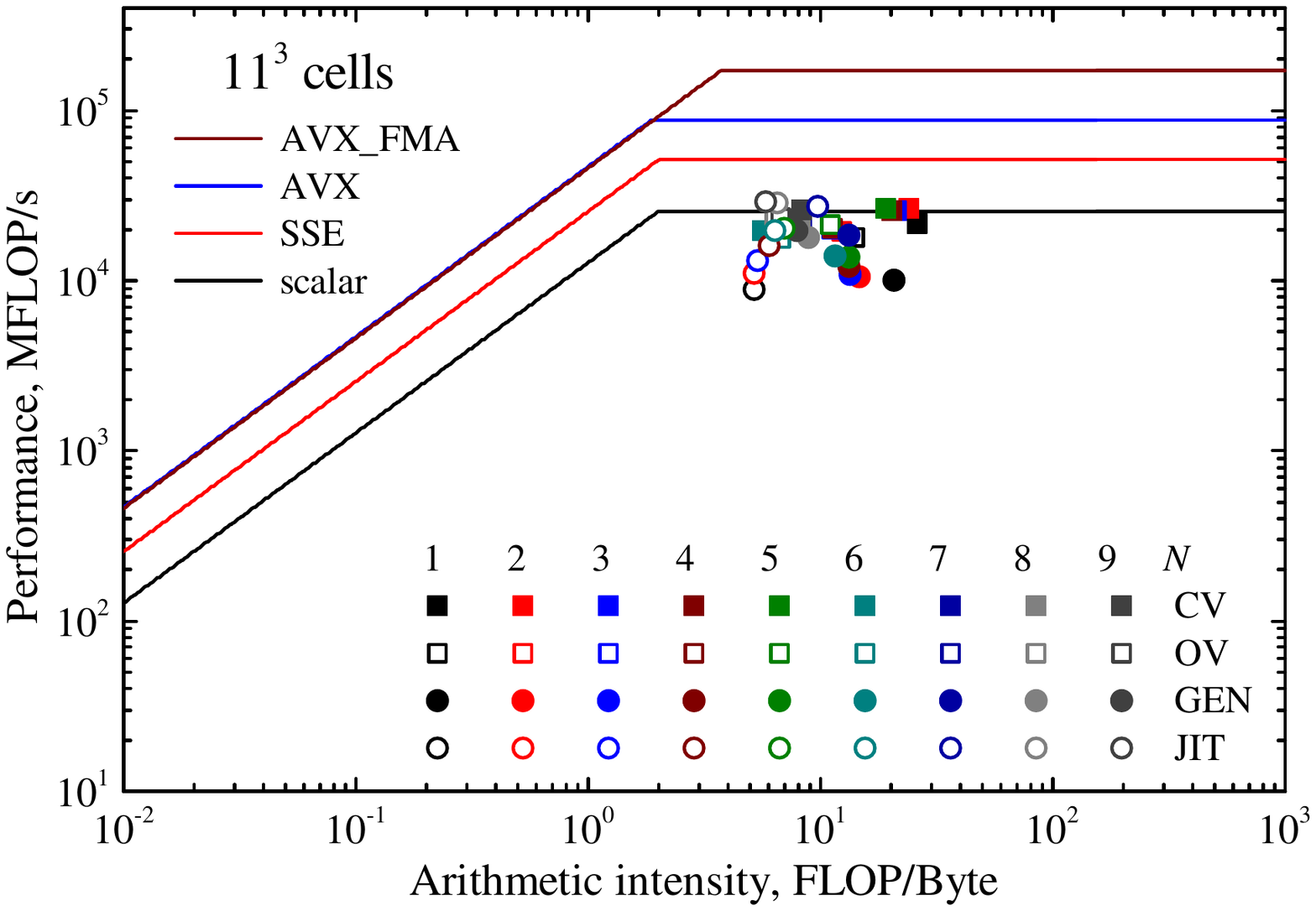}
\caption{\label{fig:roofline_DG_SCL_3d_mesh_sizes}%
Roofline models for finite-element ADER-DG-$\mathbb{P}_{N}$ method with a posteriori correction of the solution in subcells by a finite-volume ADER-WENO2 limiter for spherical explosion problem --- performance (measured in MFLOP/s) versus arithmetic intensity (measured in FLOP/Byte) on a double logarithmic scale. Peak performance values are presented for scalar operations, for operations using the SSE instruction set, using the AVX instruction set, and using the AVX-512 FMA instruction set. Data points are presented for CV, OV, general BLAS and JIT BLAS implementations. A comparison is presented for different polynomial degrees $N = 1, \ldots, 9$, at different mesh sizes --- $5^{3}$, $7^{3}$, $9^{3}$, $11^{3}$.
}
\end{figure*}

Roofline data measurements were performed using the same compilation and execution parameters as for the Section~\ref{sec:comp_costs} computational cost measurements. When executing with only one thread, the measured data significantly exceeded the roofline, which is due to the internal design of the CPU, whose scheduling system used unused computational cores in case of insignificant load. The results were obtained using execution on four threads, with approximately four tasks running simultaneously, effectively loading 16 computational cores. Figures~\ref{fig:roofline_DG_2d_degrees},~\ref{fig:roofline_DG_2d_mesh_sizes},~\ref{fig:roofline_DG_3d_degrees} and~\ref{fig:roofline_DG_3d_mesh_sizes} show the roofline models for ADER-DG-$\mathbb{P}_{N}$ method for two-dimensional and three-dimensional sine wave advection problems, with Figures~\ref{fig:roofline_DG_2d_degrees} and~\ref{fig:roofline_DG_2d_mesh_sizes} showing the results of solving a two-dimensional problem, Figures~\ref{fig:roofline_DG_3d_degrees} and~\ref{fig:roofline_DG_3d_mesh_sizes} showing the results of solving a three-dimensional problem, Figures~\ref{fig:roofline_DG_2d_degrees} and~\ref{fig:roofline_DG_3d_degrees} showing the results of the comparison for different mesh sizes at different polynomial degrees $N$, and Figures~\ref{fig:roofline_DG_2d_mesh_sizes} and~\ref{fig:roofline_DG_3d_mesh_sizes} showing the results of the comparison for different polynomial degrees $N$ at different mesh sizes. Figures~\ref{fig:roofline_WENO_2d_degrees},~\ref{fig:roofline_WENO_2d_mesh_sizes},~\ref{fig:roofline_WENO_3d_degrees} and~\ref{fig:roofline_WENO_3d_mesh_sizes} show the roofline models for finite-volume ADER-WENO($N+1$) method for two-dimensional and three-dimensional sine wave advection problems, with Figures~\ref{fig:roofline_WENO_2d_degrees} and~\ref{fig:roofline_WENO_2d_mesh_sizes} showing the results of solving a two-dimensional problem, Figures~\ref{fig:roofline_WENO_3d_degrees} and~\ref{fig:roofline_WENO_3d_mesh_sizes} showing the results of solving a three-dimensional problem, Figures~\ref{fig:roofline_WENO_2d_degrees} and~\ref{fig:roofline_WENO_3d_degrees} showing the results of the comparison for different mesh sizes at different polynomial degrees $N$, and Figures~\ref{fig:roofline_WENO_2d_mesh_sizes} and~\ref{fig:roofline_WENO_3d_mesh_sizes} showing the results of the comparison for different polynomial degrees $N$ at different mesh sizes. Figures~\ref{fig:roofline_DG_SCL_2d_degrees},~\ref{fig:roofline_DG_SCL_2d_mesh_sizes},~\ref{fig:roofline_DG_SCL_3d_degrees} and~\ref{fig:roofline_DG_SCL_3d_mesh_sizes} show the roofline models for finite-element ADER-DG-$\mathbb{P}_{N}$ method with a posteriori correction of the solution in subcells by a finite-volume ADER-WENO2 limiter for cylindrical and spherical explosion problems, with Figures~\ref{fig:roofline_DG_SCL_2d_degrees} and~\ref{fig:roofline_DG_SCL_2d_mesh_sizes} showing the results of solving a two-dimensional cylindrical explosion problem, Figures~\ref{fig:roofline_DG_SCL_3d_degrees} and~\ref{fig:roofline_DG_SCL_3d_mesh_sizes} showing the results of solving a three-dimensional spherical explosion problem, Figures~\ref{fig:roofline_DG_SCL_2d_degrees} and~\ref{fig:roofline_DG_SCL_3d_degrees} showing the results of the comparison for different mesh sizes at different polynomial degrees $N$, and Figures~\ref{fig:roofline_DG_SCL_2d_mesh_sizes} and~\ref{fig:roofline_DG_SCL_3d_mesh_sizes} showing the results of the comparison for different polynomial degrees $N$ at different mesh sizes. In all these cases, the comparison for different mesh sizes with different polynomial degrees $N$ and for different polynomial degrees $N$ with different mesh sizes is presented for the convenience of analyzing the obtained results. The roofline boundary lines presented in the figures were obtained for the four-thread case.

The results presented in Figure~\ref{fig:roofline_DG_2d_degrees} for the finite-element ADER-DG-$\mathbb{P}_{N}$ method for solving two-dimensional problems for different polynomial degrees $N$ show that in the case of small polynomial degrees $N < 5$, the data points are more than $10$ times lower than the roofline. In the case of polynomial degrees $N = 1$-$3$, for small numbers of mesh cells, the problem is a bandwidth-bound problem, but already in the case of larger meshes, the data points shift to the right and up, shifting the problem to the intermediate region and to the area of compute-bound problems. With an increase in the polynomial degrees $N$, a significant shift of data points to the right and up occurs, which indicates an increase in arithmetic intensity $I$ and computational performance $P$. At the same time, for the polynomial degrees $N \geqslant 5$ and all the numbers of mesh cells presented in the results shift into the area of compute-bound problems. In the case of small polynomial degrees $N$, with an increase in the number of mesh cells, a significant shift of data points to the right and up occurs, however, for polynomial degrees $N > 6$, this trend ceases to be strictly observed. In the case of small polynomial degrees $N$, the data points for the JIT BLAS implementation and the general BLAS implementation are significantly to the left and below the implementation of OV, and even the implementation of CV. However, in the case of a polynomial degree $N = 4$, this trend changes, and for polynomial degrees $N > 4$, JIT BLAS and general BLAS are above or to the right of the OV and CV implementations, at least for sufficiently large mesh cell counts. For sufficiently large  $N$, an interesting feature is observed for the JIT BLAS implementation and the general BLAS implementation --- with an increase in the number of mesh cells, the data points shift to the left and up, indicating a decrease in arithmetic intensity $I$, however, the data points also remain in the area of compute-bound problems. It should be noted that the obtained roofline results show that the data points are $4$-$10$ times lower than the peak performance values for AVX-512 FMA even for high polynomial degrees $N$. Therefore, purely hypothetically, there is still a certain reserve for further optimization of the implementation of the finite-element ADER-DG-$\mathbb{P}_{N}$ method in two-dimensional case, which can be implemented by a more efficient organization of the computational process in the program. However, in any case, certain computational costs will be associated with the direct and inverse calculation of flow, non-conservative and source terms, primitive variables and conservative quantities, as well as evaluation of the Riemann solver, the optimization of which for the efficient use of AVX-512 FMA can cause significant difficulties. The results for the method for different values of the number of mesh cells presented in Figure~\ref{fig:roofline_DG_2d_degrees} demonstrate the dependencies and trends described above much more clearly. It becomes clear that in the case of sufficiently large polynomial degrees, the JIT BLAS implementation has the highest performance compared to other implementations. There is also a clear tendency that in the case of a small number of mesh cells, an increase in arithmetic intensity $I$ and performance $P$ with an increase in the polynomial degrees $N$ is observed for almost all implementations, however, for a sufficiently large number of mesh cells, this trend is violated. In particular, for the JIT BLAS implementation, the arithmetic intensity of more than $200$ FLOP/Byte in the case of polynomial degree $N = 5$ and $25$-$30$ FLOP/Byte in the case of polynomial degree $N = 9$, for $12^{2}$-$16^{2}$ mesh cells. The data presented in Figures~\ref{fig:roofline_DG_2d_degrees} and~\ref{fig:roofline_DG_2d_mesh_sizes} describing comparisons of performances $P$ and arithmetic intensities $I$ for different polynomial degrees $N$ and numbers of mesh cells can explain the significant decreasing in the computational cost of the JIT BLAS implementation presented in Section~\ref{sec:comp_costs}.

The results for the finite-element ADER-DG-$\mathbb{P}_{N}$ method for solving two-dimensional problems for different polynomial degrees $N$ presented in Figures~\ref{fig:roofline_DG_3d_degrees} and~\ref{fig:roofline_DG_3d_mesh_sizes} show that for almost all studied polynomial degrees $N$ and the number of mesh cells the problem is in the region of compute-bound problems, except that the case of polynomial degree $N = 1$ is in the intermediate region. In the case of small polynomial degrees $N$, the performance $P$ is approximately $10$ times lower than the peak value $P_{\rm max}$, however, already in cases of polynomial degrees $N > 3$, the performance is only $1.5$-$2.0$ times lower than the peak value $P_{\rm max}$, which together with the high arithmetic intensity $I$ indicates a high level of use of available computing resources. It is necessary to note an interesting feature, especially clearly demonstrated in Figure~\ref{fig:roofline_DG_3d_mesh_sizes} --- with an increase in the number of mesh cells, the arithmetic intensity $I$ decreases, and for degree polynomial degrees $N \geqslant 8$ there is also a decrease in performance $P$, which is associated with the transition of problems from the region of compute-bound problems to the intermediate region. In the case of polynomial degree $N = 1$, the JIT BLAS implementation demonstrates low values of arithmetic intensity $I$ and performance $P$ compared to all other implementations, however, in the case of polynomial degrees $N > 3$, the JIT BLAS implementation is characterized by the highest performance $P$, but not the highest values of arithmetic intensity $I$. The data presented in Figures~\ref{fig:roofline_DG_3d_degrees} and~\ref{fig:roofline_DG_3d_mesh_sizes} describing comparisons of performances $P$ and arithmetic intensities $I$ for different polynomial degrees $N$ and numbers of mesh cells explain well the significant decreasing in the computational costs of the JIT BLAS implementation and other features presented in Section~\ref{sec:comp_costs}. The potential for further optimization of the finite-element ADER-DG-$\mathbb{P}_{N}$ implementation in the three-dimensional case is significantly less than in the two-dimensional case, which can be explained by the larger share of matrix-matrix operations in the total computational costs in the three-dimensional case.

The results for the finite-volume ADER-WENO($N+1$) method for solving two-dimensional problems presented in Figures~\ref{fig:roofline_WENO_2d_degrees} and~\ref{fig:roofline_WENO_2d_mesh_sizes} show that in the case of polynomial degree $N = 1$, the problems is in the region of bandwidth-bound for small numbers of mesh cells and is in the intermediate region for large values of mesh cells, while the OV and CV implementations demonstrate greater values of arithmetic intensity $I$ and performance $P$ compared to the JIT BLAS and general BLAS implementations. With an increase in the polynomial degree $N$ --- in the studied cases of polynomial degrees $N = 2$ and $3$, the data points shift to the right and up, which indicates a more efficient use of computing resources, however, from the presented results it is not possible to unambiguously conclude about the computational advantages of the JIT BLAS implementation over other implementations. In the case of the data points method, the performance values are $10$-$15$ times lower than the maximum value $I \cdot B$ in the case of bandwidth-bound problems and the peak value $P_{\rm max}$ in the case of compute-bound problems, so in this case there is a certain significant potential for further optimization. This state of the data points with respect to the roofline can be considered expected for a finite-volume ADER-WENO method, especially in the two-dimensional case, when the number of matrix-matrix operations does not constitute the largest part of the computational costs, and significant computational costs are associated with the forward and backward calculation of flow, non-conservative and source terms, primitive variables and conservative quantities, as well as solving the Riemann problem using the Riemann solver. The results presented in Figures~\ref{fig:roofline_WENO_3d_degrees} and~\ref{fig:roofline_WENO_3d_mesh_sizes} for the finite-volume ADER-WENO($N+1$) method for different polynomial degrees $N$ show that for almost all studied polynomial degrees $N$ and numbers of mesh cells, the problem is in the computer-bound region or intermediate region, which is quite interesting in relation to finite-volume ADER-WENO methods. Compared to the two-dimensional case (Figures~\ref{fig:roofline_WENO_2d_degrees} and~\ref{fig:roofline_WENO_2d_mesh_sizes}), in the three-dimensional case, the data points are significantly closer to the peak performance $P_{\rm max}$. In the case of polynomial degree $N = 1$ and, especially, a small number of mesh cells, the performance $P$ is more than $100$ times lower than the peak performance $P_{\rm max}$, however, even in this case $N = 1$, with an increase in the number of mesh cells, the performance increases significantly --- by about $5$-$15$ times lower than the peak performance $P_{\rm max}$, with a simultaneous increase in arithmetic intensity $I$. With an increase in the polynomial degrees $N$, in particular, in the case of $N = 3$, the performance $P$ is only $5$-$10$ times lower than the peak performance  $P_{\rm max}$ for a large number of mesh cells. This state of data points in relation to the roofline is quite unexpected for a finite-volume ADER-WENO method and is associated not only with a large part of matrix-matrix operations, but also with the efficient organization of computing. It can also be noted that for the finite-volume ADER-WENO method in the three-dimensional case there is potential for further optimization and increased performance, but it is significantly less than in the two-dimensional case.

The roofline for the finite-element ADER-DG-$\mathbb{P}_{N}$ method with a posteriori correction of the solution in subcells by a finite-volume ADER-WENO2 limiter is shown in Figures~\ref{fig:roofline_DG_SCL_2d_degrees} and~\ref{fig:roofline_DG_SCL_2d_mesh_sizes} for the two-dimensional case and in Figures~\ref{fig:roofline_DG_SCL_3d_degrees} and~\ref{fig:roofline_DG_SCL_3d_mesh_sizes} for the three-dimensional case. Unlike the results presented above for the finite element ADER-DG method and the finite-volume ADER-WENO method separately, in this case a different problem is solved for the performance analysis --- the explosion problem (cylindrical in the two-dimensional case and spherical in the three-dimensional case). The results clearly demonstrate some average position of the data points for the method in relation to the ADER-DG and ADER-WENO methods separately. In almost all the studied cases, except for the polynomial degree $N = 1$, the problem is in the compute-bound region. At the same time, with an increase in the polynomial degree $N$, the data points approach the peak performance $P_{\rm max}$. In the case of high polynomial degrees $N$, the data points for the JIT BLAS implementation show higher performance $P$, but not always higher arithmetic intensity $I$. In the cases of polynomial degrees $N = 7$, $8$ and $9$ in the two-dimensional case, the performance achieved for the JIT BLAS implementation is only $4$-$5$ times lower than the peak performance $P_{\rm max}$. However, despite many features similar to individual ADER-DG and ADER-WENO methods, in this case there is also a significant difference --- the obtained data points for the three-dimensional case are below the corresponding data points for the two-dimensional case. This is most likely due to the features of the explosion problem used --- in the three-dimensional case, the relative number of troubled cells is greater than in the two-dimensional case, especially for the studied numbers of mesh cells, so this feature cannot be attributed specifically to the size effect. It can also be noted that in this case, too, there is a certain potential for further optimization.

\section*{Conclusion}
\label{sec:conclusion}

In conclusion, it should be noted that implementations of finite-element ADER-DG and finite-volume ADER-WENO methods with the LST-DG predictor based on the use of the BLAS interface 
\newtext{have been} developed and proposed in this work. The use of the BLAS interface was limited to the \texttt{gemm} function only. 

The proposed approach immediately operates on AoS, which makes it possible to efficiently calculate flux, source and non-conservative terms, in this case, there is no need to carry out transposition, which must be carried out in the algorithm proposed in the work~\cite{ader_dg_eff_impl} and implemented in the work~\cite{exahype}. An important difference from existing descriptions~\cite{ader_dg_eff_impl, exahype} of the use of BLAS in the implementation of numerical methods is the use of the AoS paradigm for storage the coefficients of the discrete space-time solution $\mathbf{q}$.

The developed implementations for the LST-DG predictor are presented, in which the Picard iterative method \newtext{is} used to solve a system of nonlinear algebraic equations. \newtext{One}-step discrete finite-element ADER-DG and finite-volume ADER-WENO schemes, and implementations of the suitable piecewise-constant projection operator and the suitable high order accurate reconstruction operator used in the ADER-DG method with a posteriori correction of the solution in subcells by a finite-volume limiter are also presented. 

This paper presents detailed algorithms and tables with function parameters, which allow to develop effective software implementations of finite-element ADER-DG and finite-volume ADER-WENO methods with the LST-DG predictor. The calculated matrices are small matrices; at the same time, the proposed implementation makes it possible to effectively use existing JIT technologies, in particular, the JIT \texttt{gemm} from the \texttt{Intel MKL} library, designed for acceleration of small matrix multiplication.

The calculated computational costs demonstrate that the implementation presented in this work, outperforms the optimized vanilla implementation in computation speed by $1.5$-$2.5$ times in most cases, the implementation with general BLAS functions by $1.5$-$4.0$ times, and the completely vanilla implementation without the use of BLAS interface by 2-8 times. 

It should be noted that the simplicity of the implementation of ``optimized vanilla'' vs. ``JIT BLAS'' implementation approximately the same --- it is still necessary to explicitly use the \texttt{gemm} function, only with a simple implementation instead of creating, storing and destroying jitters in implementation with JIT BLAS. Therefore, the complexity of developing an implementation based on the approach proposed in this paper does not exceed the complexity of developing an optimized vanilla implementation.

\newtext{The paper presents the obtained results of the performance analysis using roofline. The obtained results can partly explain the observed features of the decreasing of computational costs and show that for the presented numerical methods there may still be some potential for optimization.}

Computations were carried out that showed high accuracy and convergence of numerical methods of the ADER family within the framework of this implementation with the integration of the BLAS interface, showing the correctness of the chosen implementation method.

\section*{Acknowledgments}
The reported study was supported by the Russian Science Foundation grant No.~21-71-00118:

\noindent
\texttt{https://rscf.ru/en/project/21-71-00118/}.

\noindent
The author would like to thank Popova A.P. for help in correcting the English text.

\end{document}